\documentclass[5p,authoryear,preprint]{elsarticle}

\usepackage{pdflscape}

\usepackage{graphicx}
\usepackage{url}

\usepackage[usenames]{color}
\def\red#1{\textcolor{red}{ #1}}  % flag for editing
\def\blue#1{\textcolor{blue}{ #1}}  % flag for editing
\newcommand{\be}{\begin{equation}}
\newcommand{\ee}{\end{equation}}
\newcommand{\ba}{\begin{eqnarray}}
\newcommand{\ea}{\end{eqnarray}}
\newcommand{\bc}{\begin{center}}
\newcommand{\ec}{\end{center}}

\journal{Journal of High Energy Astrophysics}

\begin{document}

\begin{frontmatter}

\title{Time-dependent modeling of TeV-detected,  young pulsar wind nebulae }
\author{D. F. Torres$^{a,b}$, A. Cillis$^{c}$, J. Mart\'in$^{a}$, \& E. de O\~na Wilhelmi$^{a}$ }
\address[1]{Institute of Space Sciences (IEEC-CSIC), Campus UAB,  Torre C5, 2a planta, 08193 Barcelona, Spain }
\address[2]{Instituci\'o Catalana de Recerca i Estudis Avan\c{c}ats (ICREA) Barcelona, Spain}
\address[3]{Instituto de Astronom\'ia y F\'isica del Espacio,  Casilla de Correo 67 - Suc. 28 (C1428ZAA), Buenos Aires, Argentina 
}

%\maketitle

\begin{abstract}
The increasing sensitivity of instruments at X-ray and TeV energies have revealed a large number of nebulae associated to bright pulsars. Despite this large data set, the observed pulsar wind nebulae (PWNe) do not show a uniform behavior and the main parameters driving features like luminosity, magnetization, and others are still not fully understood. To evaluate the possible existence of common evolutive trends and to link the characteristics of the nebula emission with those of the powering pulsar, we selected a sub-set of 10 TeV detections which are likely ascribed to 
young PWNe and model the spectral energy distribution with a time-dependent description of the nebulae's electron population. In 9 of these cases, 
a detailed PWNe model, using up-to-date multiwavelength information, is presented. 
The best-fit parameters of these nebula are discussed, together with the pulsar characteristics. We conclude that TeV PWNe are particle-dominated objects with large multiplicities, in general far from magnetic equipartition, and that relatively large photon field enhancements are required 
to explain the high level of Comptonized photons observed. We do not find significant correlations between the efficiencies of emission at different frequencies and the magnetization. The injection parameters do not appear to be particularly correlated with the pulsar properties either. We find that a normalized comparison of the SEDs (e.g., with the corresponding spin-down flux) at the same age significantly reduces the spectral distributions dispersion.
\end{abstract}


\end{frontmatter}

%%%%%%%%%%%%%%%%%%%%%%%%%%%%%%%%%%%%%%%%%%%%
%%%%%%%%%%%%%%%%%%%%%%%%%%%%%%%%%%%%%%%%%%%%
\section{Introduction}
%%%%%%%%%%%%%%%%%%%%%%%%%%%%%%%%%%%%%%%%%%%%
%%%%%%%%%%%%%%%%%%%%%%%%%%%%%%%%%%%%%%%%%%%%

During the last few years, the number of pulsar wind nebulae (PWNe) detected at TeV energies has increased from 1 (the Crab nebula, Weekes et al. 1989) to $\sim$30. The latter number of detected PWNe, mostly contributed by the H.E.S.S. survey of the Galactic plane (see, e.g., Carrigan et al. 2013 for a recent status report), is similar to the number of characterized nebulae at other frequencies. The Cherenkov Telescope Array (Actis et al. 2011) will likely increase this number to several hundreds (de O\~na Wilhelmi et al. 2013), probably providing an essentially complete account of TeV emitting PWNe in the Galaxy. 

These recent PWNe discoveries provided a basic understanding of their phenomenology:  assuming that the PWNe is maintained solely by the pulsar rotational power,  the $\gamma$-ray luminosity detected is believed to be the result of Comptonization of soft photon fields by relativistic electrons injected by the pulsar during its lifetime. This scenario can lead to TeV sources without counterparts (e.g., the first one was detected by Aharonian et al. 2002, Albert et al. 2008), when the synchrotron emission is reduced by the decay of the magnetic field. Also, it can lead to large mismatches in extension between $\gamma$ and X-ray energies, when the magnetic field is low enough that electrons emiting keV photons  actually cool faster and are more energetic 
that electrons emitting in TeV (see de Jager \& Djannati-Atai 2008 for a discussion).
The explanation of these basic properties of the behavior of PWNe does not imply that we understand the population detected in detail. 

\subsection{Pulsars with low characteristic age}

A compilation of pulsars with known rotational parameters and characteristic age of $\tau<10^4$ years is presented in { { Table 1, }} which is obtained from the updated ATNF catalog (Manchester et al. 2005) and includes the recently detected magnetar close to the Galactic Center (Mori et al. 2013, Rea et al. 2013). 
The value of the period $P$, period derivative $\dot P$, distance $D$, characteristic age $\tau$, dipolar field $B_d$, spin-down power $L_{sd}$, and $L_{sd} / D^2$ is listed. Their definitions are given below. These values are obtained directly from the catalog, neglecting some better estimations on the distances, such as those of e.g., G0.9+0.1 or pulsars at the LMC, in favor of uniformity when compiling the table. 
According to their position in the sky, we added the label H, M or V (for H.E.S.S., MAGIC or Veritas respectively) to indicate the visibility from different Cherenkov telescopes. The names of the TeV putative PWNe (or at least co-located TeV sources even if the TeV source is likely not associated to the pulsar in some cases) are also included. The majority of these pulsars, located in the inner part of the Galaxy, were in the reach of the H.E.S.S. Galactic Plane Survey (GPS), which attains a roughly uniform sensitivity of 20 mCrab (Gast et al, 2012). Some of the pulsars in the northern sky have been observed by either MAGIC or Veritas, with comparable sensitivity.

To compare the pulsar sample in Table 1 we consider their characteristic ages. Even if this is not the pulsar real age, which is usually uncertain, it can be considered a good approximation 
when the pulsar braking index is $n \sim 3$ and the initial pulsar spin-down period is much shorter than the current one. In order to give an idea of relative strength,
the spin-down power of any pulsar is compared to that of the Crab extrapolated to the corresponding characteristic age.  The last three columns in Table 1  represent, respectively, the age of Crab (assuming no change in braking index) at which it would have 
the same characteristic age as the corresponding pulsar ($T^{Crab}_{\tau}$), the Crab's spin-down power at that age ($ L_{sd}^{Crab}(T^{Crab}_{\tau}) $), and the spin-down power
of the pulsar in terms of percentage of $ L_{sd}^{Crab}(T^{Crab}_{\tau} )$, which we refer to as CFP (or Crab fractional power). When looked in this way, the Crab pulsar is no longer special.

\subsection{The influence of age in pulsars of similar spin-down}

Considering the characteristic ages provides the possibility of assessing the total power input into the nebula. Take as an example
PSR J1617--5055 and J1513--5908, and assume for the sake of the argument that both generate TeV emission via a PWN.
Both pulsars have essentially the same, and relatively high spin-down power, $1.7 \times 10^{37}$ erg s$^{-1}$.
However, one has likely been injecting this power for a much longer time, since the characteristic age of
PSR J1617--5055  is a factor of 5 larger than that of PSR J1513--5908. The electrons that populate the nebulae will sustain energy losses and live, in most conditions, 
for more than 10$^4$ years. Thus, it is reasonable to suppose that there will be more high-energy electrons with which generate 
TeV radiation in the older pulsar than in the younger one. The differences between PSR J1617--5055 and J1513--5908 
are reflected in the comparison with Crab at the moment when its characteristic age is correspondingly the same to the pulsar in question.
PSR J1617--5055 is approximately three times as luminous than Crab will be at the same characteristic age.
Instead PSR J1513--5908 spin-down corresponds to only a few percent of the one Crab will have at its characteristic age.
Thus, even when both have the same spin-down we are speaking of very different nebulae. 

\begin{figure}[t]
\centering
\includegraphics[width=84mm]{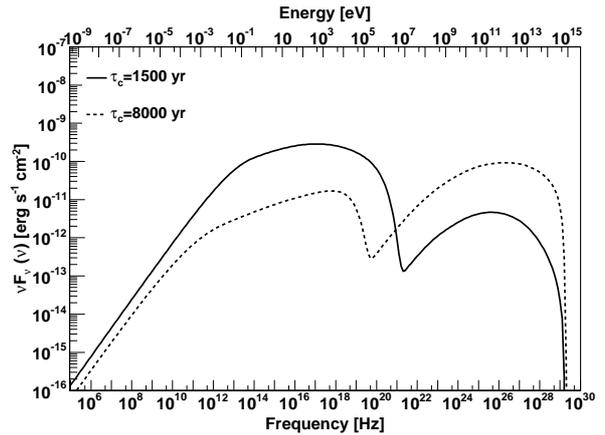}
\caption{Comparing two PWNe models that differ only in age. Parameters of these models
%, including $\tau_0$, 
are as those used for the Crab nebula, and both have the same spin-down power.} 
\label{taucomp}
\end{figure}

To exemplify further this point, consider two mock pulsars having the same spin-down evolution, magnetic fraction, injection spectrum, and photon background parameters than Crab (see below for precise definition of all these quantities) and both having also the same 
spin-down power, $1.7\times10^{37}$ erg s$^{-1}$, but two different characteristic ages of 1500 and 8000 years, respectively.  
The modeled PWNe (details of the model itself are discussed below) when every parameter is the same but just the $\tau$ and the corresponding real age vary
% that is we assume the same tau_0
turn out to be different: 
For instance, the resulting magnetic field varies from 1 to 30 $\mu$G. 
%and the size of the PWN changes from 2 to 40 pc. 
The SEDs shown in Fig. \ref{taucomp}
show that the spin-down power $L_{sd}$, or the parameter $L_{sd} / D^2$ (which is the same for the SEDs in the figure),  
unless of course when $L_{sd}$ is extremely low,
cannot by themselves blindly define dectectability of PWNe, and further considerations about the PWNe age, injection, and environment have to be taken into account.
This conclusion is emphasized when the photon background, the injection, and the magnetic fraction, among other key parameters, 
may vary from one pulsar to the next.

\subsection{Recent models and differences}

Table 1 shows that most of the young pulsars we know of were indeed surveyed for TeV emission. 
This has motivated developing detailed radiative models to tackle the complexities in each of the PWNe. 
However, whereas some of these models are time-dependent, which is essential for a proper accounting of the nebula evolution and electron losses as per the discussion above, they are different to one another, and are constructed under different approximations and assumptions. Just considering the most recent literature, one can see that some models approximate the electron population computation to obtain an advective differential equation  (e.g., Tanaka \& Takahara 2010; 2011), whereas others neglect the treatment of energy losses in full and instead replace it by the particleÕs escape time (e.g. Zhang et al. 2008; Qiao, Fang, \& Zhang 2009), and yet others do not impose any approximation at this level (e.g., Martin et al. 2012).
Some models actually assume the particle population directly and neglect any time dependence in most of the magnitudes (e.g., Abdo et al. 2010). Some assume the injection is described by a broken power law (e.g., Bucciantini et al. 2011; Tanaka \& Takahara 2010, 2011; Martin et al. 2012; Torres et al. 2013a,b), 
others consider that particle spectrum downstream of a relativistic shock can be fitted as a Maxwellian plus a power-law tail, despite the increased amount of unconstrained fitting parameters (e.g., Fang \& Zhang 2010). Some impose conservation of the total energy injected by the pulsar summing up the energy fractions distributed 
in particles and magnetic field (e.g., Tanaka \& Takahara 2011, Torres et al. 2013a,b); in others, this condition is relaxed  (e.g., Bucciantini et al. 2011). 
Some models have account of the dynamics beyond reverberation (e.g., Gelfand et al. 2009, Fang \& Zhang 2010, Bucciantini et al. 2011), while most others do that with less precision. Some take into account self-synchrotron emission (e.g.,  Tanaka \& Takahara 2011, Bucciantini et al. 2011, Torres et al. 2013b), others do not. Some consider bremsstrahlung (Martin et al. 2012), others do not, even when densities assumed are somewhat large (Li et al. 2010). Some models consider the magnetic field evolution by taking into account its work on the environment (e.g., Bucciantini et al. 2011; Torres et al. 2013a,b), others approximate it (e.g., Tanaka \& Takahara 2010, 2011). The magnitude of spectral results introduced by different underlying assumptions has been quantified only in some cases (e.g., see the impact on approximating the electron computation in Martin et al. 2012). Having a clear conversion of results from one model to another, in order to generate a uniform theoretical setting where PWNe fittings can be compared, is simply impossible.

In addition, apart from the obvious mismatches in the models per se, the nebulae that have been studied with each of them are scarce. Table \ref{thmodels} gives some examples using a certainly incomplete span of the literature. 
Our interpretation of observations is based on uncommon modeling, undermining our conclusions.

\subsection{This work}
 
The purpose of this work is to put at least a partial remedy to this situation, and provide a study of several young, TeV detected PWN. In order to do that we have improved 
our radiative model of PWNe (Martin et al. 2012) and applied it to observations. The model is one zone, leptonic, and time-dependent. 
It seeks a solution for the lepton distribution function considering the full time-energy-dependent diffusion-loss equation. The time-dependent lepton population is balanced by injection, energy losses and escape. We include losses by synchrotron, inverse-Compton (Klein Nishina inverse Compton with the cosmic-microwave background as well as with IR/optical photon fields), self-synchrotron Compton, and bremsstrahlung, devoid of any radiative approximations, and compute likewise the radiation produced by each process. We consider below in more detail the computation of the magnetic field evolution and its relation with the magnetization of the nebula. The main caveats of this model are that it contains only a free expansion dynamics (we come back to this below) and no geometry other than assuming spherical symmetry. These are clear over simplifications for some nebulae, where, for example, we know one size does not fit all frequencies. 
Still, it is a complete radiative model, and despite these caveats, it makes sense to use it for a more systematic study of the youngest nebulae.

Our sample is formed by 10 TeV detected, possibly Galactic PWNe, taken from Table 1 plus the recently detected CTA~1, which has a characteristic age slightly larger than $10^4$ years. In the Appendix of this work we comment on why we do not consider in our study the cases of HESS J1023--575, J1616--508, J1834--087/W41, and J1841--055 (in most cases, the information gathered on them imply  that the TeV emission is not univocally associated with a PWN) as well as Boomerang and HESSJ1640--465. We find that not all of the 10 cases studied are best interpreted with a PWN. In particular, we conclude that the case of HESS J1813--178 is most likely related to the SNR rather than to the PWN. The rest of this paper is organized as follows. The following Section briefly introduces the model used. Section 3 deals with each of the TeV detections in our sample, provide a PWN model when possible, and discussing the complexities of each case, surfacing caveats of our model when appropriate. Finally, Section 4 puts all our results in context, compare the modeled PWNe, and draws some conclusions from the overall population.

%%%%%%%%%%%%%%%%%%%%%%%%%%%%%%%%%%%%%%%%%%%%
%%%%%%%%%%%%%%%%%%%%%%%%%%%%%%%%%%%%%%%%%%%%
\section{Young PWNe modeling}
%%%%%%%%%%%%%%%%%%%%%%%%%%%%%%%%%%%%%%%%%%%%
%%%%%%%%%%%%%%%%%%%%%%%%%%%%%%%%%%%%%%%%%%%%

The model we use here is mostly described 
in the work by Mart\'in et al. (2012), to which we refer for details and formulae.
With respect to that model, we have 
introduced a few changes that are explicitly commented below.

%%%%%%%%%%%%%%%%%%%%%%%%%%%%%%%%%%%%%%%%%%%%
%%%%%%%%%%%%%%%%%%%%%%%%%%%%%%%%%%%%%%%%%%%%
\subsection{Spin-down and particle evolution}
%%%%%%%%%%%%%%%%%%%%%%%%%%%%%%%%%%%%%%%%%%%%
%%%%%%%%%%%%%%%%%%%%%%%%%%%%%%%%%%%%%%%%%%%%

The spin-down of the pulsar is
$
L(t)=4\pi^2 I {\dot{P}}/{P^3}
$
where
$P$ and $\dot{P}$ are the period and its first derivative and $I$ is the pulsar
moment of inertia (here assumed as $10^{45}$ g cm$^{2}$). 
The spin-down power can also be written as
$
L(t)=L_0 \left(1+{t}/{\tau_0} \right)^{-(n+1)/(n-1)},
$
using 
the initial luminosity $L_0$, the initial spin-down timescale $\tau_0$, and
the braking index $n$.
$\tau_0$ is given by (e.g., Gaensler \& Slane 2006),
$
\tau_0={P_0}/[{(n-1)\dot{P}_0}]={2\tau_c}/[{n-1}]-t_{age},
$
where
$P_0$ and $\dot{P}_0$ are the initial period and its first derivative and $\tau_c$ is the characteristic age of the pulsar.
%$
%\tau_c={P}/{2\dot{P}}. 
%$
The braking index is unknown for the great majority of pulsars, and assumed to be
$3$ when other data is lacking (corresponding to a dipole spin-down rotator). { { The above-quoted formulae
also imply that the inclination angle and the moment of inertia do not vary in time, and thus the braking index $n$ is constant.
We note that all  young pulsars with measured $n$ (see Espinoza et al. 2011, and Pons et al. 2012 and references therein) 
have $n$-values lower than 3.}}

We consider that the PWN is a sphere where the particle content is obtained from the balance 
of energy losses, injection, and escape. Thus, we solve 
\begin{eqnarray}
\label{te}
{\partial N(\gamma,t)}/{\partial t}=-{\partial}/{\partial \gamma}\left[\dot{\gamma}(\gamma,t)N(\gamma,t) \right]- \nonumber \\
{N(\gamma,t)}/{\tau(\gamma,t)}+Q(\gamma,t),
\end{eqnarray}
where 
$\dot{\gamma}(\gamma,t)$ contains
the energy losses due to  
synchrotron, (Klein-Nishina) inverse Compton, bremsstrahlung, and adiabatic expansion.
$Q(\gamma,t)$ represents the injection of particles
per unit energy (or Lorentz factor) per unit time, and
$\tau(\gamma, t)$ is the escape time (assuming Bohm diffusion). 
%e.g., as in Zhang et al. 2008, or Li et al. (2010).

Unless otherwise noted, we adopt a broken power-law for the injection of particles,
\begin{equation}
\label{injection}
Q(\gamma,t)=Q_0(t)\left \{
\begin{array}{ll}
\left(\frac{\gamma}{\gamma_b} \right)^{-\alpha_1}  & {\rm for }\, \gamma \le \gamma_b,\\
 \left(\frac{\gamma}{\gamma_b} \right)^{-\alpha_2} & {\rm for }\, \gamma > \gamma_b,
\end{array}  \right .
\end{equation}
where $\gamma_b$ is the break energy, the parameters $\alpha_1$  and $\alpha_2$
are the spectral indices. We assume that this injection is continuous along the lifetime of the PWN.
%Adopting an injection from electron's Lorentz factor of 1 and up is conservative, since assuming a new fitting parameter
%($\gamma_{min}$, the minimum energy of the injected electrons) would preferentially distribute the power to higher
%energy electrons, and thus facilitate the TeV flux production. Lacking a consistent way of determining $\gamma_{min}$
%we fix it for all nebulae to 1, and explore higher values for some cases below.

The maximum Lorentz factor of the particles is limited by requesting that the
Larmor radius $R_{L}$ is smaller than the termination shock $R_{L} = \varepsilon R_s$.
The parameter $\varepsilon$ is the so-called containment factor \citep{DeJ} (it has to
be lower than 1 in order to contain the electrons inside the acceleration region). This is
a free parameter of the model.
The Larmor Radius is
\be
\label{rl}
R_{L}=(\gamma_{max} m_e c^2)/(e B_s),
\ee
 where $B_s$ is the post-shock field strength, defined  as (see Eq. 2.1 and 2.2 of Kennel and Coroniti 1984, and 
 ($\sigma  = \eta / (1-\eta)$  or $\eta = \sigma / (1+\sigma)$)
\be
\label{bs}
B_s \sim (\chi (\eta L(t)/c)^{0.5})/R_s , \ee
with $R_s$ the termination radius. 
We have fixed $\chi$, the magnetic compression
ratio, to 3 (e.g., Venter \& de Jager 2007, 
Holler et al. 2012).
Using eq. \ref{rl} and \ref{bs} in the condition $R_{L} = \varepsilon R_s$, we find that the maximum Lorentz factor is 
\be
\gamma_{max}(t)=({\varepsilon e \chi}/{m_e c^2})\sqrt{\eta {L(t)}/{c}},
\label{g1}
\ee
where $e$ is the electron charge.

The normalization of the injection function is 
\be
(1-\eta)L(t)=\int_{\gamma_{min}}^{\gamma_{max}} \gamma m c^2 Q(\gamma,t) \mathrm{d}\gamma,
\ee
where $\eta=L_B(t)/L(t)$ is the magnetic energy fraction,  assumed constant along the evolution, with $L_B(t)$ being the magnetic power,
and $B$ is the average field in the nebula. 
Injected particles always have a fraction ($1-\eta$), and the magnetic field a fraction $\eta$, of the total power available.

%%%%%%%%%%%%%%%%%%%%%%%%%%%%%%%%%%%%%%%%%%%%
%%%%%%%%%%%%%%%%%%%%%%%%%%%%%%%%%%%%%%%%%%%%
\subsection{Magnetic field evolution}
%%%%%%%%%%%%%%%%%%%%%%%%%%%%%%%%%%%%%%%%%%%%
%%%%%%%%%%%%%%%%%%%%%%%%%%%%%%%%%%%%%%%%%%%%

%%%%%%%%%%%%%%%%%%%%%%%%%%%%%%%%%%%%%%%%%%%%
\begin{figure*}[t]
\centering
\includegraphics[width=84mm]{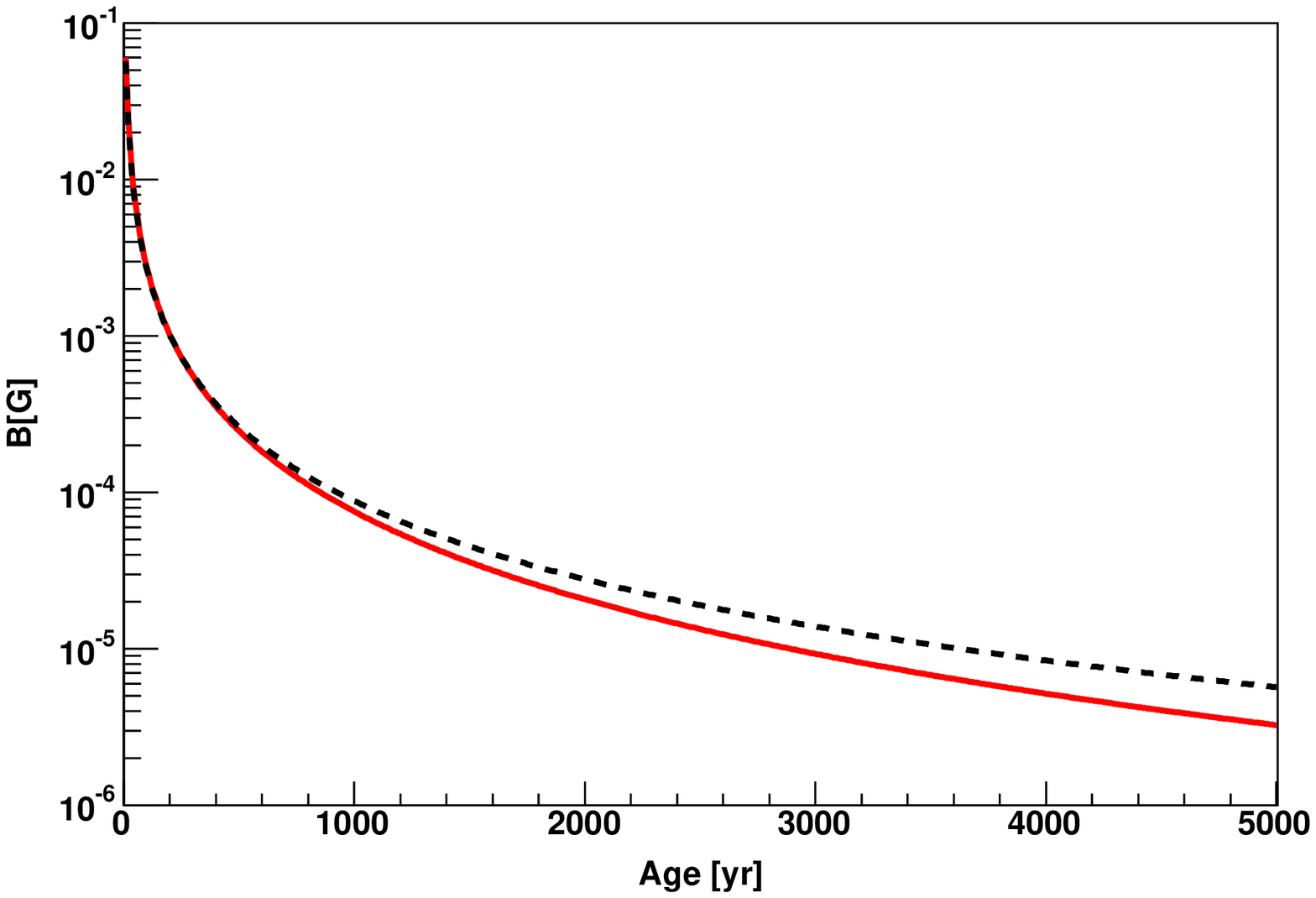}
\includegraphics[width=84mm]{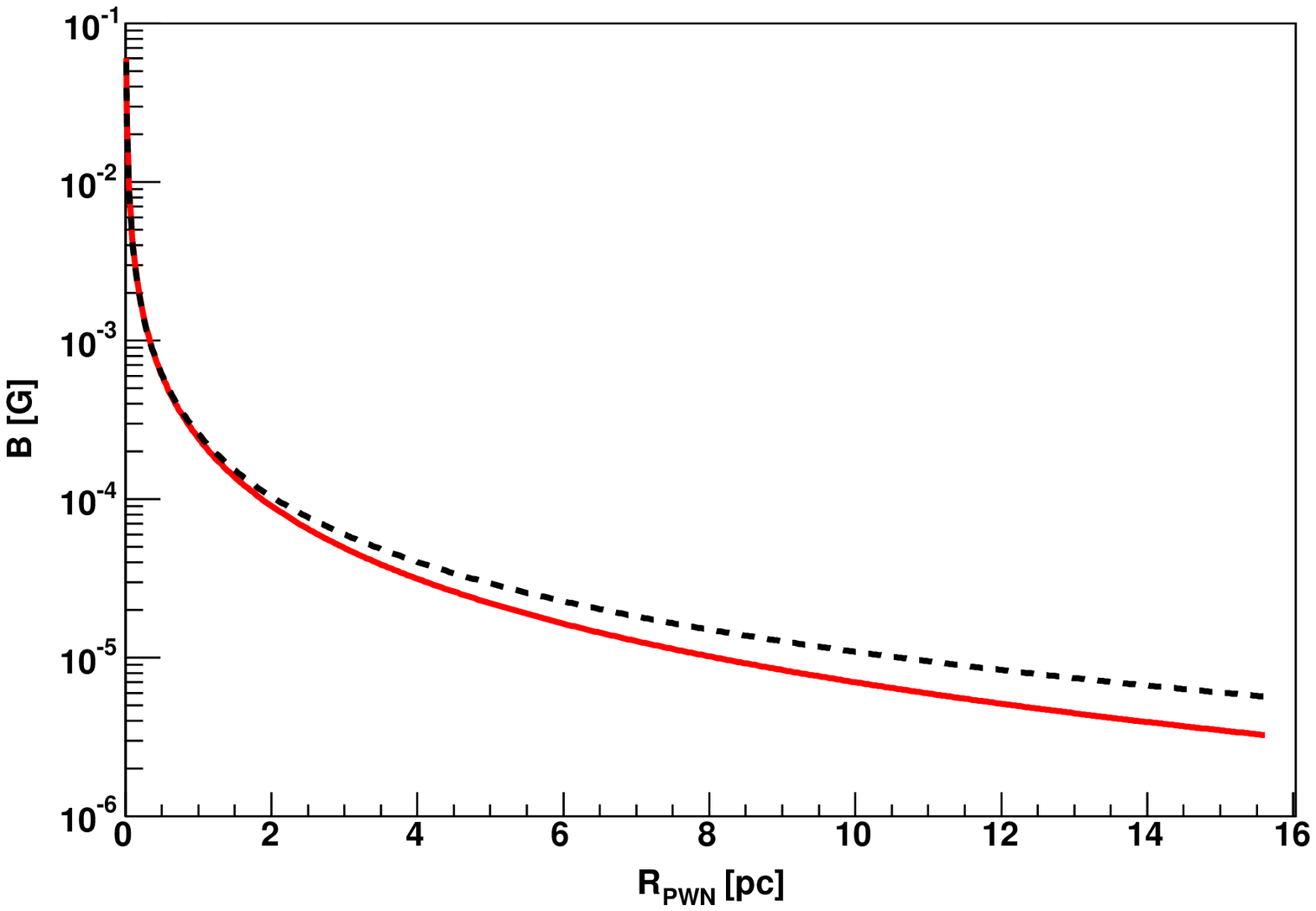}
\caption{Magnetic field as a function of time (left) and size of the PWN (right), taking the Crab nebula 
as an example. The dashed line
corresponds to Equation (\ref{T2}) whereas the solid one to Equation (\ref{new}).}
\label{field}
\end{figure*}
%%%%%%%%%%%%%%%%%%%%%%%%%%%%%%%%%%%%%%%%%%%%

The magnetic field 
$B(t)$ results from solving
\begin{equation}
\int_0^t{\eta L(t')R_{PWN}(t')dt'}=W_B\,R_{PWN},
\label{new}
\end{equation}
where 
\be W_B=\frac{4\pi}{3}R_{PWN}^3(t)\frac{B^2(t)}{8\pi}.\ee 
This equation is equivalent to 
\begin{equation}
(dW_B/dt)=\eta L-W_B(dR_{PWN}/dt)/R_{PWN},
\end{equation}
as can be seen by taking the derivative of Eq. (\ref{new}) in time.
The latter includes the adiabatic losses due to nebular expansion (e.g., 
Ostriker \& Gunn 1971,  %- Eq. 6
Pacini \& Salvati 1973, %- Eq. 2.1
%Chevalier 1977, in "Supernovae" ed. Schramm p. 53 %- Eq. 4
Reynolds \& Chevalier 1984, %- Eq. 4
Gelfand et al. 2009) %- Eq. B1);
and differs from the one adopted by Tanaka \& Takahara (2010) and subsequent literature (e.g., Li et al. 2010, Tanaka \& Takahara 2011, 2012, Mart\'in et al. 2012, and others).
In the latter case, the field is obtained from
\begin{equation}
\int_0^t{\eta L(t')dt'}=(4\pi/3)R_{PWN}^3(t)B^2(t)/(8\pi),
\label{T2}
\end{equation}
which does not take into account the energy losses due to expansion, i.e. the work done
on the surroundings. 
By comparing the left-hand side of the two definitions, one can see 
that, in order to obtain the same value for the present-time magnetic field, the
actual magnetic fraction should be $\sim 2$--3 times larger. This implies that 
models using Eq. (\ref{T2}) for the evolution of $B(t)$
without including the adiabatic losses in order to account for the present nebular field tend
to underestimate $\eta$.
To clarify on the differences we plot in Fig. \ref{field} the evolution of the two $B(t)$ mentioned above for the Crab nebula.
ÊBoth formulae for the field 
give the same power law dependence with
time, as long as $t \ll \tau_0$ ($B(t) \propto t^{-1.3}$). Instead, at later times ($t >> \tau_0$) the
resulting evolution is different (being approximately $B \propto t^{-1.8}$ in one case; $B \propto t^{-\frac{9n-4}{5(n-1)}}$ in the other  (e.g., it is $B \propto t^{-2.46}$
for $n=2.5$). 

% using L(t)~L_0 for t<<tau_0 ; L(t)~L_0(t/tau_0)^-(n+1/n-1) for t>>tau_0. 

%%%%%%%%%%%%%%%%%%%%%%%%%%%%%%%%%%%%%%%%%%%%
%%%%%%%%%%%%%%%%%%%%%%%%%%%%%%%%%%%%%%%%%%%%
\subsection{Dynamics}
%%%%%%%%%%%%%%%%%%%%%%%%%%%%%%%%%%%%%%%%%%%%
%%%%%%%%%%%%%%%%%%%%%%%%%%%%%%%%%%%%%%%%%%%%

We adopt the 
free expanding phase as in van der Swaluw et al. (2001, 2003), where the radius of the
 PWN is 
\be
 R_{PWN}(t) %\sim \left({L_0 t}/{E_0} \right)^{1/5} V_{ej} t, 
   =C \left(\frac{L_0 t}{E_0} \right)^{1/5} V_{ej} t,
   \label{rpwn}
\ee
with $V_{ej}$ determined requiring that the kinetic energy of the ejecta equals $E_0$,
$V_{ej}=\sqrt{{10 E_0}/{3 M_{ej}}}$ and where 
$E_0$ and $M_{ej}$ are the energy of the supernova explosion and the ejected mass, respectively. 
The constant C is 
 \begin{equation}
C=\left(\frac{6}{15(\gamma_{PWN}-1)}+\frac{289}{240} \right)^{-1/5},
\end{equation}
with $\gamma_{PWN}=4/3$ since we consider the PWN material as a relativistically hot gas. 
The velocity of expansion can be obtained doing the derivative
of equation (\ref{rpwn}).  
The swept-up mass resulting from these parameters is
$M_{sw} = M_{ej} (R_{PWN}/V_{ej}t)^3$.
We consider that the systems we study are not
in the reverberation phase and beyond (see e.g., Gelfand et al. 2009). But some of they could perhaps be beyond reverberation, 
%for instance, this could be the case for CTA~1 or, perhaps, of G0.9+0.1. 
When (if) so, our model is just a simplification of the latests stages of the nebula evolution.
%
% We checked that the time that the reverse shock would take to reach the bubble (see e.g., Reynolds and Chevalier 1984, eq. 29) is larger than the age of the PWN we analyze.
%
The size of the nebula (as given above in Eq. \ref{rpwn}) is used to model the spectrum at all frequencies. 
This non-dependent size assumption, in the essence of all one-zone models quoted in the introduction,
and probably similarly to the use of a single $B$-field, is inadequate for, e.g., the Crab nebula
(we discuss more on this below).
Having different sizes for, e.g., the synchrotron nebula, does not necessarily render the spectral 
model results in question, unless the size of the synchrotron emitting ball is such that it creates
a different balance of contributions by significantly modifying  the relative importance of the energy densities.

%%%%%%%%%%%%%%%%%%%%%%%%%%%%%%%%%%%%%%%%%%%%
%%%%%%%%%%%%%%%%%%%%%%%%%%%%%%%%%%%%%%%%%%%%
\subsection{Photon backgrounds}
%%%%%%%%%%%%%%%%%%%%%%%%%%%%%%%%%%%%%%%%%%%%
%%%%%%%%%%%%%%%%%%%%%%%%%%%%%%%%%%%%%%%%%%%%

The local conditions of the  interstellar radiation field (ISRF) around each PWNe are  highly uncertain.
We assume that the ISRF has three components. Permeating all nebulae, there is the CMB. Additionally,  the
spectra in the infrared and optical bands are assumed as diluted blackbodies, each of them characterized by a given temperature and energy density.
The dependence of the results on the temperatures of the IC/FIR ($T_{IR} \sim  20-100$ K; i.e., the infrared or far-infrared component) and the
NIR/OPT ($T_{NIR} \sim 3000-5000$ K; i.e., the optical or near infrared component) is relatively weak. 
%El aumento de temperatura hace mover el pico hacia las frecuencias menores, pero hace disminuir el flujo. 
We compare our densities 
with models of Galactic backgrounds (Porter et al. 2006)  in the conclusions.

%%%%%%%%%%%%%%%%%%%%%%%%%%%%%%%%%%%%%%%%%%%%
%%%%%%%%%%%%%%%%%%%%%%%%%%%%%%%%%%%%%%%%%%%%
\section{Individual modeling results}
%%%%%%%%%%%%%%%%%%%%%%%%%%%%%%%%%%%%%%%%%%%%
%%%%%%%%%%%%%%%%%%%%%%%%%%%%%%%%%%%%%%%%%%%%

%For the following spectral energy distribution (SED) plots, we use as a benchmark for GeV data --if no specific upper limits or detections are available in this regime-- the 3 years {\it Fermi}-LAT sensitivity, as obtained from the LAT performance webpage.\footnote{{\url http://www.slac.stanford.edu/exp/glast/groups/canda/lat_Performance.htm}} It is a Galactic, P7SOURCE$\_$V6 differential sensitivity plot for three years in 4 bins per energy decade. It assumes a point source with a power-law spectrum with index $-2$ and uniform background around it. 

%%%%%%%%%%%%%%%%%%%%%%%%%%%%%%%%%%%%%%%%%%%%
%%%%%%%%%%%%%%%%%%%%%%%%%%%%%%%%%%%%%%%%%%%%
\subsection{Crab nebula}
%%%%%%%%%%%%%%%%%%%%%%%%%%%%%%%%%%%%%%%%%%%%
%%%%%%%%%%%%%%%%%%%%%%%%%%%%%%%%%%%%%%%%%%%%

\begin{figure*}[t]
\centering
\includegraphics[width=84mm]{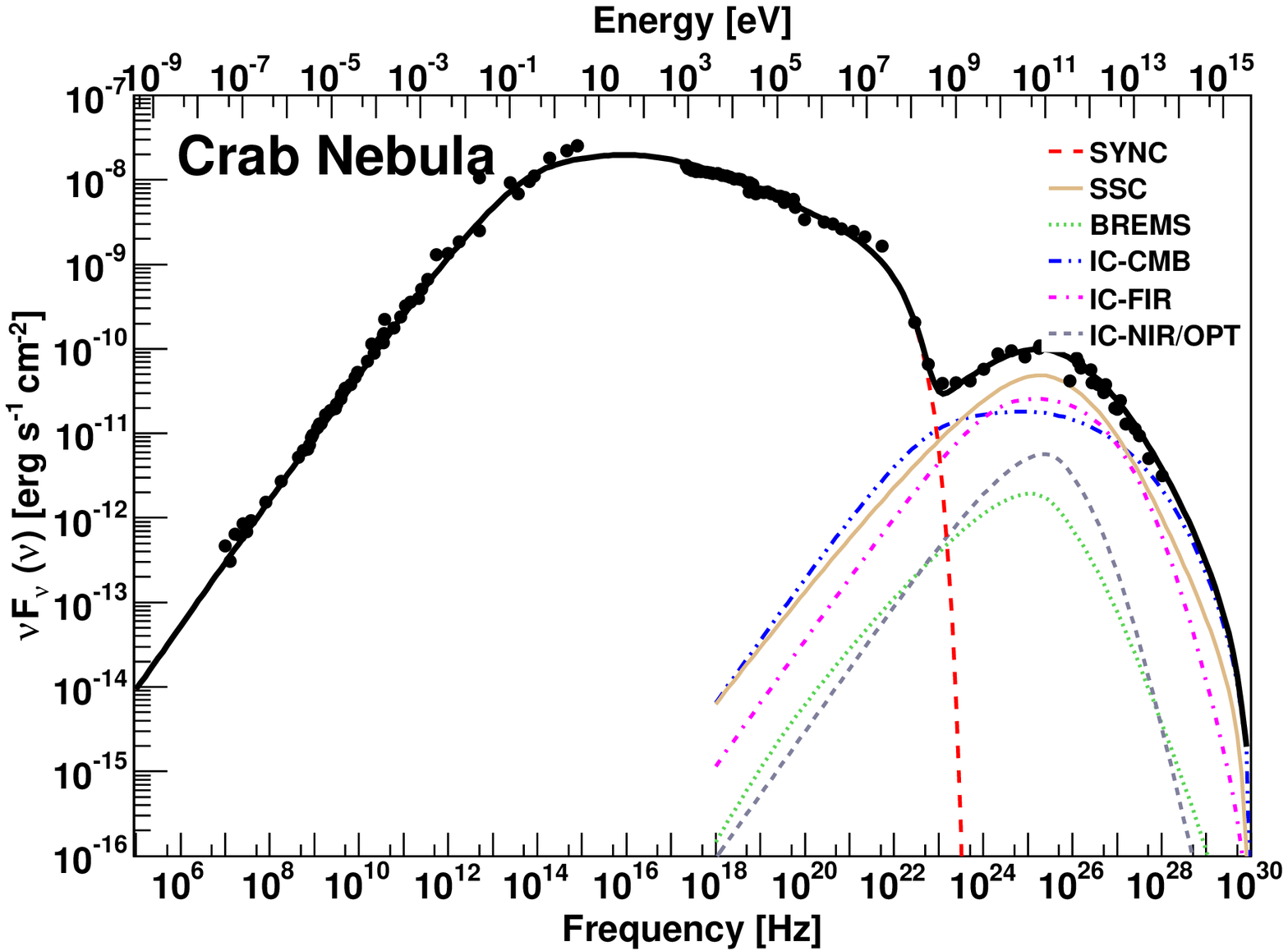}
\includegraphics[width=84mm]{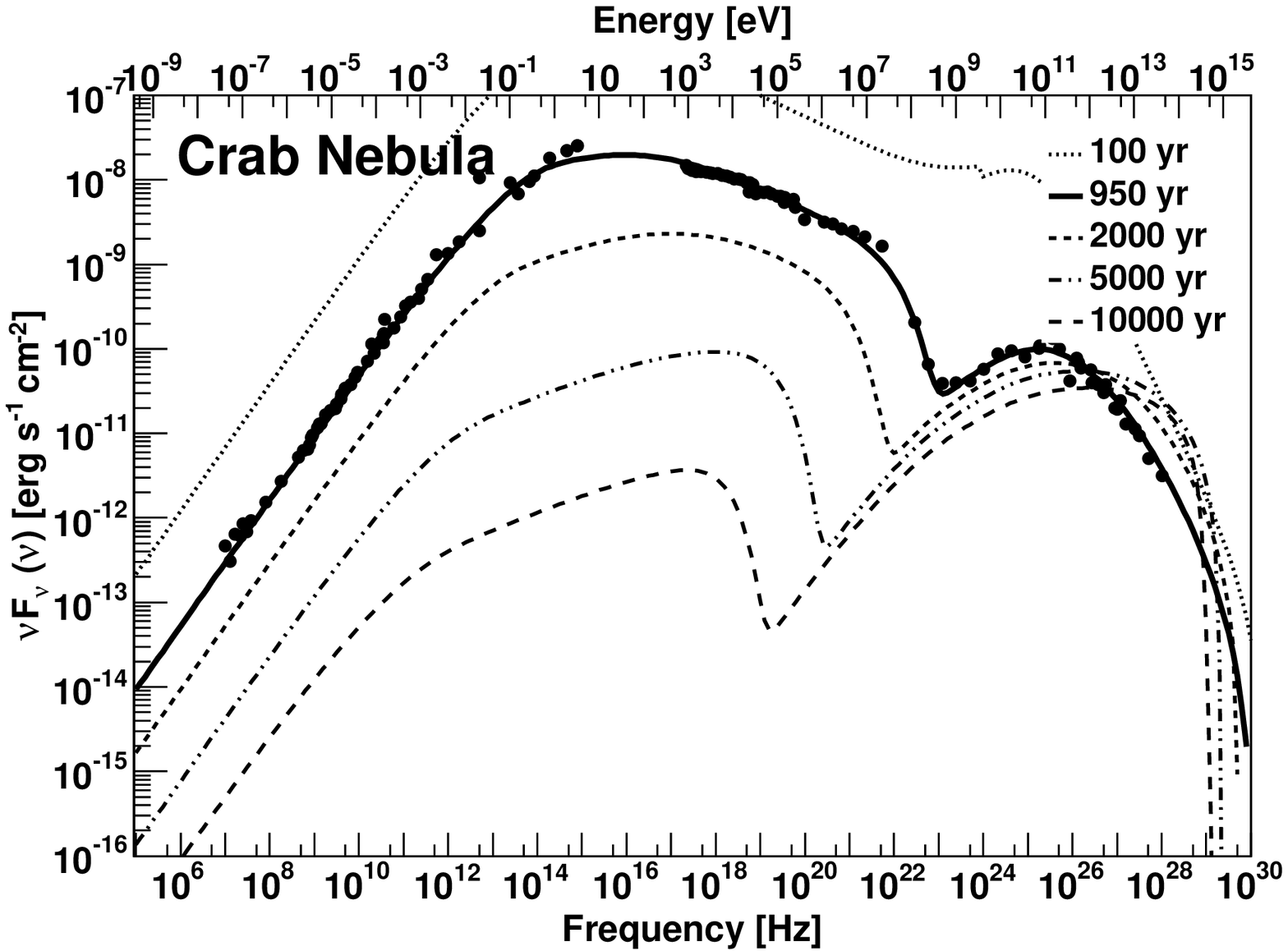}
\includegraphics[width=84mm]{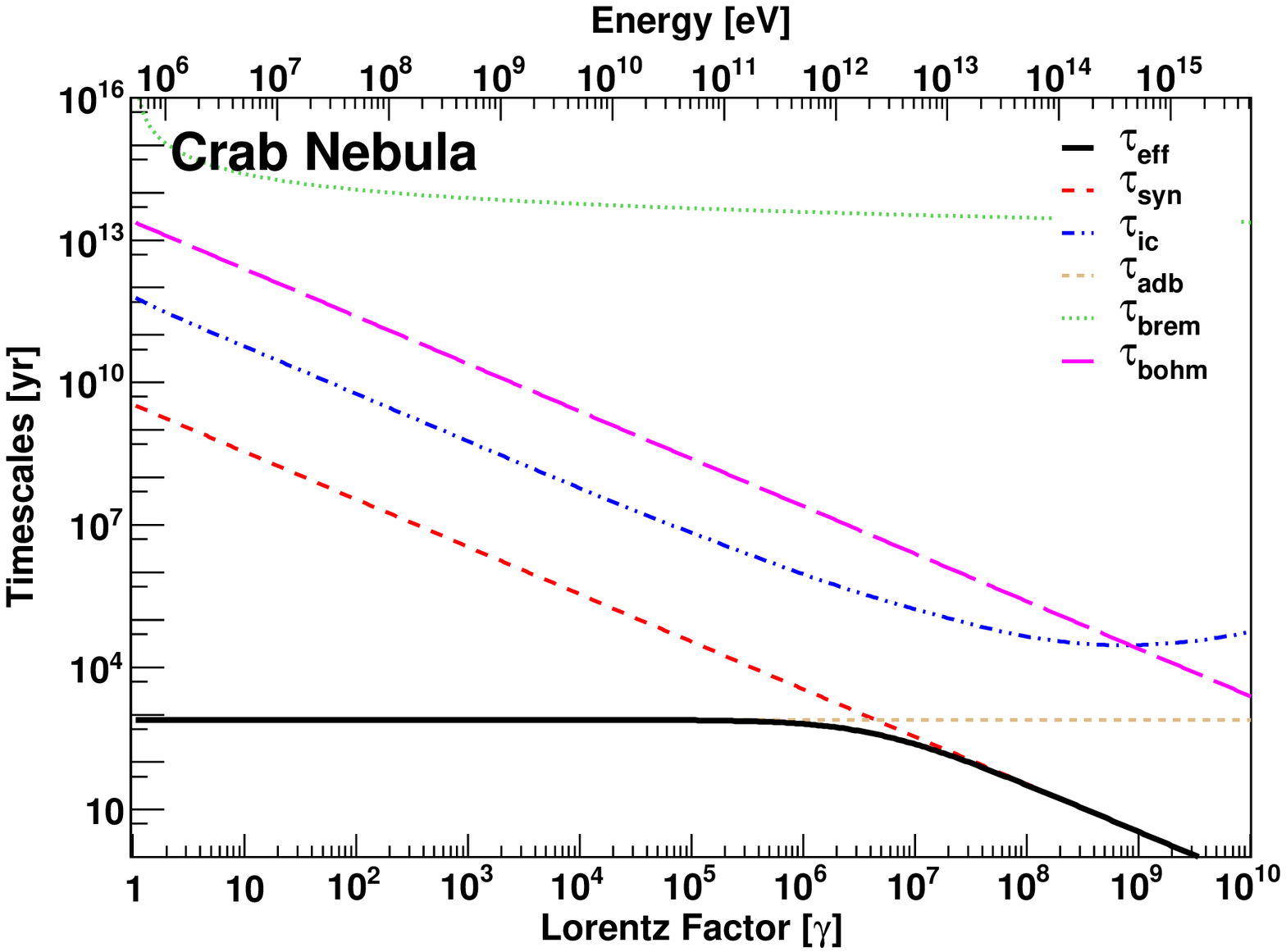}
\includegraphics[width=84mm]{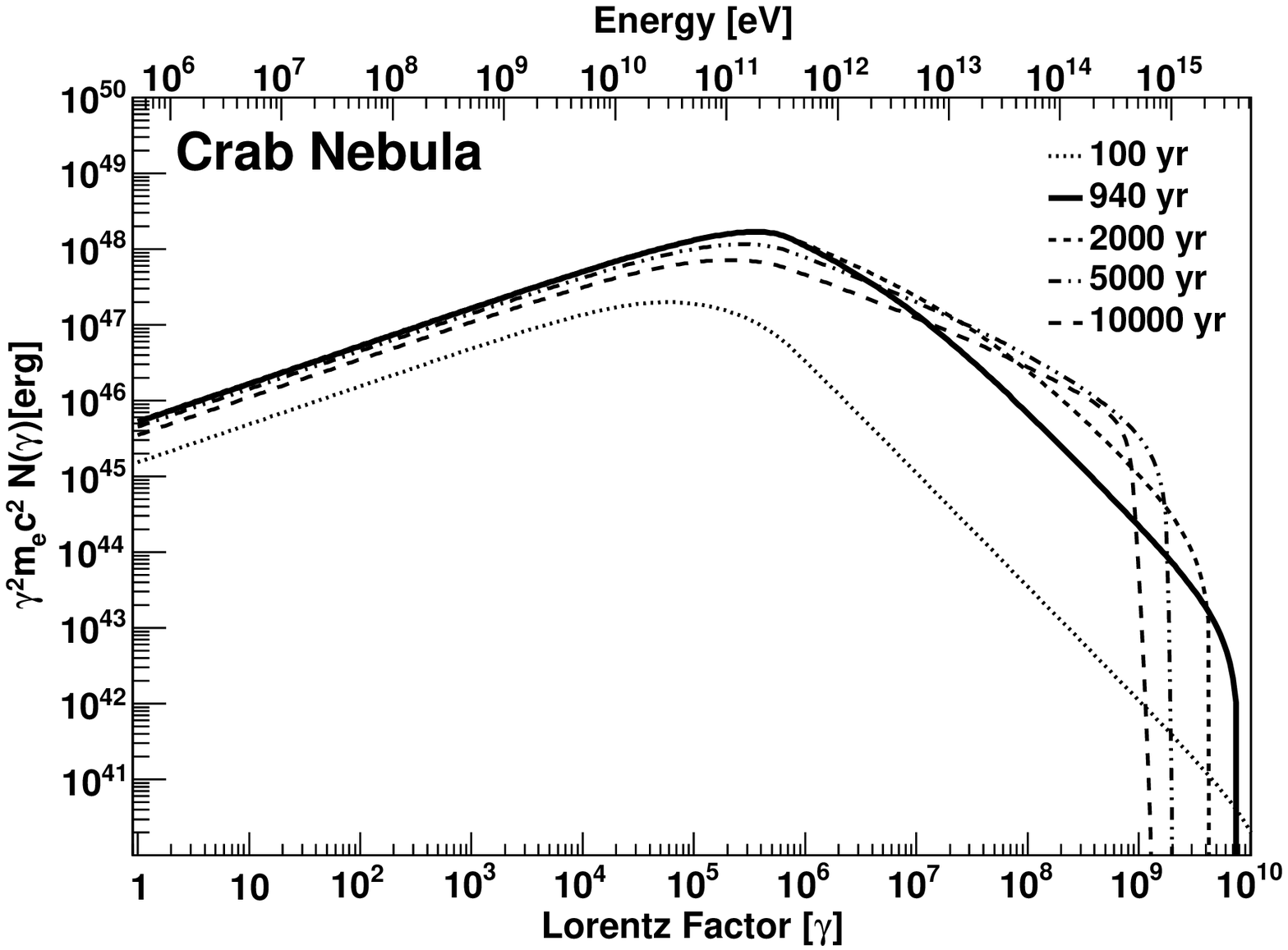}
\caption{The Crab nebula
as fitted by our model. The top left panel shows the SED at the adopted age (i.e., today),  
whereas the top right panel does it along the time evolution.
The bottom panels represent the timescales for the different losses today (the effective timescale for the losses is represented with a bolder curve)
and the evolution of the electron spectra in time. }
\label{CrabN}
\end{figure*}

Table \ref{param} presents all the fit parameters and assumed physical magnitudes of the model fitting the Crab nebula.
Our results for the Crab nebula are shown in 
Fig. \ref{CrabN}. The top left panel shows the SED at the adopted age (i.e., today),  
whereas the top right panel does it along the time evolution.
The bottom panels represent the timescales for the different losses today (the effective timescale for the losses is represented with a bolder curve)
and the evolution of the electron spectra in time. We plot the resulting SED today and the 
electron population as grey curves in all the corresponding plots of other nebulae, for comparison. For more details see Mart\'in et al. (2012), 
and Torres et al.  (2013). 

%%%%%%%%%%%%%%%%%%%%%%%%%%%%%%%%%%%%%%%%%%%%
%%%%%%%%%%%%%%%%%%%%%%%%%%%%%%%%%%%%%%%%%%%%
\subsection{VER J1930+188 (G54.1+0.3)}
%%%%%%%%%%%%%%%%%%%%%%%%%%%%%%%%%%%%%%%%%%%%
%%%%%%%%%%%%%%%%%%%%%%%%%%%%%%%%%%%%%%%%%%%%

%%%%%%%%%%%%%%%%%%%%%%%%%%%%%%%%%%%%%%%%%%%%
\begin{figure*}[t!]
\centering\includegraphics[width=84mm]{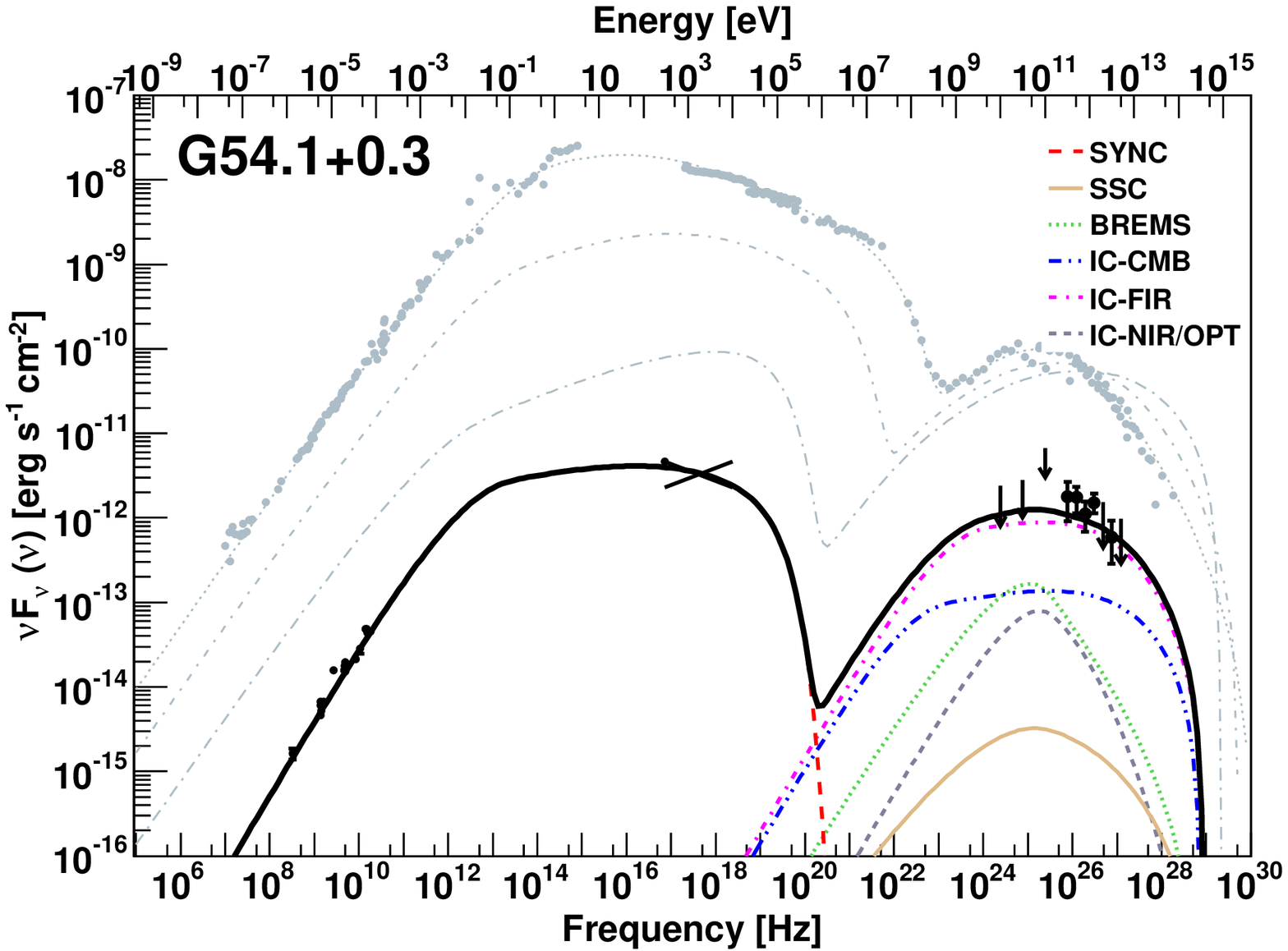}
\includegraphics[width=84mm]{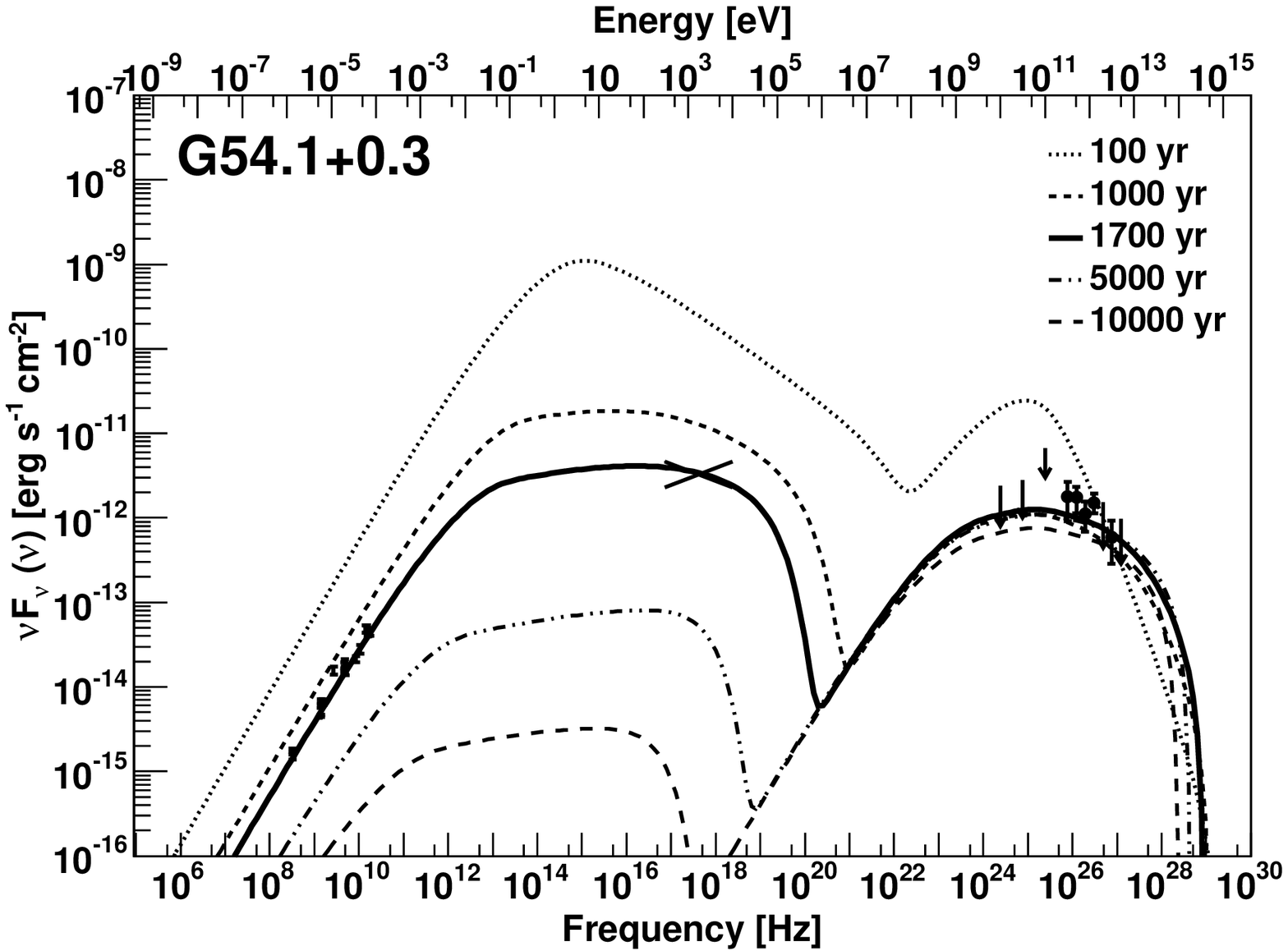}
\includegraphics[width=84mm]{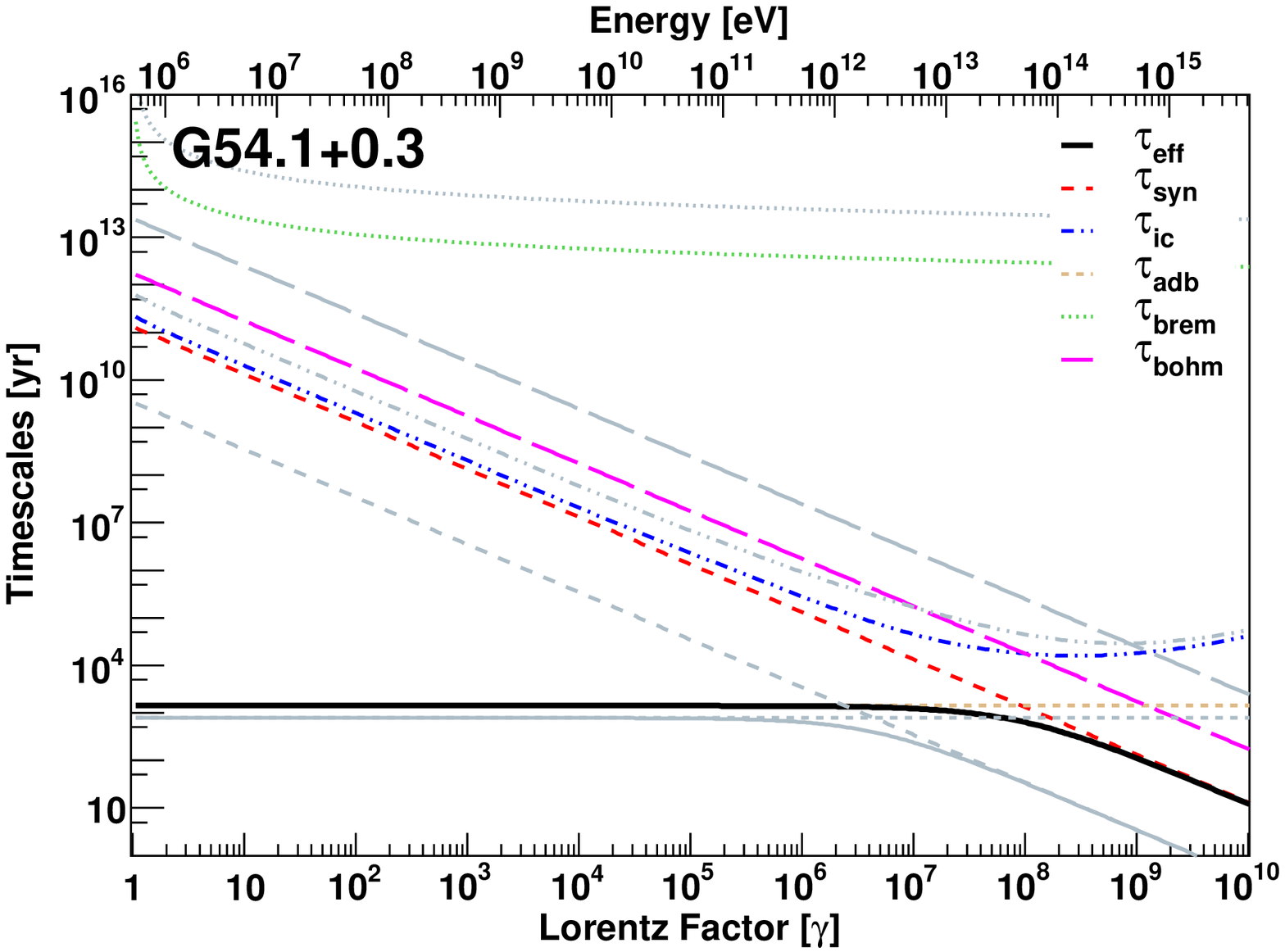}
\includegraphics[width=84mm]{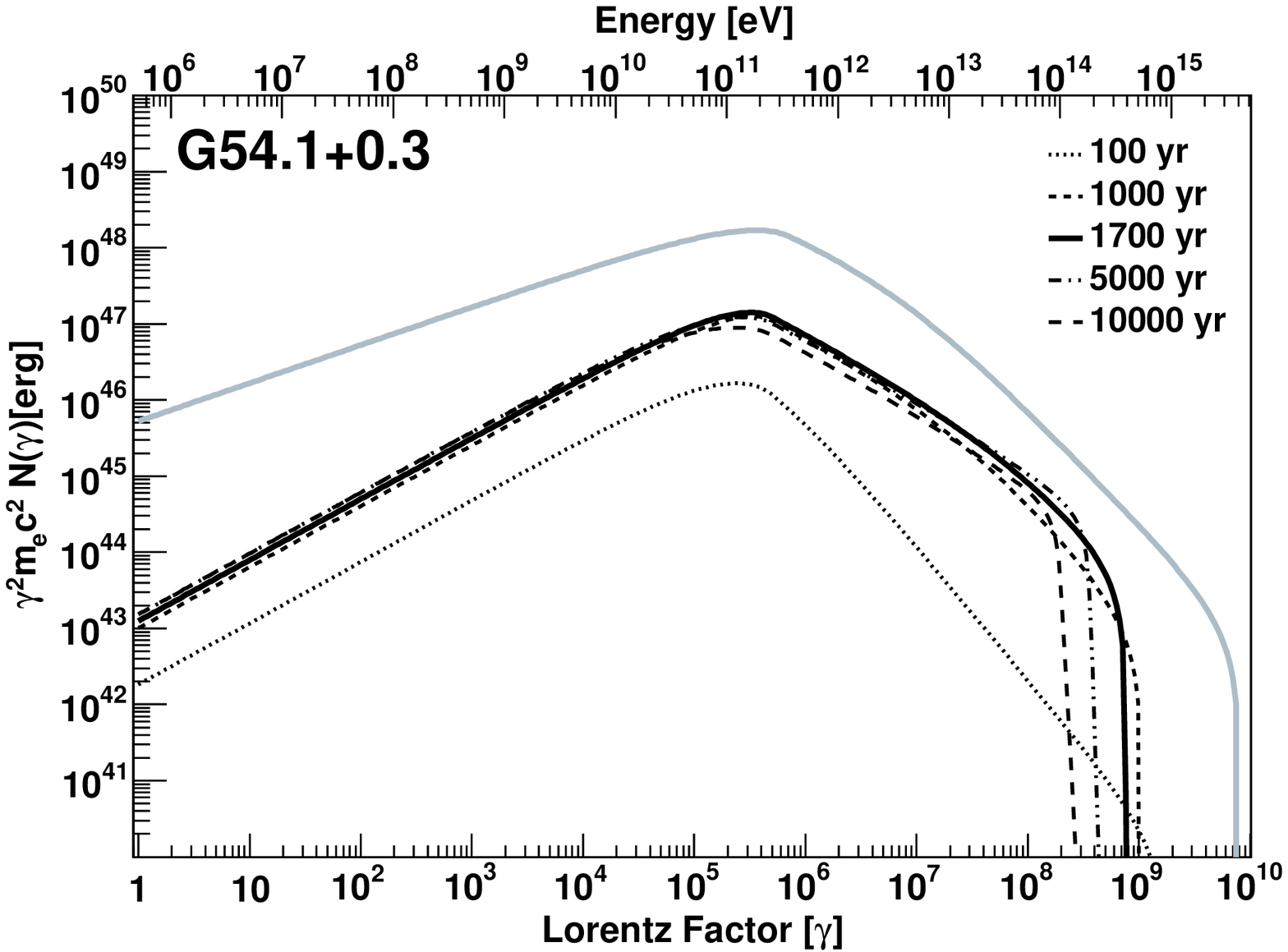}
\caption{Details of the SED (black bold line) of G54.1+0.3 
as fitted by our model. The top left panel shows the SED at the adopted age (i.e., today),  
whereas the top right panel does it along the time evolution.
The bottom panels represent the timescales for the different losses today (the effective timescale for the losses is represented with bolder curves, both for G54.1+0.3 and the Crab nebula)
and the evolution of the electron spectra in time. Here and in the figures that follow, 
we use the results of the Crab nebula model 
as a comparison. In the top-left panel, we plot (in grey, from top to bottom) three curves corresponding to the Crab nebula's SED at 940,  2000, and 5000 years.
In the bottom left panel we compare the losses of G54.1+0.3 to each of the processes with those of Crab (in grey). In the case of the electron distribution we compare with the electron population resulting from the Crab nebula model at its current age. 
For details regarding the observational data and a discussion of the fit, see the text.}
\label{G54}
\end{figure*}
%%%%%%%%%%%%%%%%%%%%%%%%%%%%%%%%%%%%%%%%%%%%

The central pulsar in G54.1+0.3 (PSR J1930+1852) is observed in radio and X-rays to have 
a period of 136 ms, and a period time derivative of $7.51\times10^{-13}$ s s$^{-1}$, implying a characteristic age of $\tau_c \sim 2.9$ kyr \citep{Camilo2002}. The
braking index is unknown, we assume it to be $3$.
Considering a possible range of braking indices and initial spin periods, Camilo et al. (2002) estimated the age of G54.1+0.3 to be between 1500 and 6000 yr. 

The PWN was first discovered by Reich et al. in 1985 in radio wavelengths. The later observation by Lu et al. in 2001 and 2002 revealing the X-ray non-thermal spectrum and the ring and bipolar jet morphology confirmed the source as a PWN. From the equations describing the PWN evolution in the model by Chevalier (2005), Camilo et al. (2002) calculated an age of 1500 yr and an initial spin period of 100 ms. Based on HI line emission and absorption measurements, the distance to G54.1+0.3 was reported to be in the 5--9 kpc range
\citep{Weisberg2008,Leahy2008}, while the pulsar dispersion measure implied 
a distance less than or equal to 8 kpc \citep{Camilo2002, Cordes2003}.  Leahy et al. (2008) 
suggested a morphological association between the nebula and a CO molecular cloud at a distance of 6.2 kpc. However, the absence of X-ray thermal emission and the lack of evidence for an interaction of the SNR with the cloud are caveats in this interpretation.  
According to Temim et al. (2010), who also assumes a distance of 6 kpc, the size of the PWN is 2 $\times$ 1.3 arcmin. Extrapolating these magnitudes to the spherical case by matching the projected area of the nebula to that of a circle,  the radius for the nebula assumed in our model is  $\sim $ 1.4 pc at 6 kpc.
We also assume Tenim et al.'s (2010) estimation of the mass of the ejecta ($\sim 20$ M$_\odot$).  
Since the SNR shell has not been detected, the particle density in the nebula is more uncertain. 
Tenim et al. (2010) have derived a density of 30 cm$^{-3}$ at one IR knot that appears to be interacting with one of the {\it jets} of the PWN.
To be conservative (see the discussion on the 
influence of the bremsstrahlung component in the SED below) we will adopt a lower, average density of 10 cm$^{-3}$.
%The radio nebula could be larger than the formerly quoted X-ray measured size.

The observations against which we fit the theoretical model are collected from different works.
Radio observations are obtained from Altenhoff et al. (1979)
Reich et al. (1984, 1985), Caswell \& Haynes (1987),
Velusamy \& Becker (1988),  Condon et al. (1989),  Griffith et al. (1990), and Hurley-Walker et al. (2009).
X ray data come from Temim et al. (2010), where we have considered the fluxes given in their table 2 except the one corresponding to the central object. 
%their regions 2 to 8.
For the spectral slope span, we have adopted the limiting cases of $-1.8$ and $-2.2$, also from Tenim et al. (2010).
We note that the
X-ray observations of Lu et al. (2002) and Lu, Aschenbach, \& Song (2011)
(used for instance in Lang et al. 2010, Li et al. 2010, and Tanaka et al. 2011) also took 
into account the central source (region 1 of Tenim et al. (2010); leading to a higher flux, 
and did not account for pileup effects (see Tenim et al. 2010 for a discussion). Use of these X-ray flux values are thus disfavored for modeling the PWN.
Finally, 
TeV observations represent the results of the VERITAS array \citep{Acciari2010}. {\it Fermi}-LAT did not detect
G54.1+0.3 (Acero et al.  2013).

For the ISRF, the region around G54.1+0.3 has been observed in the
infrared by Koo et al. (2008), and Temim et al. (2010). These observations 
suggest that the ISRF around G54.1+0.3 is larger than
that  resulting from Galactic averages as obtained, for instance, from CR propagation models. We concur (see Table \ref{param}). 
%These results are comparable to those used by Tanaka \& Takahara (2011) in their model 2, and Li et al. (2010).\footnote{We caveat on a direct comparison with these models since not only the magnetic field definition, but also the radiative content and the way in which the electron population is computed are different, in some cases leading to spurious changes in the time evolution that result only from the approximations made (see Mart\'in et al. 2012).}
Considering further additional components in the ISRF, as for instance Li et al. (2010)
did with the optical/UV contribution from nearby YSOs, does not yield to any significant changes in the fit.

% 0.3 solar masses of swept up mass
% 3342 km/s of V_ej
% 1700 years of age
% 1.4 pc of R_PWN

%%%%%%%%%%%%%%%%%%%%%%%%%%%%%%%%%%%%%%%%%%%%
%%%%%%%%%%%%%%%%%%%%%%%%%%%%%%%%%%%%%%%%%%%%
%\subsubsection{Discussion}
%%%%%%%%%%%%%%%%%%%%%%%%%%%%%%%%%%%%%%%%%%%%
%%%%%%%%%%%%%%%%%%%%%%%%%%%%%%%%%%%%%%%%%%%%

Table \ref{param} and Fig. \ref{G54} present the fitting result of our model of G54.1+0.3. Radio and X-ray data can be fitted very well with a synchrotron 
component driven by a low magnetic field of only 14 $\mu$G.  
We found a very small parameter dependence for differences in the value of the shock radius fraction; for instance for values of $\epsilon  = 0.5, 0.3, 0.2$, other parameters are only slightly changed.
The magnetic fraction  
in our model is 0.005 (half of a percent).  This turns out to be a factor of 6 smaller than that of Crab nebula. Clearly, 
G54.1+0.3 is a particle dominated 
nebula.

At high energies, 
the influence of the SSC, and the NIR/OPT IC contribution is negligible, with the FIR-IC contribution clearly dominating and the CMB-IC
and bremsstrahlung contributing at the same level at $\sim$100 GeV 
(albeit both do so at one order of magnitude lower than the dominant component).  The bremsstrahlung contribution is linear with the uncertain particle density.
Then, the selection of 10 cm$^{-3}$ as the average particle density against which we compute the bremsstrahlung contribution 
may be subject to further discussion. We note that it is a factor of 3 lower than that measured in the
IR knots (see, e.g., Tenim et al. 2010). However, the average density of the medium is probably lower than that found in such IR enhancements, and in addition, relativistic electrons may not be able to fully penetrate into the knots. Other authors, e.g., Li et al. (2011), used the IR-knot measured 30 cm$^{-3}$
as average particle density, but did not compute the bremsstrahlung luminosity in his leptonic models. For such densities, the bremsstrahlung would overcome the IC-CMB contribution to the SED in a narrow range of energies.
In agreement with observations, G54.1+0.3 should not be seen by {\it Fermi}-LAT in the framework of this model.

One interesting difference with the results of the work by Tanaka \& Takahara (2011) is the value of the high-energy index ($\alpha_2$). In our model, it results in 2.8 where it is 2.55 for Tanaka \& Takahara (2011). Contributing to this difference is likely 
the fact that in the latter model the maximum energy of electrons is fixed all along the evolution of the nebula, whereas in ours it evolves in time in agreement with the rest of the physical magnitudes. Having a fixed maximal electron energy hardens the needed slope to fit the data.

Li et al. (2010) have argued for a lepto-hadronic origin of the TeV radiation from G54.1+0.3.
The main reason argued for this case is that a leptonic-only model would produce a low magnetic field, as indeed we find.
This would result, these authors claim, very low in comparison with estimates of 
an equipartition magnetic field of 38$\mu$G, obtained from the radio luminosity of the 
PWN or a magnetic field of 80--200 $\mu$G from the lifetime of X-ray emitting particles as discussed by Lang et al. (2010).
But there is no indication that the PWN is in equipartition (in fact, models such as ours, including 
a proper calculation of losses) show that it is not  necessary to include 
any significant relativistic hadron contribution to fit the SED. 

Finally, we have also considered uncertainties in parameters that lead to degeneracies in the fit quality. One such is the age. Indeed, considering ages around 1700 years would still make possible to produce a good fit to the spectral data if changes to the photon backgrounds are allowed. For instance, the FIR energy density would need to shift from 2 to 3 eV cm$^{-3}$ in order to have a good fit when the age is 1500 yrs. Another aspect of note is the degeneracy in $\gamma_b$, which, within a factor of a few, can lead to equal-quality fits requiring a smaller magnetic field (and magnetic fraction) or small changes in the FIR density.

%\begin{figure}
%\includegraphics[width=84mm]{./G54/G54-alternative-fermi.eps}
%\caption{Details of the SED of G54.1+0.3 as fitted by the alternative model with lower break energy. See Table \ref{param} for the parameters used.}
%\label{late-G54}
%\end{figure}

%%%%%%%%%%%%%%%%%%%%%%%%%%%%%%%%%%%%%%%%%%%%
%%%%%%%%%%%%%%%%%%%%%%%%%%%%%%%%%%%%%%%%%%%%
\subsection{HESS J1747-281 (G0.9+0.1)}
%%%%%%%%%%%%%%%%%%%%%%%%%%%%%%%%%%%%%%%%%%%%
%%%%%%%%%%%%%%%%%%%%%%%%%%%%%%%%%%%%%%%%%%%%

%%%%%%%%%%%%%%%%%%%%%%%%%%%%%%%%%%%%%%%%%%%%
\begin{figure*}[t!]
\centering\includegraphics[width=84mm]{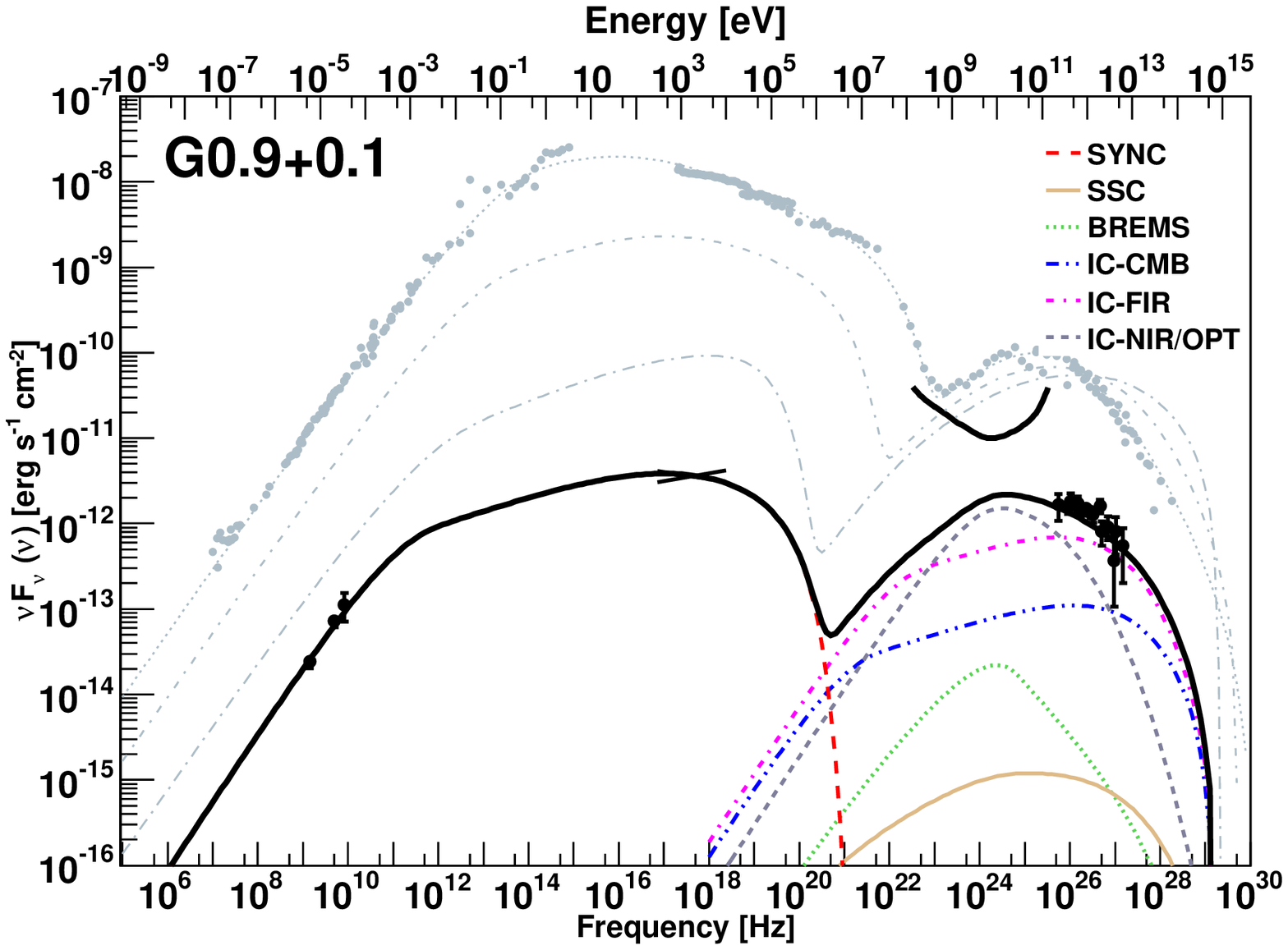}
\includegraphics[width=84mm]{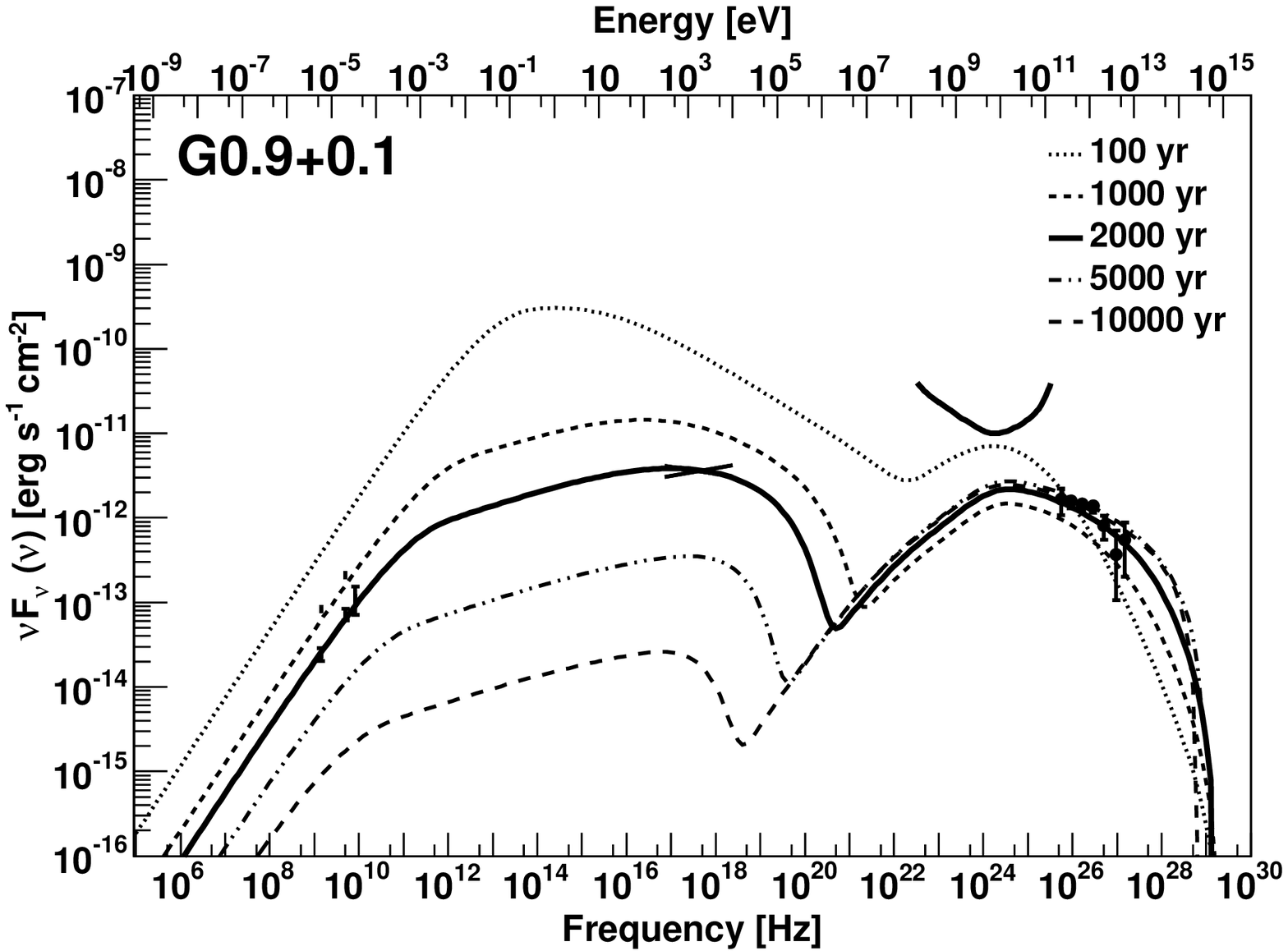}
\includegraphics[width=84mm]{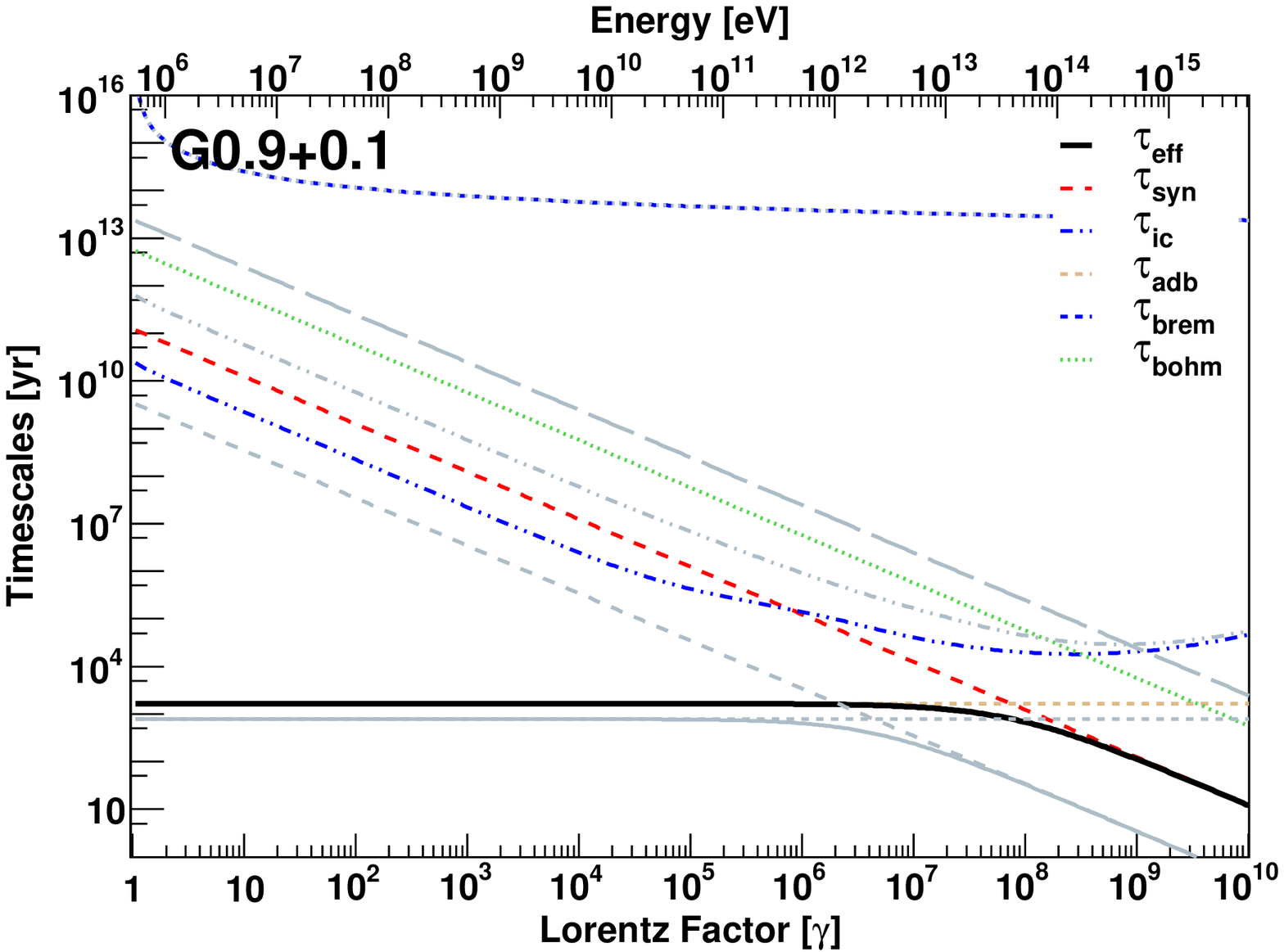}
\includegraphics[width=84mm]{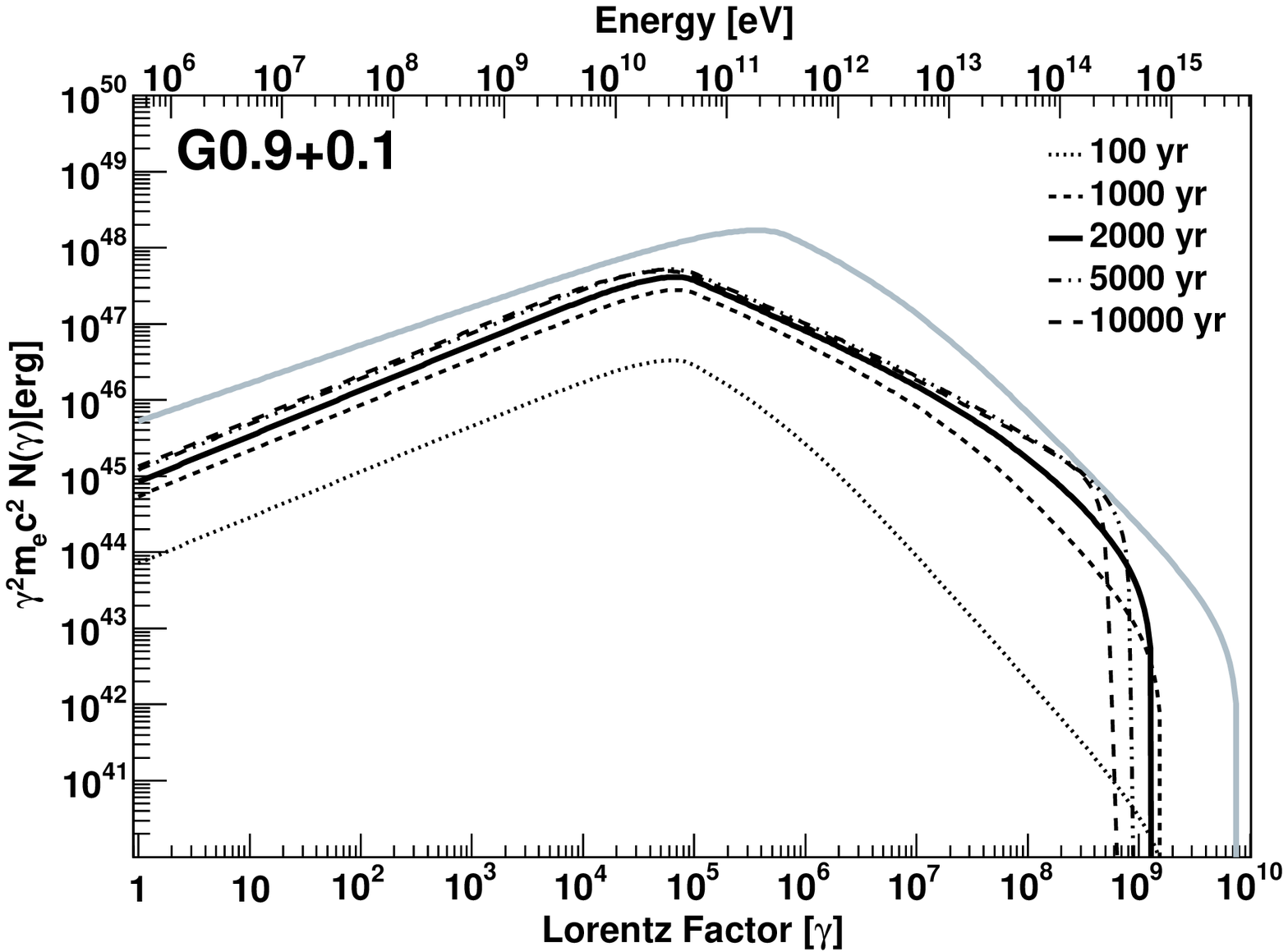}
\caption{Details of the SED of G0.9+0.1
as fitted by our model. The panels are as in Fig.~\ref{G54}.
For details regarding the observational data and a discussion of the fit, see the text.}
\label{G09}
\end{figure*}
%%%%%%%%%%%%%%%%%%%%%%%%%%%%%%%%%%%%%%%%%%%%

The PWN G0.9+0.1 
 was first identified in radio emission \citep{Helfand1987}, and then detected in X-rays \citep{Mereghetti1998, Sidoli2000}. Its central pulsar, PSR J1747-2809, was detected years later \citep{Camilo2009}. The period of this pulsar is
52.2 ms, with a period derivative of $1.56\times10^{-13}$ s s$^{-1}$, leading to a characteristic age $\tau_c$ = 5300 kyr, and a spin-down luminosity of $4.3\times10^{37}$erg s$^{-1}$ \citep{Camilo2009}, one of the largest among Galactic pulsars. 
The braking index of PSR J1747-2809 is unknown, and we assume $n = 3 $. 
The actual age of G0.9+01 is also unknown. Camilo et al. (2009) estimated an age between 2000 and 3000 yr, which is compatible with the properties of the composite SNR in radio and in X-rays  \citep{Sidoli2000}. The average radius of the PWN in radio  is $\sim$ 1 arcmin  \citep{Porquet2003}. G0.9+01 is close to the Galactic Center.
Because of that a distance of  8.5 kpc is usually adopted
 \citep{Aharonian2005, Dubner2008}. Camilo et al. (2009) estimated a distance of 13 kpc according to the dispersion measure and the NE2001 electron model \citep{Cordes2003}, but this estimation can be especially faulty towards the inner Galactic regions, and only a range between 8 and 16 kpc can be reliably suggested.

The observational data against which we fit the theoretical models come from different sources.
We use new high-resolution radio images from observations at 4.8 GHz and at 8.4 GHz carried out with the Australia Telescope Compact Array, and from reprocessed archival VLA data at 1.4 GHz  \citep{Dubner2008}. 
%The first column of Table 3 of  the mentioned paper, lists the total SNR emission, the second column corresponds to the core component alone, the third one has the same but with the underlying shell emission subtracted, and the fourth one lists the shell component alone. We use the value of the PWN with the shell subtracted in our fit.  
The X-rays observations we use were done by XMM (Porquet et al. 2003), and have an unabsorbed flux in the range 2--10 keV of $5.78 \times 10^{-12}$ erg s$^{-1}$ cm$^{-2}$, with a power-law index $1.99 \pm{0.19}$. This corresponds to an X-ray luminosity of $\sim 5 \times 10^{34}$ erg s$^{-1}$, if located at 8.5 kpc. The lack of non-thermal X-ray emission from the shell of G0.9+0.1 argue against the TeV radiation being leptonically originated there.
TeV observations are as in Fig.  3 of \cite{Aharonian2005}. 

The values needed of FIR and NIR/OPT energy densities for the nebula to be detected in the TeV range, which 
we found by fitting -see Table \ref{param}-, are
higher than what is found in the model by GALPROP (Porter et al. 2006). This discrepancy is not surprising at the central Galactic region.

%%%%%%%%%%%%%%%%%%%%%%%%%%%%%%%%%%%%%%%%%%%%
%%%%%%%%%%%%%%%%%%%%%%%%%%%%%%%%%%%%%%%%%%%%
%\subsubsection{Discussion}
%%%%%%%%%%%%%%%%%%%%%%%%%%%%%%%%%%%%%%%%%%%%
%%%%%%%%%%%%%%%%%%%%%%%%%%%%%%%%%%%%%%%%%%%%

It is interesting to note that different authors 
have used alternative set of observations for their fits. 
Aharonian et al. (2005) used the XMM data \cite{Porquet2003} like us, but for the radio data they used the work by Helfand \& Becker (1987) since their paper is prior to that of Dubner et al. (2008).  The latter authors argue for an overestimation of the radio flux of the PWN given by Helfand \& Becker (1987).
On the other hand, Tanaka \& Takahara (2011) used the data by Dubner et al. (2008) for radio, but {\it Chandra}  observations for X-ray data (Gaensler et al. 2001), 
a choice making the X-ray spectrum higher in the SED, see the discussion in Porquet et al. (2003). 
These differences in the assumed multi-wavelength spectra of the PWN  reflect in the fits, and have to be taken care of when analyzing results.

Due to the uncertainties in the distance, age, and ejected mass, we consider two cases in our fit:  
In Model 1 (to which the plots in Fig.~\ref{G09} correspond)
we assume a distance of 8.5 kpc, and an age of 2000 yrs. We consider that the PWN is a sphere with a physical radius of 2.5 pc.
In Model 2 we assume a larger distance of 13 kpc,  and an age of 3000 yrs, leading to a physical radius of 3.8 pc. 
We assume a value of 11 $M_\odot$ (Model 1) and  17 $M_\odot$ (Model 2) for the ejected mass. 
In both models we assume a density of 1 cm$^{-3}$. 
There are no significant differences (beyond the defining values for the dynamics and location) between these two models. 
The magnetic field obtained from our fits is  low $\sim$15 $\mu$G, and 
the magnetic fraction is in the order of 1--2\%.
The spectral break in the electron distribution is equal to $1\times 10^5$ for Model 1 and $0.5\times 10^5$ for Model 2. 
The spectral indices for the two cases are given in Table \ref{param} and they are very similar for the two models as well. 
This similarity gives an idea of the importance of knowing the age and distance of the PWN in fixing model parameters.

We have also analyzed the case in which the injected spectrum is a single power-law; but in practice, this required increasing the minimum energy
of the electrons in the nebula up to the break energy. The values obtained for the energy densities in FIR and NIR/OPT in order to fit the data change accordingly. The SED distribution of all
of these models (Models 1 and 2, both described in Table \ref{param}, and their analogous with a single power-law) is essentially exactly the same as the one plotted in Fig. \ref{G09}, implying that the degeneracy will be hard to break without precise measurements or modeling of the ISRF backgrounds. 

In order to reduce the FIR and NIR/OPT densities the only solution is of course to have more high-energy 
electrons in the nebula. This can be achieved for instance assuming an injection of electrons in the form of 
a single power-law with a fixed maximum and minimum energy, as in the case of 
Tanaka \& Takahara (2011). 
However, there are no particular reasons to choose given values for the latter parameters. Other differences with the assumptions in the
 Tanaka \& Takahara (2011) model is that
their nebula is 4500 years-old (instead of 2000--3000 yrs) and located slightly closer, at  8 kpc (instead of 8.5 kpc). At this adopted age/distance, which seems not particularly preferred by any observation, the total power would be $\sim 1$ order of magnitude larger than that in our Model 1; 
what explains the lesser need of target photon backgrounds to achieve the same
TeV fluxes. 
This set of assumptions for the injection and age does not appear preferable or particularly justifiable when confronted
with the possibility of having larger local background in the Galactic Center environment.
Fang \& Zhang (2010) also studied the spectral evolution of G0.9+0.1; but under the assumption that the particle distribution at injection is given by a relativistic Maxwellian distribution plus a single power-law distribution. The latter produces a distinctive feature in the SED at about $10^{-9}$ MeV for which there is no observational need yet. Even when different assumptions and modeling techniques are used, a low magnetic field is also singled out by their study.

In  agreement with our prediction in all the models analyzed, {\it Fermi}-LAT did not detect this PWN, and because of the Galactic Center location, it has been impossible to impose useful upper limits either (Acero et al. 2013). 
The SED fit in Fig. \ref{G09} shows only a guiding-curve for the 3-years {\it Fermi}-LAT sensitivity.

%%%%%%%%%%%%%%%%%%%%%%%%%%%%%%%%%%%%%%%%%%%%
%%%%%%%%%%%%%%%%%%%%%%%%%%%%%%%%%%%%%%%%%%%%
\subsection{HESS 1833$-$105 (G21.5$-$0.9)}
%%%%%%%%%%%%%%%%%%%%%%%%%%%%%%%%%%%%%%%%%%%%
%%%%%%%%%%%%%%%%%%%%%%%%%%%%%%%%%%%%%%%%%%%%

%%%%%%%%%%%%%%%%%%%%%%%%%%%%%%%%%%%%%%%%%%%%
\begin{figure*}[t!]
\centering\includegraphics[width=84mm]{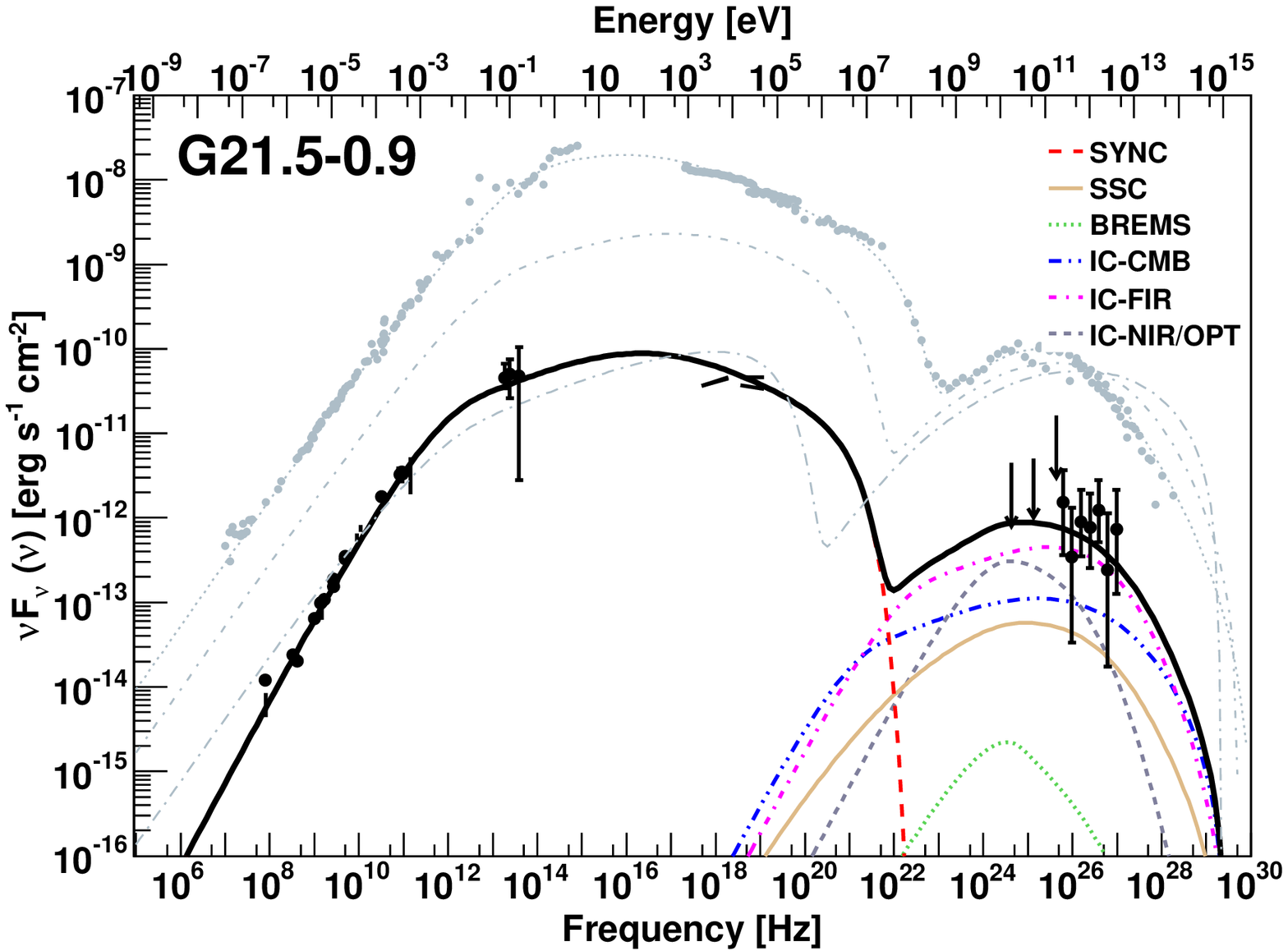}
\includegraphics[width=84mm]{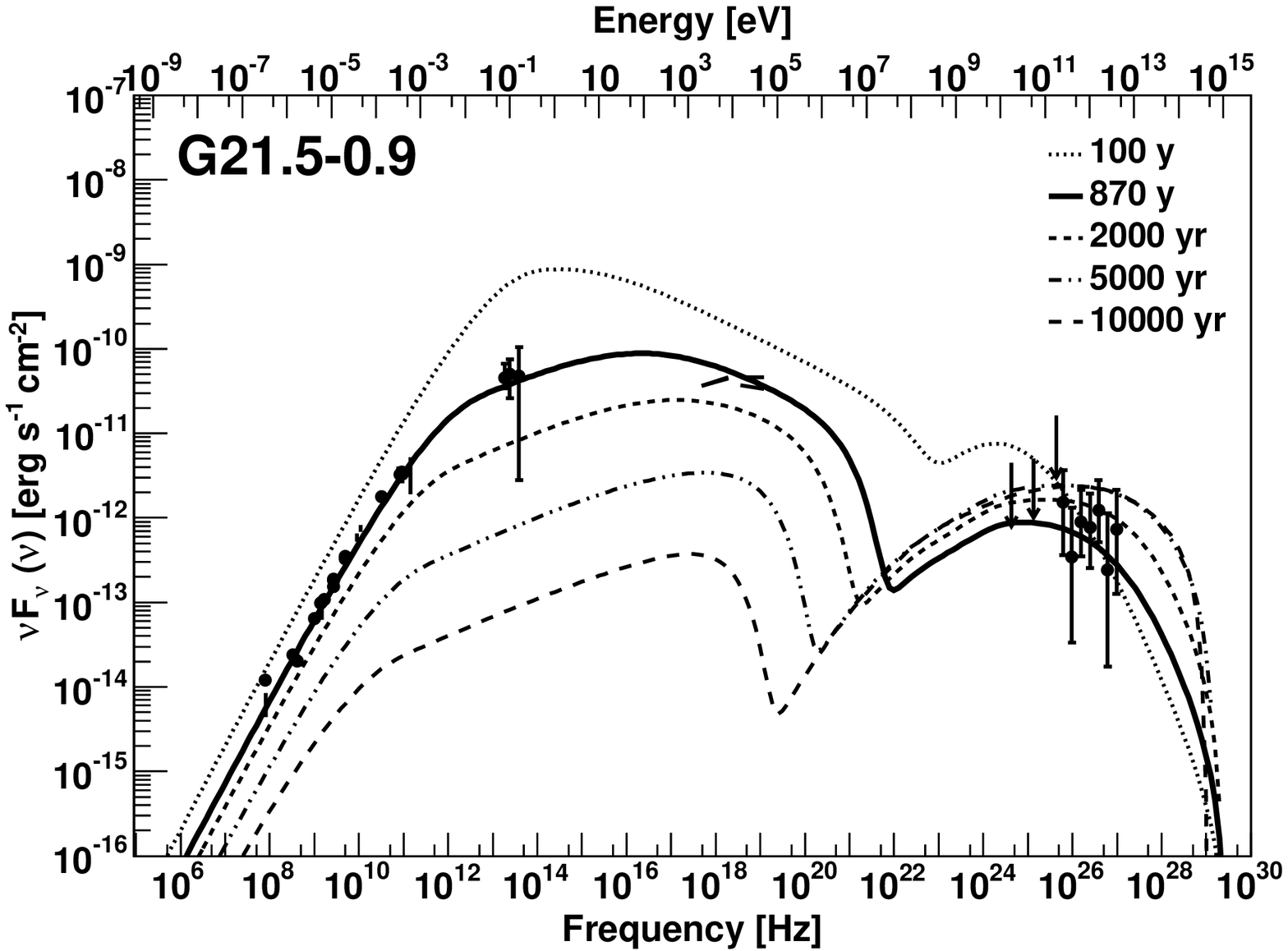}
\includegraphics[width=84mm]{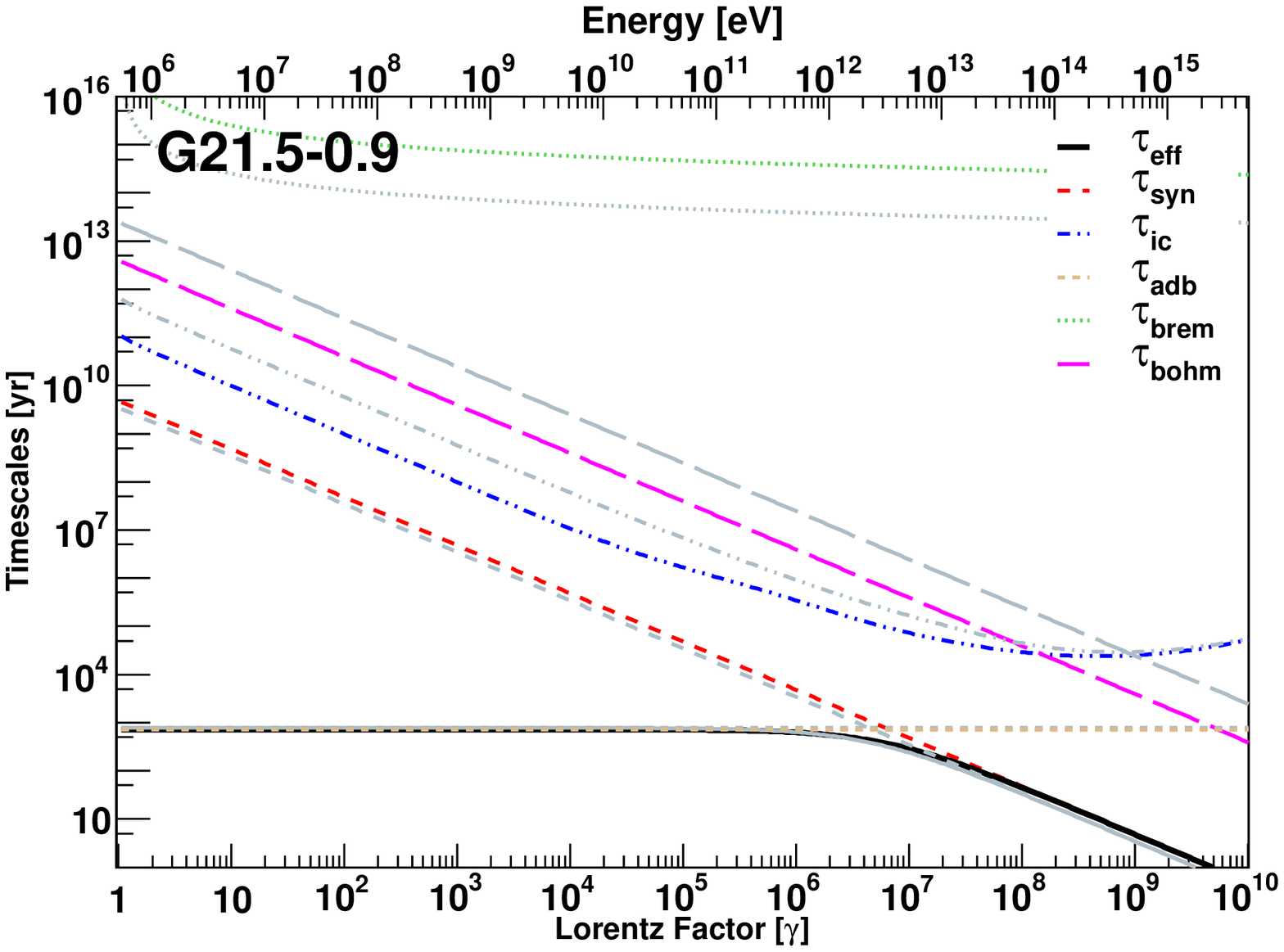}
\includegraphics[width=84mm]{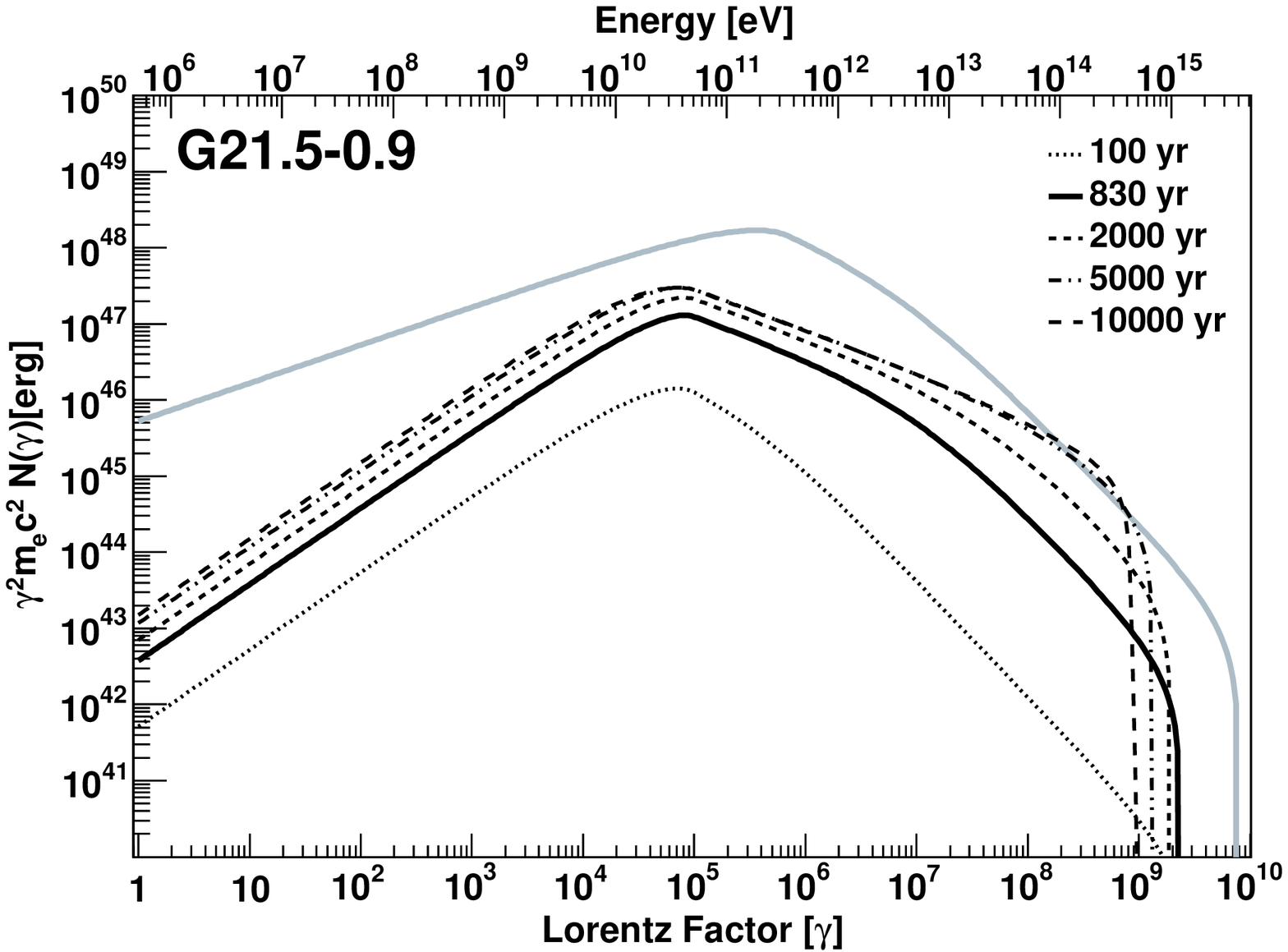}
\caption{Details of the SED model of G21.5$-$0.9.The panels are as in Fig.~\ref{G54}. For details regarding the observational data and a discussion of the fit, see the text. }
\label{G21}
\end{figure*}
%%%%%%%%%%%%%%%%%%%%%%%%%%%%%%%%%%%%%%%%%%%%

G21.5$-$0.9 is a plerionic SNR with an
approximately circular shape having a radius of $\sim$ 40'' in radio, infrared and X-ray. The pulsar is at its center. 
The central pulsar of G21.5$-$0.9, PSR J1833-1034, was observed in radio having a period of 61.8 ms, and a period derivative of  $2.02\times10^{-13}$~s~s$^{-1}$, yielding  a characteristic age $\tau_c$ $=$ 4860 yrs \citep{Camilo2006}.  It was not possible to measure the braking index, and we take $n = 3$.
PSR J1833-1034 was also observed pulsating in GeV by {\it Fermi}-LAT \citep{Abdo2010}, but not in X-rays  (see for example, Camilo et al. 2006). 

The pulsar is one of the youngest in the galaxy. A recent age estimate based on measuring the PWN expansion rate in the radio band gives an age of 870 yr \citep{Bietenholz2008}.
In case of decelerated expansion, this real age could be even lower. However,  Wang et al. (1986) 
suggested that G21.5-0.9 might be the historical supernova of 48 BC. Uncertainty remains in this point. We assume the 870 years of age in our model.
The distance to the system was estimated, based on HI and CO measurements, to be 4.7$\pm$0.4 kpc \citep{Camilo2006}.
The same value (within errors) was obtained by other authors \citep{Tian2008}. We approximate the nebula as an sphere of radius $\sim$1 pc.
We assumed a mass of 8 $M_{\odot}$ for the ejected mass. Matheson \& Safi-Harb (2005) derived an upper limit for the upstream density of $\sim$ $0.1-0.4$ cm$^{-3}$. For our fitting procedure, then, we assumed that the PWN expands in a low density media with a value of $0.1$ cm$^{-3}$.
 
G21.5$-$0.9 has been observed at different frequencies. 
In our analysis we have used the radio data obtained in the works by Salter et al. (1989), Morsi \& Reich (1987),
Wilson \& Weiler (1976), and Becker \& Kundu (1976).
We have also used the infrared observations performed by Gallant \& Tuffs (1998,1999). There are additional X-ray and IR data that we are not using in the fit
 \citep{Zajczyk2012} and corresponding to the compact nebula only, a region of 2 arcsec surrounding the central pulsar.

G21.5$-$0.9 is usually taken as a calibration source for X-ray satellites, see for example the works by Slane et al. (2000), Warwick et al. (2001),
Safi-Harb et al. (2001), 
Matheson \& Safi-Harb (2005, 2010), and
De Rosa et al. (2009).
We have used  
the joint calibration of {\it Chandra}, {\it INTEGRAL},  {\it RXTE}, {\it Suzaku}, {\it Swift}, and {\it XMM-Newton} done by Tsujimoto et al. (2011) when considering the X-ray spectrum.
The latter shows an spectral softening with radius \citep{Slane2000, Warwick2001}.
{\it Chandra} data showed for the first time evidence for variability in the nebula, a similar behavior that occurs in Crab and Vela \citep{Matheson2010}. 
{\it Fermi}-LAT data come from Acero et al. (2013).
Finally, at TeV energies, the data comes from H.E.S.S. observations, which detected the PWN as the source 
HESS 1833$-$105 \citep{Gallant2008,Djannati2007}.

G21.5$-$0.9 was the first PWN discovered to be surrounded by a
%($ r \sim$150 arcsec) 
low-surface brightness X-ray halo that was suggested to be associated with the SNR shell;  its spectrum being non-thermal  \citep{Slane2000}. The halo was not observed in radio wavelengths.
Slane et al. (2000) argued that the halo may be the evidence of the expanding ejecta and the blast wave formed in the initial explosion.  Warwick et al. (2001) posed that the halo may be an extension of the central synchrotron nebula. But deep {\it Chandra} observations revealed  limb-brightening in the eastern portion of the X-ray halo and wisp-like structures, with the photon index being constant across the halo \citep{Matheson2005}.
%These features may indicate that the halo is unlikely an expansion of the nebula \cite{Matheson2005}.  
Another interpretation of the origin of the halo is that it could be composed by diffuse extended emission due to the dust scattering of X-ray from the plerion \citep{Bocchino2005}. Spectroscopy analysis done by Matheson \& Safi-Harb (2010) with {\it Chandra} data revealed a partial shell on the eastern side of the SNR.  Safi-Harb et al. (2001) could not find evidence for line emission in any part of the remnant.

%%%%%%%%%%%%%%%%%%%%%%%%%%%%%%%%%%%%%%%%%%%%
%%%%%%%%%%%%%%%%%%%%%%%%%%%%%%%%%%%%%%%%%%%%
%\subsubsection{Discussion}
%%%%%%%%%%%%%%%%%%%%%%%%%%%%%%%%%%%%%%%%%%%%
%%%%%%%%%%%%%%%%%%%%%%%%%%%%%%%%%%%%%%%%%%%%

Table \ref{param} summarizes the values of the parameters and the result of the fit. The latter is shown in Fig. \ref{G21}, which has the same panels as in the previously analyzed PWNe. It is particularly interesting to note that the electron losses in our model (see bottom left panel of Fig. \ref{G21}) are almost exactly the same as those of Crab, and has $\sim$10\% of its spin-down power. Table \ref{param} gives further account of this similarity as regards of age and energy densities of the photon backgrounds.
%
%The main uncertainty in the observational data come from the IR measurements (see Gallant \& Tuffs 1998, 1999). 
%But we have found essentially no change in the resulting parameters by removing the IR data constraints in the fit altogether (the most noticeable is a difference of 10 $\mu$G in the magnetic field, which is in the range of  60-70 $\mu$G). In both cases, 
G21.5-0.9 is a particle dominated nebula,
 with a magnetic fraction of $0.03-0.04$.
 This value is higher than that the one obtained by \cite{Tanaka2011}, in correspondence with the different equation used for the 
 definition of magnetic field, as described above. Otherwise, the resulting model parameters are very similar, which is probably due to a significant domination of the FIR
 component, almost one order of magnitude above the CMB contribution to the inverse Compton yield at 1 TeV.
We fixed the temperature of FIR and NIR/OPT photon distributions at the same values obtained from GALPROP. In order to be detected in the TeV range as has been, the value for the energy density in the FIR is $\sim 1.4$ eV cm$^{-3}$. The Comptonization of these photons dominates the spectrum at the highest energies. There is some degeneracy in the precise determination of the FIR and NIR densities and temperatures. For instance, we have checked that our fits would be very similar with temperature of 70 and 5000 K, and densities of 2 eV cm$^{-3}$ in the FIR and NIR, respectively. 
We have analyzed the impact of having a smaller braking index (e.g., 2.5), and a different shock fraction (from 0.1 to 0.3), 
but did not find any significant differences in our fits due to the change in these parameters.

%\begin{figure}
%\includegraphics[width=84mm]{G21/G21-SEDtoday_Model2.eps}
%\caption{Spectrum of G21.5$-$0.9 fitted by our model, Model 2 }
%\end{figure}

%%%%%%%%%%%%%%%%%%%%%%%%%%%%%%%%%%%%%%%%%%%%
%%%%%%%%%%%%%%%%%%%%%%%%%%%%%%%%%%%%%%%%%%%%
\subsection{HESS 1514--591 (MSH 15--52)}
%%%%%%%%%%%%%%%%%%%%%%%%%%%%%%%%%%%%%%%%%%%%
%%%%%%%%%%%%%%%%%%%%%%%%%%%%%%%%%%%%%%%%%%%%

%%%%%%%%%%%%%%%%%%%%%%%%%%%%%%%%%%%%%%%%%%%%
\begin{figure*}[t!]
\centering\includegraphics[width=84mm]{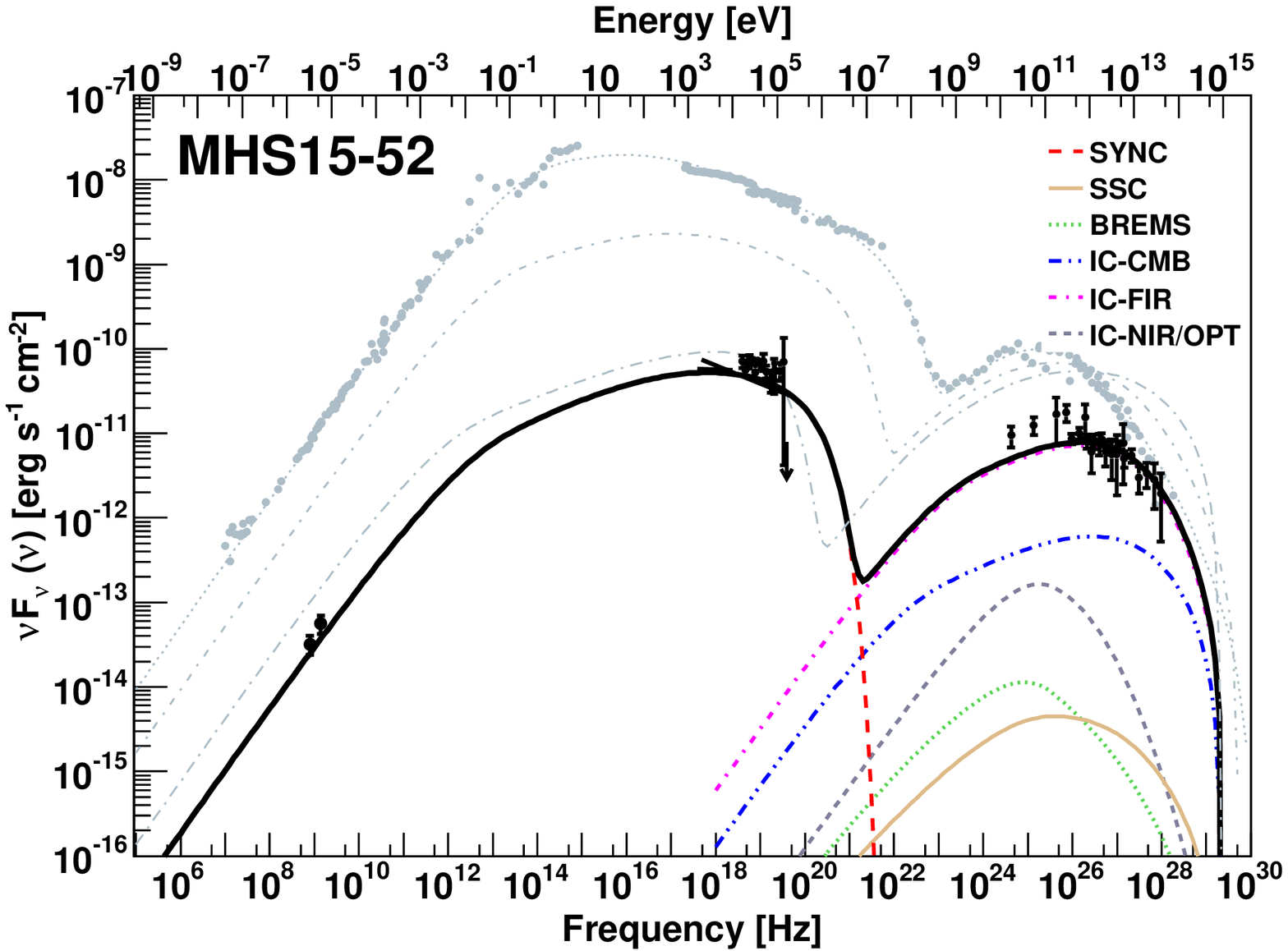}
\includegraphics[width=84mm]{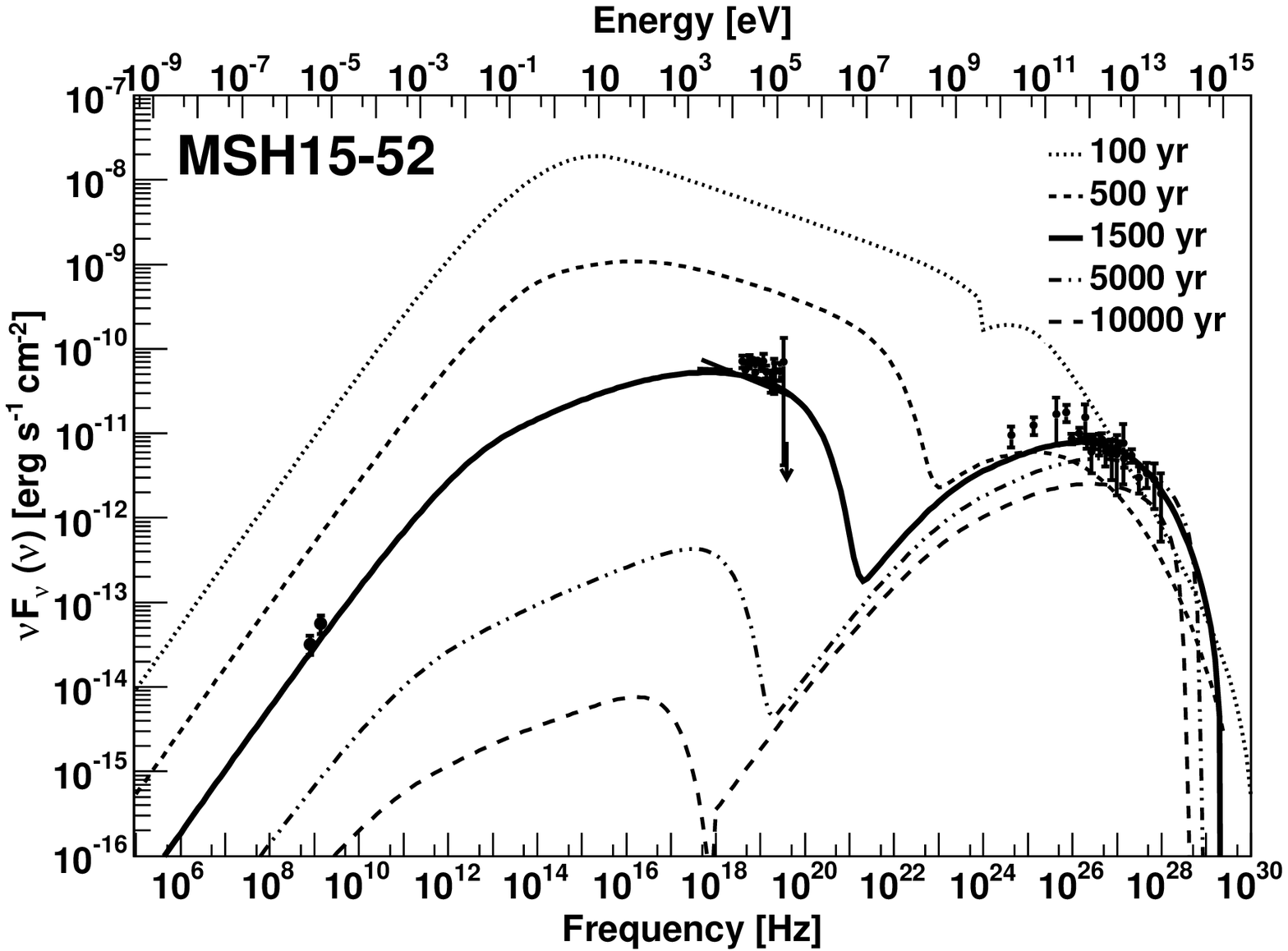}
\includegraphics[width=84mm]{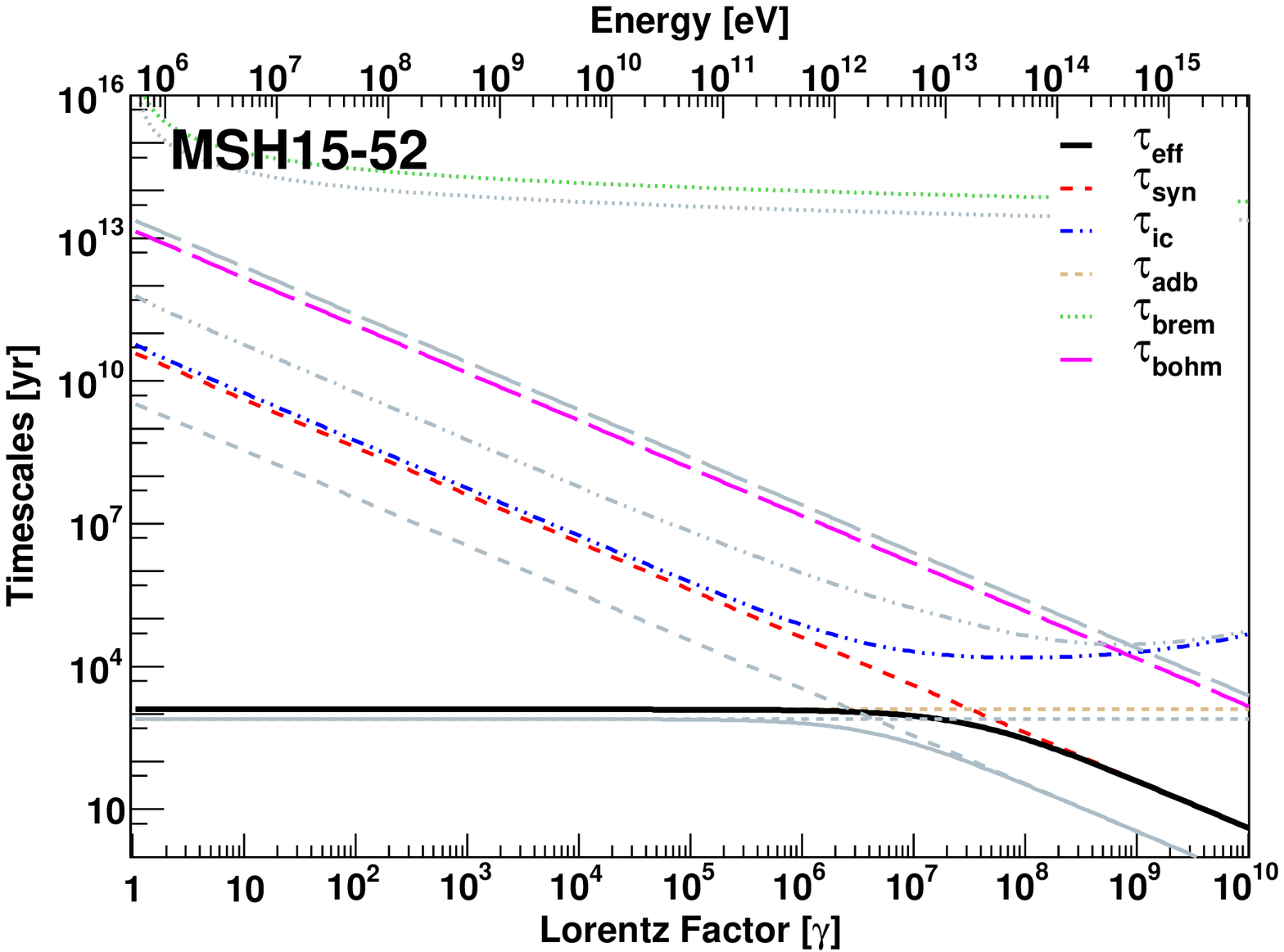}
\includegraphics[width=84mm]{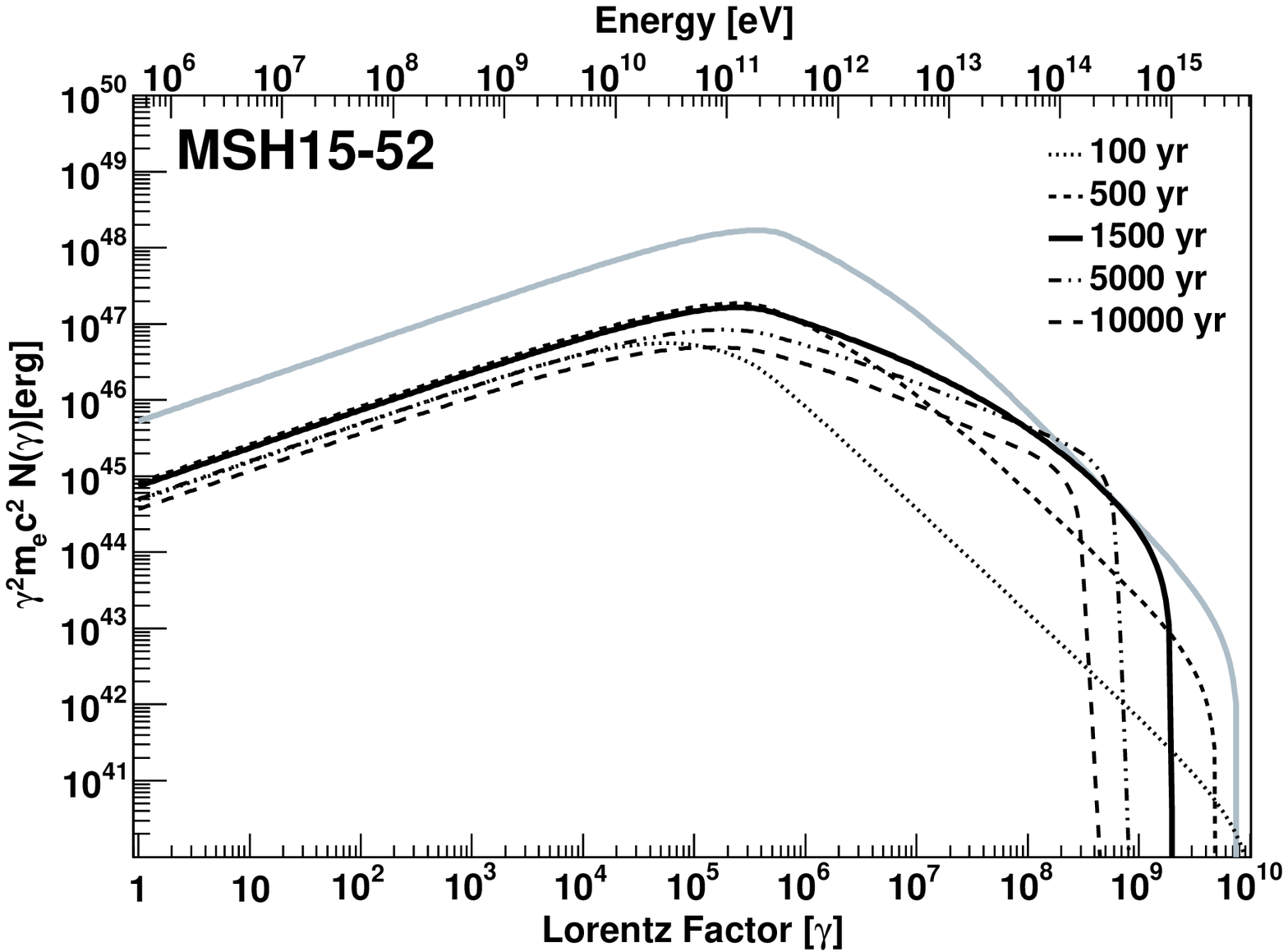}
\caption{Details of the SED model of MSH15--52.The panels are as in Fig.~\ref{G54}.
For details regarding the observational data and a discussion of the fit, see the text.}
\label{MSH1552}
\end{figure*}
%%%%%%%%%%%%%%%%%%%%%%%%%%%%%%%%%%%%%%%%%%%%

The composite SNR G320.4--1.2 / MSH 15--52 \citep{Caswell1981} is associated with the
 radio pulsar PSR B1509-58.  This pulsar is one of the youngest and most energetic known, with a 150 ms 
rotation period. It was discovery by the {\it Einstein} satellite \citep{Seward1982}, and was also detected at radio
 frequencies by Manchester et al. (1982). It has a period derivative of $1.5 \times 10^{-12}$ s s$^{-1}$, and a characteristic age of $\sim$1600 yrs, leading to a spin-down power of
 $1.8 \times10^{37}$ erg s$^{-1}$.
It is one of the pulsars with measured braking index \citep{Kaspi1994,Livingstone2005}; and we adopt for it 
the value of 2.839.  The pulsar was detected also in gamma-rays using
{\it Fermi}-LAT \citep{Abdo2010c}. 
The central non-thermal source of the system has been interpreted as a PWN powered by the pulsar 
\citep{Seward1984, Trussoni1996}.
The distance to the system was estimated using HI absorption measurements \citep{Gaensler1999} 
to be 5.2 $\pm$ 1.4 kpc, which is consistent with the value obtained by \cite{Cordes2003}  
from dispersion measure estimates, 4.2 $\pm$ 0.6 kpc.

The dimension of the PWN as observed by ROSAT \citep{Trussoni1996} and H.E.S.S. \citep{Aharonian2005}
are $10 \times 6$, and $6.4 \times 2.3$ arcmin respectively. 
The dimensions obtained in the TeV data, corresponds to a radius of a circle of $\sim$ 3 pc, 
%if we considered that the ellipse and the circle have the same surface and 
at a distance of 5.2 kpc. 

The measured braking index of the pulsar implies a young age,  lower than $\sim$ 1700 yr.
%However, as is also the case of G292.2$-$0.5 analyzed below, the braking index could not be the same for all the pulsar 
%evolution, allowing for an older system.  
According to the standard parameters of the 
ISM, the age of the system was estimated to be in the range 6--20 kyr, an order of the magnitude larger than 
the age estimated by the pulsar parameters. A plausible explanation for this discrepancy is that 
the SNR has expanded rapidly into a low-density cavity, what can also explain  
the unusual SNR morphology, the offset of the pulsar from the apparent center of the SNR, and the 
faintness of the PWN at radio wavelengths \citep{Gaensler1999,Dubner2002}. 
The south-southeastern half of the SNR seems to have expanded 
across a lower density environment of $\sim$ 0.4 cm$^{-3}$. And the north-northwestern radio limb 
has instead encountered a dense HI filament. In our models we adopt a density of 0.4 cm$^{-3}$.
However, the morphology of MSH 15--52 is complex and not taken into account in our model (similarly to other analysis alike
e.g., Tanaka and Takahara, 2011, Abdo et al. 2010, Zhang et al. 2008, Nakamori et al. 2008).

To perform our multi-wavelength fit we acquired the observational data as follows: Radio 
observations were obtained from Gaensler et al. (1999, 2002).
Observations of the nebula in the hard X-rays come from  {\it Beppo-SAX} \citep{Mineo2001}, and {\it INTEGRAL}-IBIS 
telescopes \citep{Forot2006}. COMPTEL and EGRET measurements \citep{Kuiper1999} combine the pulsar 
and the PWN measurement, so we did not consider them in our fit.
The PWN was detected and its 
spectral distribution in GeV energies was obtained by {\it Fermi}-LAT during the first year of operation of 
this instrument \citep{Abdo2010}. {\it Fermi}-LAT  observations used in our work come from subsequent work by 
Acero et al. (2013). At even higher energies,
Cangaroo III observations are in agreement with the previous H.E.S.S. observations. Both data sets were 
used below.
In the models presented here an ejected mass of 10 $M_{\odot}$ is assumed.

%%%%%%%%%%%%%%%%%%%%%%%%%%%%%%%%%%%%%%%%%%%%
%%%%%%%%%%%%%%%%%%%%%%%%%%%%%%%%%%%%%%%%%%%%
%\subsubsection{Discussion}
%%%%%%%%%%%%%%%%%%%%%%%%%%%%%%%%%%%%%%%%%%%%
%%%%%%%%%%%%%%%%%%%%%%%%%%%%%%%%%%%%%%%%%%%%

We consider different scenarios  to fit the multiwavelength data. 
In the model presented in Fig. \ref{MSH1552} we assume that the age of the 
system is 1500 yrs, close to the characteristic age of the pulsar.
We also assume a broken power-law injection. In order to fit the measured 
GeV and TeV data we use a FIR photon field 
of 
5 eV cm$^{-3}$, at a  
temperature of 20 K.  
This component is 
dominating the IC yield, while the contribution of 
the optical photon field is much lower in comparison (see Table \ref{param}). 
The other parameters resulting from the fit are 
$\alpha_1$=1.5, $\alpha_2$=2.4, a break 
Lorentz Factor of $5\times$ 10$^5$, a maximum Lorentz Factor of $1.9\times10^9$, a 
nebula magnetic field of 21 $\mu$G, and a magnetic fraction of 0.05. 
It would seem that the {\it Fermi}-LAT data is not perfectly well reproduced. This can be cured by choosing higher densities
and temperatures of the photon backgrounds, but we have not been able to find a perfect match in these conditions.

It was already proposed that the local photon background for this PWN could be higher than the average Galactic 
value, in particular in the FIR \citep{Aharonian2005}.
Nakamori et al. (2008)  and Du Plessis et al. (1995) suggested that the SNR itself could be the origin of the excess 
of the IR photon field.  As in the work of Bucciantini et al. (2011), we have also 
investigated the possibility of performing our fit assuming a contribution of a local IR photon 
field with a temperature of $\sim$ 400 K. 
This possibility is presented in our Model 2.  
Indeed, we have found that we could 
fit the observational data with a temperature (energy density) of the IR photon field of 20 K 
(4 eV cm$^{-3}$), and local IR photon field with a temperature (energy density) of 400 K 
(20 eV cm$^{-3}$). The quality and final SED 
corresponding to these assumptions (leaving all other parameters unscathed) is better matching also to the 
{\it Fermi}-LAT data, and both M1 and M2 models are compared in 
Fig.~\ref{MSH1552}. 
As the result of the M2 fit we obtained $\alpha_1$=1.5, $\alpha_2$=2.4, a break 
Lorentz Factor of $5\times$ 10$^5$, a maximum Lorentz Factor of $2.3\times10^9$, a 
nebula magnetic field of 25 $\mu$G, and a magnetic fraction of 0.07. 

\begin{figure}[t!]
\centering
\includegraphics[width=84mm]{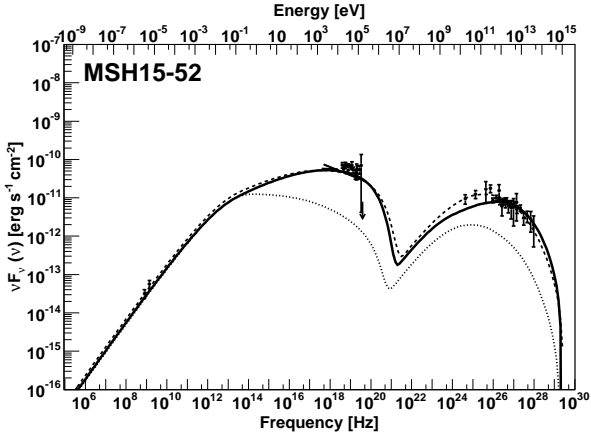}
\caption{SED of MSH 15--52 fitted with the parameters of the model described in Table \ref{param} (solid line), together with a comparison with the resulting fit using as
photon temperatures and corresponding energy densities (20 K and 4 eV cm$^{-3}$ for the FIR, and 400 K and 20 eV cm$^{-3}$ for the NIR; leaving all other parameters the same, dashed line). We also compare with the current SED results
if the parameters of Abdo et al. (2010) are assumed (dotted line). }
\label{MSH1552-2}
\end{figure}

Previous to {\it Fermi}-LAT observations,  Aharonian et al. (2005)  presented a fit of the X ray and VHE data 
using a static IC model \citep{Khelifi2002}. Using this model they reproduced the VHE spectrum of 
the whole nebula assuming a power-law energy spectrum for the population of the accelerated 
electrons with an spectral index of 2.9. The energy density of the dust component is more than a factor of 2 
higher than the nominal value given by GALPROP, similar to ours.
Abdo et al. (2010) also performed a fit of the observational data, including radio, X-ray, {\it Fermi}-LAT,   
and TeV observations using the one-zone, static model described by Nakamori et al. (2008).  According to 
their model the gamma-ray emission is dominated by the IC of the FIR photons from the interstellar 
dust grains with a radiation density fixed at 1.4 eV cm$^{-3}$ which actually is the nominal value of GALPROP at the position of MSH 15--52.
The energy densities in the model by Aharonian et al. (2005) are similar to those assumed by Abdo et al. (2010) when presenting 
{\it Fermi}-LAT results. In these
works, no time evolution is considered in any of the quantities.
We tried performing a fit with the same parameters used in \cite{Abdo2010};  i.e., assuming their spectral 
indexes, break in the spectrum of the injected particles, magnetic field, and energy 
densities of the photon fields  (see table 4 of the mentioned paper). 
We compare the results of the fits of Model 1 and 2 with the resulting model having the same parameters of Abdo et al. (2010) in Fig. \ref{MSH1552-2}. The main difference between Abdo et al. (2010) model and ours reside, apart that the latter is static, is the assumed lower photon field densities and the steeper high-energy slope of the injected electrons. These changes make for a significant underestimation of the TeV emission.
The nebula magnetic field obtained in our model (of the order of 20--25 $\mu$G) is however similar to the one 
obtained by \cite{Aharonian2005} and \cite{Abdo2010} (17 $\mu$G). Previous estimations \cite{Gaensler2002} 
gave a lower limit of the field  (8 $\mu$G), which is also compatible.

%%%%%%%%%%%%%%%%%%%%%%%%%%%%%%%%%%%%%%%%%%%%
%%%%%%%%%%%%%%%%%%%%%%%%%%%%%%%%%%%%%%%%%%%%
\subsection{HESS J1119$-$614 (G292.2$-$05)}
%%%%%%%%%%%%%%%%%%%%%%%%%%%%%%%%%%%%%%%%%%%%
%%%%%%%%%%%%%%%%%%%%%%%%%%%%%%%%%%%%%%%%%%%%

%%%%%%%%%%%%%%%%%%%%%%%%%%%%%%%%%%%%%%%%%%%%
\begin{figure*}[t!]
\centering\includegraphics[width=84mm]{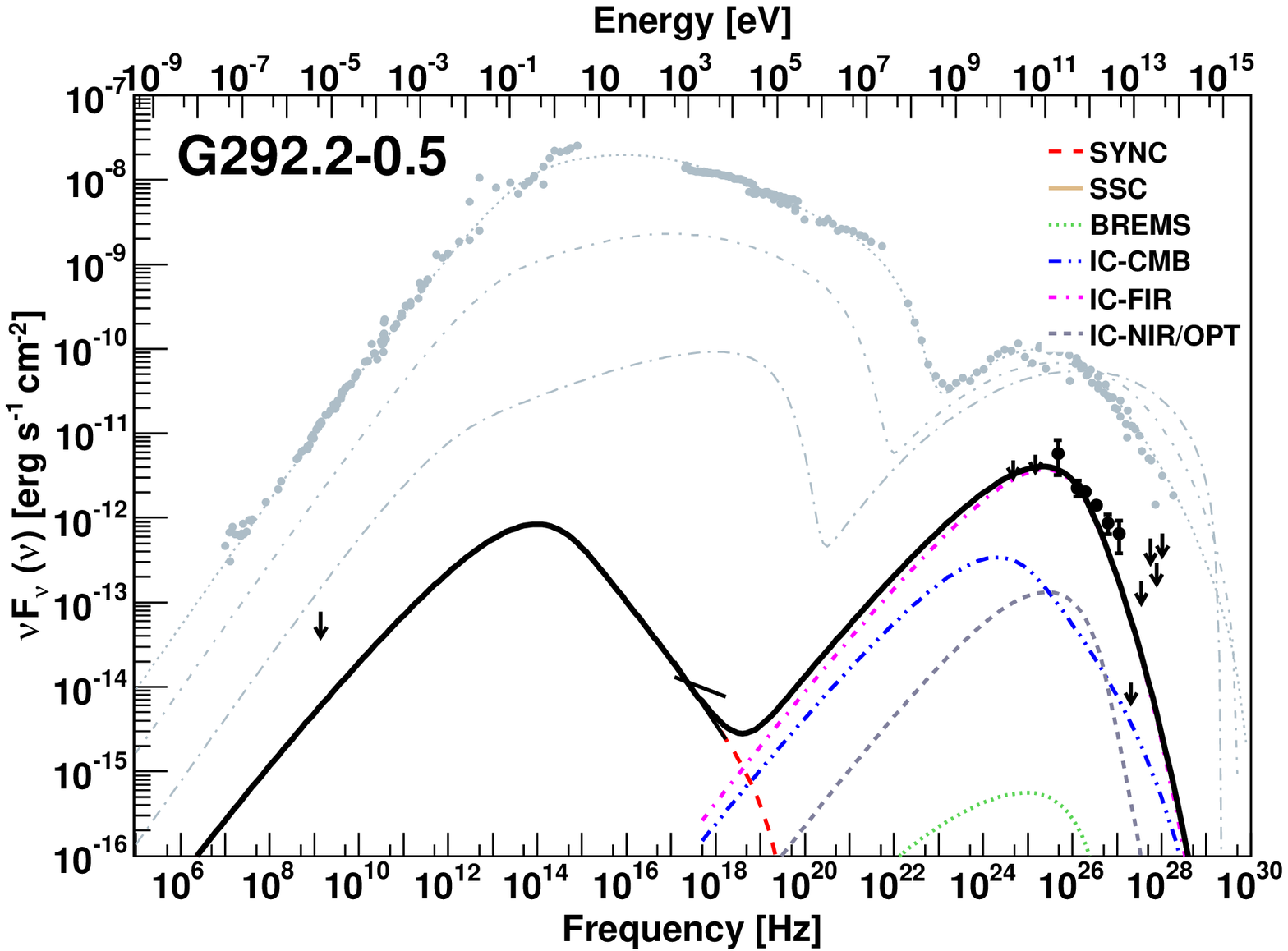}
\includegraphics[width=84mm]{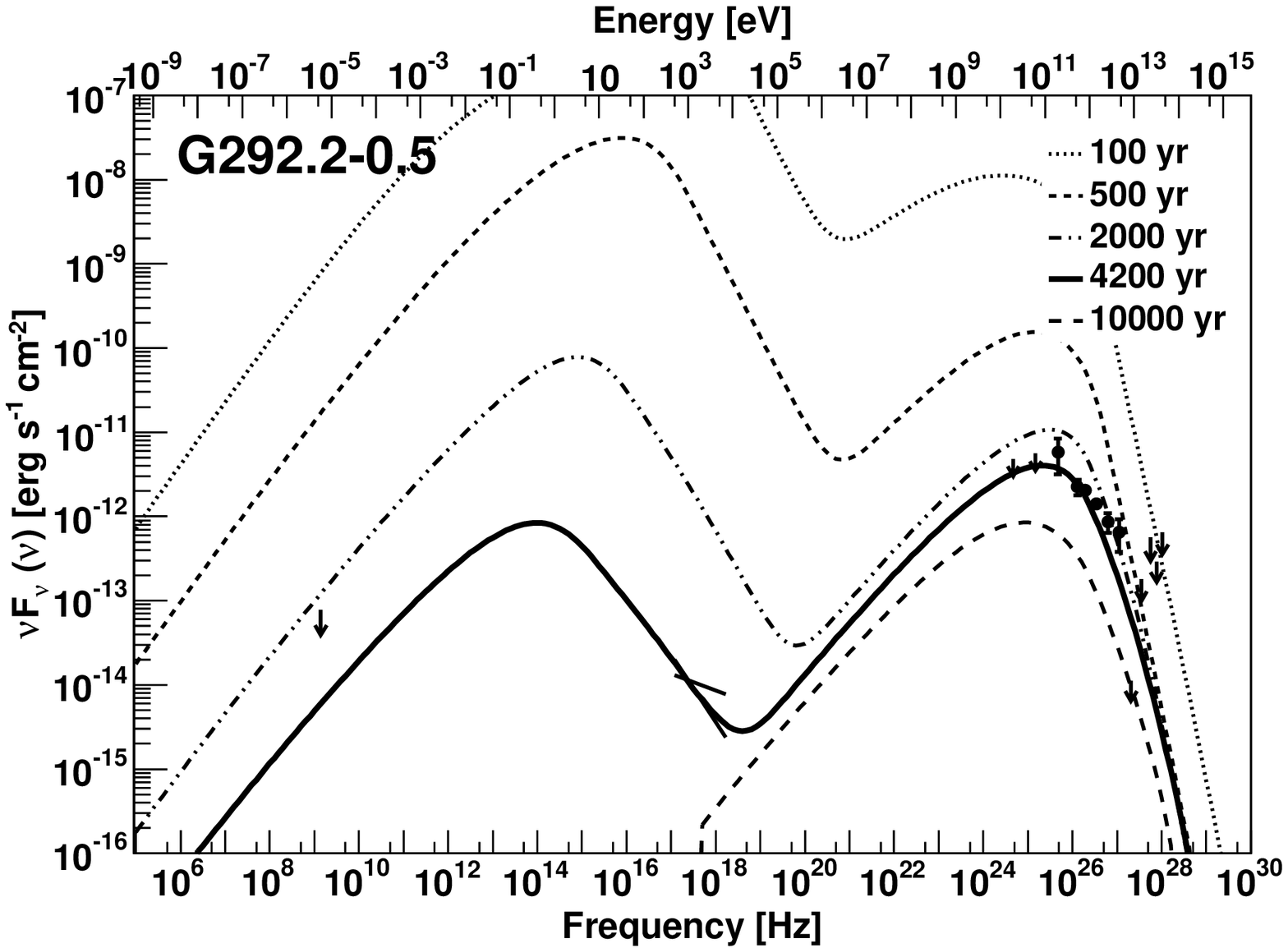}
\includegraphics[width=84mm]{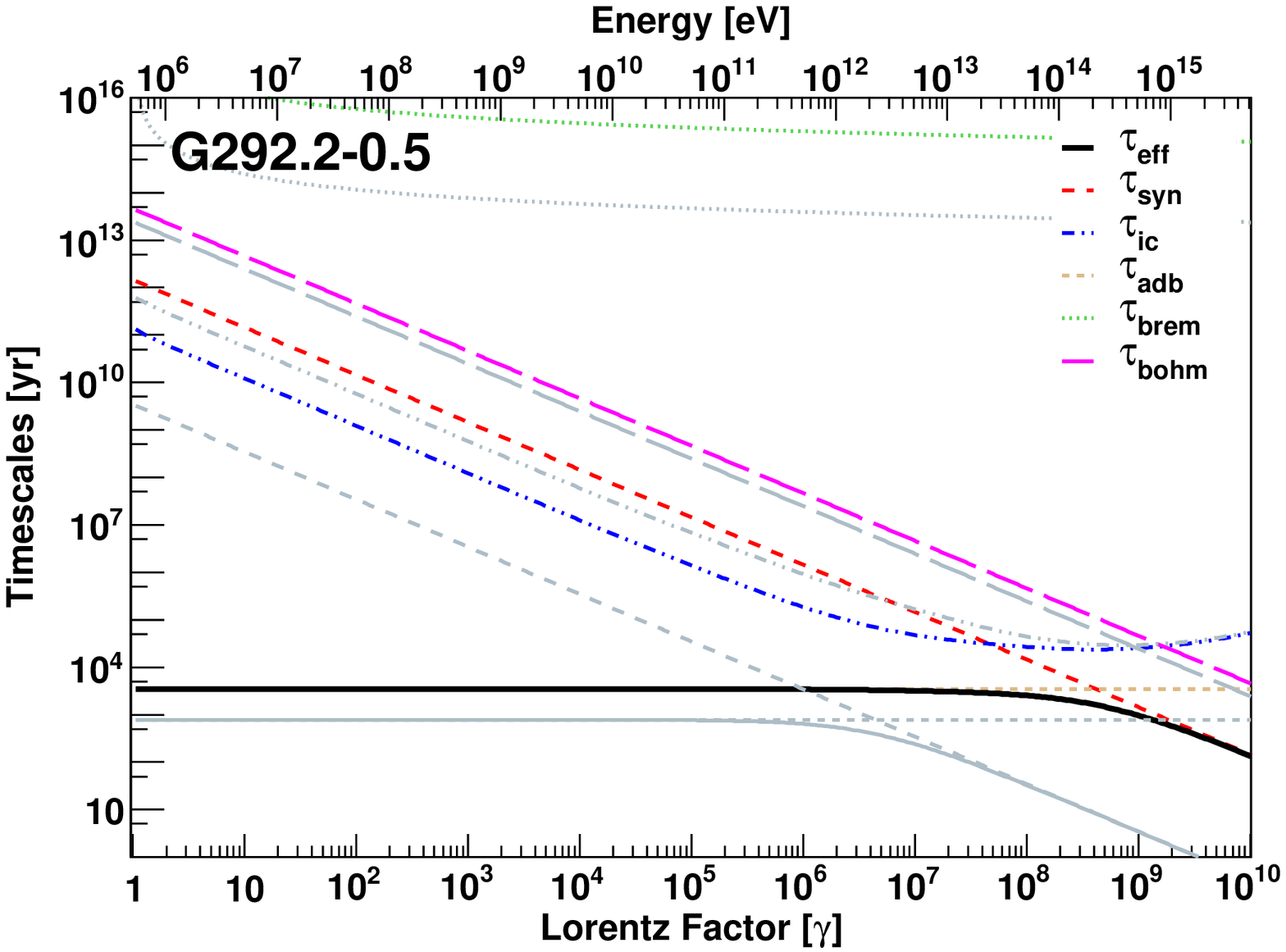}
\includegraphics[width=84mm]{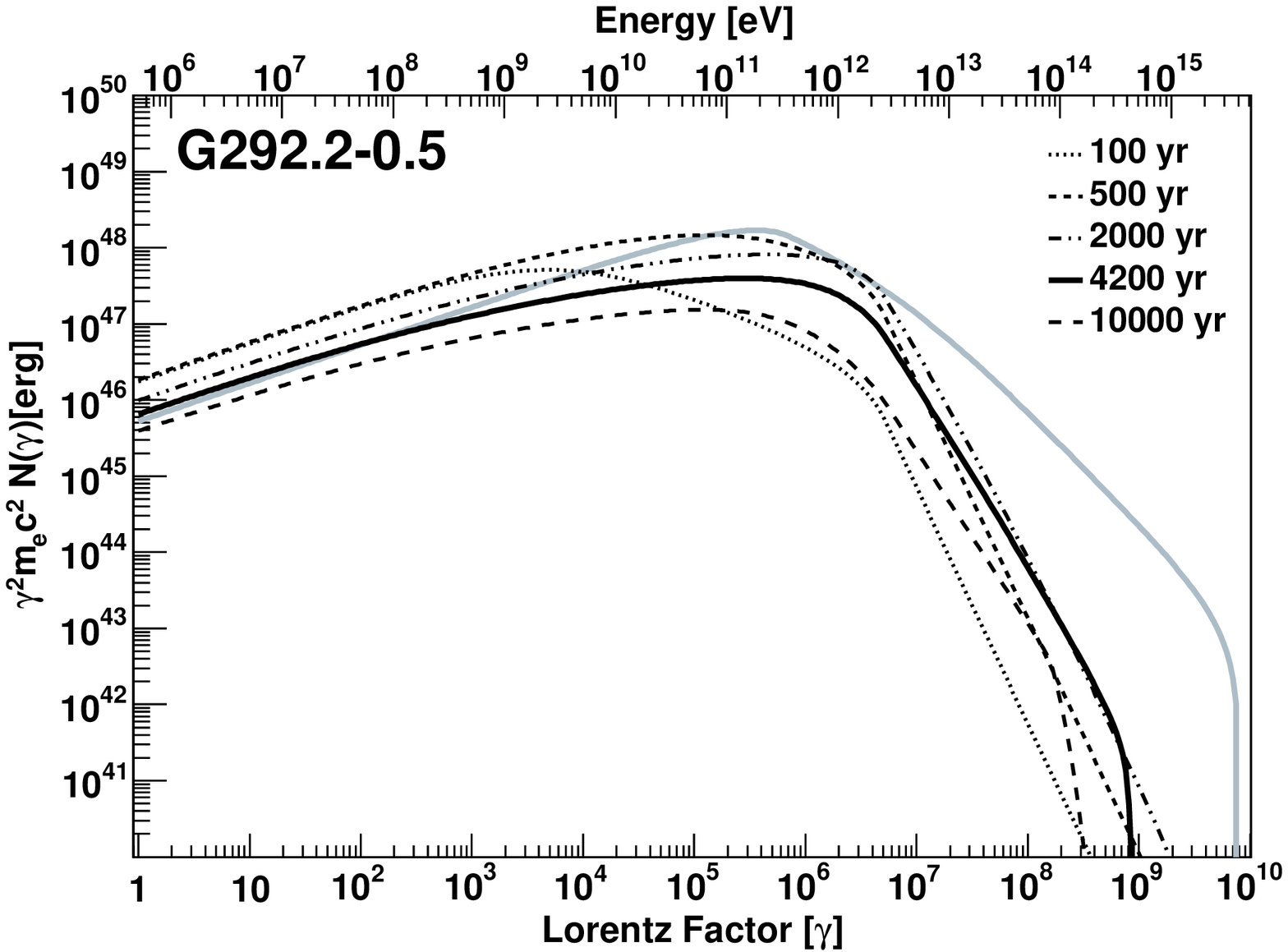}
\caption{Details of the SED model of G292.2--0.5.The panels are as in Fig.~\ref{G54}.
For details regarding the observational data and a discussion of the fit, see the text. }
\label{G292}
\end{figure*}
%%%%%%%%%%%%%%%%%%%%%%%%%%%%%%%%%%%%%%%%%%%%

G292.2$-$0.5 is a SNR associated with the high-magnetic field radio pulsar 
J1119-6127, which was discovered in the Parkes multibeam pulsar survey \citep{Camilo2000}. 
The pulsar was also detected in X-rays \citep{Gonzalez2005} and gamma-rays \citep{Parent2011}. 
It has a rotational period of 408 ms, and a period derivative of $4\times10^{-12}$ s s$^{-1}$, leading to 
a characteristic age of $\sim$1600 yr, and a spin-down luminosity of $2.3\times10^{36}$ erg~s$^{-1}$. 
The braking index was measured for the first time by Camilo et al. (2000), but this value was recently 
refined using more than 12 years of radio timing data to  $2.684\pm0.002$ \citep{Weltevrede2011}.
The high value of the pulsar magnetic field, $\sim 4.1\times10^{13}$~G places PSR J1119$-$6127 between typical radio pulsars and usual
magnetars.  

A faint PWN surrounding the pulsar was detected in X-rays \citep{Gonzalez2003,Safi2008}. The X-ray unabsorbed flux 
between $0.5$ and 7 keV was measured to be $1.9 \times 10^{-14}$~erg~cm$^{-2}$~s$^{-1}$ for the compact nebula, and 
$2.5\times10^{-14}$~erg~cm~$^{-2}$~s$^{-1}$ for the associated {\it jet}, with spectral indices 
of 1.1 $\pm$ $^{0.9}_{0.7}$ and 1.4 $\pm$ $^{0.8}_{0.9}$, respectively. 
These are extremely low values in comparison to other PWNe, 
G292--0.5 is a very faint PWN in X-rays, which remains so even in the case of adding the southern {\it jet} flux. 
 The PWN was also detected at high energies by {\it Fermi}-LAT  \citep{Acero2013} and at very high energies by H.E.S.S. \citep{Mayer2010,Djanati2009}.\footnote{We remark that these are not official
 claims of the H.E.S.S. collaboration; they are not confirmed, but not ruled out either. We entertain the possibility that the final TeV data may differ from the current available spectrum.}
TeV measurements have shown a flux of 4\% of the Crab nebula and a steeper spectrum (with slope larger than 2.2) compared with other young PWNe. 
The luminosity in TeV gamma-rays (at 8.4 kpc, see below) is 
$3.5 \times 10^{34}$ erg s$^{-1}$, which makes for an efficiency of 1.5\% in comparison of the current pulsar spin-down. Thus, the ratio of 
$L_X$/$L_\gamma$ is $\sim10^{-3}$, which would imply a low magnetic field.
%mirar por formula de Djannati atai

The mass of the progenitor of the SN explosion is large \citep{Kumar2012}; these authors
%  $\sim 30 M_{\odot}$ \cite{Kumar2012}
inferred that the expansion occurred in a very low-density medium. We 
assumed in our calculations that the ejected mass had a value between 30 and 35 $ M_{\odot}$, and that the density of the medium was $0.02$ cm$^{-3}$.
The kinematic distance to the system was suggested to be 8.4 $\pm$ 0.4 kpc based on HI absorption measurements \citep{Caswell2004}. 
According to Safi-Harb \& Kumar (2008), the size of the compact PWN in X-rays is  6$\times$15  arcsec,
with the jet corresponding to a faint structure of 6 $\times$20 arcsec. 
For a distance of 8.4 kpc, this size corresponds to $\sim$ 0.5 pc.  In the TeV range, the source is extended and the size is larger,  its diameter is of the order of  $\sim$30 pc \citep{Kargaltsev2010,Djanati2009}. 

Kumar et al. (2012) estimated the age of the SN in a range between 4200 yrs
(for a free expansion phase, assuming an expansion velocity of 5000 km s$^{-1}$) and 7100 yr (for a Sedov phase). 
This estimation is larger than the one obtained using the pulsar parameters, of 1900 yr. 
In our Model we propose 
a fit of the data assuming an age of 4200 yr (and $n=1.7$), and compare it with the results of assuming an age of 1900 yrs (and $n=2.7$) in alternative fittings.

To compute the fit we then consider the H.E.S.S. measurements \citep{Kargaltsev2010,Djanati2009}; together with the X-ray flux quoted above
 \cite{Safi2008}. These are both crucial assumptions, which, as we shall see, reflect in a very steep injection at high energies. We comment more on them below.
 ATCA deep measurements revealed only a 15 arcmin SNR shell \citep{Crawford2001}, but no radio emission from the PWN. The latter authors interpreted the absence of a radio PWN as being the result of the pulsar's high magnetic field; which would lead to a short time of high energy electron injection (due to a large spin-down)
%
%see Bhattacharya (1990)
%from Crawford et al. 2011: even if a high magnetic field pulsar is born spinning rapidly, it will quickly slow down to a long period because of severe magnetic braking. At later times, there is therefore no significant energy injection into the PWN from the pulsar, and the energetics of the nebula are dominated by the significant losses it experiences because of expansion into the ambient medium. The net result is that a PWN associated with a high field pulsar should have an observable lifetime that is very brief, corresponding to the reduced period for which the pulsar provides significant amounts of energy to the nebula.
%
What they see is a limb brightening elliptical shell (in fact designated thereafter as G292--0.5) of dimensions 14' x 16'  with a 1.4 GHz flux density of 5.6 $\pm$ 0.3 Jy. At 2.5 GHz, the measured flux density of G292.2--0.5 is 1.6 $\pm$ 0.1 Jy (but this should likely be taken as a lower limit since the shell is larger than the largest scale to which the interferometer used is sensible). We shall take this SNR flux measurement at 1.4 GHz as a safe upper limit for the PWN radio emission.

%%%%%%%%%%%%%%%%%%%%%%%%%%%%%%%%%%%%%%%%%%%%
%%%%%%%%%%%%%%%%%%%%%%%%%%%%%%%%%%%%%%%%%%%%
%\subsubsection{Discussion}
%%%%%%%%%%%%%%%%%%%%%%%%%%%%%%%%%%%%%%%%%%%%
%%%%%%%%%%%%%%%%%%%%%%%%%%%%%%%%%%%%%%%%%%%%

We consider first an age of 4200 years, as derived by Kumar et al. (2012) based on SNR properties. 
To reconcile the pulsar age with the supernova, 
Kumar et al. (2012) suggested that the braking index has to be smaller than 2 for most of the pulsar lifetime.  We assume it to be 1.7.
With this age, a fit can be obtained with the FIR dominating the IC yield, with relatively low energy densities. However, the injected 
electron spectrum at high energy needs to be steep (4.1) to achieve good agreement with observational data. This is an interesting result, since it is by far steepest $\alpha_2$ we shall see in the whole sample, and it is quite constrained by the observations of both GeV and TeV emission from this source. Another interesting difference in this case is that the spectral break of the injected electron is higher, in the three models, that the one obtained for other PWNe.  However, the extra degree of freedom given by  the lack of a detection  of the synchrotron at low frequencies peak is a caveat.
The resulting model parameters under this age assumption are given in Table \ref{param} and Fig.~\ref{G292}.

\begin{figure}
\centering
\includegraphics[width=84mm]{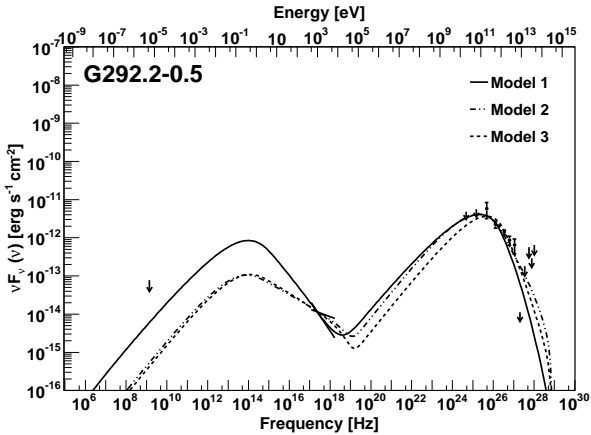}
\caption{Spectrum of the three different models for  G292.2--0.5. See the text for a discussion of the differences and caveats underneath each of these models.}
\label{G292comp}
\end{figure}

We have also explored models in which 
the age of the PWN is lower, as resulting from the estimate of the pulsar period, period derivative, and braking index \citep{Weltevrede2011}.  We have found that it is especially difficult to find models that could consistently fit the whole set of observations, with the more constraining range being the GeV gamma-rays.
In order to fit 
the MW observational data for lower pulsar ages, either  we assume that 
the energy densities of the FIR and NIR/OPT components are significantly larger 
(10 and 130 eV cm$^{-3}$, respectively),  or we assume that there is 
a contribution of a local IR field at 400 K, similar to the alternative model considered above 
for MSH 15-52; which, in any case, would need  a large energy density (33 eV cm$^{-3}$).
These values of NIR/OPT densities  would 
make the corresponding IC component to significantly contribute, or overcome the FIR IC yield.
Both of these models take the measured value of $n \sim 2.7$, 
show a radius of about 6 pc, and similar magnetic field, magnetization, injection slopes, and break energies that the corresponding 
ones shown in Table \ref{param}, but 
are less satisfying due to the large energy densities involved without a clear a priori justification. 
In any case, degeneracies in modeling can be broken at radio and optical frequencies (see Fig. \ref{G292comp}).

Interestingly, the three models show a very low magnetic field for the nebula, which is
consistent with the expectations coming from the extremely low value of the ratio of X-ray and gamma-ray luminosities, and
about one order of magnitude lower than the 
one estimated earlier by \cite{Mayer2010}, of 32 $\mu$G.
A lower ejected mass or a higher energy explosion (that can make the size of the nebula larger) will make the magnetic field even lower than the ones obtained in the models presented here.

Another point of discussion in this case is the size of the nebula.  
Whereas the different sizes could be explained due to the larger losses of X-ray generating electrons, this PWN has one of the largest mismatches.
Electrons generating keV photons have, for the resulting $B$ field, a very high energy, in excess of 70 TeV,
% Ee= 70 TeV B_5^-1/2 E_kev^1/2 
much larger than the energy of electrons generating TeV photons.
In the model of Table \ref{param}, we obtain a radius of $\sim 13$ pc and use it for all frequencies. However, the X-ray and TeV emission regions are probably not the same, and a more detailed model could be needed
for a more proper accounting. 
%

%%%%%%%%%%%%%%%%%%%%%%%%%%%%%%%%%%%%%%%%%%%%
%%%%%%%%%%%%%%%%%%%%%%%%%%%%%%%%%%%%%%%%%%%%

%%%%%%%%%%%%%%%%%%%%%%%%%%%%%%%%%%%%%%%%%%%%
%%%%%%%%%%%%%%%%%%%%%%%%%%%%%%%%%%%%%%%%%%%%
\subsection{HESS J1846--029 (Kes 75)}
%%%%%%%%%%%%%%%%%%%%%%%%%%%%%%%%%%%%%%%%%%%%
%%%%%%%%%%%%%%%%%%%%%%%%%%%%%%%%%%%%%%%%%%%%

%%%%%%%%%%%%%%%%%%%%%%%%%%%%%%%%%%%%%%%%%%%%
\begin{figure*}[t!]
\centering\includegraphics[width=84mm]{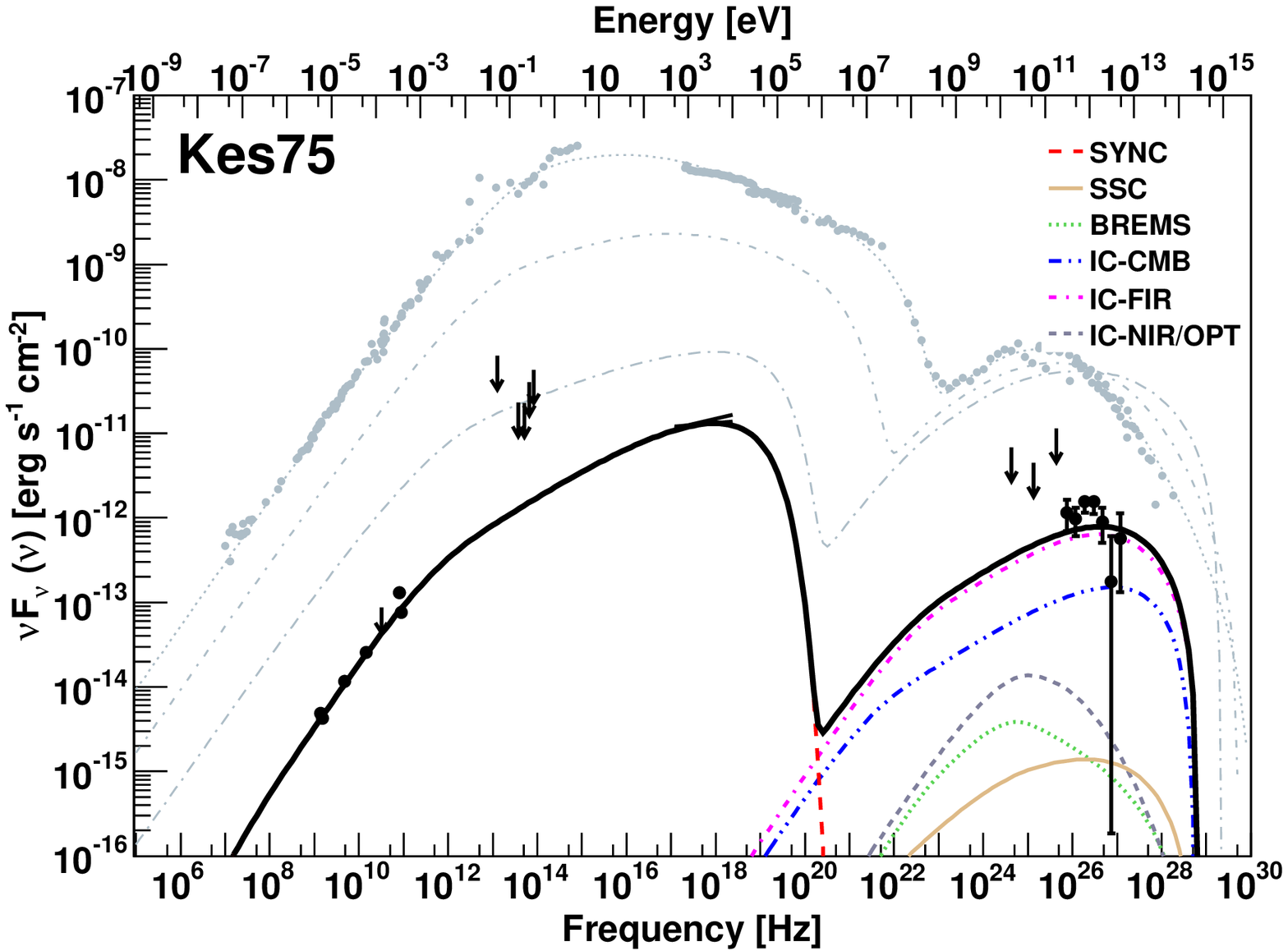}
\includegraphics[width=84mm]{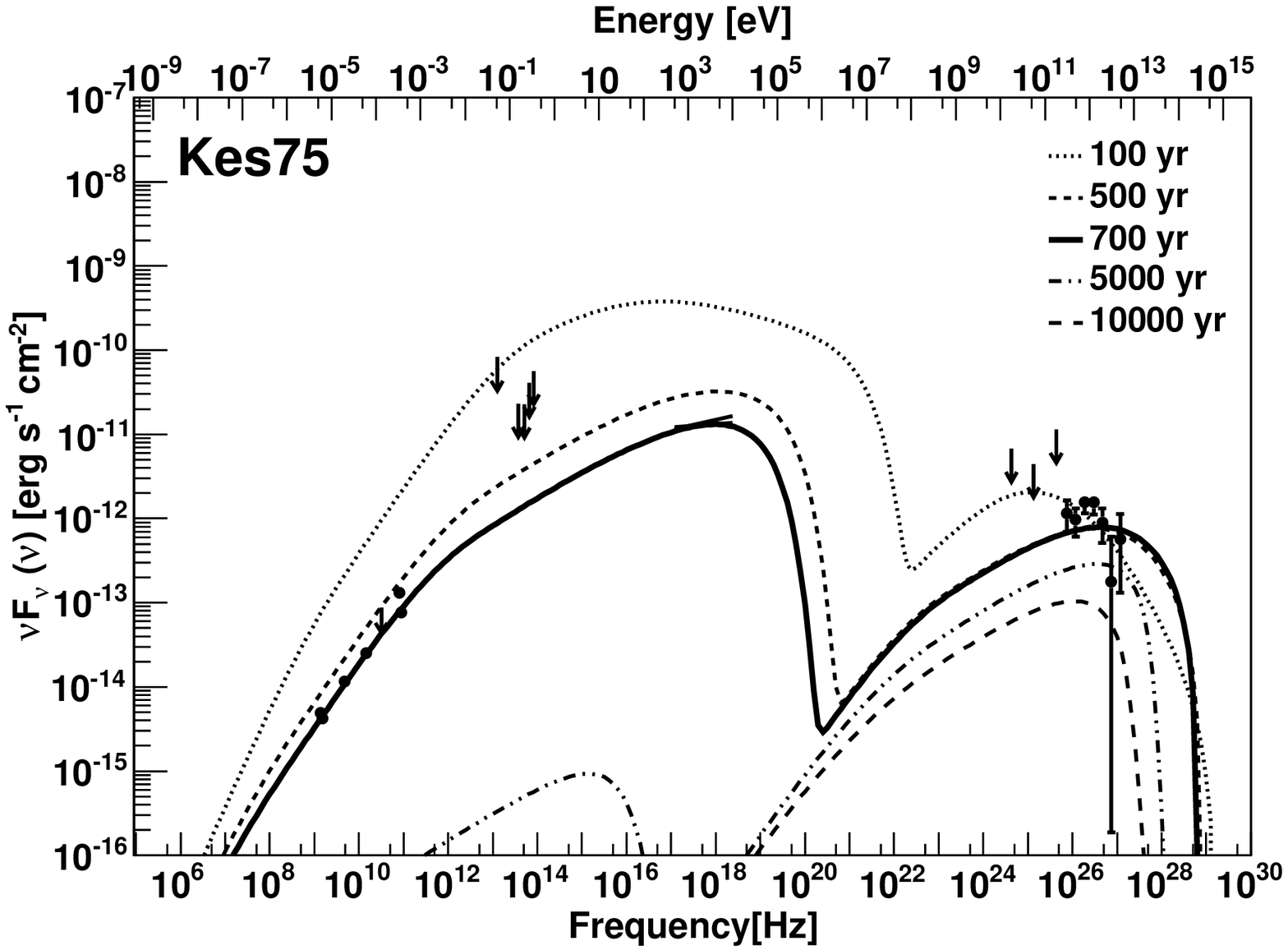}
\includegraphics[width=84mm]{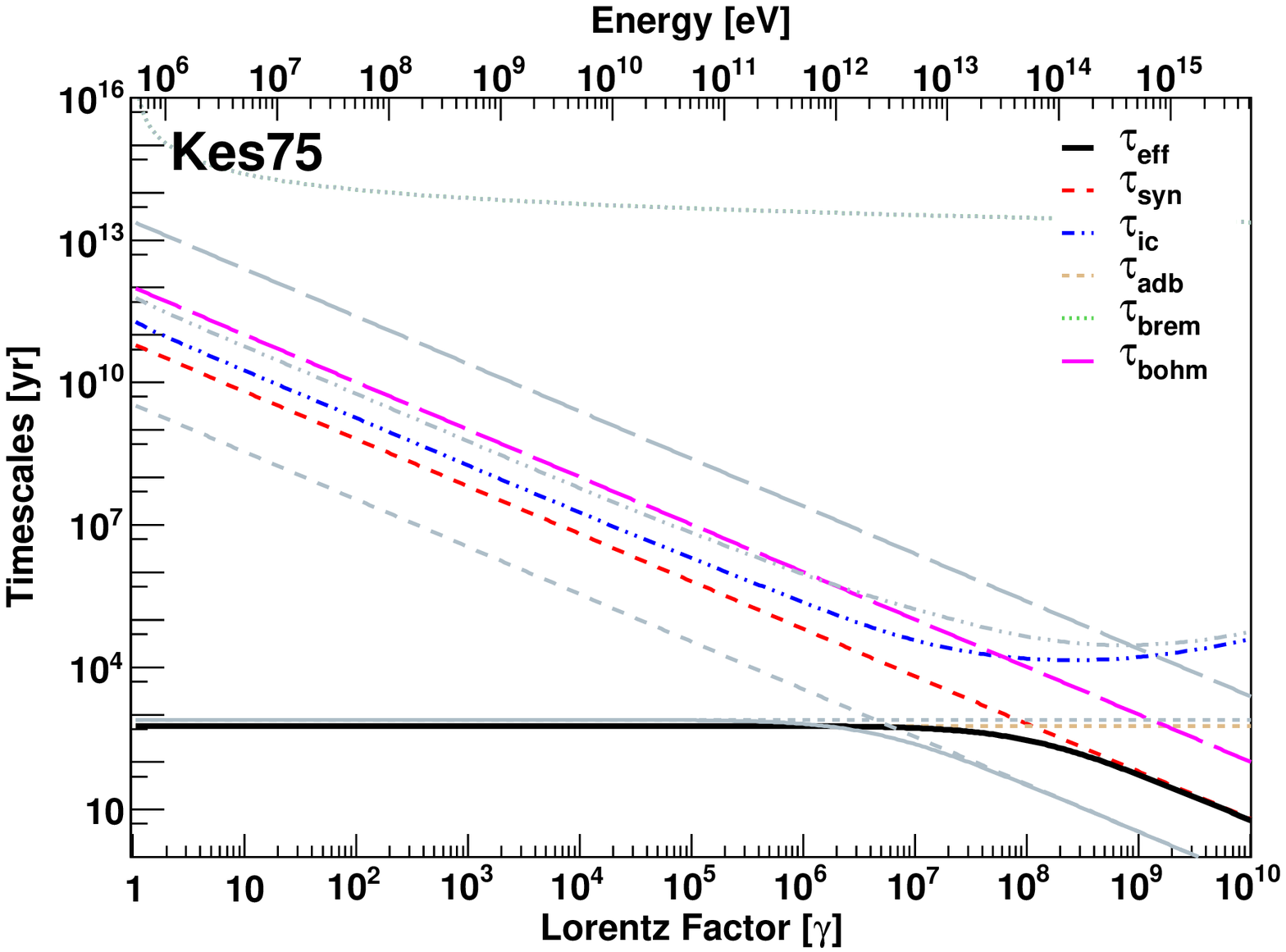}
\includegraphics[width=84mm]{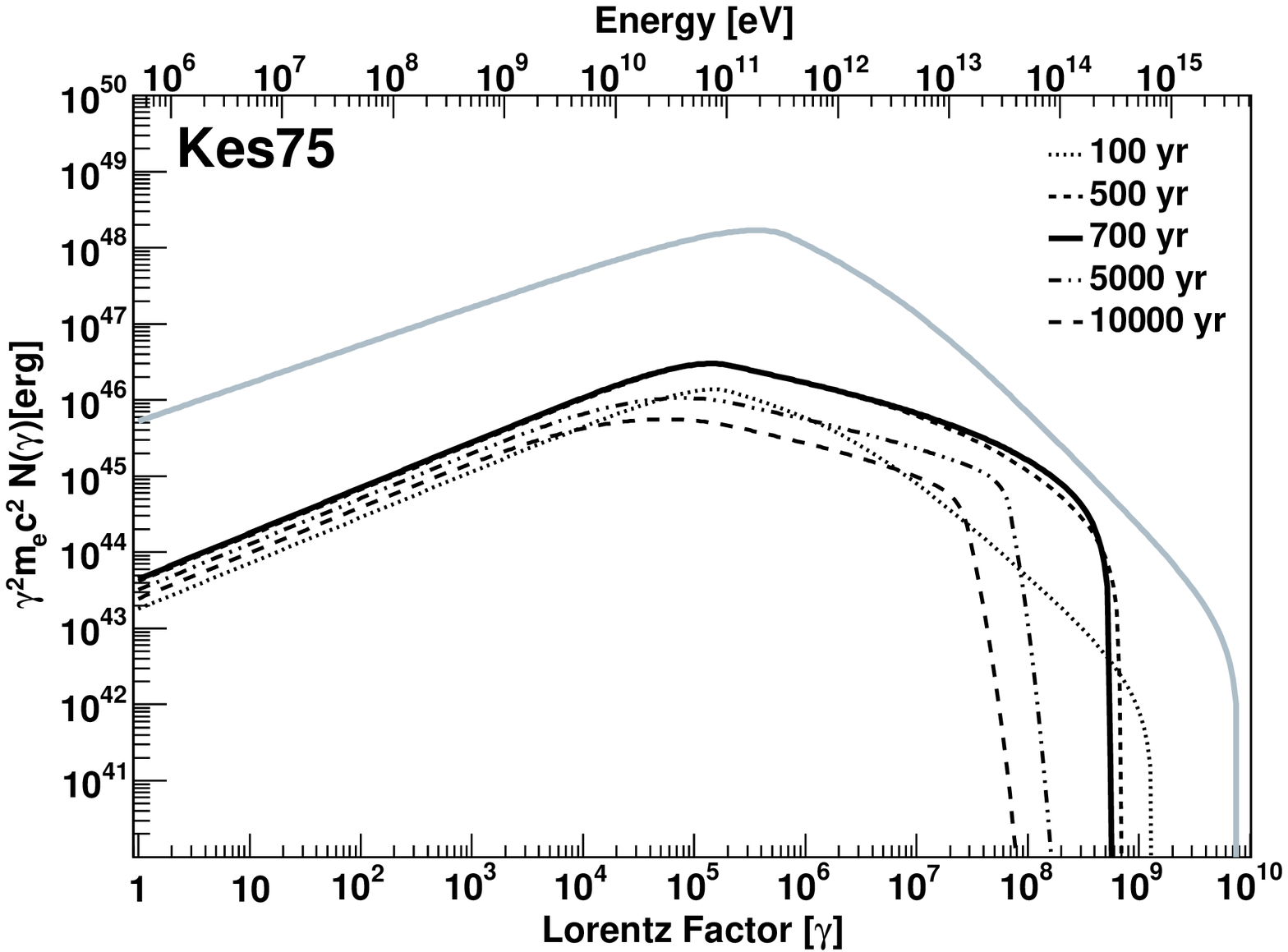}
\caption{Details of the SED of Kes 75
as fitted by our model. The panels are as in Fig.~\ref{G54}.
For details regarding the observational data and a discussion of the fit, see the text.}
\label{Kes75}
\end{figure*}
%%%%%%%%%%%%%%%%%%%%%%%%%%%%%%%%%%%%%%%%%%%%

Kes 75 (also known as G29.7--0.3) is a shell-type supernova remnant with a central core whose observed properties suggest an association with a PWN.  
The pulsar associated with this system, PSR J1846-0258, was discovered in a timing analysis of the X-ray data from {\it RXTE} and {\it ASCA} \citep{Gotthelf2000}. The pulsar has not been detected in the radio band, perhaps due to beaming. {\it Fermi}-LAT  did not detect the pulsar at high energies either. 
PSR J1846-0258 has a spin period of $\sim$324 ms, and a spin-down age of 7.1 $\times 10^{-12}$ s s$^{-1}$, implying a large spin-down luminosity of 8.2 $\times10^{36}$ ergs s$^{-1}$, a high surface magnetic field of $\sim$ 5 $\times10^{13}$ G, and a small characteristic age $\sim$ 720 yr \citep{Kuiper2009}. This pulsar exhibited a 
magnetar-like outburst with a large glitch in 2006 \citep{Gavriil2008,Kumar2008,Livingstone2011}.
The pulsar's braking index was measured  using {\it RXTE} observations \citep{Livingstone2006}. The latter authors found a value of 2.65 $\pm$ 0.01, which implies a spin-down age of 884 years,  placing this pulsar among the youngest in the Galaxy.
During the magnetar-like outburst and the large glitch of 2006, the pulsar presented 5 very short X-ray bursts, changes in the spectra, timing noise, increase in the flux  (6 times larger than in the quiescent state), and softening of the spectral index \citep{Ng2008,Gavriil2008,Kumar2008}. After that episode the braking index decreased, and has now a value of 2.16 $\pm$ 0.13 and
the pulsar and the PWN came back to the previous flux and spectral index  \citep{Livingstone2011}.
It was proposed that the
PWN variability observed in 2006 is most likely unrelated to the outburst and is probably similar in origin to the variation of  
small-scale features seen in other PWNe \citep{Livingstone2011}. Detailed studies of the variability of the PWN using deep {\it Chandra} observations were 
also presented by Ng et al. (2008).
While fitting the multiwavelength emission from Kes 75, 
we have assumed a value of 2.16 for the braking index, and 
analyzed the differences in the predictions entailed by changing the value of $n$ to that valid before 
the outburst. 
 
The morphology of the nebula in X-rays is similar to the one observed in radio wavelengths. It is highly structured and it has a dimension, 
according to high-resolution {\it Chandra} images, 
of 26 $\times$ 20 arcsec$^2$.  A detail of the complex morphology of the nebula according to {\it Chandra} observations is presented by Ng et al. (2008).
The first estimation of the distance to the system based on neutral hydrogen absorption measurements was 19 kpc \citep{Becker1984}. More recently \cite{Leathy2008} estimated 
a new distance between 5.1 and 7.5 kpc from HI and $^{13}$CO maps. 
However, Su et al. (2009) also estimated a new distance to the system of 10.6 kpc based on the association between the remnant and the molecular shells.
There is then a significant uncertainty in the distance to this PWN, and thus we have assumed two different models; with 
a distance of 6 kpc in our Model 1 and a distance of 10.6 kpc in our Model 2. 

To perform the multiwavelength fit  presented below, we took radio observations \citep{Salter1989, Bock2005}, and infrared upper limits  \citep{Morton2007}. 
The X-ray spectra, resulting from {\it Chandra} observations, was taken from Helfand et al. (2003). 
{\it Fermi}-LAT  upper limits in the photon flux corresponding to three energy bands 
are presented in Acero et al. (2013). In all of these energy bins, the significance (TS value) is very low (5 in the range 10--31 GeV, and 0 in the ranges of 31--100 GeV and 100--316 GeV). To obtain the upper limits in energy we multiplied the photon flux in each bin by the energy of the center of the bin.  
%These values are the ones plotted in Fig. \ref{Kes75}.
At very high energies the nebula was detected by H.E.S.S. \citep{Djannati2007} with an intrinsic extension compatible with a point-like source and a position in good agreement with the pulsar associated to the nebula. 

%%%%%%%%%%%%%%%%%%%%%%%%%%%%%%%%%%%%%%%%%%%%
%%%%%%%%%%%%%%%%%%%%%%%%%%%%%%%%%%%%%%%%%%%%
%\subsection{Discussion}
%%%%%%%%%%%%%%%%%%%%%%%%%%%%%%%%%%%%%%%%%%%%
%%%%%%%%%%%%%%%%%%%%%%%%%%%%%%%%%%%%%%%%%%%%

We present the results of our fit to the multiwavelength observations of Kes 75 assuming that the age and distance to the system are 700 yr and 6 kpc for Model 1, 
and 800 yr and 10.6 kpc for Model 2. In both models, we have assumed a braking index of 2.16 \citep{Livingstone2011} and a density of the medium of 1 cm$^{-3}$ \citep{Safi2012}. The ejected mass for Model 1 was assumed to be 6 $M_{\odot}$ and 7.5 $M_{\odot}$ for  Model 2. These models span the range of the uncertainties in distance.

To fit the TeV data we assume a temperature (energy density) of 25 K (2.5 eV cm$^{-3}$) for the FIR and  5000 K (1.4 eV cm$^{-3}$) for the NIR/OPT photon field in Model 1.  In Model 2 (corresponding to the slightly larger age and farther distance) we need to double the energy density in the FIR to fit the observational data. We comment more on this below.
In both of these models, the IC with the FIR photon field is the most important component, being the IC with CMB the second contributor to the total yield. The full set of assumed and fitted parameters are shown in Table \ref{param}, whereas the results for Model 1 are presented in Fig.  \ref{Kes75}.
 
The {\it Spitzer} upper limits do not constrain the parameters of the models in any significant way.
The break in the spectrum between the radio and X-ray bands appears at $\sim$100 GeV for Model 1 and $\sim$50 GeV for Model 2 in our fit. These low breaks are in agreement with the results presented by Bock et al. (2005). 
The average magnetic field obtained for the nebula was 19 $\mu$G in Model 1 and 33 $\mu$G in Model 2. In both cases the magnetic fraction is  low  and comparable to other PWNe.  The average magnetic field obtained are similar to the ones obtained by \cite{Tanaka2011}. Djannati-Atai et al. (2007) also suggested a low magnetic field for this nebula of the order of $\sim$10 $\mu$G. The first spectral index, $\alpha_1$, of the injected spectrum are both also in agreement with the ones obtained by Tanaka et al. (2011), but as in other cases, our second spectral index, $\alpha_2$ are lower than the ones obtained in their fits; which may result from a different treatment of the radiative losses. 
The final SED results for Models 1 and 2 are quite similar, showing a problematic degeneracy which cannot be broken by the data now at hand. 
In fact, other degeneracies resulting from the uncertainty in age can be accommodated by modifying the high energy slope of the injected power law, or the magnetic field.
Changes are not severe, though, and do not affect the main conclusions.

We could also fit the observational data assuming a braking index of 2.65 (with an age of 700 yrs).
For instance, for
an ejected mass of 6 $M_{\odot}$, at a distance of 10.6 kpc, a nebula magnetic field of 40 $\mu$G with a magnetic fraction of  0.055, and spectral indices of 1.4 and 2.2 for the injected particle spectrum with a break Lorentz factor at 2$\times10^5$ would fit the spectrum equally well, for energy densities and temperatures of photon backgrounds similar to those assumed in Models 1 and 2 presented in Table \ref{param}.

All in all, Kes 75 is a difficult case to model in detail: in particular, we find difficult to provide an overall (along all frequencies) significantly 
better fit than the one we show in Fig.  \ref{Kes75}, which we see a bit dissatisfying at the largest energies. There, the fall out 
of the TeV emission is plausibly steeper than in the model we show, what should be studied with future datasets. The VHE energy data seems to {\it peak}
around 1 TeV. However, since this is not clear within the reach of the present dataset, we have not tried to model a peak. 
We have considered models with larger break energies,
different photon background and injection parameters, but they do not provide significant improvements. We explored 
increasing the NIR density, i.e., increasing the IC contribution at energies of 10$^{11}$ eV so that the curve at the highest energies flattens. 
With $\omega_{FIR} = 2$ eV cm$^{-3}$ at a central temperature of 100 K and $\omega_{NIR} = 20-25$ eV cm$^{-3}$ at 3000 K
the contribution of IC-NIR becomes comparable to that of IC-FIR but peaking at lower energies, thus flattening or even steppening 
the high-energy yield.

%%%%%%%%%%%%%%%%%%%%%%%%%%%%%%%%%%%%%%%%%%%%
%%%%%%%%%%%%%%%%%%%%%%%%%%%%%%%%%%%%%%%%%%%%
\subsection{HESS J1356--645 (G309.9--2.51)}
%%%%%%%%%%%%%%%%%%%%%%%%%%%%%%%%%%%%%%%%%%%%
%%%%%%%%%%%%%%%%%%%%%%%%%%%%%%%%%%%%%%%%%%%%

%%%%%%%%%%%%%%%%%%%%%%%%%%%%%%%%%%%%%%%%%%%%
\begin{figure*}[t!]
\centering\includegraphics[width=84mm]{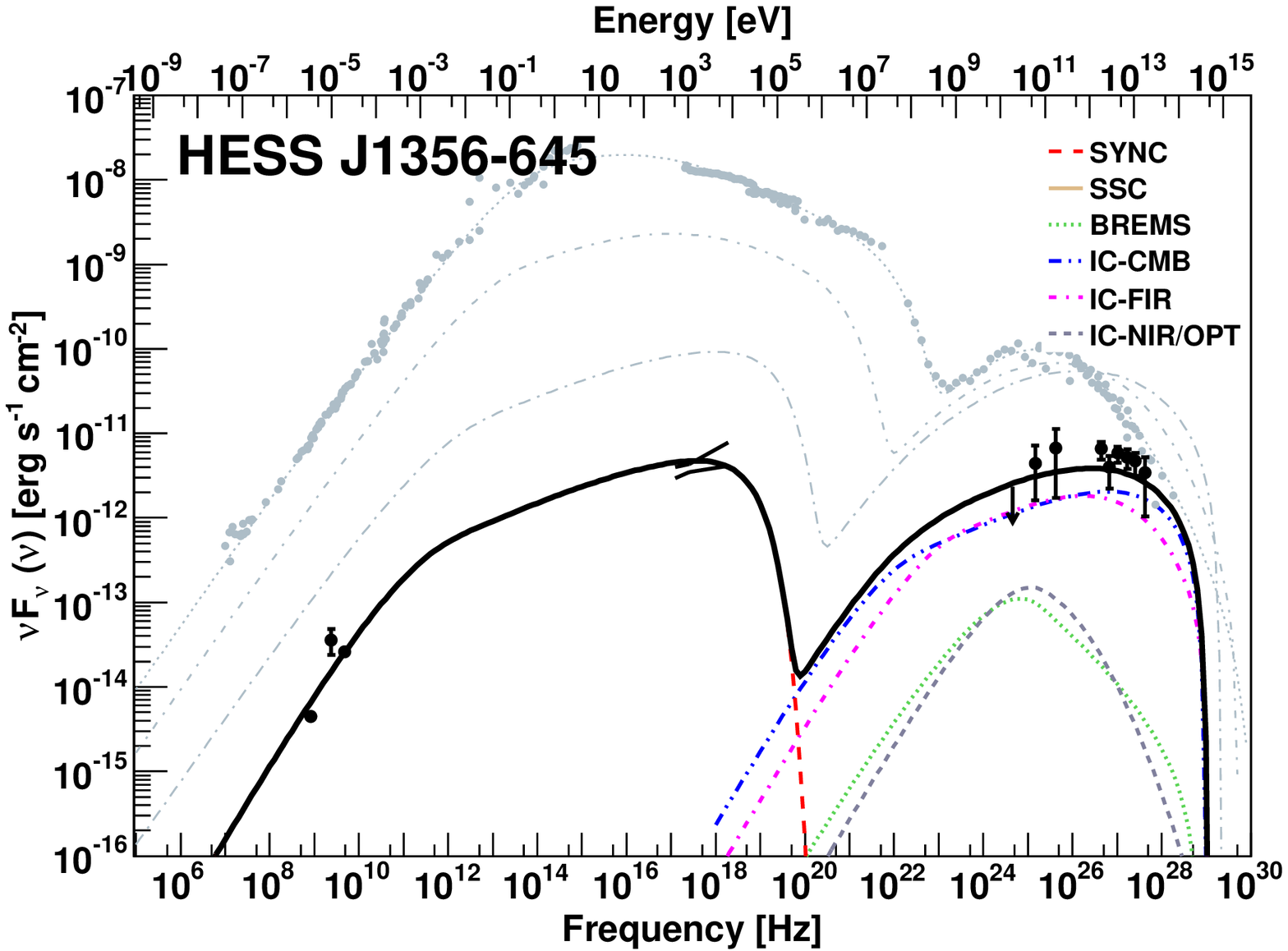}
\includegraphics[width=84mm]{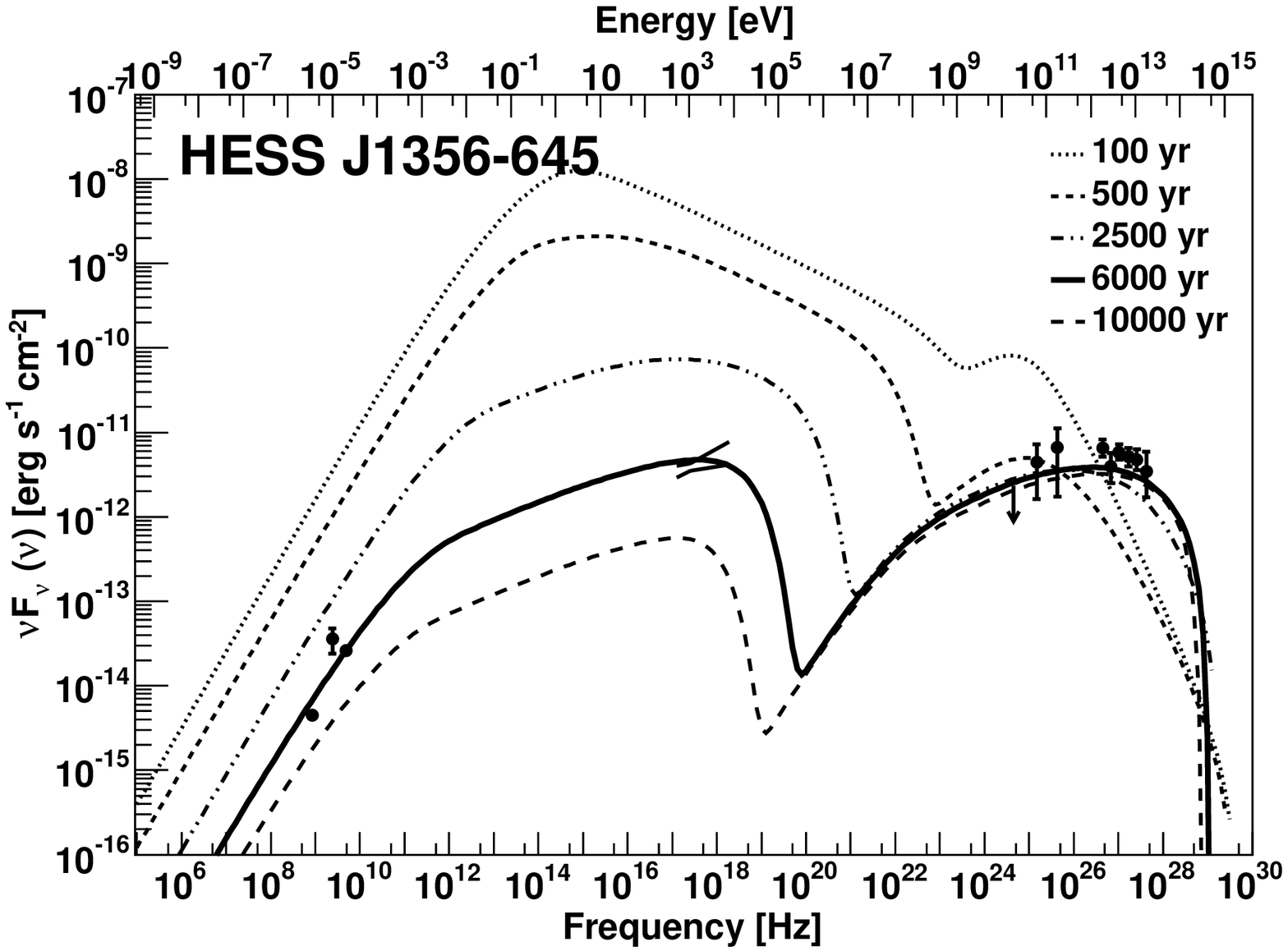}
\includegraphics[width=84mm]{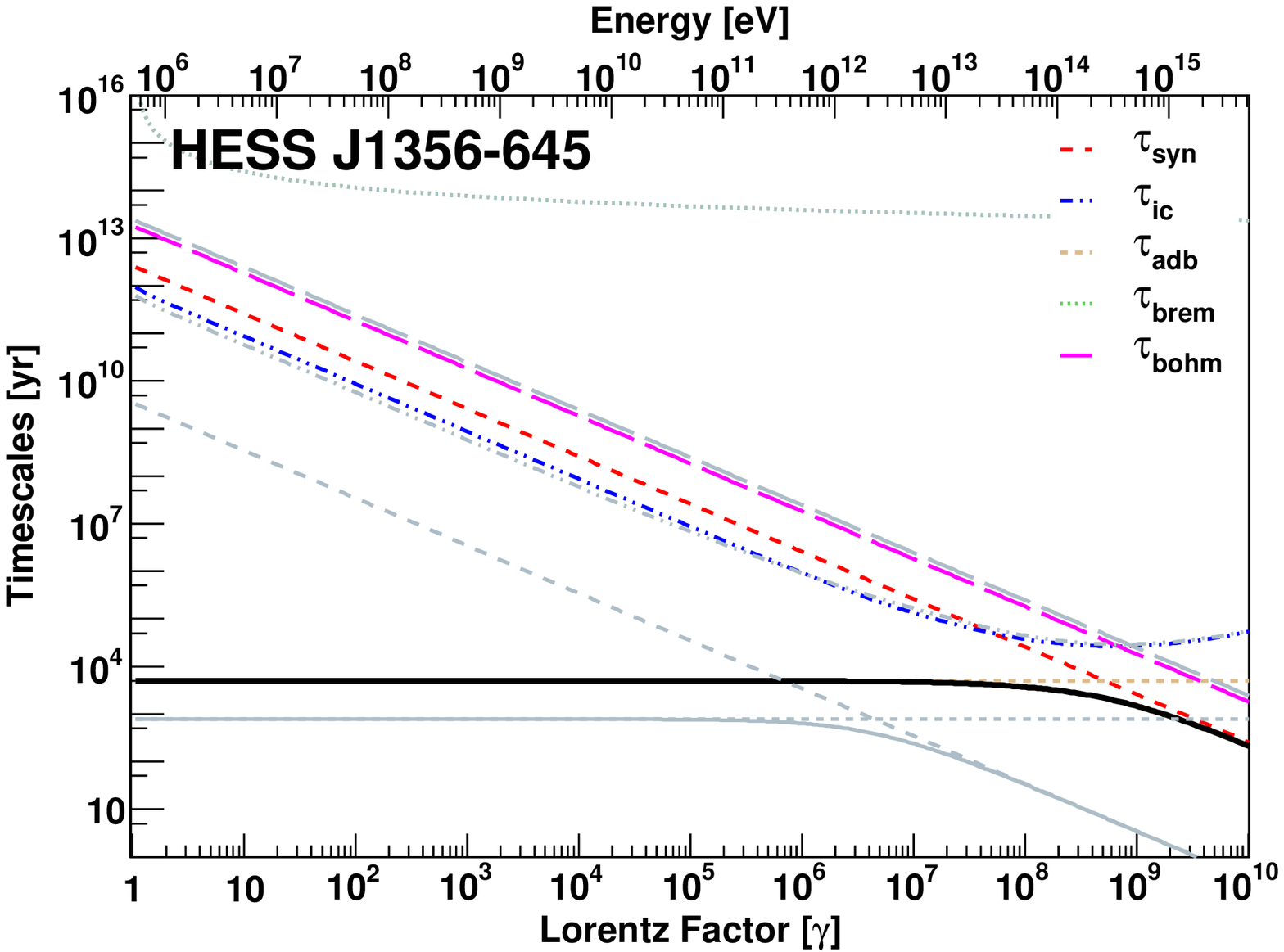}
\includegraphics[width=84mm]{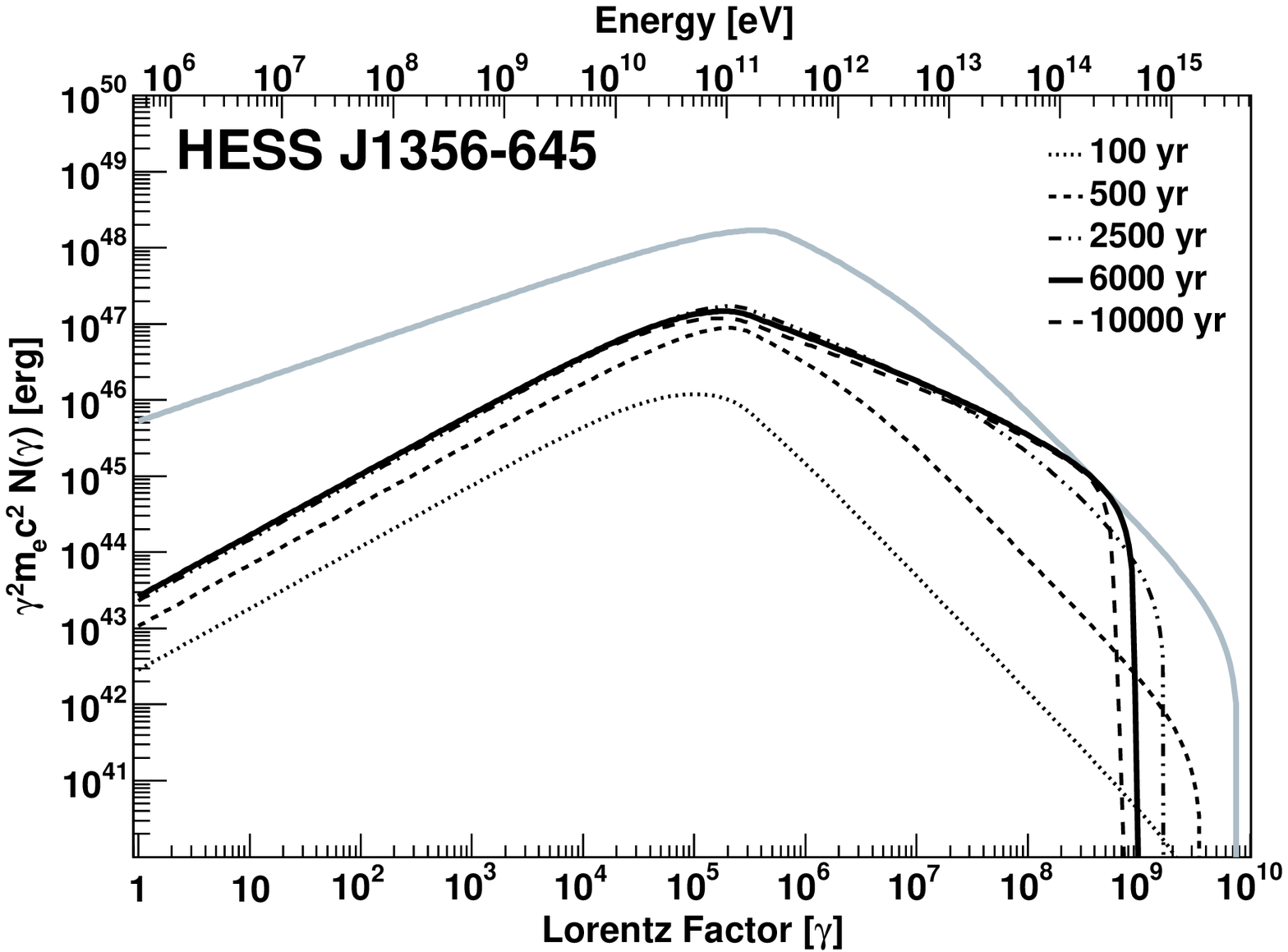}
\caption{Details of the SED model of HESS J1356--645.The panels are as in Fig.~\ref{G54}.
For details regarding the observational data and a discussion of the fit, see the text. }
\label{HESS1356}
\end{figure*}
%%%%%%%%%%%%%%%%%%%%%%%%%%%%%%%%%%%%%%%%%%%%

HESS J1356--645 is localized at $\sim$5 pc from the pulsar PSR J1357--6429, if at the same distance, and has an intrinsic Gaussian width of (0.2 $ \pm$  0.02) deg \cite{HESS2011}. 
%The most plausible scenario for the origin of the VHE emission is, according to studies at different wavelengths, an evolved and offset PWN powered by the mentioned pulsar \cite{HESS2011,Chang2012,Acero2013}.
%
PSR J1357--6429 is a young pulsar with a $\tau_c$=7.3 kyr, a spin-down luminosity of $3.1\times 10^{36}$ 
erg s$^{-1}$, and a period of 166 ms. It was discovered during the Parkes multibeam survey of the 
Galactic Plane \citep{Camilo2004}. Lemoine-Goumard et al. (2011)  detected pulsations using data from 
{\it Fermi}-LAT and {\it XMM-Newton} observations. A possible optical counterpart was also reported 
\citep{Danilenko2012}. Several authors pointed out the similarities of this pulsar with Vela \citep{Esposito2007, HESS2011, Acero2013}.
Particularly, 
they both have a low X-ray efficiency, presence of thermal X-ray photons, and a similar ratio of the compact to diffuse sizes of the nebula. 
The distance to the pulsar was estimated, based on its dispersion measure, to be 2.4 kpc \citep{Camilo2004}.

The first upper limit of the X-rays emission of the PWN of this pulsar was established by Esposito et al. (2007).  
Later, the H.E.S.S. collaboration studied {\it ROSAT} and {\it XMM-Newton} images and reported the X-ray spectra of the nebula \citep{HESS2011}.  Radio and X-ray data, although faint, are coincident in extension with the VHE emission, which provides arguments for the association between the HESS source and the nebula \citep{HESS2011}. The morphology of the PWN was also recently 
studied in detail by Chang et al. (2012), 
who also arrived to the same conclusion about the possible association of the nebula with the very high energy source.
{\it Fermi}-LAT detected a faint counterpart to the nebula after 45 months of observations \citep{Acero2013}. The spatial and spectral coincidences between {\it Fermi}-LAT and HESS emission also suggests that they are coming from the same source.

To perform our fit we then take the radio, X-ray, and TeV data as quoted in the discovery paper by H.E.S.S. (Abramowski et al. 2011): Radio data comes from the Molonglo Galactic Plane Survey at 843 MHz, Parkes 2.4 GHz, and Parkes-MIT-NRAO (PMN) at 4.85 GHz.  The X-ray spectral shape comes from {\it XMM-Newton} observations.  {\it Fermi}-LAT observations were taken from Acero et al. (2013).

%%%%%%%%%%%%%%%%%%%%%%%%%%%%%%%%%%%%%%%%%%%%
%%%%%%%%%%%%%%%%%%%%%%%%%%%%%%%%%%%%%%%%%%%%
%\subsection{Discussion}
%%%%%%%%%%%%%%%%%%%%%%%%%%%%%%%%%%%%%%%%%%%%
%%%%%%%%%%%%%%%%%%%%%%%%%%%%%%%%%%%%%%%%%%%%

To fit the observational data, we have assumed an age of 6000 years, a braking
index of 3, an ejected mass of 10~$M_{\odot}$, and a distance of 2.4 kpc (see Table \ref{param}).
We could fit the data with a broken power-law injection having a hard low-energy spectral index $\alpha_1$=1.2,  and a high-energy slope of    
 $\alpha_2$=2.52. 
 We found no need of adding a constraint on $\gamma_{min}$ in this model. 
 The break in the spectrum happens at a Lorentz factor of  $3 \times 10^5$. We found HESS J1356-645 to be a particle dominated nebulae too, with a magnetic fraction
 of 0.06. 
 The FIR and NIR/OPT photon fields of the model have temperatures of 25 K and 5000 K, and
 energy densities of 0.4 and 0.5 eV cm$^{-3}$, respectively. These values are quite low in comparison with other 
 PWNe we have studied, and near the estimations obtained from GALPROP (see below).
 The average magnetic field we obtain is also very low 
 $\sim$3.1 $\mu$G.  A magnetic field higher than $\sim 4$ $\mu$G would make it impossible to fit the data, 
 even varying other parameters.
 The SED today, its evolution over time, the electron population, and the losses  are plotted in Fig. \ref{HESS1356}.
At high and very high energies, the most important contributions are coming from the IC with the
 CMB and FIR, almost in an equal extent, being the contributions to the IC coming from the NIR/OPT photons, as well as from
 bremsstrahlung, negligible in comparison.  
 For comparison, the HESS Collaboration \citep{HESS2011}
have modeled the source assuming a static one-zone leptonic scenario, with an electron population injected with an
 exponential cutoff power-law of index 2.5 and cutoff energy of 350 TeV.  They also assumed photon fields
 with temperatures of $\sim$35 K and 350 K and optical photon field of temperature of  $\sim$4600 K. 
%They obtained an average magnetic field for the nebula of  $\sim$3.5 $\mu$G. 
We do not find the need of incorporating 
an additional component to the IR distribution at 350 K in order to fit the data.

\begin{figure}[t!]
\centering
\includegraphics[width=84mm]{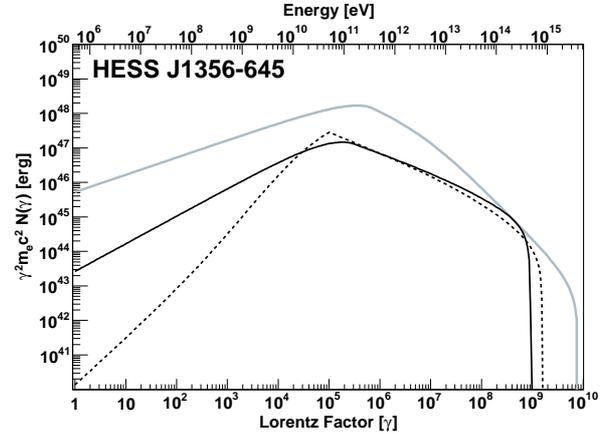}
\caption{Comparison of the electron distributions for the two models considered for HESS J1356--645. The solid (dashed) line corresponds
to Model 1 (2), with the parameters given in Table \ref{param}. Recall that the age in these two models is different. The grey solid line is the Crab nebula
electron distribution today.}
\label{1356-elec}
\end{figure}

We have found that it is also possible to have a good fit to the data with a single power law in the spectrum of injected electrons (with slope 2.6), if electrons are energetic enough. 
To allow for this possibility the braking index is reduced to 2, so that the initial spin-down age is increased by about a factor of $\sim$5 (up to 6622 years). With such an spin-down age, the pulsar is injecting more electrons along most of its lifetime. An slightly larger age (assumed to be 8000 years) and
 magnetic fraction (0.08) would allow for an equally good SED fit.  
 Finally, the  $\gamma_{min}$  value is here constrained to be larger than $10^5$. In practice,  electrons injected are assumed to 
 be above the break energy of the prior model, and losses populate lower levels in electron energy. 
 These parameters are summarized in Table \ref{param}, quoted as Model 2. Fig. \ref{1356-elec}
 compares the two resulting electron distribution at the corresponding current age.  By compensating with a longer injection age and more energetic electrons, 
 the electron distribution can be made similar in both models, leading to equally acceptable SEDs. This degeneracy still remains, although preference for model 1 can be argued:
  the alternative model 2 referred above requires more contrived assumptions to work and would make the nebula an outlier in comparison with others.  
%

%%%%%%%%%%%%%%%%%%%%%%%%%%%%%%%%%%%%%%%%%%%%
%%%%%%%%%%%%%%%%%%%%%%%%%%%%%%%%%%%%%%%%%%%%

%%%%%%%%%%%%%%%%%%%%%%%%%%%%%%%%%%%%%%%%%%%%
%%%%%%%%%%%%%%%%%%%%%%%%%%%%%%%%%%%%%%%%%%%%
\subsection{VER J0006+727 (CTA~1)}
%%%%%%%%%%%%%%%%%%%%%%%%%%%%%%%%%%%%%%%%%%%%
%%%%%%%%%%%%%%%%%%%%%%%%%%%%%%%%%%%%%%%%%%%%

%%%%%%%%%%%%%%%%%%%%%%%%%%%%%%%%%%%%%%%%%%%%
\begin{figure*}[t!]
\centering\includegraphics[width=84mm]{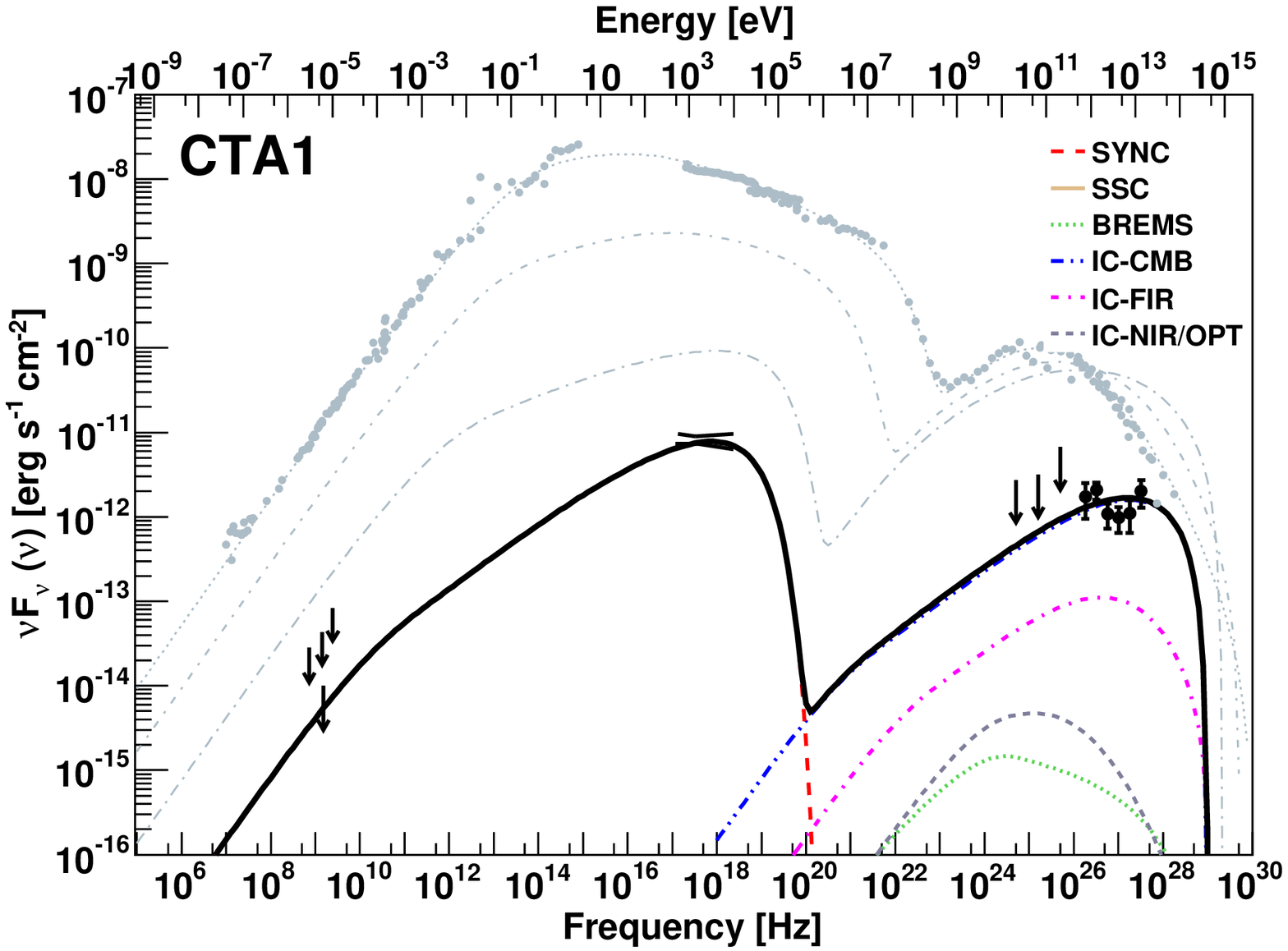}
\includegraphics[width=84mm]{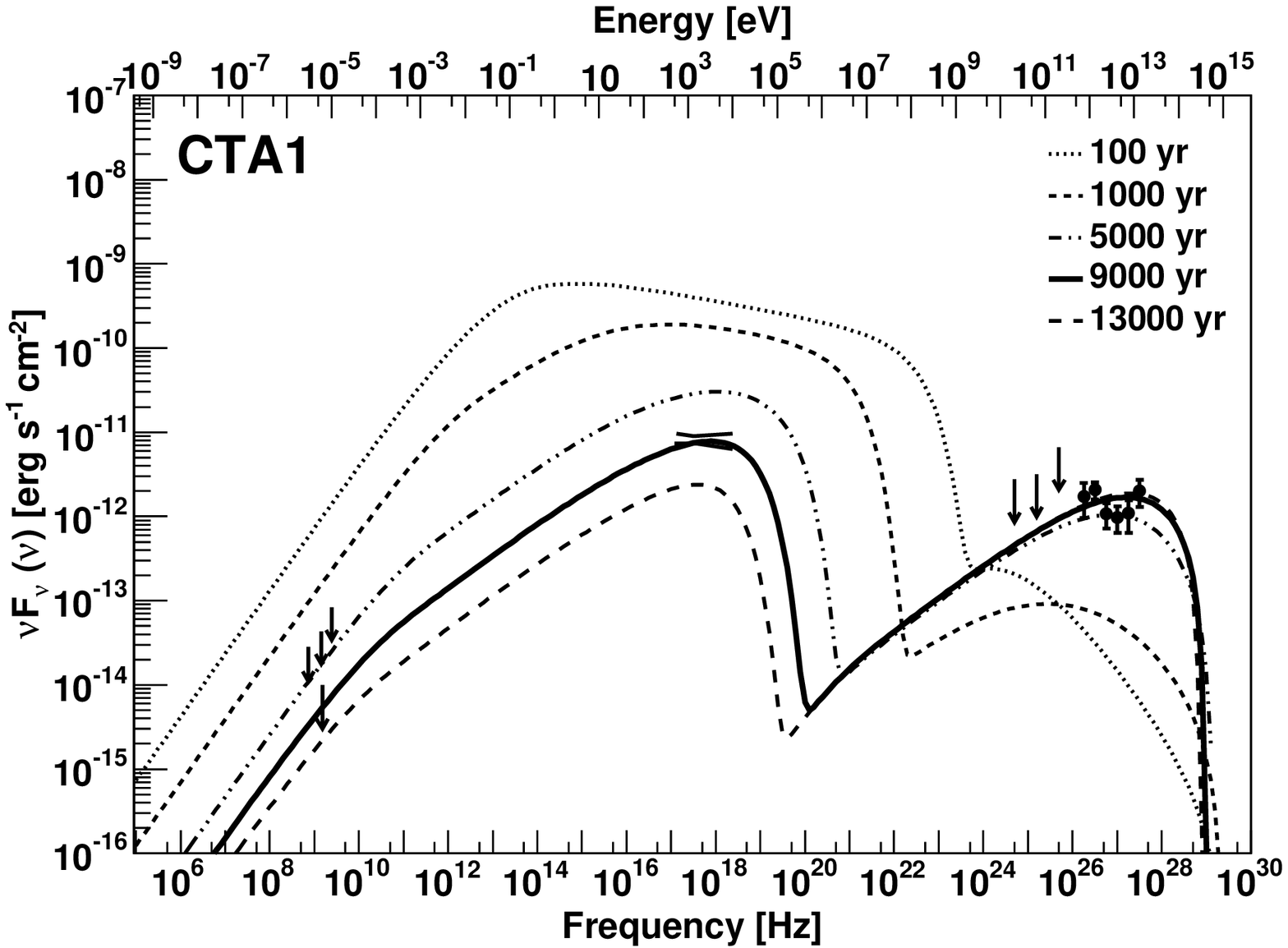}
\includegraphics[width=84mm]{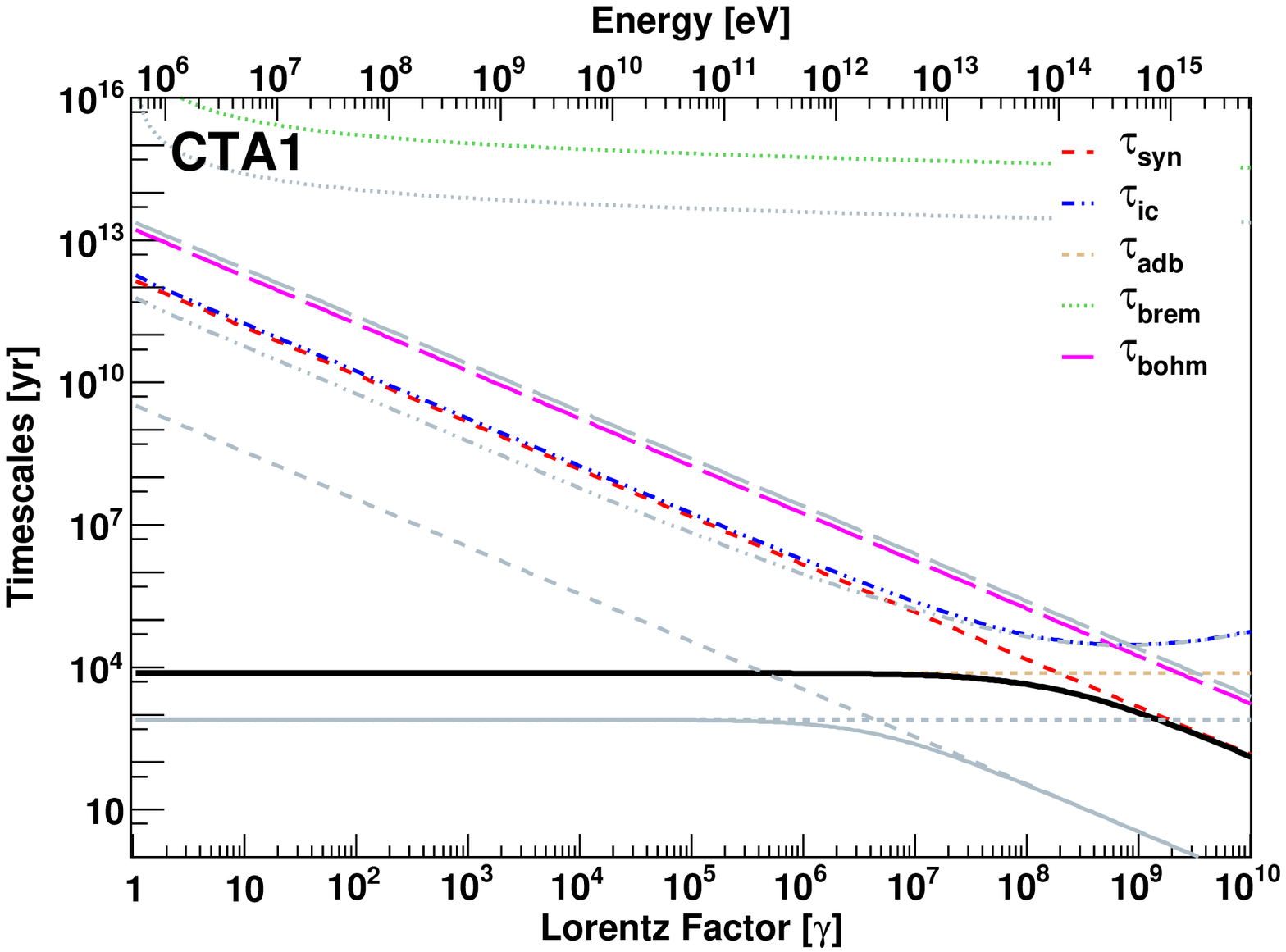}
\includegraphics[width=84mm]{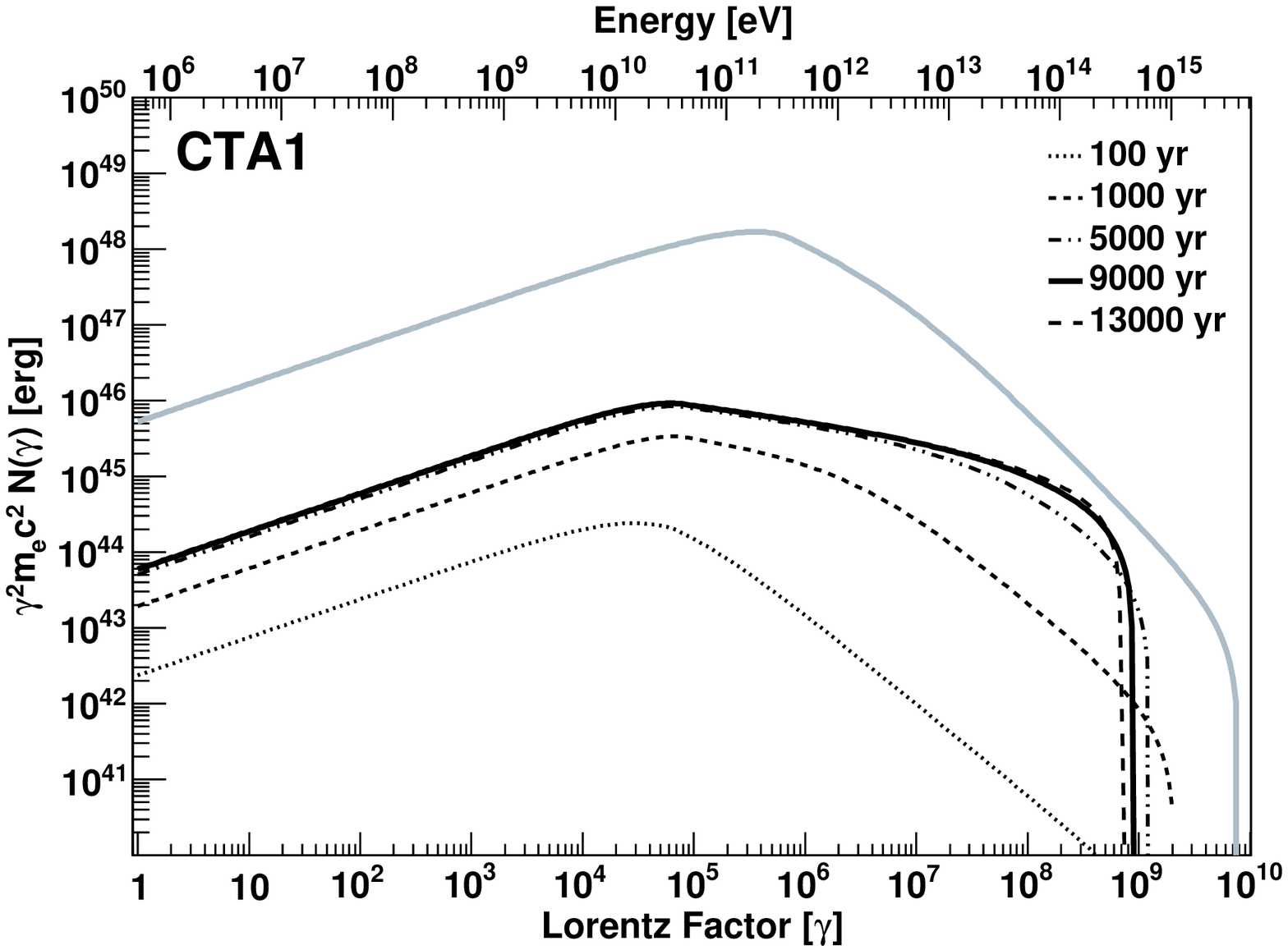}
\caption{Details of the SED model of CTA~1.The panels are as in Fig.~\ref{G54}.
For details regarding the observational data and a discussion of the fit, see the text. }
\label{CTA1}
\end{figure*}
%%%%%%%%%%%%%%%%%%%%%%%%%%%%%%%%%%%%%%%%%%%%

The extended radio source CTA~1 (G119.5+10.2) 
was first proposed as a SNR by Harris \& Roberts (1960).
The SNR was first detected in X-rays by ROSAT by Seward et al. (1995). The authors also reported the presence of a faint compact source, RXJ 0007.0+7302, located within the central region. Slane et al. (1997) confirmed the non-thermal nature of the central emission using ASCA data. These early detections were indicative of the presence of a synchrotron nebula powered by an active neutron star, for which the most plausible candidate was the source RX J0007.0+7302.
Further studies performed with the XMM-Newton and ASCA satellites towards 
RX J0007.0+7302 have resolved the X-ray emission into a point-like 
source and a diffuse nebula of 18 arcmin in size \citep{Slane2004}. 
Using the {\it Chandra} observatory \cite{Halpern2004} have found a point source, RX J0007.0+7302, embedded in a compact 
nebula of  $3 ^{\prime\prime}$ in 
radius, and a jet like extension.
At high energies, Mattox et al. (1996) proposed that the EGRET source 
3EG J0010+7309 (which lies in spatial coincidence with RX J0007.0+7302), 
was a potential candidate for a radio-quiet gamma-ray pulsar. 
Brazier et al. (1998) also pointed out that this source was pulsar-like, but a search for gamma-ray pulsation using 
EGRET data failed \citep{Ziegler2008}.
During the commissioning phase 
 of the {\it Fermi} satellite, a radio-quiet pulsar in CTA~1 was finally discovered 
\citep{Abdo2008}.  
X-rays pulsations from this source were finally detected by XMM-Newton \citep{Lin2010, Caraveo2010}. 
The pulsar in CTA~1 has a period of $\sim$316 ms and a spin-down power of $\sim$4.5$\times10^{35}$ erg s$^{-1}$.
No radio counterpart to RX J0007.0+7302 was identified, most likely due to beaming. No optical counterpart
is known either (Mignani et al. 2013).

%Using 16 months of Fermi/LAT data \cite{Ackermann2011} 
%noted a candidate for an off-pulse emission from PSR J0007+7303, but it was 
%below the detection threshold. With more accumulation of data 
Abdo et al. (2011) 
reported the detection of an extended source in the off-pulse emission at $\sim6 \sigma$ level using 2 years 
of Fermi/LAT data. Acero et al. (2013) improved on this result (which we use for modeling).
The VERITAS Collaboration also detected an extended source of 0.3 $\times$ 0.24 deg at 
5 min from the Fermi gamma ray pulsar PSR J0007+7303 \citep{Aliu2013}.

CTA~1 characteristics in radio and X-rays suggest an age between 5000 and 15000 
yrs \citep{Pineault1993,Slane1997,Slane2004} for the SNR, which is in agreement with the 
spin-down age of the pulsar ($\sim$14000 yr).
Pineault et al. (1993) derived a kinematic distance of 1.4 $\pm$ 0.3 kpc based on associating
an HI shell found northwestern part of the SNR.
In order to perform our fit we take the radio upper limits
 from Aliu et al. (2013) --where the authors have used a 1.4
 GHz image to estimate the flux upper limit within 20 arcmin radius around the pulsar and extrapolated
 this upper limit to lower and higher frequencies assuming respectively a radio spectral index of 
0.3 and 0. The other UL we use, at 1.5 GHz, was obtained from  
a new VLA image \citep{Giacani2013} considering a size for the nebula of 20 arcmin in radius. 

%\section{Discussion}

We performed our fit considering a distance to the system of 1.4 kpc, an ejected mass between 6 and 10 $M_{\odot}$,
 a braking index equal to 3, and a density of the media of 0.07 similar to the one proposed by the Veritas Collaboration 
\citep{Aliu2013}. We explored the possibility of different ages for the nebula, between 9000 and 12000 yrs.
The best fit of the data was obtained with an age of 9000 yrs and 10 $M_{\odot}$ of ejected mass.
The injected spectrum was assumed to follow a power-law with slopes  
$\alpha_1$ = 1.5 and $\alpha_2$=2.2.
The magnetic field obtained for the model presented in Table 2 was of 4.1 $\mu$G, with an extension of 
the nebula of 8 pc in radius. 
For this nebula the main contribution to the flux at high and very high energies comes from the IC 
with the CMB, being the IC with the FIR and NIR/OPT components almost negligible. 
Compared to the other PWNe analyzed in this work, the magnetic fraction of this nebula is much higher,
 $\eta$=0.4. 
 %To test this result, we have 
 %tried fitting the nebula with lower values of $\eta$, for example 0.2 (nevertheless high in comparison to other PWNe), like the value obtained by 
 %the Veritas Collaboration \citep{Aliu2013} using  the model by \cite{Gelfand2009}, but
 %we could not fit the observational data at TeV energies.
A low $\eta$ value, like the one obtained with our model for Crab nebula ($\eta$=0.03), 
over-estimates the flux values at TeV energies compared to the observations of Veritas.

Previous to Veritas observations, Zhang et al. (2009) over-predicted the value of the flux at high energies. 
%They developed a model of the unpulsed emission with a one-zone model with six free parameters to fit to the
%radio and X-ray data, with emission coming from synchrotron radiation and inverse
%Compton scattering of CMB and diffuse IR photons (see Figures 2-4 of their paper).  
To model the radio upper limits these authors assumed that all the emission obtained from the 
images  of Pineault et al. (1997) was coming from the PWN, which caused also an over-estimation of the radio flux.
In the model presented in Fig. \ref{CTA1},  Fermi upper limits are higher (by about a factor of 8) than the predictions of our model at those energies. 

%A possible explanation is that the subtraction performed by the Fermi  Collaboration of the pulsar emission could not be totally accurate. Another, but less likely possibility in order to fit the GeV data is that the IC-NIR/OPT component plays an important role.  If this is the case, we found that the energy density of this component has to be larger than 15 eV cm$^{-3}$). }}

%%%%%%%%%%%%%%%%%%%%%%%%%%%%%%%%%%%%%%%%%%%%
%%%%%%%%%%%%%%%%%%%%%%%%%%%%%%%%%%%%%%%%%%%%
\subsection{HESS J1813--178 (G12.8--0.0)}
%%%%%%%%%%%%%%%%%%%%%%%%%%%%%%%%%%%%%%%%%%%%
%%%%%%%%%%%%%%%%%%%%%%%%%%%%%%%%%%%%%%%%%%%%

HESS J1813--178 is a TeV source discovered at high energies in the inner galaxy survey done by H.E.S.S. (Aharonian et al. 2006). It was also observed by MAGIC (Albert et al. 2006), obtaining its differential $\gamma$-ray spectrum as
(3.3$\pm$0.5) $\times 10^{-12}$(E/TeV)$^{2.1\pm0.2}$ cm$^{-2}$ s$^{-1}$ TeV$^{-1}$.
The angular extension of the source is 2.2'. With a distance of 4.8 kpc \citep{halp2012}, this gives 3.1 pc of diameter.
The associated central source is the pulsar PSR J1813--1749, which has a period of 44.6 ms \citep{gott2009} and a period derivative
of 1.26 $\times 10^{-13}$ s s$^{-1}$ \citep{halp2012}. The spin-down power nowadays is 5.59 $\times 10^{37}$ erg s$^{-1}$, and its
characteristic age is 5600 yr.

Brogan et al. (2005) discovered a radio shell (SNR G12.8-0.0) 
%using {\it VLA} (Very Larga Array) data 
coincident with the position of
HESS J1813-178, having an angular diameter of $\sim$2.5'. The flux density spectrum was fitted with a power law with an index of 0.48 between 
3 cm to 90 cm wavelength. In X-rays, {\it ASCA} detected the source AX J1813-178 also coincident with the position of the SNR and the
H.E.S.S. source, but the pointing uncertainty was too large to distinguish if the origin of the emission is the center of the 
remnant or from the shell. Helfand et al. (2007) resolved the X-ray central source and the PWN using observations from {\it Chandra}.
The flux of the PWN was fitted with a power law with an index of 1.3 and an absorbed flux of 5.6 $\times 10^{-12}$ erg cm$^{-2}$ s$^{-1}$
between 2 and 10 keV. A distance of 4.5 kpc was assumed and they inferred a luminosity for the PWN of 1.4 $\times 10^{34}$ erg s$^{-1}$.
The pulsations of the central source in X-rays were discovered two years later  using data from {\it XMM-Newton}
 \citep{gott2009}.  Concerning the age of the system, if the SNR shell
were expanding freely, the dynamic age of the system would be about 285 yr whereas in a Sedov expansion, the age increases 
until 2520 yr \citep{brogan2005}. We adopt an intermediate case of  $\sim$1500 yr here, similarly to other analysis.
{\it XMM-Newton} also observed this source and could resolve the PWN with an spectral index of 1.8 and a flux between 2 and
10 keV of 7 $\times 10^{-12}$ erg cm$^{-2}$ s$^{-1}$ \citep{funk2007}, which is similar to the one obtained by Helfand et al. (2007), but 
softer. Ubertini et al. (2005) observed a soft gamma source with INTEGRAL with an spectral index between 20 and 100 keV of 1.8, as in the
XMM-Newton data. They inferred a luminosity of 5.7 $\times 10^{34}$ erg s$^{-1}$ assuming a distance of 4 kpc.

The origin of the emission in the TeV energy range is
not clear and we shall use our model to assess the possibility that a PWN produces it.
Other authors have considered this problem before. For instance, Funk et al. (2007) considered two scenarios, one in which the VHE and 
the X-ray emission are produced leptonically, by electrons in a PWN; and another, in which the VHE and the radio emission 
are generated in the SNR shell. They considered two alternatives for the leptonic scenario producing both the X-ray and the VHE photons:
a normal FIR and NIR background with a single power law (with slope 2.4) electron spectrum (model 1); and 
a significant excess of NIR photons (a factor of 1000 beyond the expected from GALPROP) subject to 
an injection spectrum described by a hard, single power law (model 2). 
In both of these alternatives one is forced to require that the maximal energy of the electrons is beyond 1.5 PeV, that the minimal energy
is also high ($\gamma_{min}$ of the order of $5\times 10^4$)
and that the magnetic fields are low (a few $\mu$G). The high value needed for $\gamma_{min}$ would convert this PWN in an outlier
with respect to the rest of the population. 
In any case, 
these models are both unsatisfying. Model 1 is barely a good fit 
to the TeV data, significantly overproducing the measurements  at the highest energies. Model 2 
has an extremely high photon background, even considering the contribution of the nearby star forming region W 33 (Funk et al. 2007). We have built
similar models, and whereas the results cannot be directly compared due to the different treatments, we essentially find the same trends in the case $\gamma_{max}$ is indeed allowed to reach high values.
PWN are capable of accelerating electrons to PeV energies (see Table \ref{param}). However, in the framework of our 
model (and in a real physical situation), the maximum Lorentz factor that electrons can achieve is not a free parameter. Here it is set
by requesting that the
Larmor radius be smaller than the termination shock (Eq. \ref{g1}). 
Even assuming that the 
fractional size of the radius of the shock is 1, we would attain lower values than 1 PeV, leading --leaving all other parameters the same--
to a bad fitting in both alternatives presented by Funk et al. (2007). For our analogous to their model 1, the redistribution of the power to lower electron energies 
would not allow for a good fit to the X-ray peak and the radio emission will increase, being close or beyond the upper limits. For our analogous to their model 2, we
would  significantly overproduce the spectral points at all energies. We need a much lower NIR density of about 55 eV cm$^{-3}$, nevertheless very high,
to match the spectrum better. However,  particularly at high TeV energies, it would become impossible to comply with all observational constraints
in the case $\gamma_{max}$ is allow to reach a high value and the slope of the injection power-law is 2, so as to provide a good fit to the X-ray part: the electrons interacting with the CMB would already overproduce the highest energy data.  \cite{fang2012} also studied models for HESS J1813--178, and although the injection is different from a simple power-law, the general trend is maintained: they cannot attain a good fit to the VHE and X-ray part of the SED with a PWN model either.

Taking into account all of the former, it seems more natural to suppose that HESS J1813--178 VHE emission is generated at the shock of a SNR, or in the interaction of accelerated protons with the environment (as in Gabici et al. 2009, or Torres et al. 2010). We shall not consider this source further in our sample.

%%%%%%%%%%%%%%%%%%%%%%%%%%%%%%%%%%%%%%%%%%%%
%%%%%%%%%%%%%%%%%%%%%%%%%%%%%%%%%%%%%%%%%%%%
\section{Discussion}
%%%%%%%%%%%%%%%%%%%%%%%%%%%%%%%%%%%%%%%%%%%%
%%%%%%%%%%%%%%%%%%%%%%%%%%%%%%%%%%%%%%%%%%%%

%%%%%%%%%%%%%%%%%%%%%%%%%%%%%%%%%%%%%%%%%%%%
%%%%%%%%%%%%%%%%%%%%%%%%%%%%%%%%%%%%%%%%%%%%
\subsection{A comment on the model limitations: the Crab nebula }
%%%%%%%%%%%%%%%%%%%%%%%%%%%%%%%%%%%%%%%%%%%%
%%%%%%%%%%%%%%%%%%%%%%%%%%%%%%%%%%%%%%%%%%%%

We have already noted that the model used here (as do essentially all of the other models quoted in the introduction)
contains no morphology nor energy-dependent 
size information. That is, 
the size of the synchrotron ball is the size of the nebula itself, at all frequencies. The model  
focuses on reproducing radiative properties of PWNe 
assuming a 1D system, a one-zone emission region, and a uniform magnetic field and magnetization therein. The relative simplicity of 
these assumptions contrasts with the goodness of the fits that one is able to obtain for instance for the Crab nebula, for which data points are numerous along all bands.
It is then important to remark what are we missing in this kind of approach: for the 
Crab Nebula we know that the size of the synchrotron emitting region 
increases towards the optical frequencies, being always smaller than what one gets for the size of nebula from the use of a dynamical
free expansion solution. For instance, Hillas et al. (1998), see also Meyer et al. (2010), use a radius of approximately 0.4 pc up to 0.02 eV, and slightly smaller for larger energies. 
For the unique case of the Crab nebula,
where the self-synchrotron contribution dominates, this is especially relevant. If we were to assume to Hillas et al. (1998) parameterization of the decreasing sizes of the emitting regions, and still maintain 
the same magnetic field all across, we would be unable to fit the data well. This is understandable, given that  assuming 
different sizes of the emitting ball is likely inconsistent with field uniformity. 
It might also be  inconsistent to actually use the same sizes as a function of frequency 
along the whole time evolution of the PWN, although we would lack information to model it otherwise.

%%%%%%%%%%%%%%%%%%%%%%%%%%%%%%%%%%%%%%%%%%%%
%%%%%%%%%%%%%%%%%%%%%%%%%%%%%%%%%%%%%%%%%%%%
\subsection{SED component dominance}
%%%%%%%%%%%%%%%%%%%%%%%%%%%%%%%%%%%%%%%%%%%%
%%%%%%%%%%%%%%%%%%%%%%%%%%%%%%%%%%%%%%%%%%%%

Table \ref{models}  shows which components dominate
the SED at TeV energies (the first and second contributors are given in the first two columns). It also provides the ratio (integrating our models in the range 
1--10 TeV) between the two largest contributions to the SED at very-high energies (third column).
The radio (at 1.4 GHz), X-ray (1--10 keV), and gamma-ray luminosities (1--10 TeV), and their corresponding efficiencies (when compared with the pulsar spin-down), $f_r$, $f_X$, and $f_\gamma$, are also shown in Table \ref{models}. To obtain the luminosities we use the distances to each nebulae according to Table \ref{param}, and obtained them from an integration on our fits. This
allows to uniformize the energy range,
introducing no change in the conclusions given that all fits are reasonably good descriptions of the observational data when such exist.

We first see that for all the sources studied, only the Crab nebula is SSC dominated. Given the age, power, and photon backgrounds
of the PWNe studied, this is an expected result (Torres et al. 2013b).
It is interesting to see that in the setting of a leptonic model, all the remaining PWNe except for HESS J1356-645 and CTA~1 are IC-FIR dominated. 
The dominance of the FIR contribution to IC is always large in these cases, and the ratio with the second contributor to the SED at 1 to 10 TeV energies 
spans from 1.3 to $\sim 10$, with the outlying PWN G292--0.5, for which the ratio is 31.
The efficiencies of emission are consistently grouped as follows: $\sim 10^{-6 \div 7}$ in radio, $\sim 10^{-2 \div 3}$ in X-rays, 
and $\sim 10^{-3 \div 4}$ in gamma-rays, except for 
G292.2--0.5, which shows a very low X-ray efficiency in comparison with the others.

%%%%%%%%%%%%%%%%%%%%%%%%%%%%%%%%%%%%%%%%%%%%
%%%%%%%%%%%%%%%%%%%%%%%%%%%%%%%%%%%%%%%%%%%%
\subsection{Slopes of injection \& electron population}
%%%%%%%%%%%%%%%%%%%%%%%%%%%%%%%%%%%%%%%%%%%%
%%%%%%%%%%%%%%%%%%%%%%%%%%%%%%%%%%%%%%%%%%%%

We have considered a broken or a single power law for the injection distribution of electrons. Other injections can be tried.
However, if we use, e.g.,
the injection model based on the particle in cell (PIC) simulations done by Spitkovsky (2008), we would have several additional --and observationally unconstrained-- parameters. This kind of injection is not devoid of significant extrapolations when considered in a PWN setting (e.g., the maximal PIC simulated Lorentz factor is far from the maximal electron energies considered in the PWNe). Thus a priori it would seem that the power-law distributions are a more reasonable choice for the time being, due to their simplicity. Their ability to produce good fits in all cases give a posteriori support. 

We have found that the energy distribution of the electron population is well described almost in all cases by a broken power law. The  high energy slope
is found to be in the range 2.2 -- 2.8 except for one outlier, G292.2--0.5, for which $\alpha_2=4.1$. The low energy part is instead much harder, in the range 1.0 -- 1.6. These results are consistent with previous studies of part of the sample we have treated, see, for instance, Bucciantini et al. (2011). The breaks, on the other hand, appear at a Lorentz factor in the range 10$^5$ -- $10^{6.7}$, and for most of the models are actually concentrated in a narrower range around $5 \times 10^5$. 
These very small ranges of values of the slopes and break energies for modeling sources that appear so different at first sight suggests that the processes at the pulsar wind termination shock are common. The only models that are exceptional to these trends are G292.2--0.5, and the Model 2 of HESS J1356--645. 
For the PWN likely associated with HESS J1356--645, a broken power law with parameters in agreement with the previous trends produces a good fit to the data; and the single power law was explored only as an alternative to give account of ignorance or degeneracies in parameters such as age and pulsar braking index.  G292.2--0.5 is also 
outlier to other phenomenology discussed in this Section. The spectral break of the injected electron needed in G292.2--0.5 is the  highest of all PWNe studied. 
Despite the obvious caveats in trying to model a spatially complex region with a one zone radiative model,
we note that we are also uncomfortable with the large ejected mass that would be needed in our model to have a good fit to G292.2--0.5 radiative data. It may well be the case that this PWN is just different in their acceleration properties (the pulsar has one of the largest magnetic field in our sample, in excess of 10$^{13}$, the other one being Kes 75), or that the model fails due to a large influence of more advanced dynamical states. In fact, the PWN is offset with respect to the position of the pulsar,  what could be originated 
if the nebula has been displaced after being crushed by an asymmetric reverse shock caused by the presence of the dark cloud in the vicinity.
%in the northeast
Finally, it may also be that 
the steepness of the G292.2--0.5 spectrum points towards an alternative origin, related to the SNR, a possibility discussed, but not favored, by Kumar et al. (2012).
All in all, due to the more uncertain origin of the radiation at the highest energies, the case of G292.2--0.5  requires special attention when looking at the overall properties of the population. 
We also note that G292.2--0.5 and the Model 2 of HESS J1356--645 are the only two cases in which we have braking indices of 2 or lower.

%%%%%%%%%%%%%%%%%%%%%%%%%%%%%%%%%%%%%%%%%%%%
%%%%%%%%%%%%%%%%%%%%%%%%%%%%%%%%%%%%%%%%%%%%
\subsection{Pair multiplicity and bulk Lorentz Factor}
%%%%%%%%%%%%%%%%%%%%%%%%%%%%%%%%%%%%%%%%%%%%
%%%%%%%%%%%%%%%%%%%%%%% %%%%%%%%%%%%%%%%%%%%%

We now consider the PWN injection rates resulting from our models.
We will compare the injection rate with the electrodynamics minimum suggested by Goldreich \& Julian~(1969),
\begin{eqnarray}
\dot N &=& \left( \frac{cI \Omega \dot \Omega}{e^2} \right)^{1/2} \nonumber \\
&=&7.6 \times 10^{33} \left( \frac {I_{45} \dot P} {P_{33}^3 \, 4\times 10^{-13} } \right)^{1/2} {\rm s}^{-1} 
\end{eqnarray}
where $P$ and $\dot P$ of Crab have been used for normalizing ($P_{33}=P/33$ ms), $\Omega = 2 \pi / P$ is the angular velocity,
and $I_{45} = I / 10^{45}$ g cm$^{2}$.
We can directly compute the injection rate by integrating Eq. (\ref{injection});
\be
Q=\int Q(\gamma,t) d\gamma , 
\ee
from where the multiplicity follows
\be
\kappa = \frac{Q}{\dot N}.
\label{kappa2}
\ee
The values of $\kappa$ for all the PWNe in our sample are shown in Table \ref{kappa}. 
Multiplicities are large in all cases, although they should be taken as upper limits.
We have found that at the level of the SED, the lower limit value of $\gamma_{min}$ (critical in defining the value of $\kappa$)
remains unconstrained in most cases.
For instance, for the Crab nebula, $\gamma_{min}$ values larger than $10^4$ would make very difficult to realize a proper description of the synchrotron part of the SED, but instead, the SED is essentially unchanged for lower values.
The same happens in other cases, for instance, with a $\gamma_{min} = 1\times 10^5$ it is already difficult to fit well the radio spectrum of G0.9+0.1 and G21.5--0.9. The same happens with G54.1+0.3 for which  $\gamma_{min}$ values up to 1000 would require no change in any of the parameters, and up to $5\times 10^4$, similarly good fits can be obtained  with slight variations of 
the injection slopes.
The only case in which we need a large value of $\gamma_{min}$ is in fact in the Model 2 of HESS J1356--645, the particularities of which were discussed above.

If the wind is characterized by a single value of the Lorentz factor $\gamma_w$, we may write
the average energy per particle in the spectrum as
\begin{eqnarray}
 <E>   =  \frac{(1-\eta) L(t) }{ \int Q(\gamma, t) d\gamma } 
 %\equiv
%\frac{\int_{\gamma_{min}}^{\gamma_{max}} \gamma m c^2 Q(\gamma,t) \mathrm{d}\gamma}
%{\int_{\gamma_{min}}^{\gamma_{max}} Q(\gamma,t) \mathrm{d}\gamma,} 
%\nonumber \\
\equiv 
\gamma_w m_e c^2.
\end{eqnarray}
The values of $\gamma_w$ are given in Table \ref{kappa}. To compute these values we have used the 
$\gamma_{min}$,  $\gamma_{max}$,  
and $\gamma_{b}$ values, as well as the slopes $\alpha_{1,2}$
when broken power laws are a good representation of the
electron spectra, for each of the nebulae.
We see that in all cases, $\gamma_b$ is larger than $\gamma_w$ by up to several orders of magnitude.
This can be understood from the mean energy definition above,
which can be analytically computed. 
This formula is time-independent and $\gamma_w$ is fully characterized by 5 parameters: $\gamma_{min}$, $\gamma_{max}$,
$\gamma_b$, $\alpha_1$ and $\alpha_2$. To get a better idea on the dependence of $\gamma_w$ on each parameter, we can simplify the
expression taking into account that normally $1<\alpha_1<2$, $\alpha_2>2$ and $\gamma_{min}<\gamma_b<\gamma_{max}$. With this
assumptions, we can simplify it to yield,
\begin{equation}
\gamma_w \simeq \left[\frac{\frac{1}{2-\alpha_2}-\frac{1}{2-\alpha_1}}{\frac{1}{1-\alpha_1}\left(\frac{\gamma_b}{\gamma_{min}} \right)^{\alpha_1-1}+\frac{1}{1-\alpha_2}} \right]\gamma_b,
\end{equation}
with the order of magnitude being $\gamma_w \sim \gamma_b (\gamma_b/\gamma_{min} )^{(1-\alpha_1)}$.
Physically, 
the population of low energy electrons
is more numerous, and it is responsible for the radio to IR emission of the nebulae.

%%%%%%%%%%%%%%%%%%%%%%%%%%%%%%%%%%%%%%%%%%%%
%%%%%%%%%%%%%%%%%%%%%%%%%%%%%%%%%%%%%%%%%%%%
\subsection{ISRF values compared with a Galactic model}
%%%%%%%%%%%%%%%%%%%%%%%%%%%%%%%%%%%%%%%%%%%%
%%%%%%%%%%%%%%%%%%%%%%%%%%%%%%%%%%%%%%%%%%%%

\begin{figure*}[t!]
\centering
\includegraphics[width=85mm]{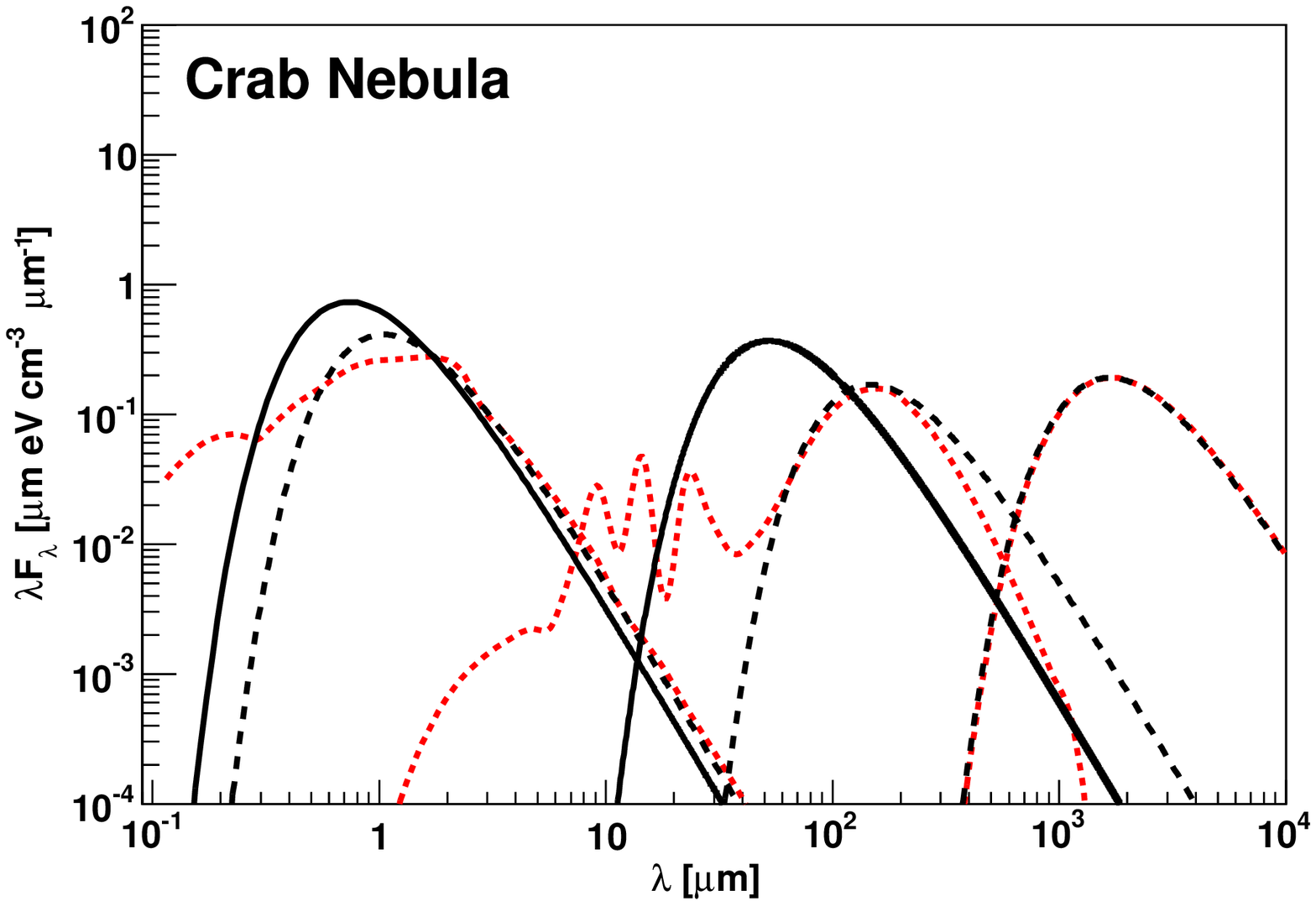}
\includegraphics[width=85mm]{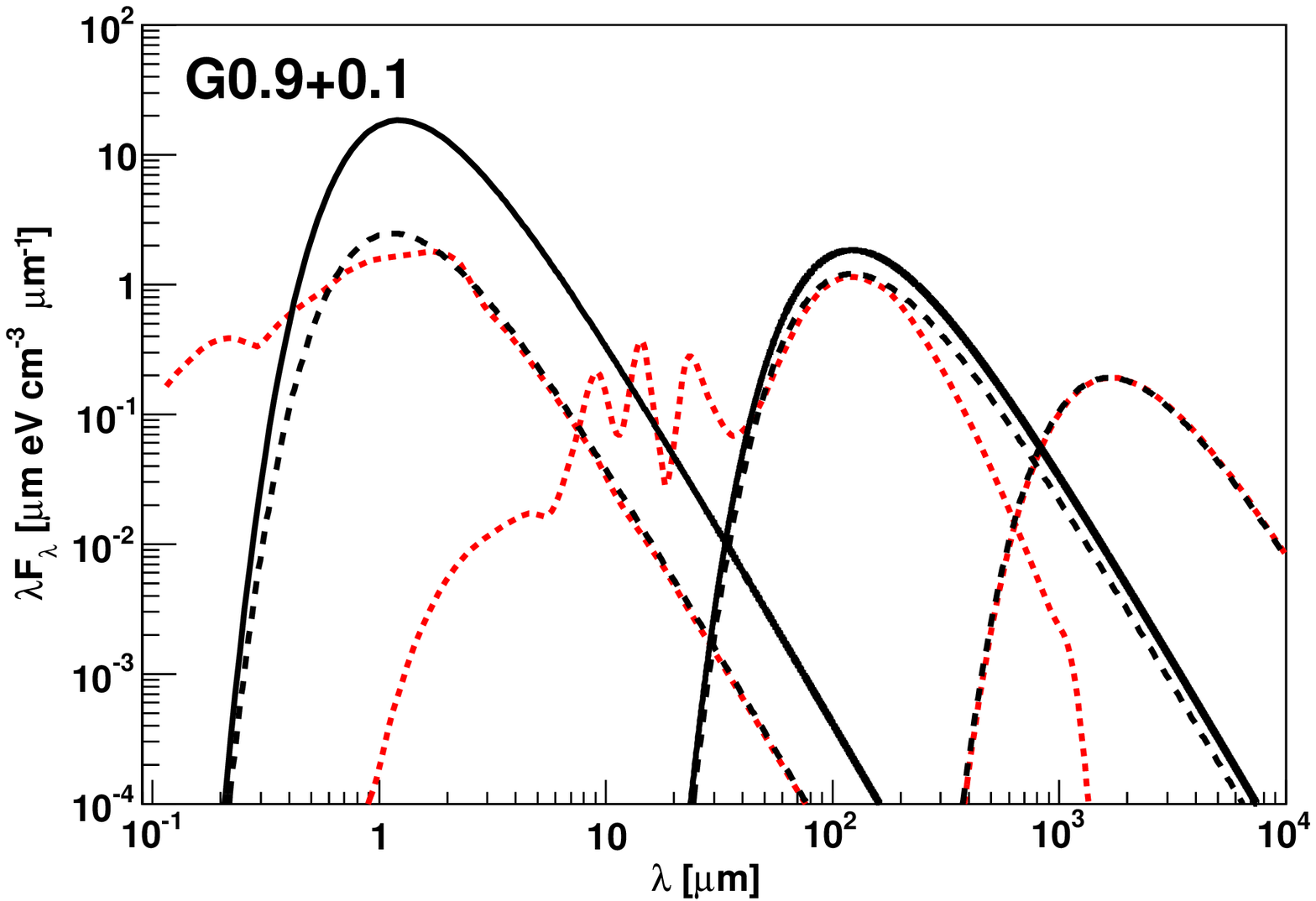}
\includegraphics[width=85mm]{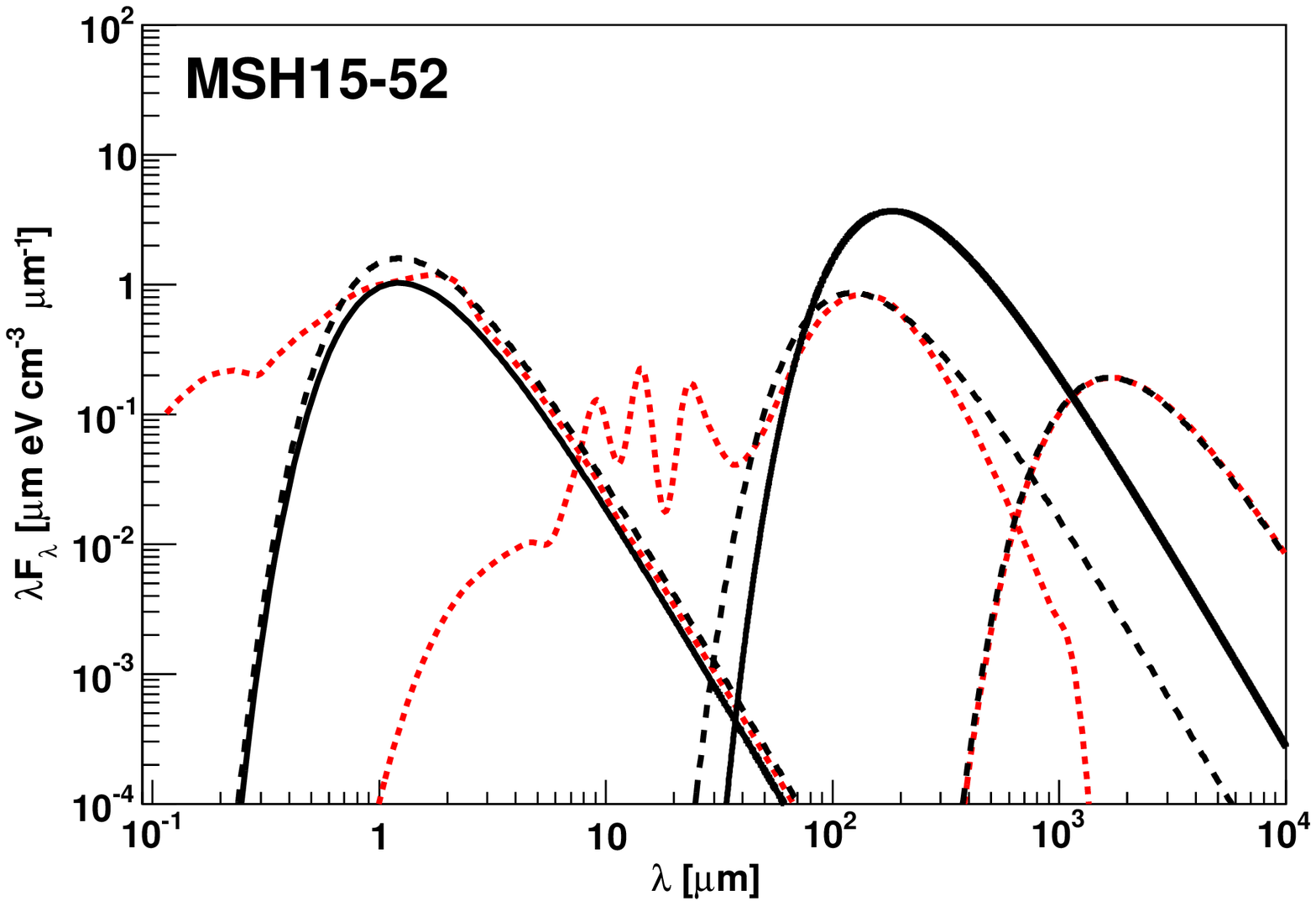}
\includegraphics[width=85mm]{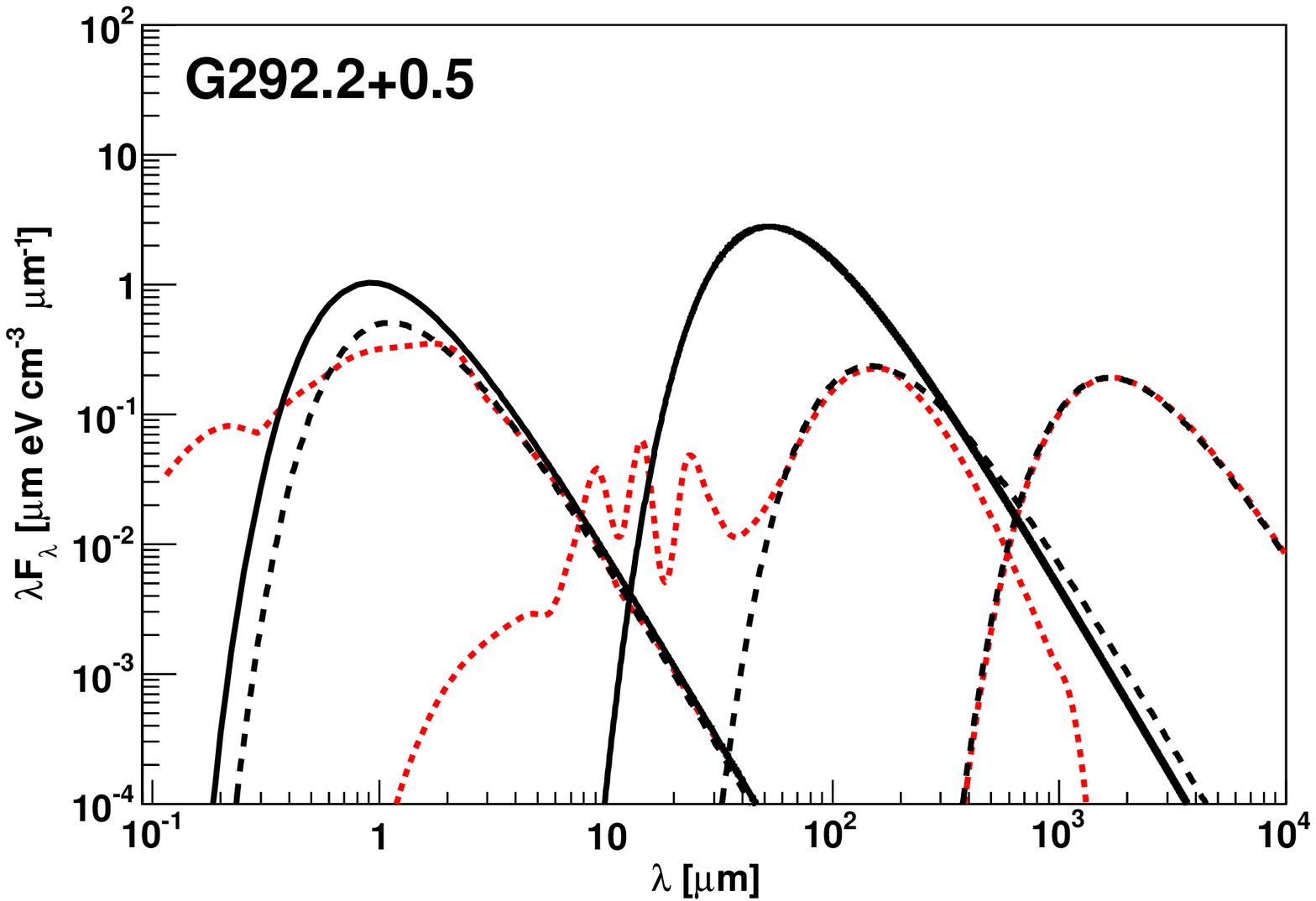}
\caption{Example of the comparison between the ISRF obtained from the GALPROP code (Porter et al. 2006)
and the assumptions made to fit the PWNe models. We show the FIR and NIR diluted blackbodies (with the parameters of
Table \ref{param} in bold black curves), in comparison with the GALPROP raw results (in red) and fits to these results using diluted blackbodies (in black thin lines, 
and as given in Table \ref{ISRF}.) The rightmost component stands for the CMB in all panels.} 
\label{isrf-comp}
\end{figure*}

Table \ref{ISRF} compares the energy densities 
used to fit the observational data of each of the PWNe studied 
with those obtained from the GALPROP code (Porter et al. 2006). In order to do this, we have obtained the ISRF 
from GALPROP and fitted three diluted blackbodies, for which the energy densities and temperatures are referred
to as $w^G$ and $T^G$, respectively. 
As shown in Table \ref{ISRF}, the values of the FIR energy densities 
obtained from GALPROP are generally lower (by up to a factor of a few) than what we found is needed to fit 
the PWN high-energy emission. Fig. \ref{isrf-comp} shows four examples.  

The use of GALPROP ISRFs all along the Galaxy is known to be subject to local uncertainties. Galactic locations in which freshly accelerated electrons target overdensities of FIR photons contributed by nearby stars, star-forming regions, or the supernova remnants themselves, could produce these local variations.
As mentioned above in some of the individual PWNe studied, the need of larger energy densities than those found in GALPROP when time-dependent models have been used has been spotted in the past, but for scattered PWN. The possibility of finding relatively high energy densities in the background photon fields nearby PWNe is interesting from a couple of perspectives: On the one hand, it would imply that CTA could be mapping PWNe also at averaged (and thus lower) Galactic photon backgrounds, ultimately helping determine the latter. On the other hand, detailed studies of the IR emission around PWNe should reveal significant sources. This is in general true, as examples, one could quote the case of G54.1+0.3 in which Temim et al. (2010) proposed that the SN dust is being heated by early-type stars belonging to a cluster in which the SN exploded; or MSH 15--52
where there is an O star 13 arcsec away from the corresponding pulsar (Arendt et al. 1991,  Koo et al. 2011). A statistical study of the correlation between mass (traced by CO and dust) and TeV sources has been recently performed by Pedaletti et al. (2013), finding that there are hints of a positive correlation with IR excess at the level of 2--3$\sigma$, which still needs to be confirmed.

%%%%%%%%%%%%%%%%%%%%%%%%%%%%%%%%%%%%%%%%%%%%
%%%%%%%%%%%%%%%%%%%%%%%%%%%%%%%%%%%%%%%%%%%%

%%%%%%%%%%%%%%%%%%%%%%%%%%%%%%%%%%%%%%%%%%%%
%%%%%%%%%%%%%%%%%%%%%%%%%%%%%%%%%%%%%%%%%%%%
\subsection{Magnetization of the nebulae}
%%%%%%%%%%%%%%%%%%%%%%%%%%%%%%%%%%%%%%%%%%%%
%%%%%%%%%%%%%%%%%%%%%%%%%%%%%%%%%%%%%%%%%%%%

\begin{figure*}[t!]
\centering\includegraphics[width=63mm]{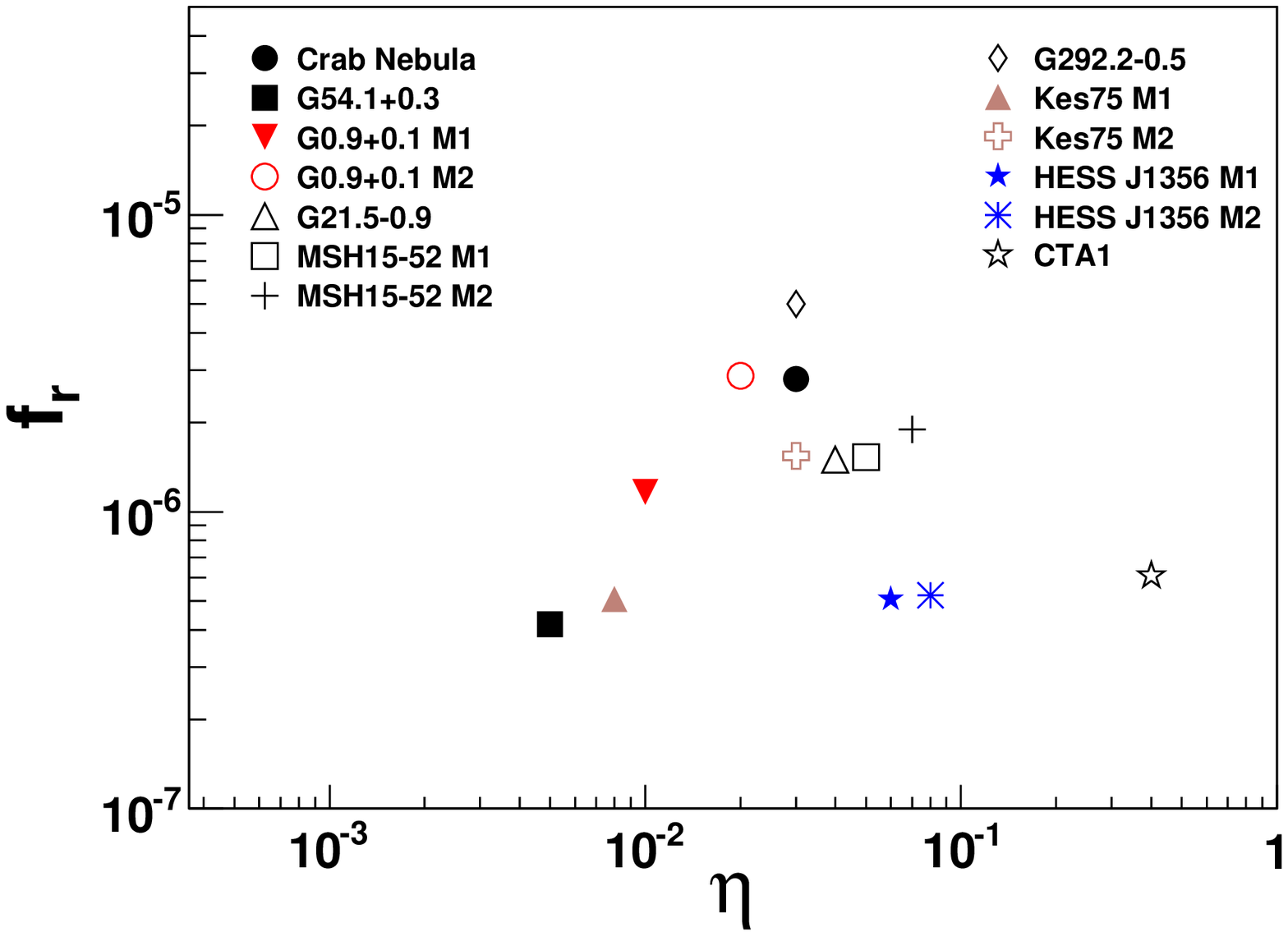} \hspace{-.6cm}
\includegraphics[width=63mm]{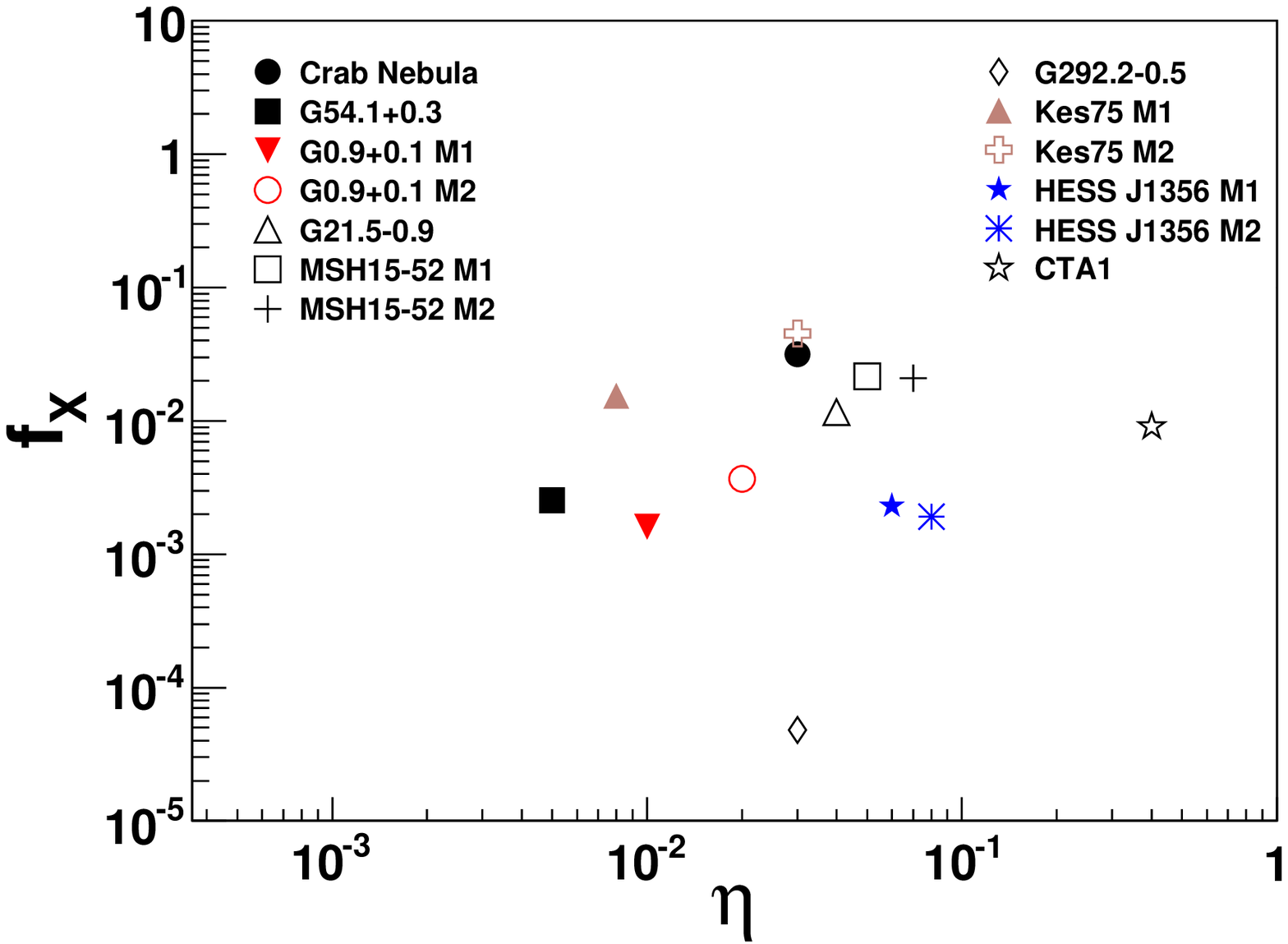} \hspace{-.6cm}
\includegraphics[width=63mm]{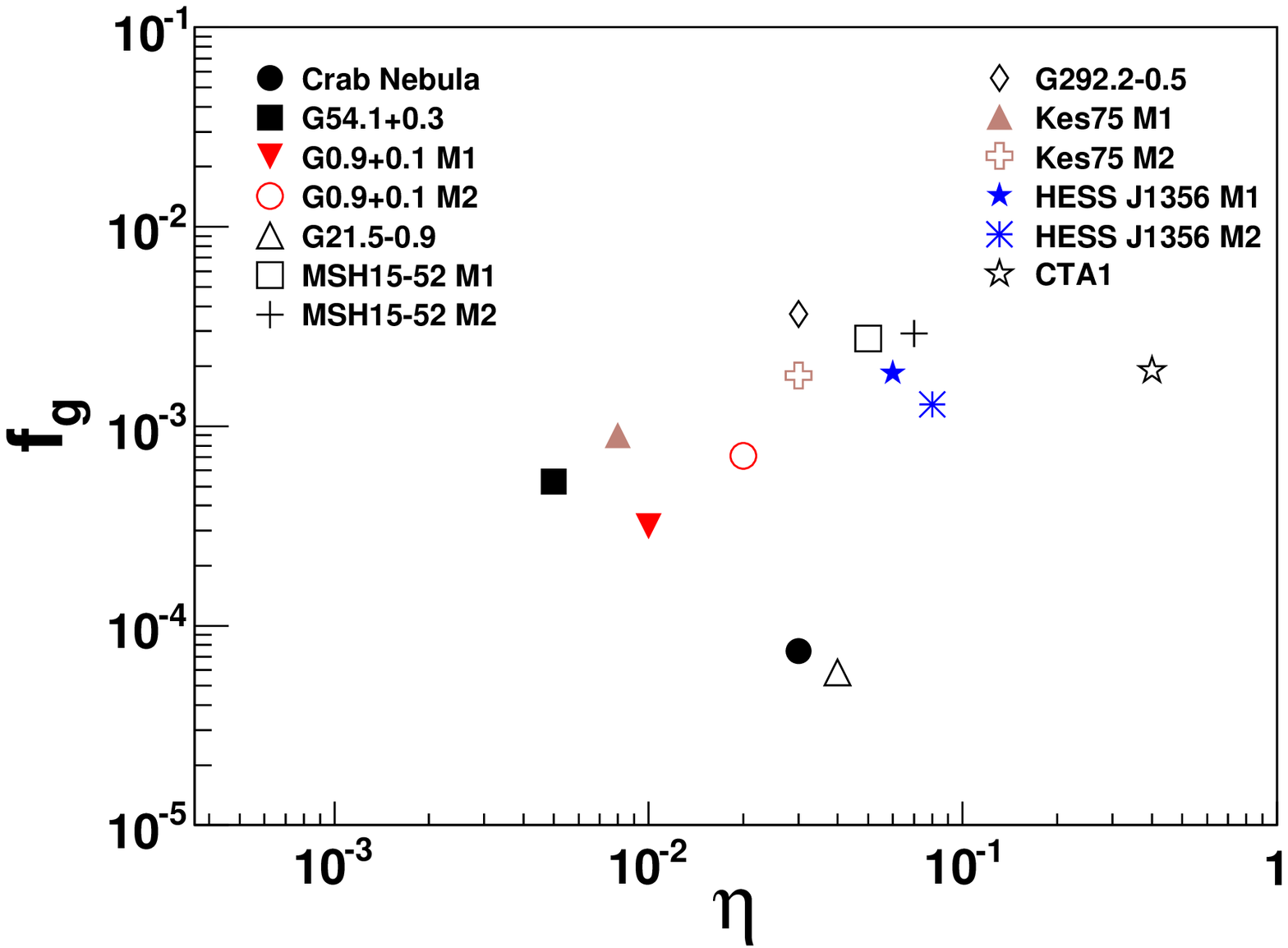}\\
\includegraphics[width=63mm]{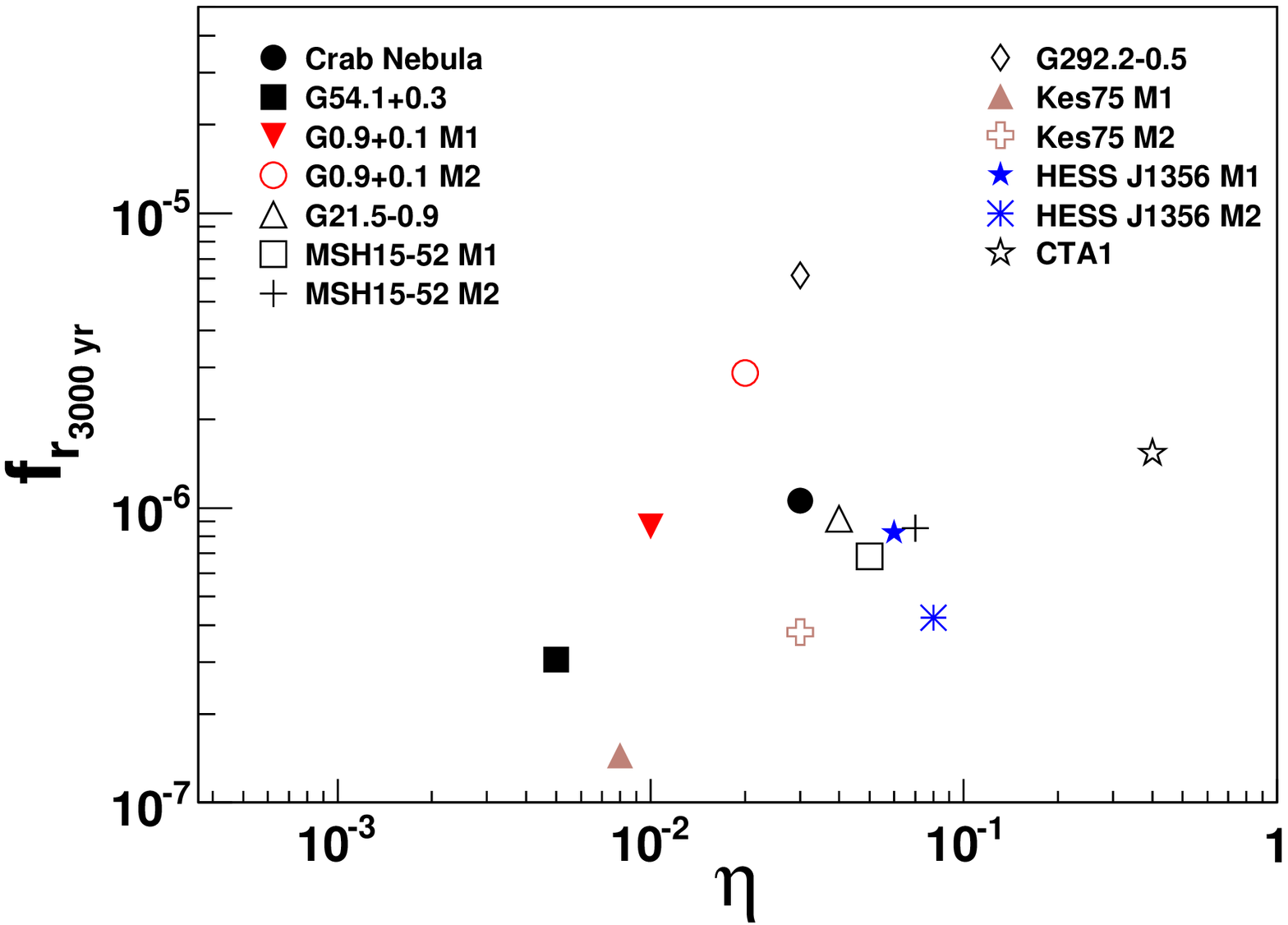} \hspace{-.6cm}
\includegraphics[width=63mm]{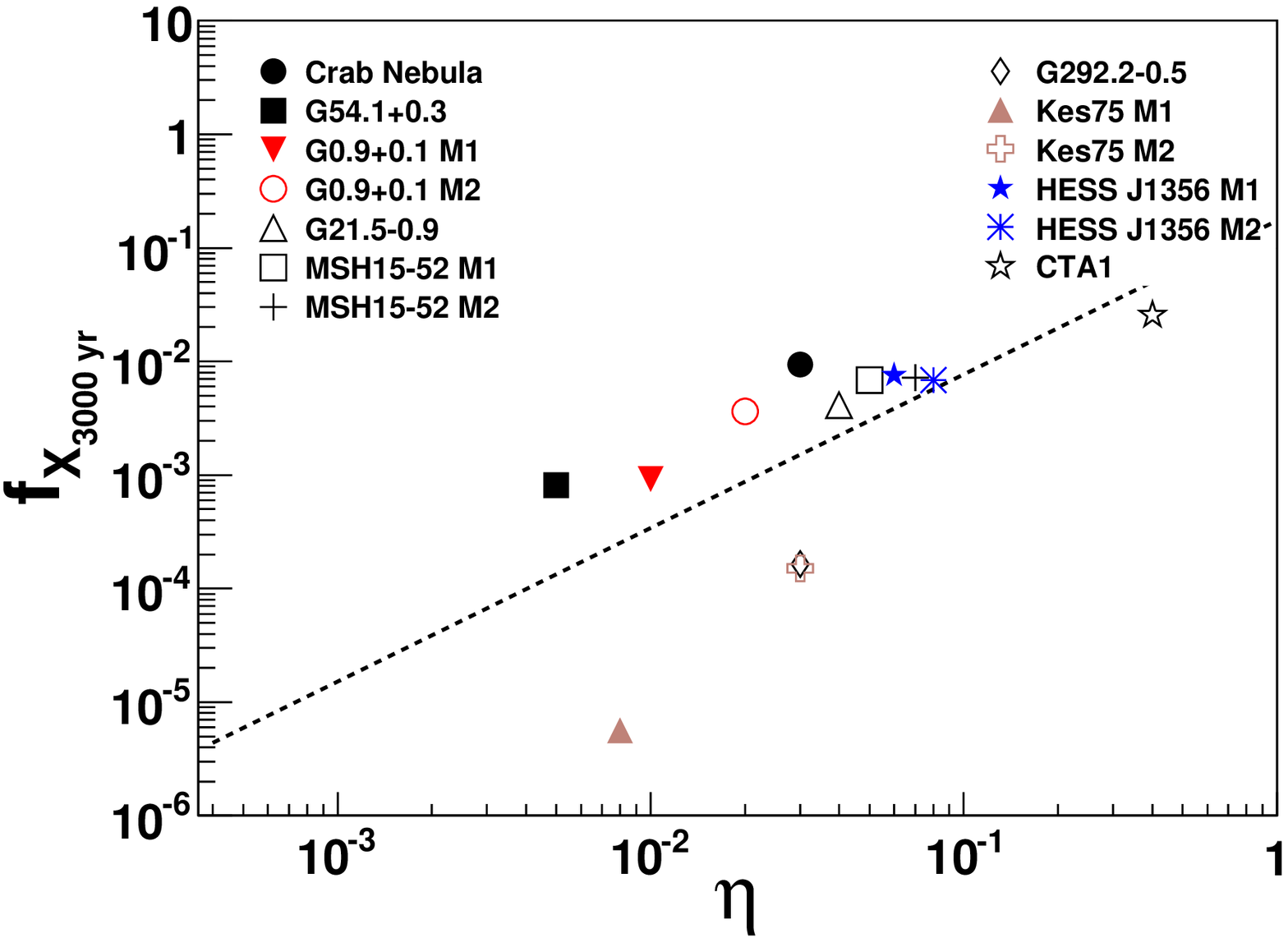}\hspace{-.6cm}
\includegraphics[width=63mm]{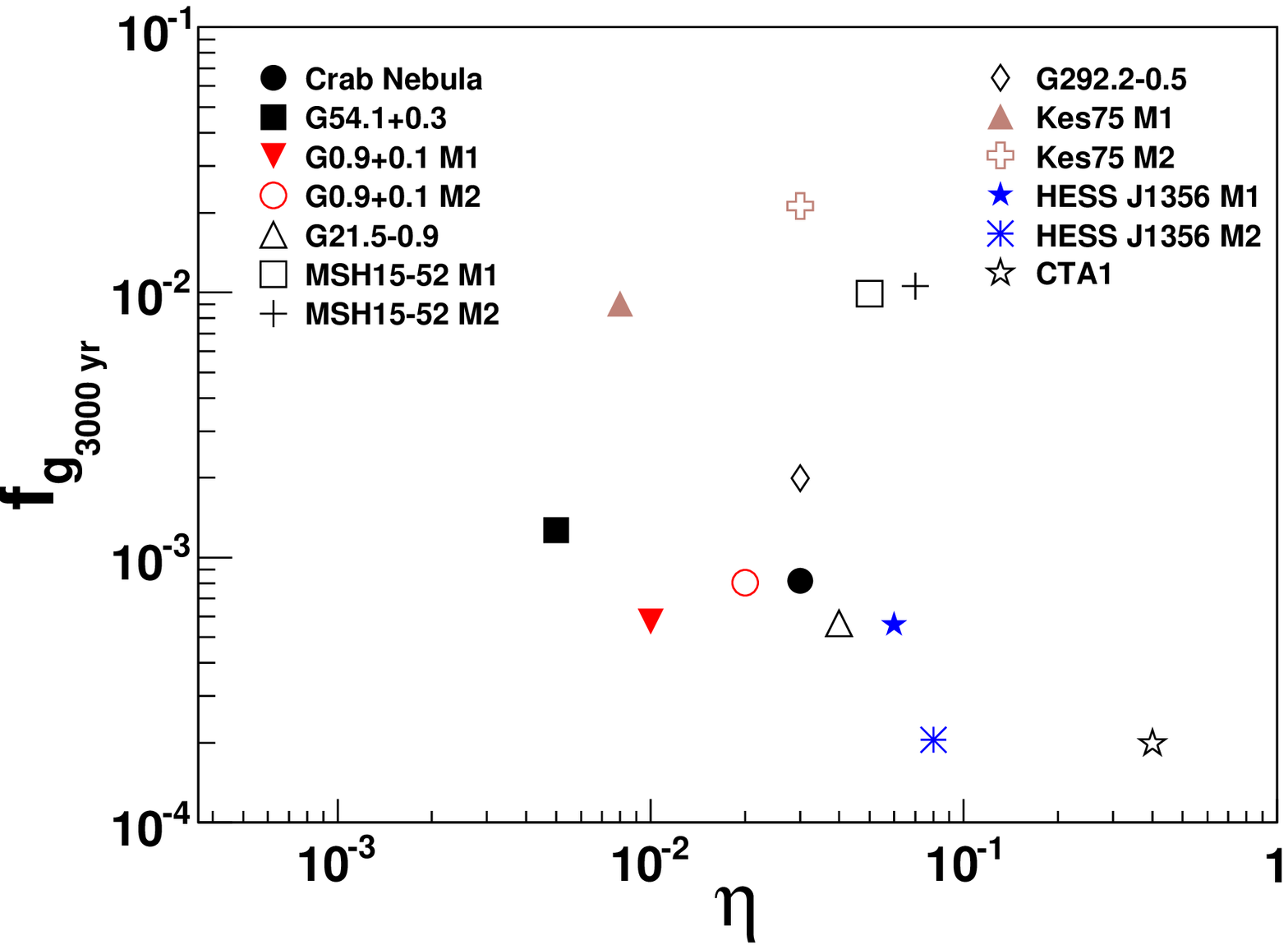}
\caption{Magnetization of PWNe as a function of the radio, X-ray, and gamma-ray efficiency. In the first row, all luminosity fraction values correspond to those today; in the second row, to the values they have from the evolution of each of the PWN when considered at 3000 years. }
\label{Magnetization}
\end{figure*}

From Table \ref{param} we see that all young nebulae detected at TeV are particle dominated, with
magnetic fractions that in all cases except CTA~1, never exceed a few percent. 
Fig. \ref{Magnetization}  shows the values of the obtained radio, X-ray, and gamma-ray efficiencies
as a function of the magnetic fraction of the nebulae (which in our model 
is constant along the evolution).
The two sets of panels distinguish the values of the efficiencies obtained today (at different ages for each of the nebulae considered) from those obtained at the same age, fixed at 3000 years.

To consider whether there is a correlation in any of these (and subsequent) magnitudes we use a Pearson test. 
The Pearson $r$ estimator is computed using 9 PWN models (unless otherwise clarified). When more than one model was considered
plausible for a given PWN we use M1, although we have verified that considering the alternatives would not introduce a significant change to the results. We do not emphasize here the search for precise fit parameters (unless an obvious connection would appear), but of plausible correlations.
The latter will be hinted in
those cases in which the Pearson coefficient for the pair of magnitudes considered
yields to a non-directional probability of incorrectly rejecting the 
null hypothesis (i.e., no correlation) smaller than 0.05.
In these cases, we quote the fit parameters in Table \ref{fits}, as well as we show the fit in the corresponding figure.

There is no apparent correlation of the efficiencies with the magnetization except when we consider the X-ray efficiency
$f_x$ of the nebulae normalized at the same age. In that case, the Pearson coefficient yields to a probability of 0.043 of incorrectly rejecting the null hypothesis, but the coefficients of a linear fit are poorly determined because of the dispersion of the data. The significance of the correlation barely meets our cut. The
radio and gamma-ray efficiencies computed at the same age present significances of the order of 10\%. 
The fact that we do not see a correlation of the gamma-ray efficiency with the magnetization implies that $\eta$ is neither the only nor the dominant order parameter to impact the luminosities. The fact that we see essentially very similarly magnetized PWNe from a magnetic point of view reduces the $\eta$-distinguishing power further.

%%%%%%%%%%%%%%%%%%%%%%%%%%%%%%%%%%%%%%%%%%%%
%%%%%%%%%%%%%%%%%%%%%%%%%%%%%%%%%%%%%%%%%%%%

%%%%%%%%%%%%%%%%%%%%%%%%%%%%%%%%%%%%%%%%%%%%
%%%%%%%%%%%%%%%%%%%%%%%%%%%%%%%%%%%%%%%%%%%%
\subsection{Is there a low-magnetization observational bias?}
%%%%%%%%%%%%%%%%%%%%%%%%%%%%%%%%%%%%%%%%%%%%
%%%%%%%%%%%%%%%%%%%%%%%%%%%%%%%%%%%%%%%%%%%%

The only high-magnetization nebula we found in the sample we study is CTA~1, for which $\eta=0.4$, is close to equipartition. Should $\eta$ be much lower than this value we would find TeV fluxes in excess of what has been detected. The possibility that CTA~1 is beyond free expansion could play a role here; a compression of the nebula due to reverberation could lead to an increase of the magnetization. Note that in the model of CTA~1 by Aliu et al. (2013), where a reverberation has been taken into account, the magnetization was also found to be in the high end, more than an order of magnitude larger than in Crab nebula. It is to note that the highest magnetized nebula in the sample is showing one of the lowest magnetic fields (see Table \ref{param}), something which has also been found with other models (e.g., Aliu et al. 2013). 
However, the conclusion that all the other nebulae are heavily particle-dominated is not affected by uncertainties 
in the modeling. 
To prove this we have tried to fit these nebulae data with an {\it ad-hoc} increase of $\eta$ up to 0.5
(equal distribution of the power between particles and field) and explored the range of parameters, if any, which would allow for a good fit.
Models with larger $\eta$ allow us to investigate whether we would have detected the 
nebulae should they have an increased magnetic fraction. Earlier,
we have concluded that if the injection and environment of PWNe were as those of Crab, 
only  in the case of a large, Crab-like, spin-down power feeding into a nebula located at 2 kpc or less, 
a H.E.S.S.-like telescope would detect magnetically-dominated nebula beyond $\eta \sim 0.5$ (Torres et al. 2013b).
Different to our earlier study, we here consider the 
injection and environmental properties specifically derived for each nebulae. 

\begin{figure*}[t!]
\centering
\includegraphics[width=84mm]{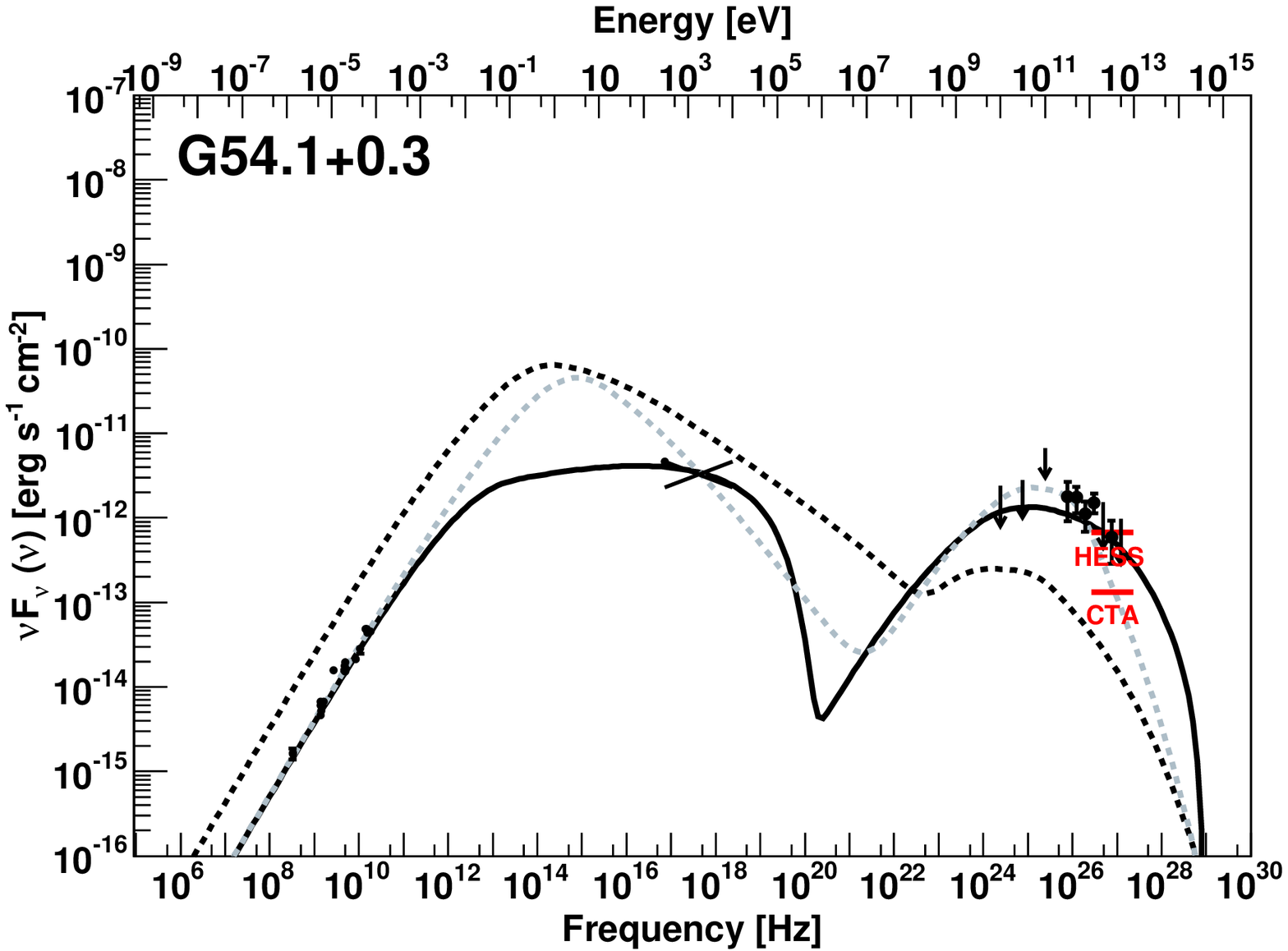}
\includegraphics[width=84mm]{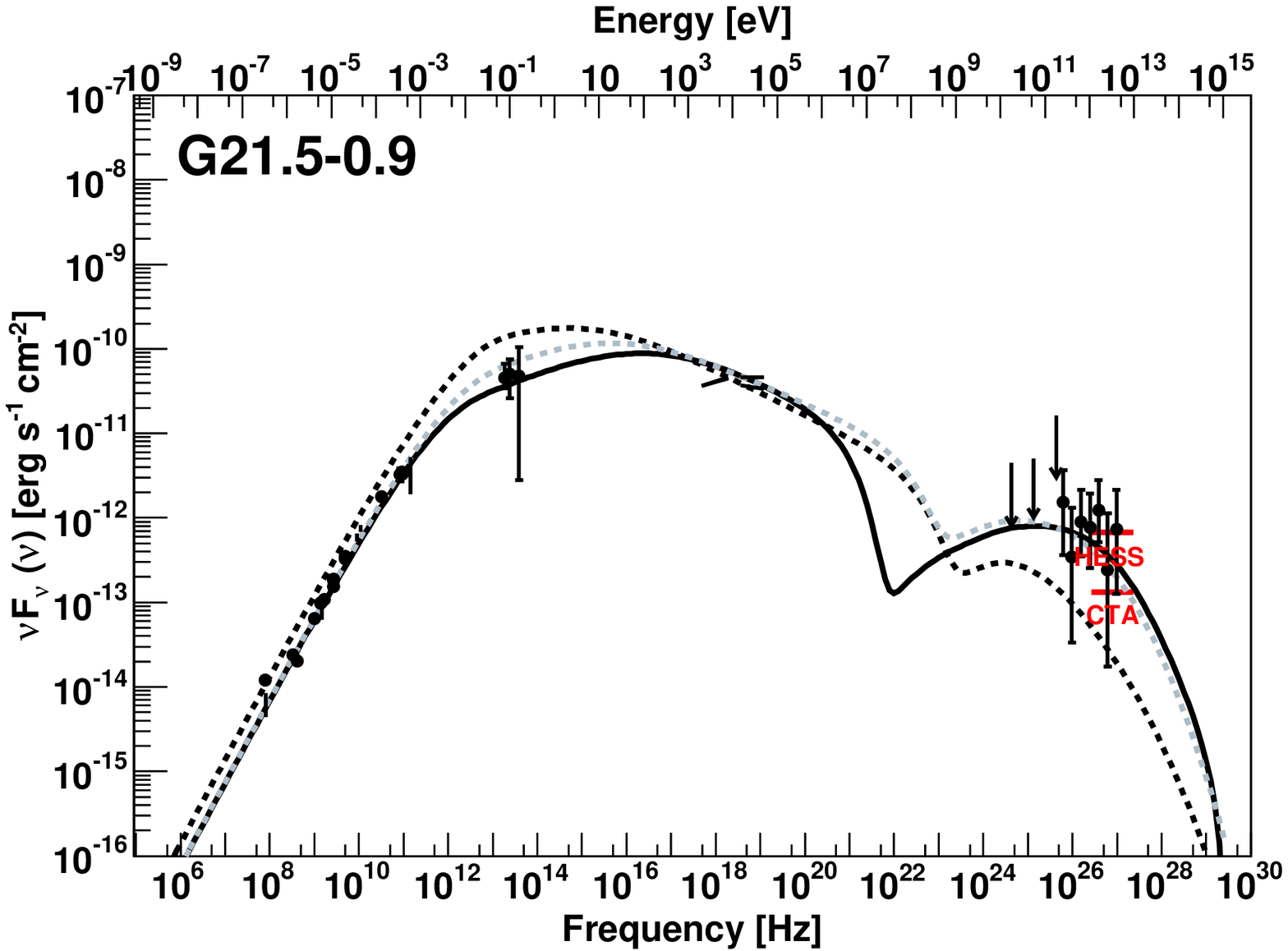}
\caption{G54.1+0.3 (left) and G21.5--0.9 (right) modeled with an imposed equipartition of the energetics ($\eta=0.5$) as compared with the adopted (particle dominated) models. The solid line represents the fitted model of Table \ref{param}, the dashed black line represents a model with $\eta=0.5$ and no changes in other parameters with respect to the fitted model of Table  \ref{param}, and
the dashed grey line stands for an equipartition model where other parameters are adjusted ad-hoc so that a relatively good fit is attained. For a discussion of the caveats of latter models see the text. 
The sensitivity of a H.E.S.S.-like telescope and of CTA are marked by the horizontal lines. 
}
\label{etabias}
\end{figure*}

Fig. \ref{etabias} shows two examples, for PWN G54.1+0.3 and G21.5--0.9, when modeled with imposed equipartition of the energetics keeping other parameters
the same (e.g., with the same FIR/NIR densities).
The increase in $\eta$ implies enlarging it by a factor of $\sim 100$ and $\sim 10$ in the fitted $\eta$-value, respectively. 
The predicted TeV emission fits the data badly, and the TeV fluxes are below the sensitivity of CTA. 

We have also searched for a fit in case the PWNe are in equipartition but all other parameters are allowed to vary. The solutions we found require 
extreme values of other parameters and are thus not preferred. For instance, in the case of G21.5--0.9, a relatively good fit (albeit of poorer quality than the one we show in Fig. \ref{G21})
can be found by increasing the FIR density to 6 eV cm$^{-3}$ (a factor of 6 larger than the GALPROP outcome at the position) and reducing the ejected mass by a factor of 2 (what enlarges the nebula size in our model and contributes to dilute the magnetic field energy). It is clear that there is no preference for these stretched parameters over the ones shown in our earlier fit. 
The case of G54.1+0.3 is similar, although requires even larger changes in the FIR and NIR densities, and the ejected mass in order to yield to a fit which is not even close to all data points, particularly those at high energies. In particular, Fig. \ref{etabias} shows a model with $\eta = 0.5$ a FIR (NIR) density of 4 (40) eV cm$^{-3}$, and an ejected mass more than a factor of 3 smaller --implying a factor of $\sim 2$ larger nebula. It is clear that no equipartition model can be sustained in this case either.
These conclusions are similarly obtained in the analysis of other PWNe. 
The finding of CTA~1, however, shows that the fact that most of the PWNe we see are particle dominated cannot be fully ascribed 
to an observational bias; at least in some cases (but not in the majority) we would be able to detect them with the current generation of
telescopes. 
%

%%%%%%%%%%%%%%%%%%%%%%%%%%%%%%%%%%%%%%%%%%%%
%%%%%%%%%%%%%%%%%%%%%%%%%%%%%%%%%%%%%%%%%%%%

%%%%%%%%%%%%%%%%%%%%%%%%%%%%%%%%%%%%%%%%%%%%
%%%%%%%%%%%%%%%%%%%%%%%%%%%%%%%%%%%%%%%%%%%%
\subsection{Searching for a more meaningful SEDs and electron population comparison}
%%%%%%%%%%%%%%%%%%%%%%%%%%%%%%%%%%%%%%%%%%%%
%%%%%%%%%%%%%%%%%%%%%%%%%%%%%%%%%%%%%%%%%%%%

\begin{figure*}[t!]
\centering
\includegraphics[width=88mm]{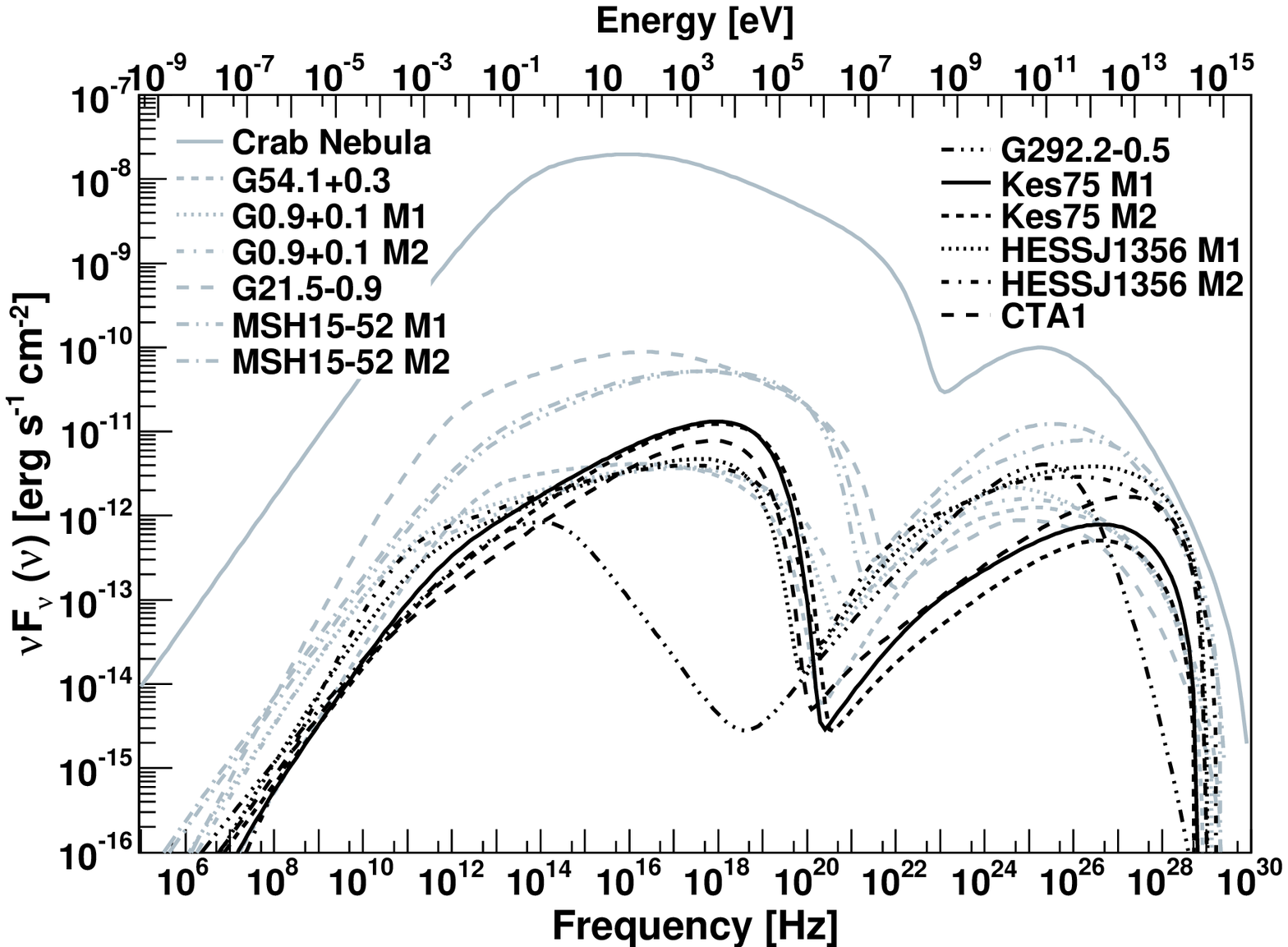}
\includegraphics[width=88mm]{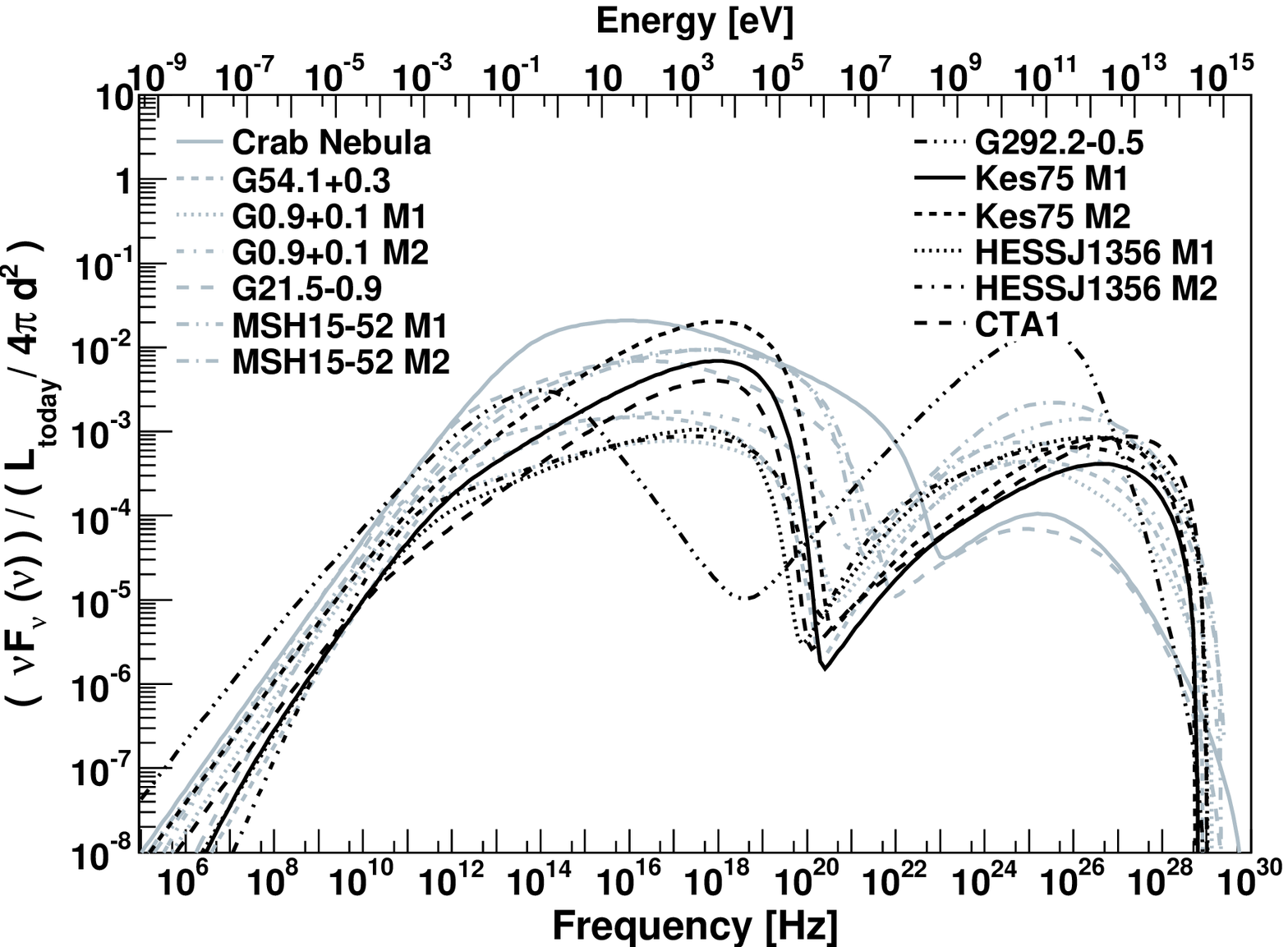}

\includegraphics[width=88mm]{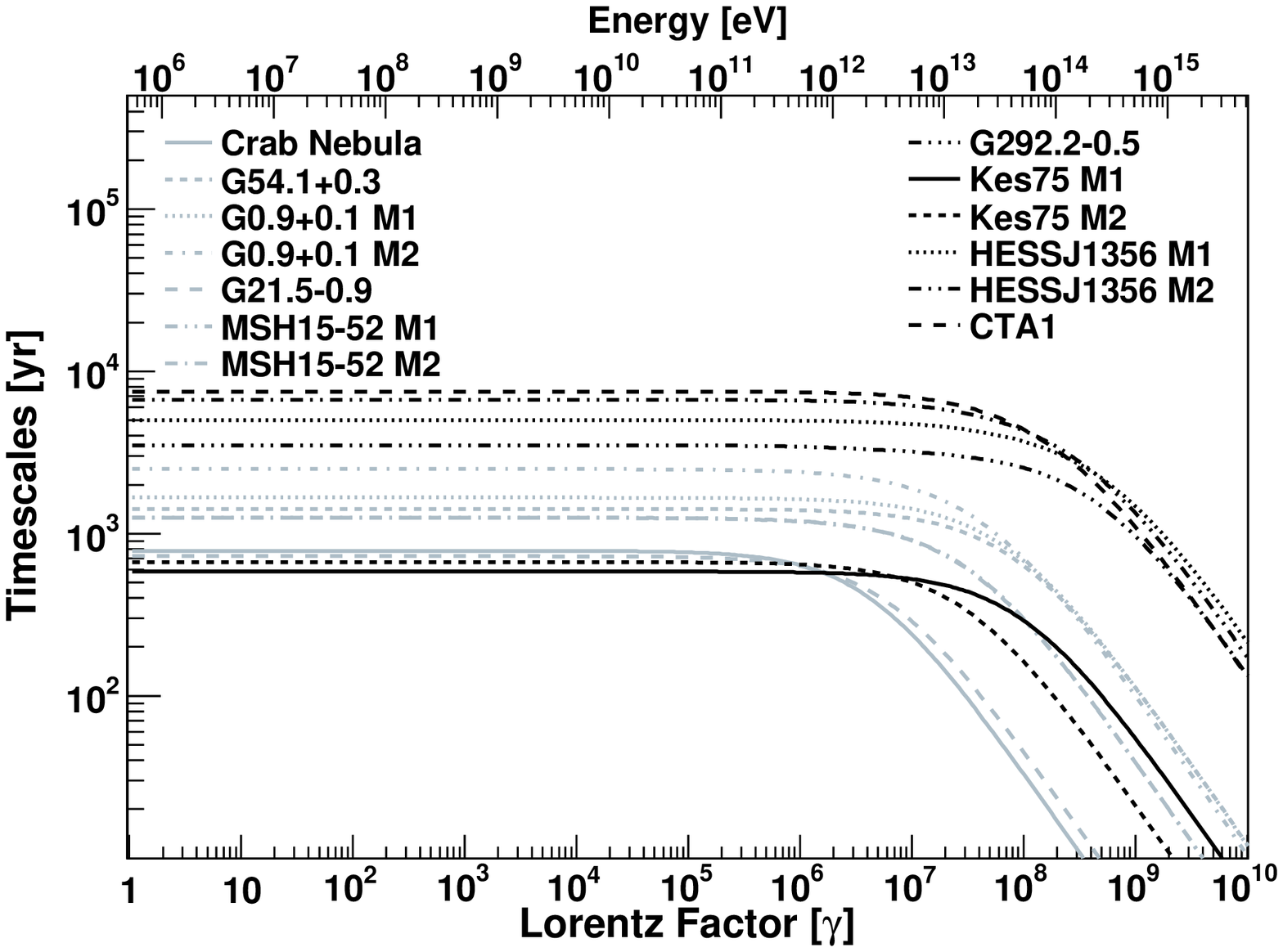}
\includegraphics[width=88mm]{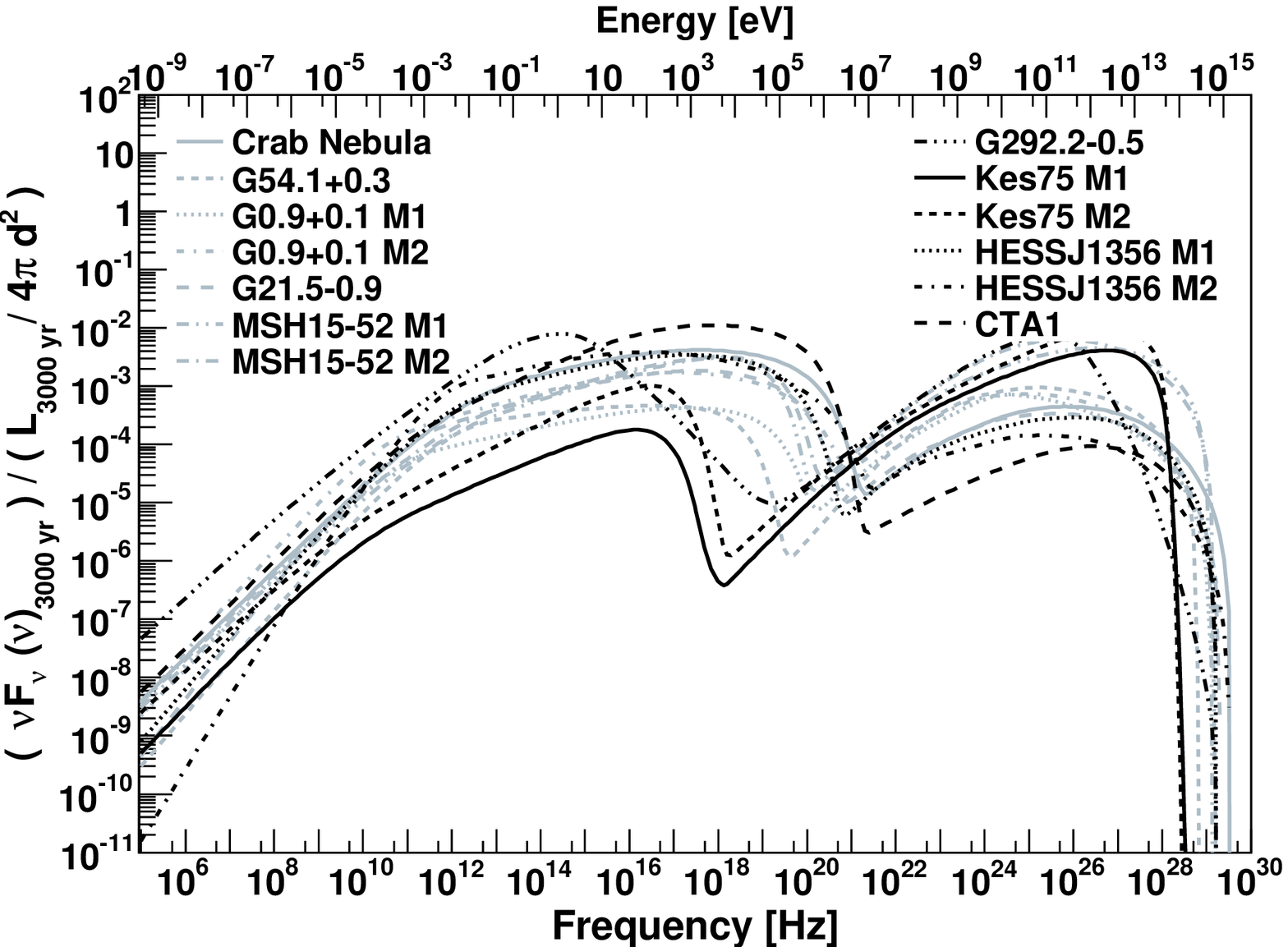}

\includegraphics[width=88mm]{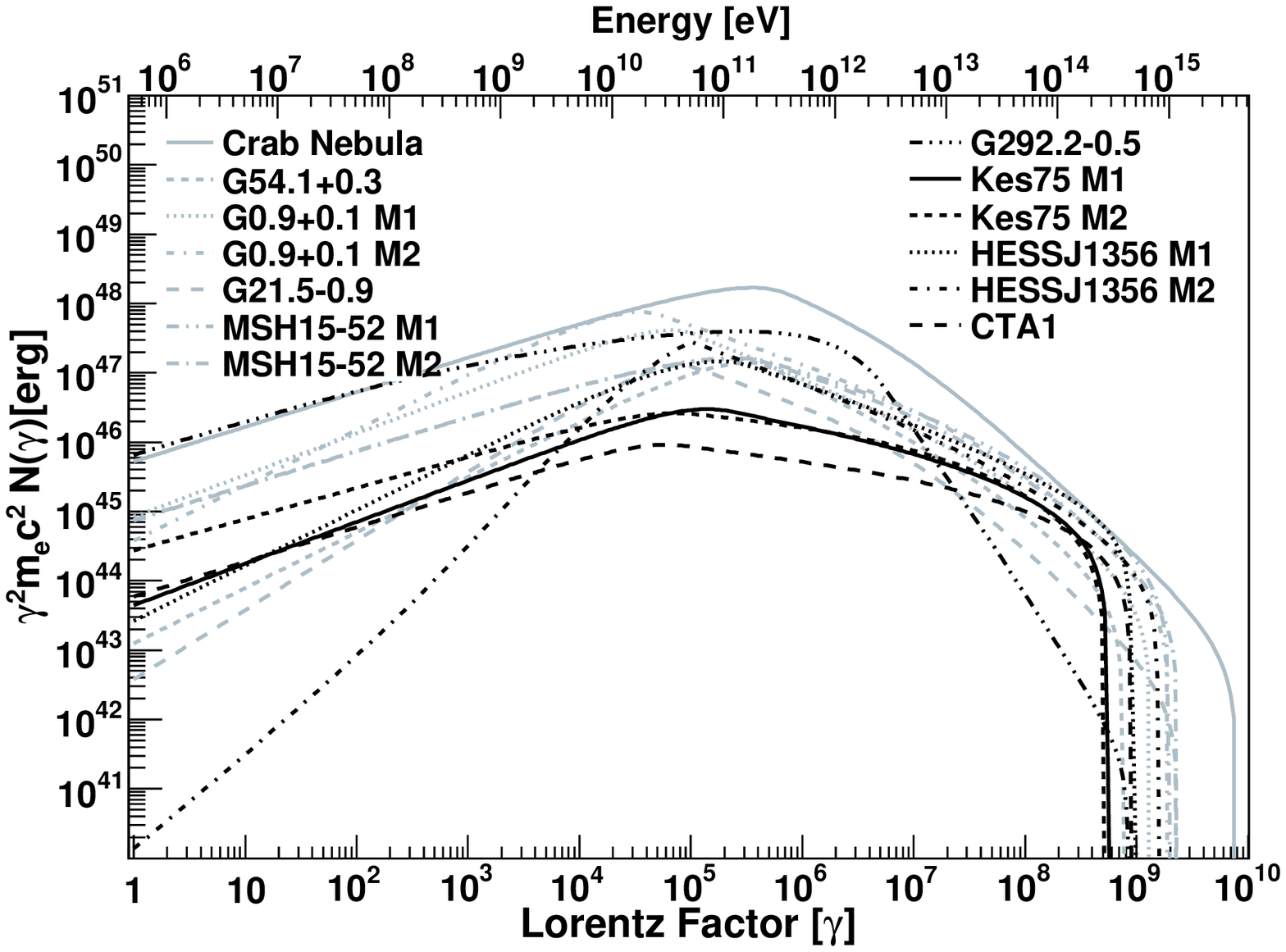}
\includegraphics[width=88mm]{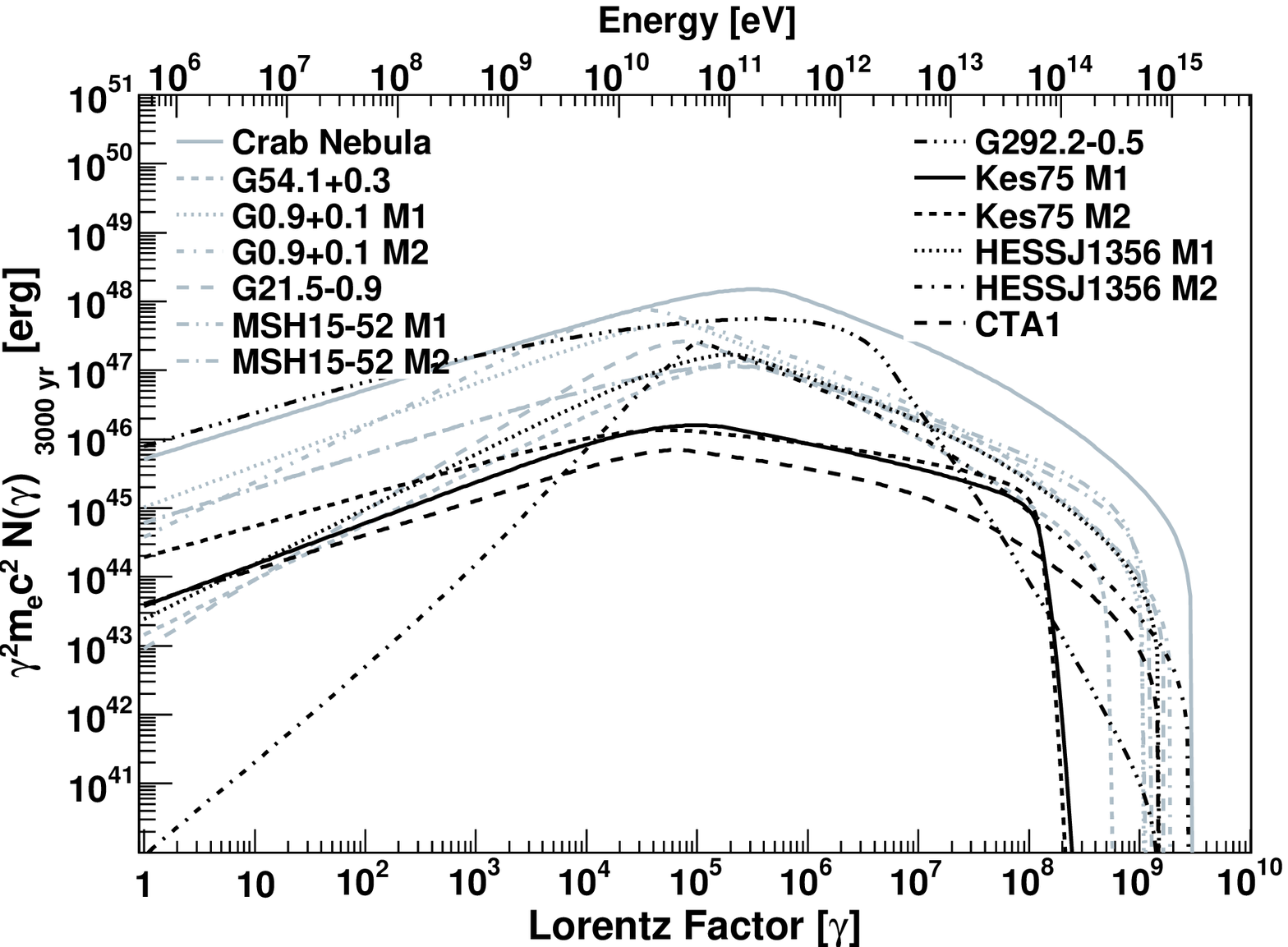}
\caption{Comparison of 
PWNe results.  Left panels: from top to bottom,  SEDs, electron losses, and  electron distributions today.
Right panels: from top to bottom, SEDs normalized by the corresponding spin-down flux   ($L_{sd}/4\pi D^2$, as obtained from Table \ref{param}), SEDs at 3000 years normalized
with the spin-down flux that each pulsar would have at that age, and electron populations 
at 3000 years.
}
\label{common}
\end{figure*}

Fig. \ref{common}
put together the currently observed 
SEDs, the corresponding electron losses, and the electron populations. Whereas this is an interesting figure to gather the variety of the sources detected, a direct comparison of the multi-frequency emission (as it is usually done)
has to be taken with care: we are looking at objects at different ages and powered by pulsars of different spin-down. 
The variety we found at the SED level (top left panel) contrasts with the little dispersion (one order of magnitude) in the timescales for the losses that are operative in all the PWN. From the SED results today, the two outliers from the bulk of models are the Crab nebula and G292.2--0.5. Whereas the former can be understood due to the large difference in spin-down power, the reason for the latter discrepancy is less clear (see the discussion above).

In order to search for a more meaningful comparison we explore two normalizations of the SEDs. On the one hand, we normalize the SED of each PWN by its corresponding spin-down flux ($F_{sd}= L_{sd}/4\pi D^2$, as obtained from Table \ref{param}) each pulsar has at its current age. 
On the other hand, we compute the SEDs at the same age (arbitrarily chosen to be 3000 years) for all pulsars, and normalize them with the spin-down flux that each pulsar would have at that age  ($L_{sd}^{3000}/4\pi D^2$). These normalized SEDs are shown in the right panels of Fig. \ref{common}. The bottom-right panel of 
Fig. \ref{common} shows the electron populations of all PWNe at the same age (3000 years).

It is interesting to compare the Crab nebula's SED with respect to the others when one normalize it with the corresponding spin-down power and/or look at all PWNe at the same age: the Crab nebula becomes an unnoticeable member of the same population of sources. 
It is also interesting to notice that the other outlier, G292.2--0.5, is now also in the bulk of models (see second panel, right column).
The population is only distinguished by differences in the electron content, where slight variations in the position of the breaks and cutoffs is retained even when looked at the same age.

 %%%%%%%%%%%%%%%%%%%%%%%%%%%%%%%%%%%%%%%%%%%%

%%%%%%%%%%%%%%%%%%%%%%%%%%%%%%%%%%%%%%%%%%%%
%%%%%%%%%%%%%%%%%%%%%%%%%%%%%%%%%%%%%%%%%%%%
\subsection{PWN versus PSR properties: $L_{sd}$ and $\tau$}
%%%%%%%%%%%%%%%%%%%%%%%%%%%%%%%%%%%%%%%%%%%%
%%%%%%%%%%%%%%%%%%%%%%%%%%%%%%%%%%%%%%%%%%%%

\begin{figure*}[t!]
\centering
\includegraphics[width=63mm]{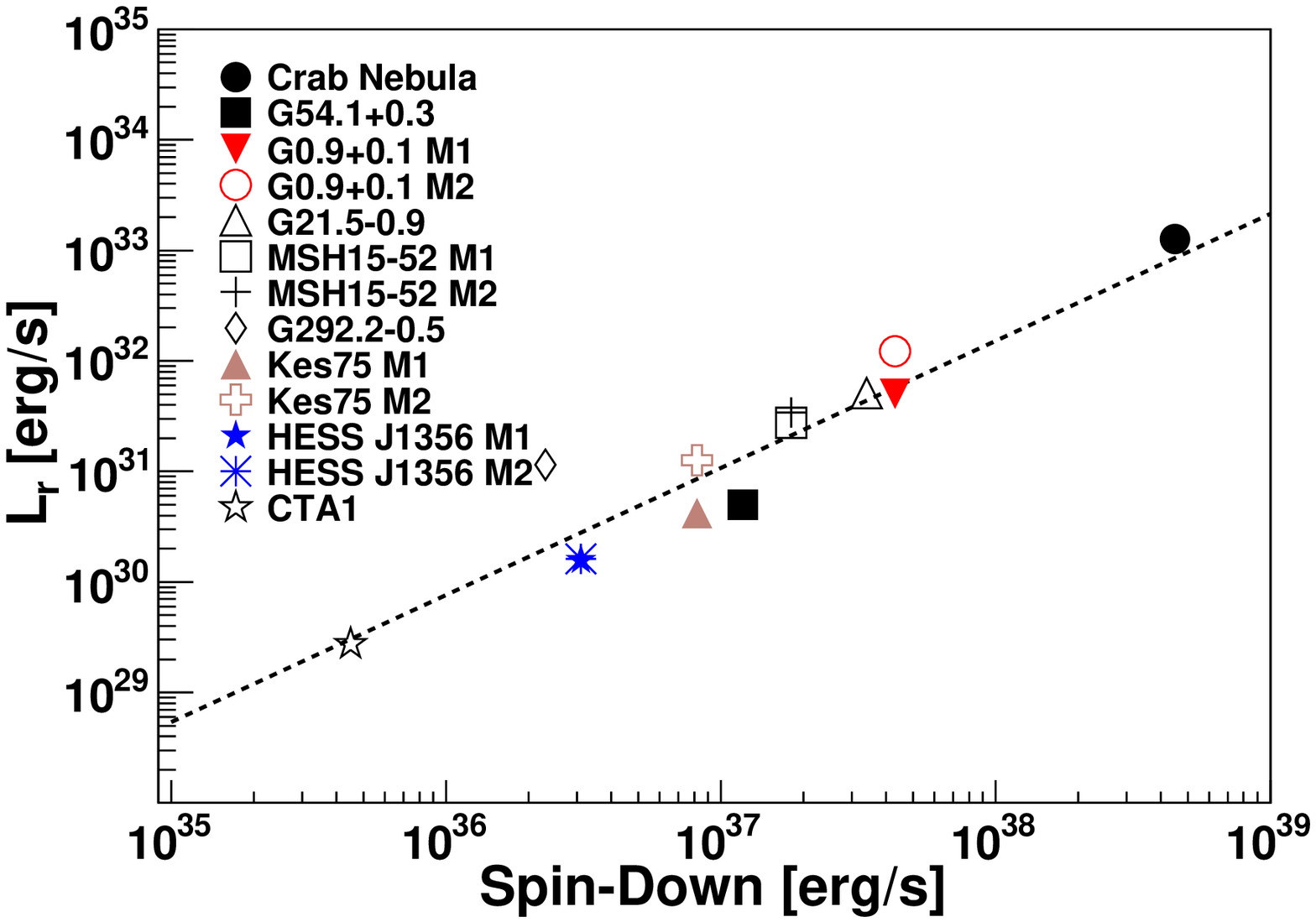} \hspace{-.6cm}
\includegraphics[width=63mm]{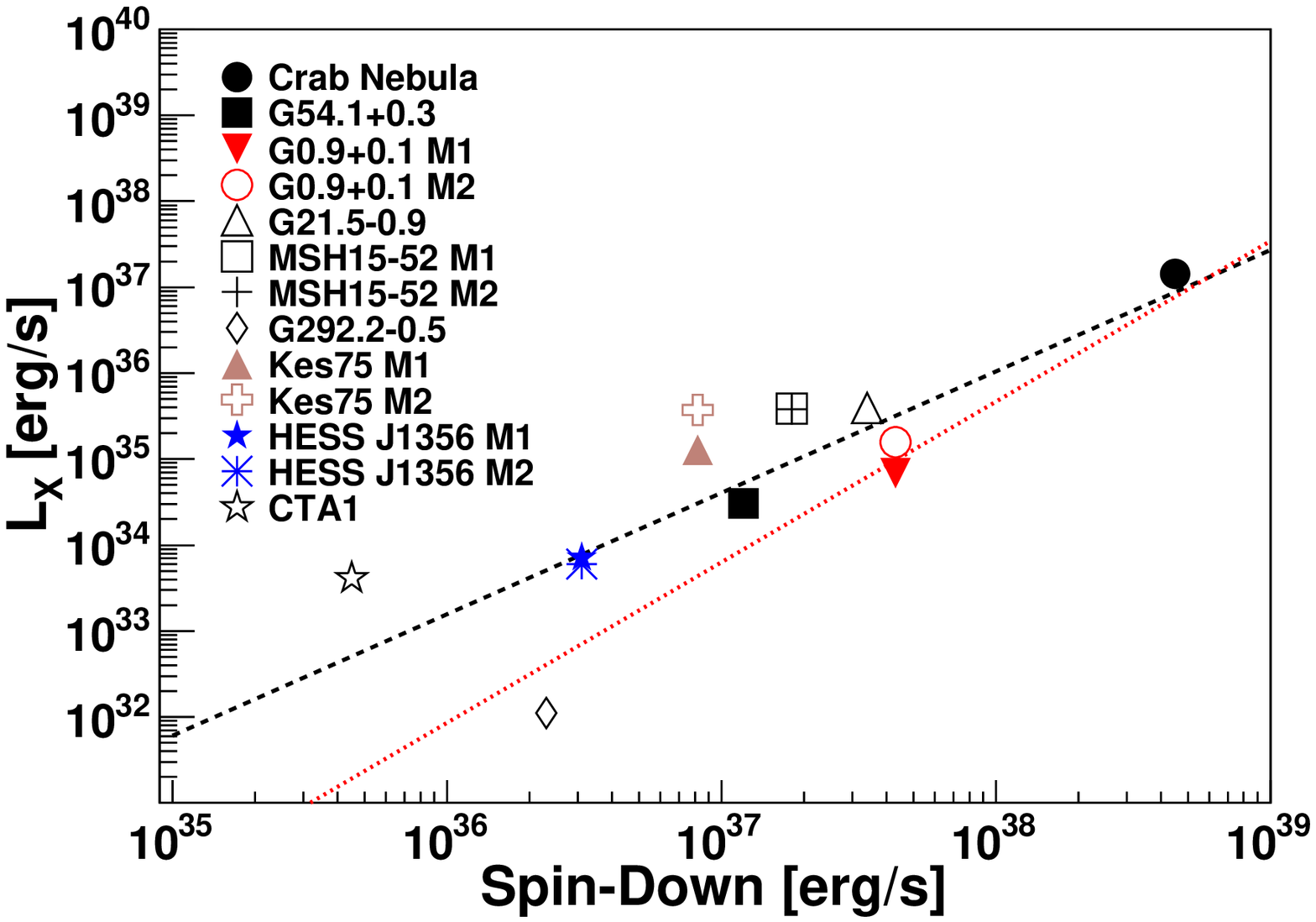} \hspace{-.6cm}
\includegraphics[width=63mm]{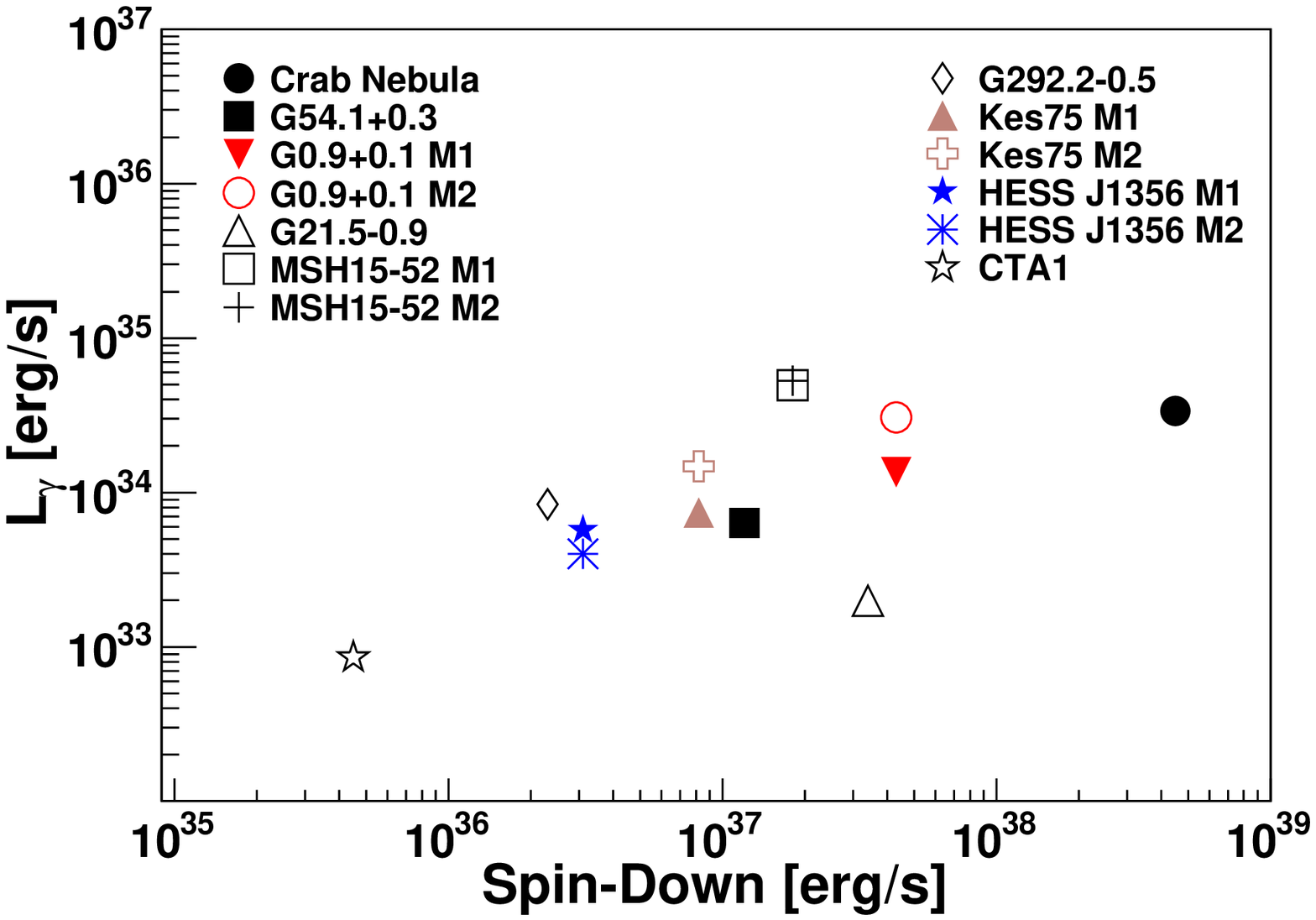}
\includegraphics[width=63mm]{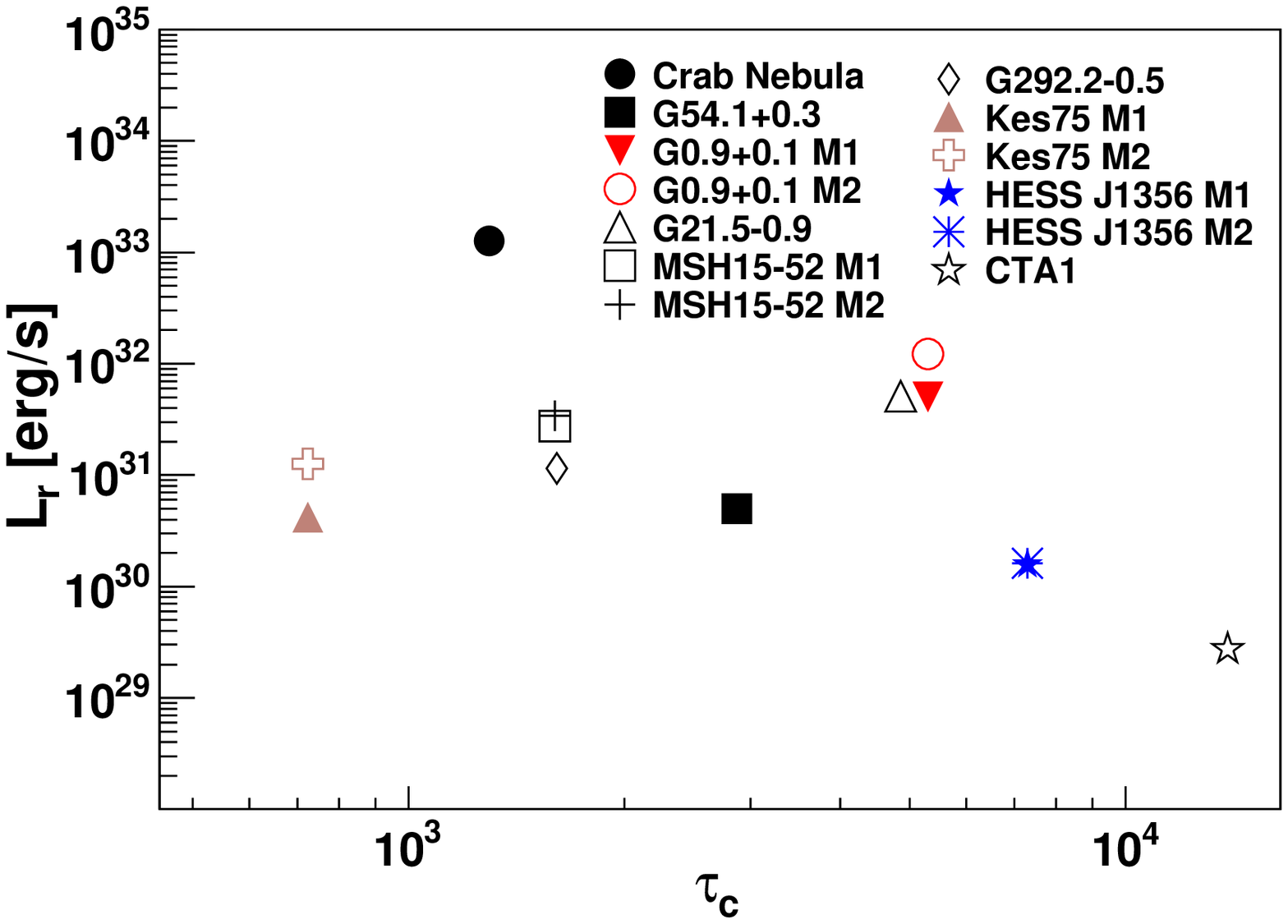}\hspace{-.6cm}
\includegraphics[width=63mm]{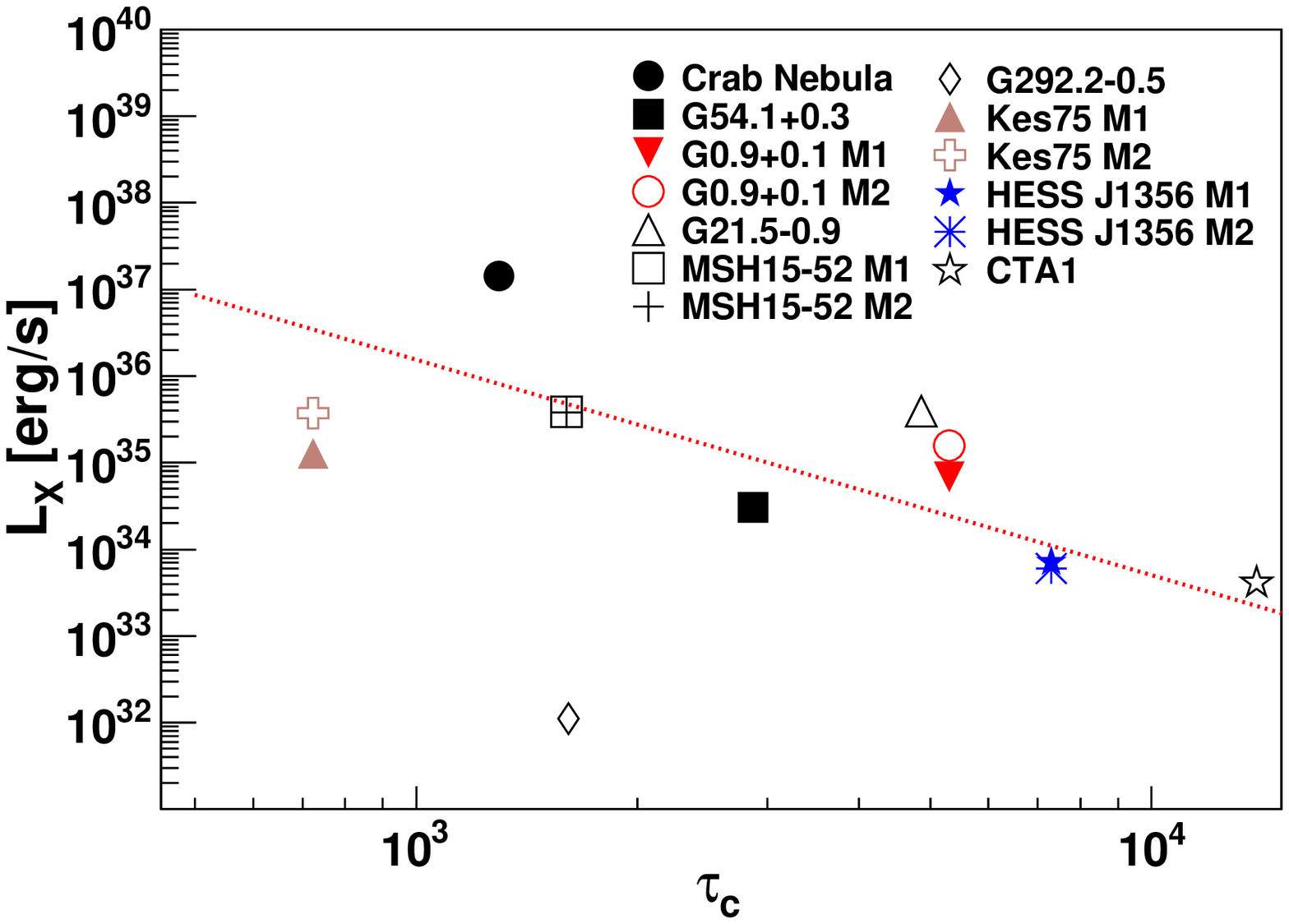} \hspace{-.6cm}
\includegraphics[width=63mm]{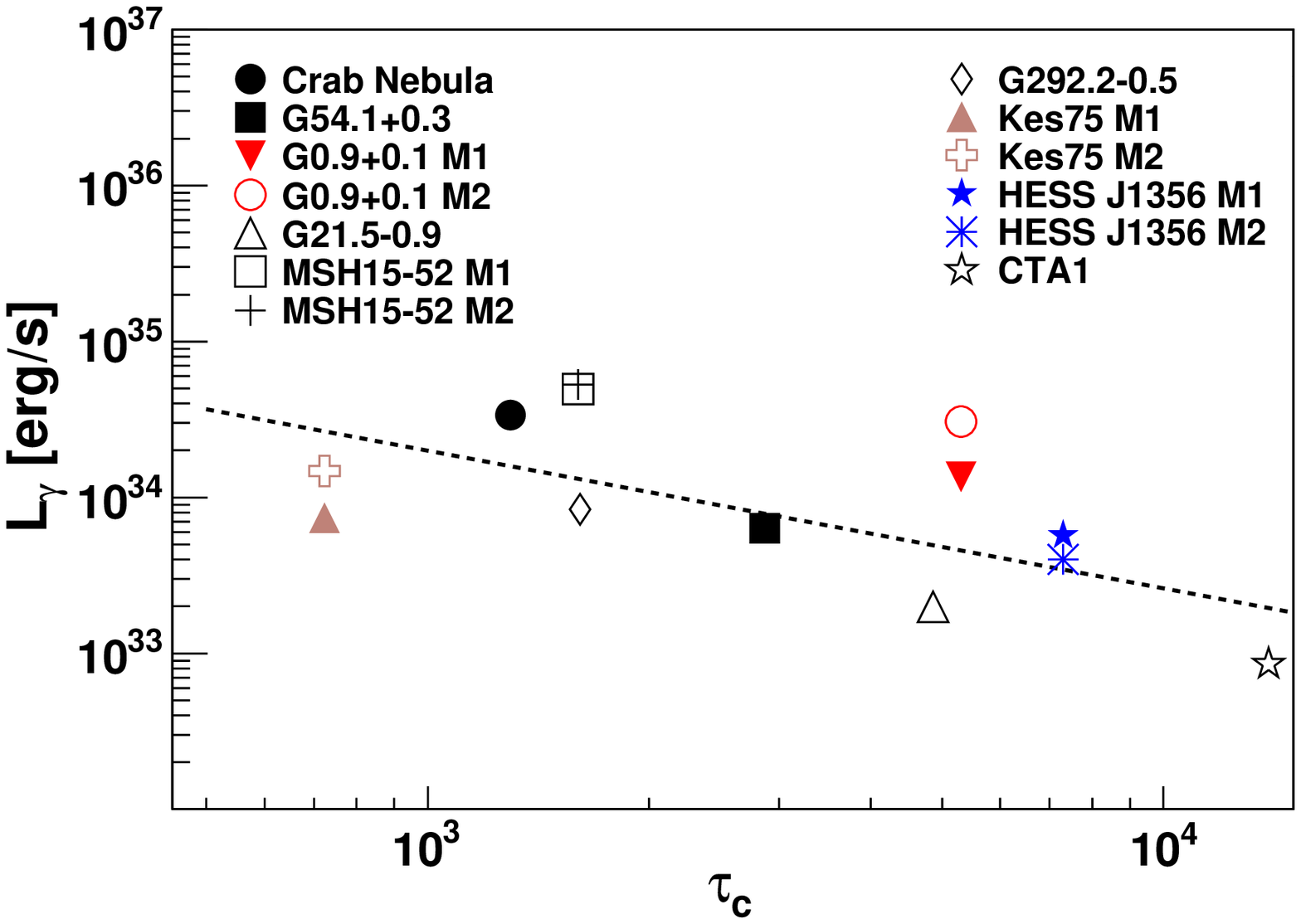}
\includegraphics[width=63mm]{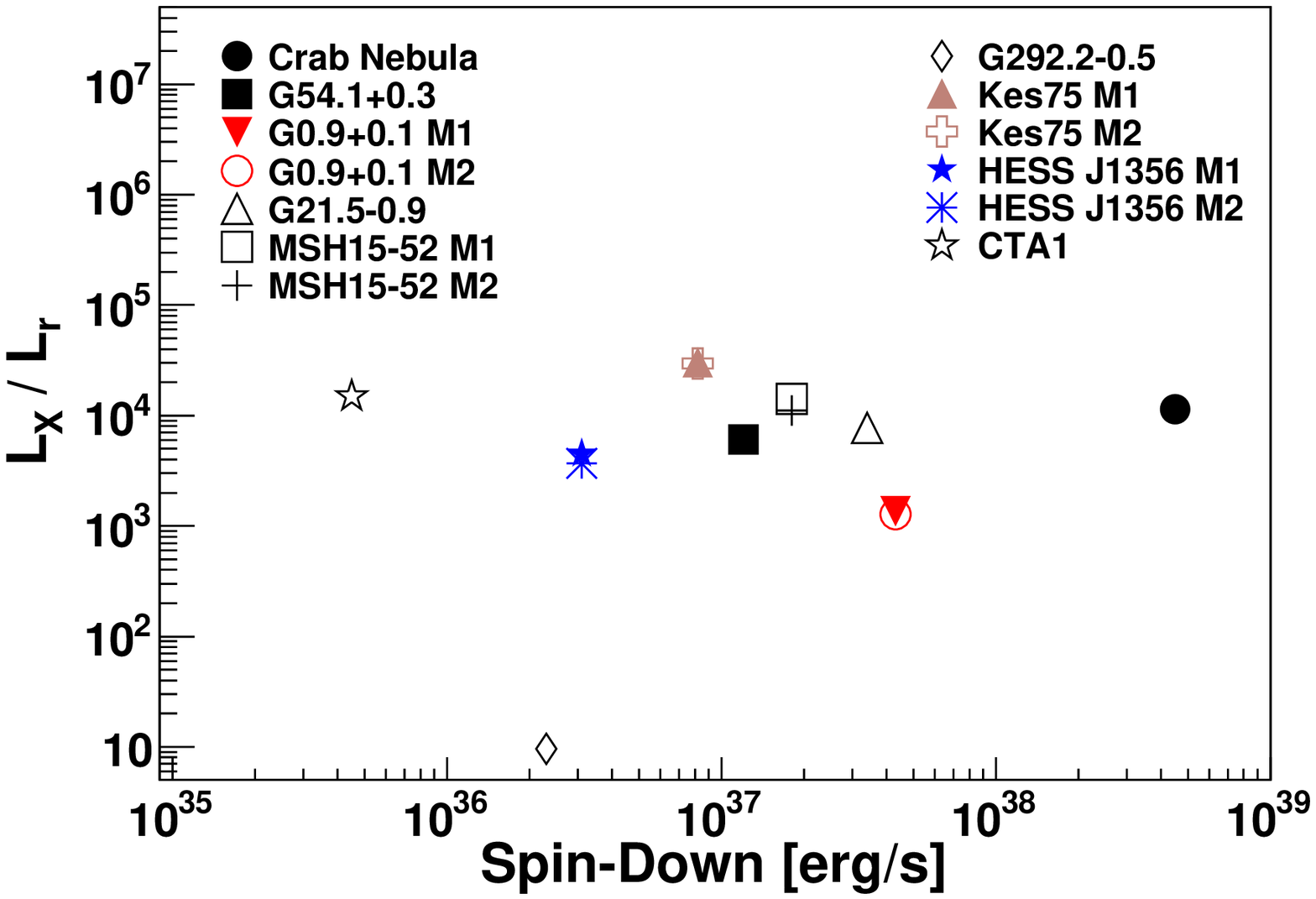}\hspace{-.6cm}
\includegraphics[width=63mm]{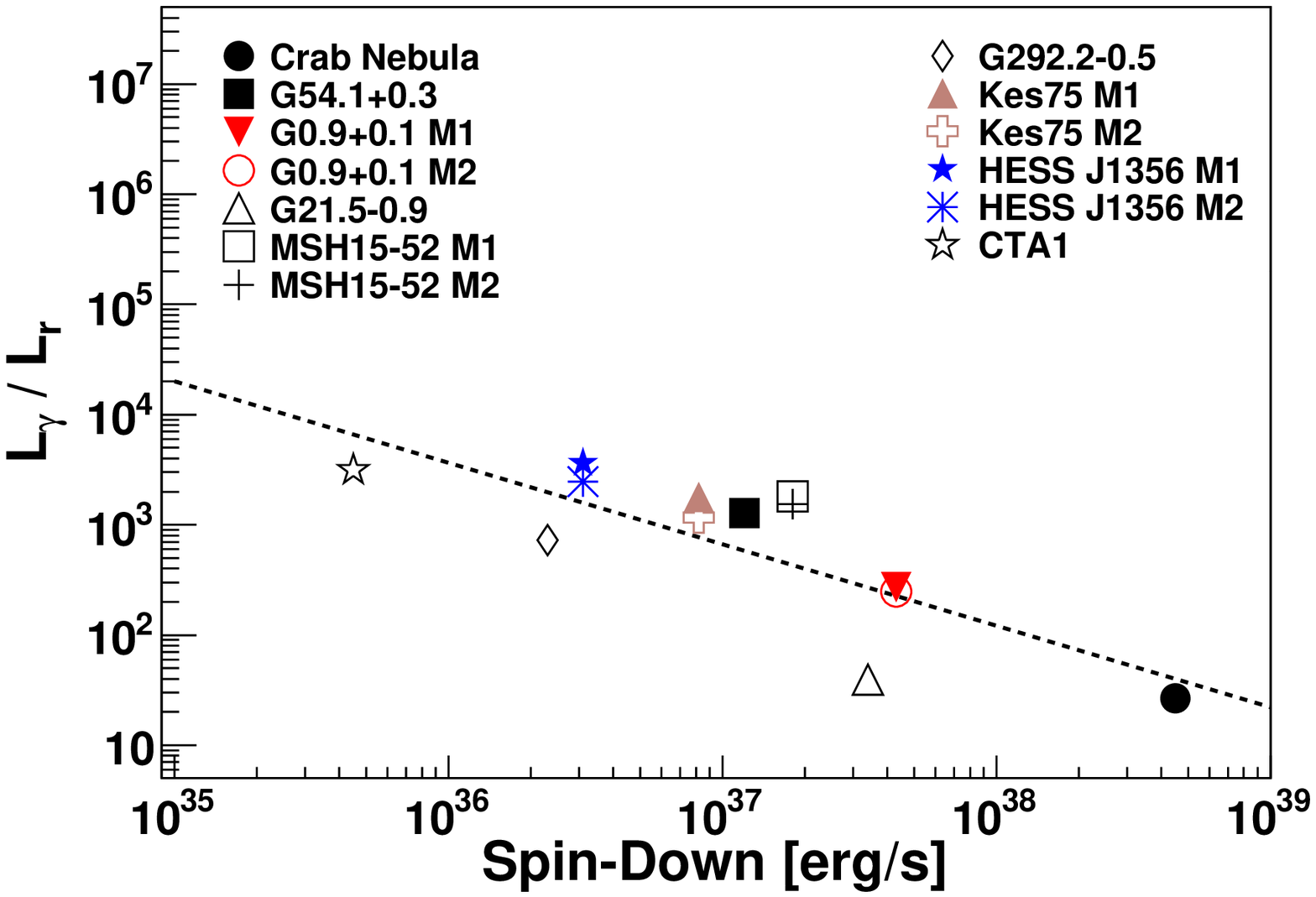}\hspace{-.6cm}
\includegraphics[width=63mm]{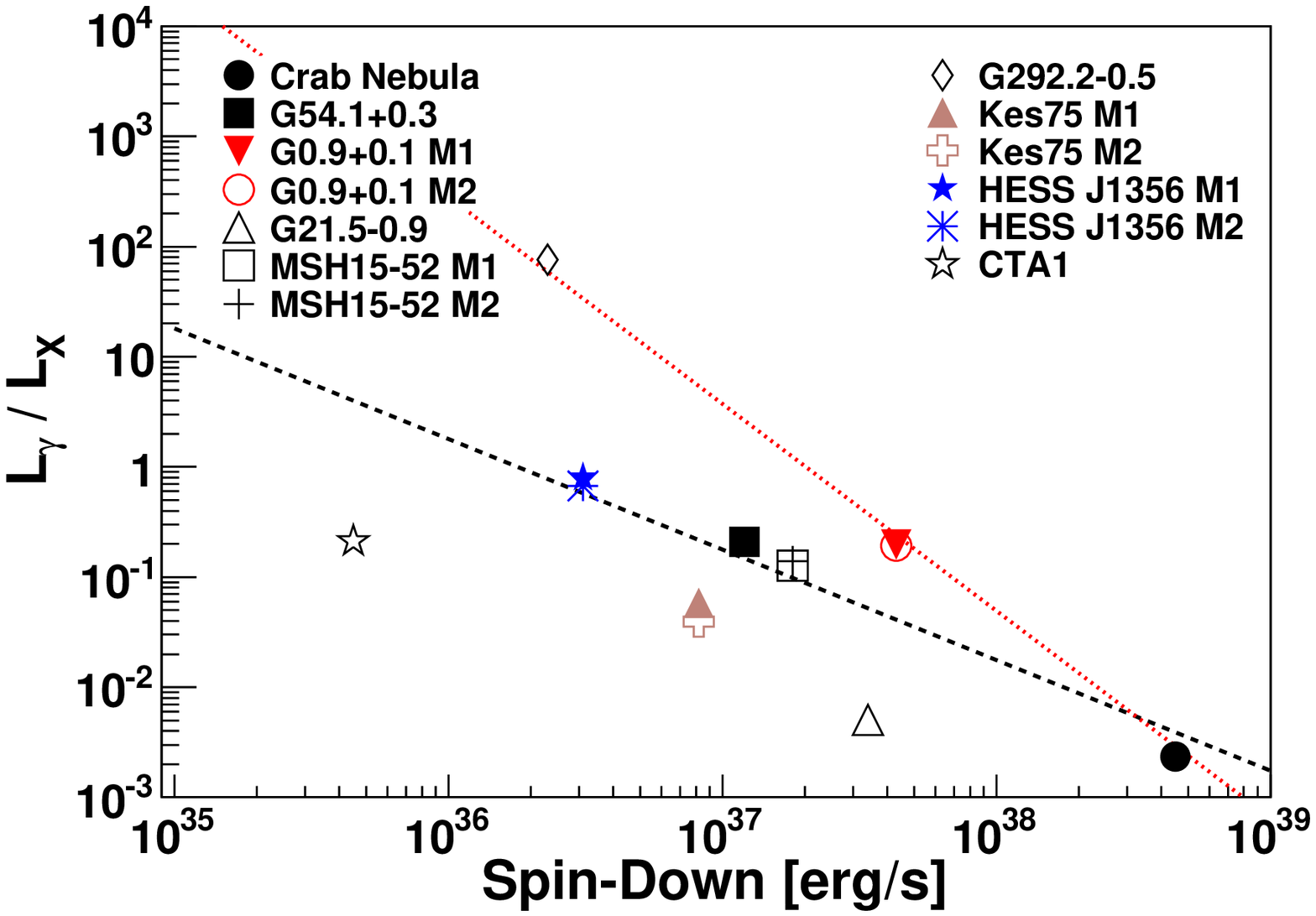}
\caption{Radio, X-ray, and gamma-ray luminosities of young, TeV-detected PWNe as a function of spin-down power and characteristic ages
of their pulsars. Linear fits to the data (black dashed lines) are also shown for magnitudes with a 
high Pearson coefficient (see text for details). Red dashed lines stand for fits
presented in Mattana et al. (2009) using observational data on pulsars of up to 10$^5$ years of age. The bottom row shows 
the ratios between the X-ray and radio, gamma-ray and radio, and gamma-ray and X-ray luminosities. } 
\label{mattana}
\end{figure*}

Possible correlations between the luminosities obtained from our models and two of the 
main features of the central pulsars, their spin-down power and characteristic age, are explored in Fig. \ref{mattana}.
It shows the distribution of radio, X-ray, and gamma-ray luminosities, and their ratios  (see Table \ref{models}) 
as a function of spin-down power and characteristic ages. A line is added (and parameters are shown in Table \ref{fits}) when the Pearson coefficient is such that the correlation is significant to better than 95\% of confidence, as above. A red line is added to those panels for which Mattana et al. (2009) provided a fit when considering observational values of TeV-detected PWNe up to 10$^5$ years of age.
 
The possible correlation of the PWN luminosities with the PSR 
characteristic ages (second row in Fig. \ref{mattana}) is not clear for young PWNe; for $L_r$ and $L_x$ we actually do not find them at the
confidence cut imposed. At the latter case, however, the fit by Mattana et al. (2009) is in agreement with the overall (visual) trend of our sample. The only correlation barely surviving  our 95\% confidence cut is the one between $\tau$ and $L_\gamma$ (see Table \ref{fits}), which Mattana et al. (2009) did not find. We see that the larger the characteristic
age the lower the gamma-ray luminosity. This trend is opposite to the example made in the introduction, where we find more gamma-ray luminosity for pulsars with larger $\tau$
when all other parameters were the same, and thus requires a careful look. On the one hand,  we have in our sample cases of similar spin-down power and $\tau$, for 
G21.5--0.9 and G0.9+0.1; but different real (or assumed real) age (the age assumed for G0.9+0.1 is a factor of 2 to 4 
larger than that of G21.5--0.9).  In this case, one should also expect variance in $L_\gamma$ (being smaller for the youngest, as found) even if all other parameters influencing the gamma-ray production are the same (which usually are not). On the other hand, 
CTA~1 (at the extreme of the distribution) has the largest magnetization and lowest spin-down power of the sample, what reduces
its gamma-ray luminosity despite its larger $\tau$.

The possible correlation of the luminosities  with the spin-down power is visually apparent 
for all three luminosities considered (see top row of Fig. \ref{mattana} and Table \ref{fits}), although in the case of the $\gamma$-ray luminosity
the confidence cut is not met (the resulting probability for no correlation  
is $P=6.2 \times 10^{-2}$).
This is compatible with Mattana et al. (2009) results.
The scaling between X-ray luminosities and spin-down power was also noted by Seward \& Wang (1988) and Becker \& Tr\"umper (1998); in the form $L_x \sim 10^{-3} L_{sd}$,  see also Kargaltsev et al. (2009). 
The radio luminosity / spin-down power correlation is the best in the sample we study. 

We have also found correlations in two of the ratios of luminosities explored, $L_\gamma / L_r$ and $L_\gamma / L_x$. That is, when we compare the IC gamma-ray luminosity with the synchrotron generated ones, we find that the larger the spin-down, the smaller their ratio. We have seen above that all three luminosities  apparently 
increase with the spin-down, with the luminosity of the synchrotron components increasing faster. 
The larger the spin-down power, the more particles are in the nebulae and the larger is the maximum energy they attain. However, the timescale for cooling of electrons via radiating synchrotron emission is faster than for IC, and whereas the radio emission is greatly enhanced, the gamma-ray emission grows at slower rate.

We have considered what happens to these correlations when all the systems are evolved to the same pulsar age, at 3000 years. We see that the correlations between the luminosities and the spin-down power (both at 3000 years of age) still appear at our confidence cut level, but their significances worsen with respect to the one pointed out above. This worsening makes for the correlation of the ratio of the luminosities to disappear in this case.

%%%%%%%%%%%%%%%%%%%%%%%%%%%%%%%%%%%%%%%%%%%%
%%%%%%%%%%%%%%%%%%%%%%%%%%%%%%%%%%%%%%%%%%%%

\subsection{PWN versus PSR properties: other parameters}

We now consider possible correlations between other PWN properties resulting
from our fits and those of the central pulsar. 
We compute for each pulsar the surface magnetic field, the potential difference at the polar cap, the light cylinder, and the magnetic field at the light cylinder (assuming the neutron star is a dipole). The definitions used for these quantities are summarized in Table \ref{def2}, as well as the values obtained for all pulsars in our study. These quantities relate to each other and to the spin-down power, all being functions of $P$ and $\dot P$; thus, it is to expect that if we find a correlation of any magnitude with the spin-down power, we would also find it with the potential difference at the polar cap, and the magnetic field at the light cylinder. The spin-down -- surface magnetic field dispersion can introduce different correlations, depending on the values of $P$ and $\dot P$.

%%%%%%%%%%%%%%%%%%%%%%%%%%%%%%%%%%%%%%%%%%%%
\begin{figure*}[t!]
\centering
\includegraphics[width=46mm]{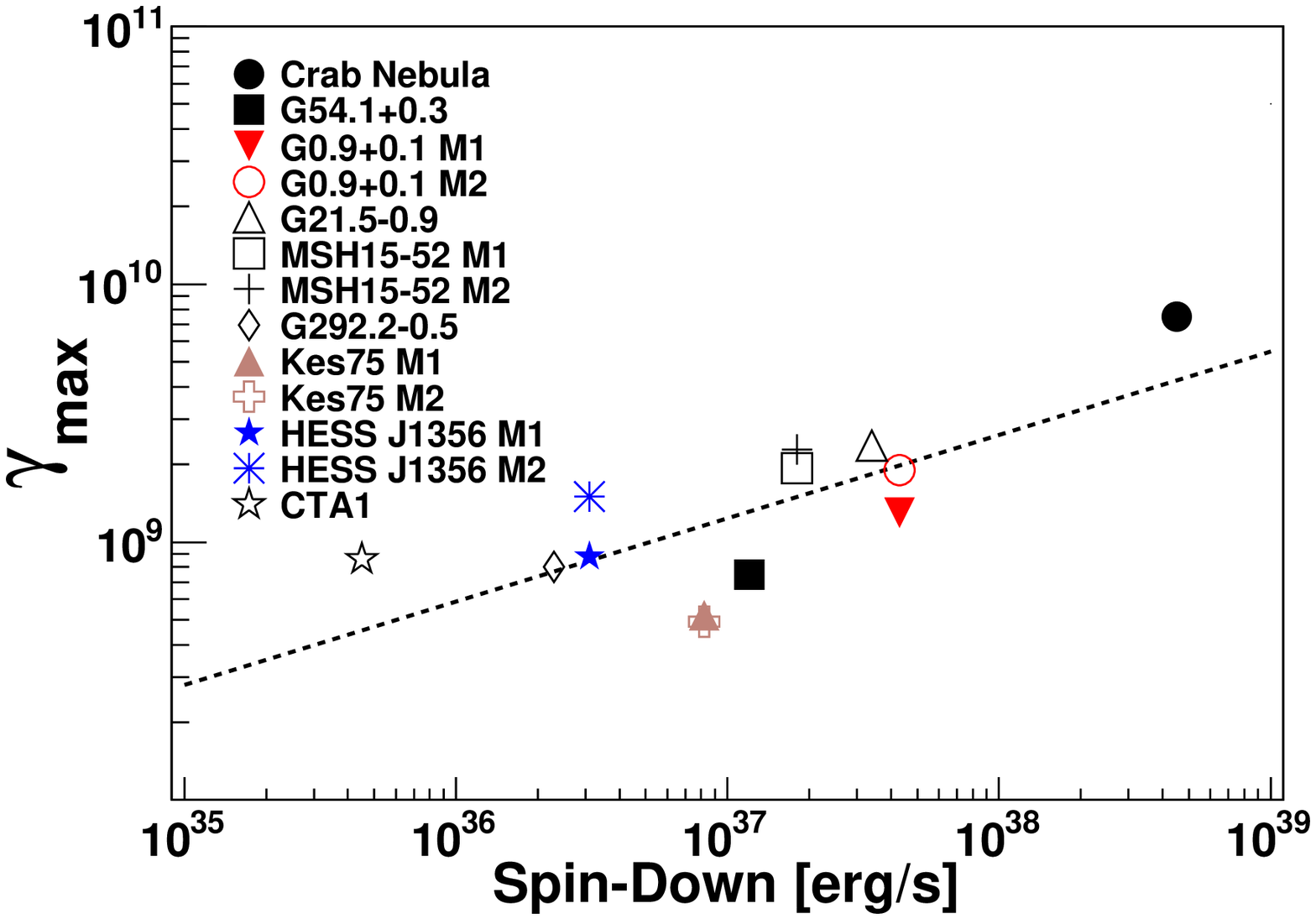} \hspace{-0.3cm}
\includegraphics[width=46mm]{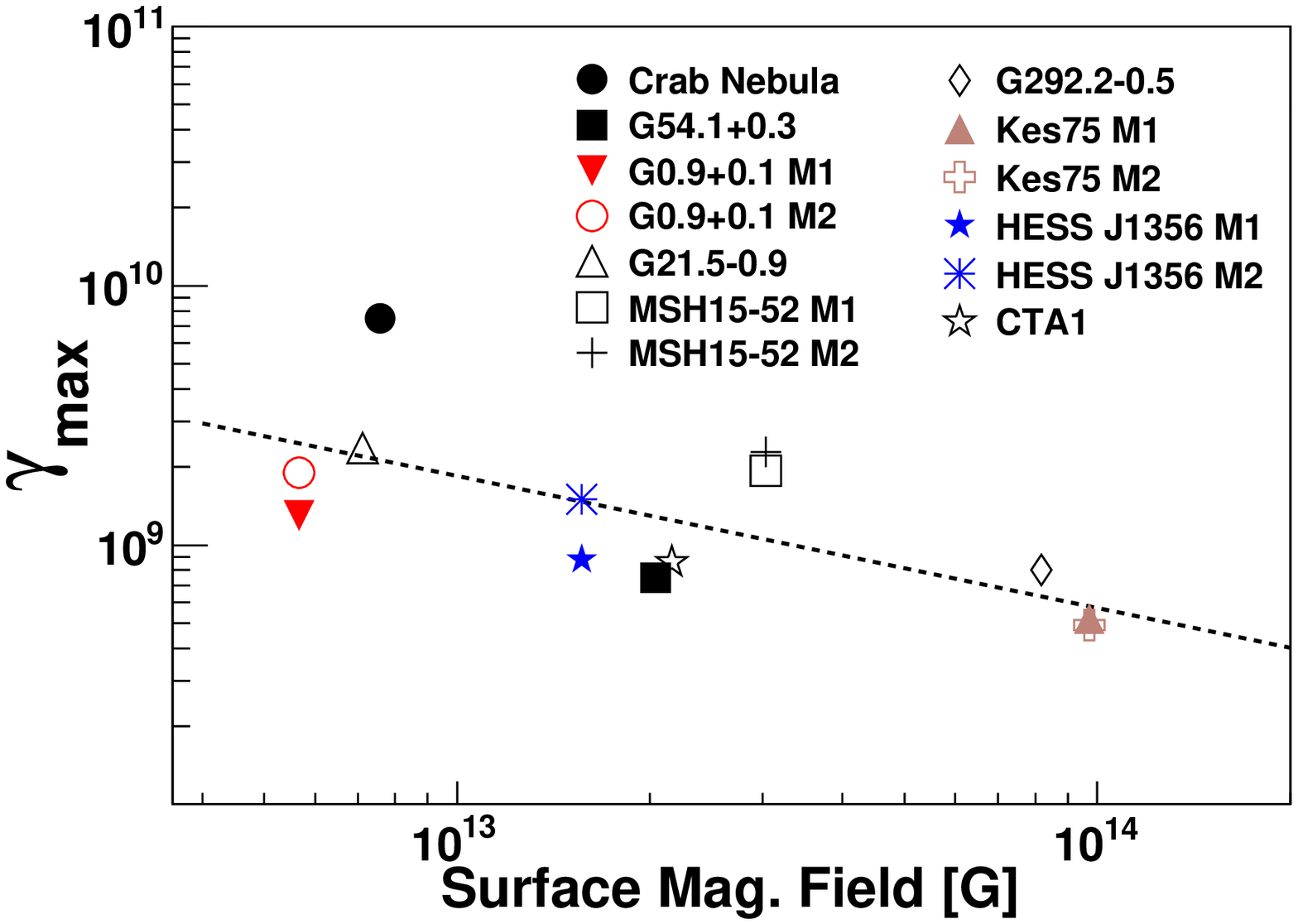} \hspace{-0.3cm}
\includegraphics[width=46mm]{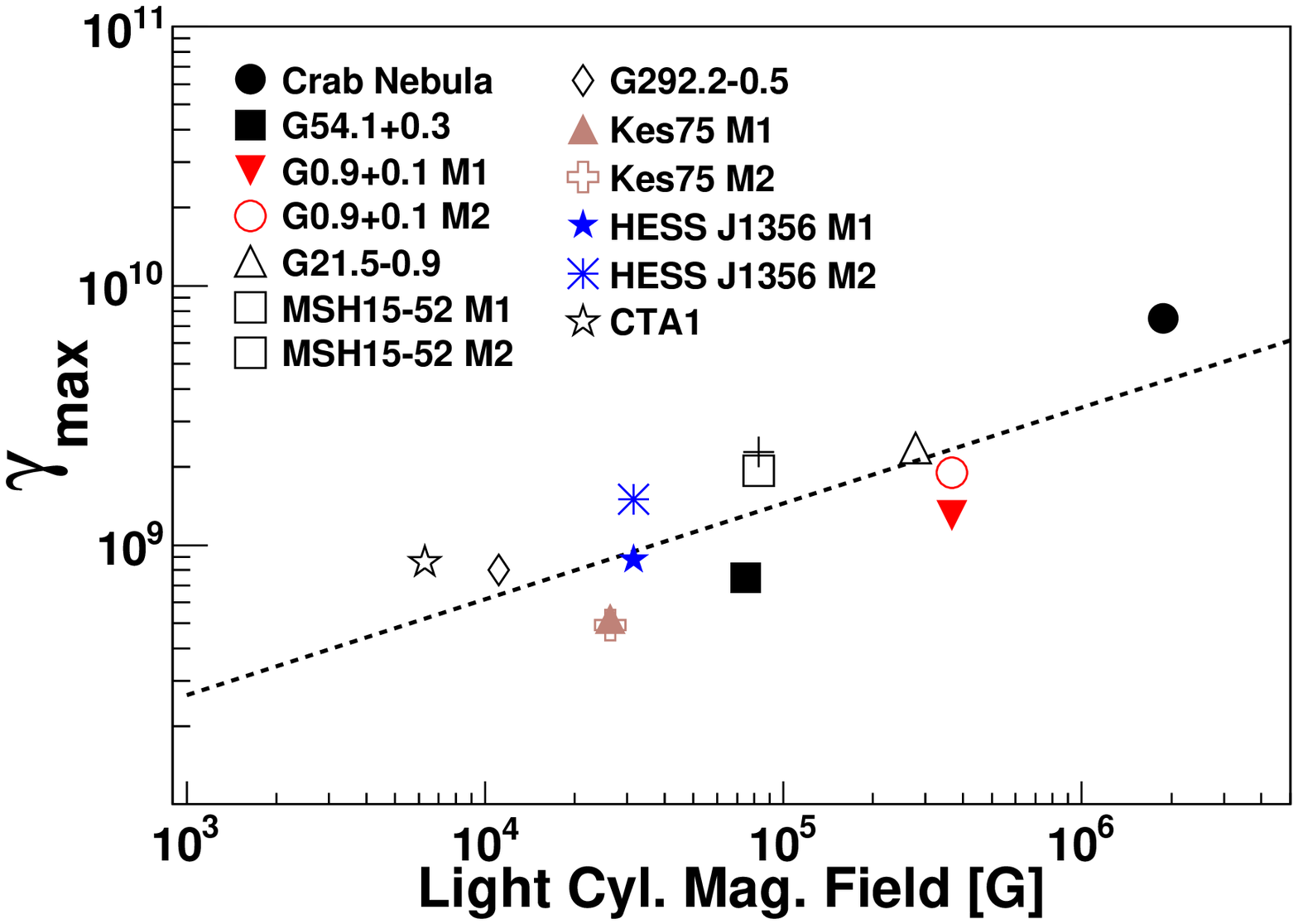}\hspace{-0.3cm}
\includegraphics[width=46mm]{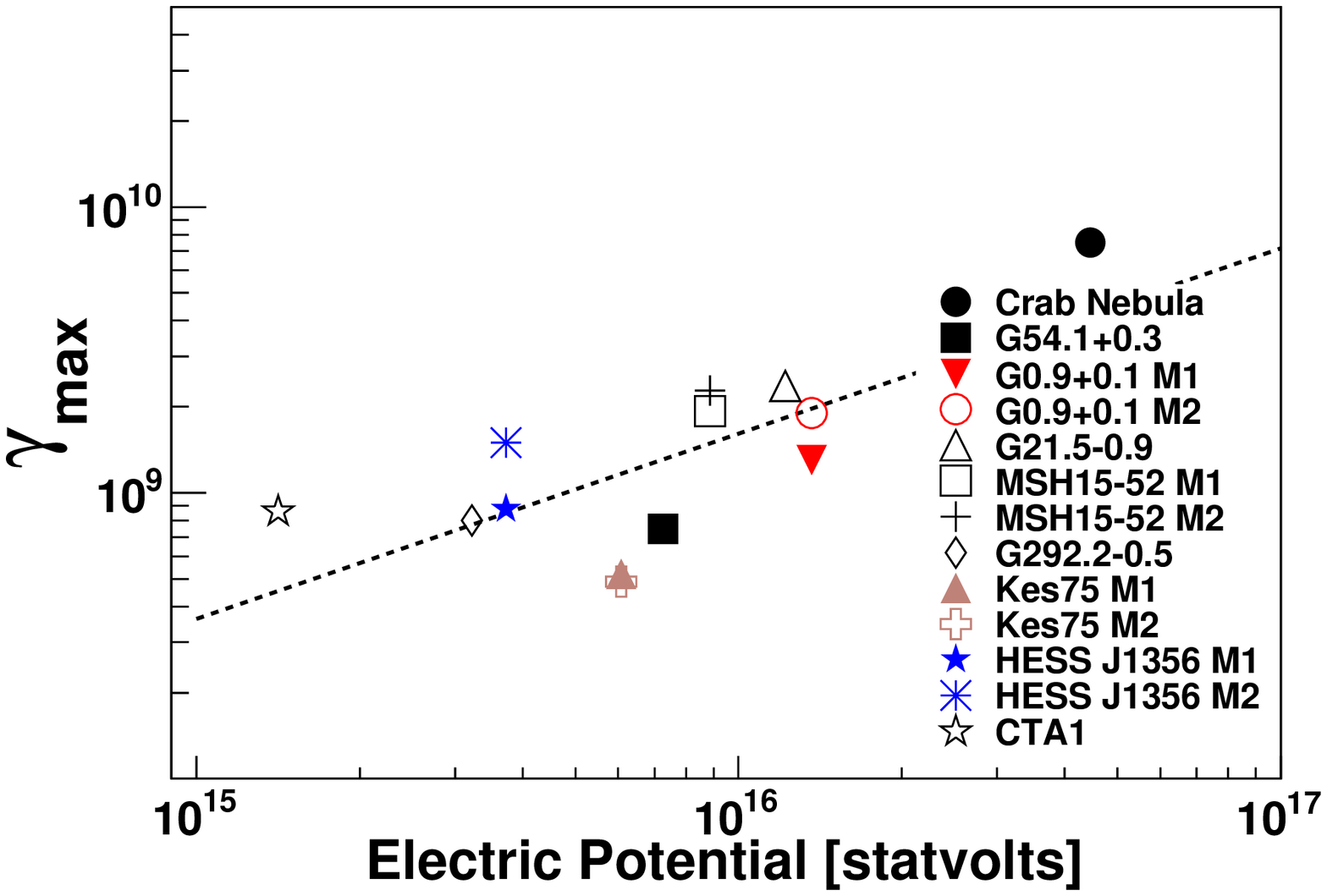} \\
\includegraphics[width=46mm]{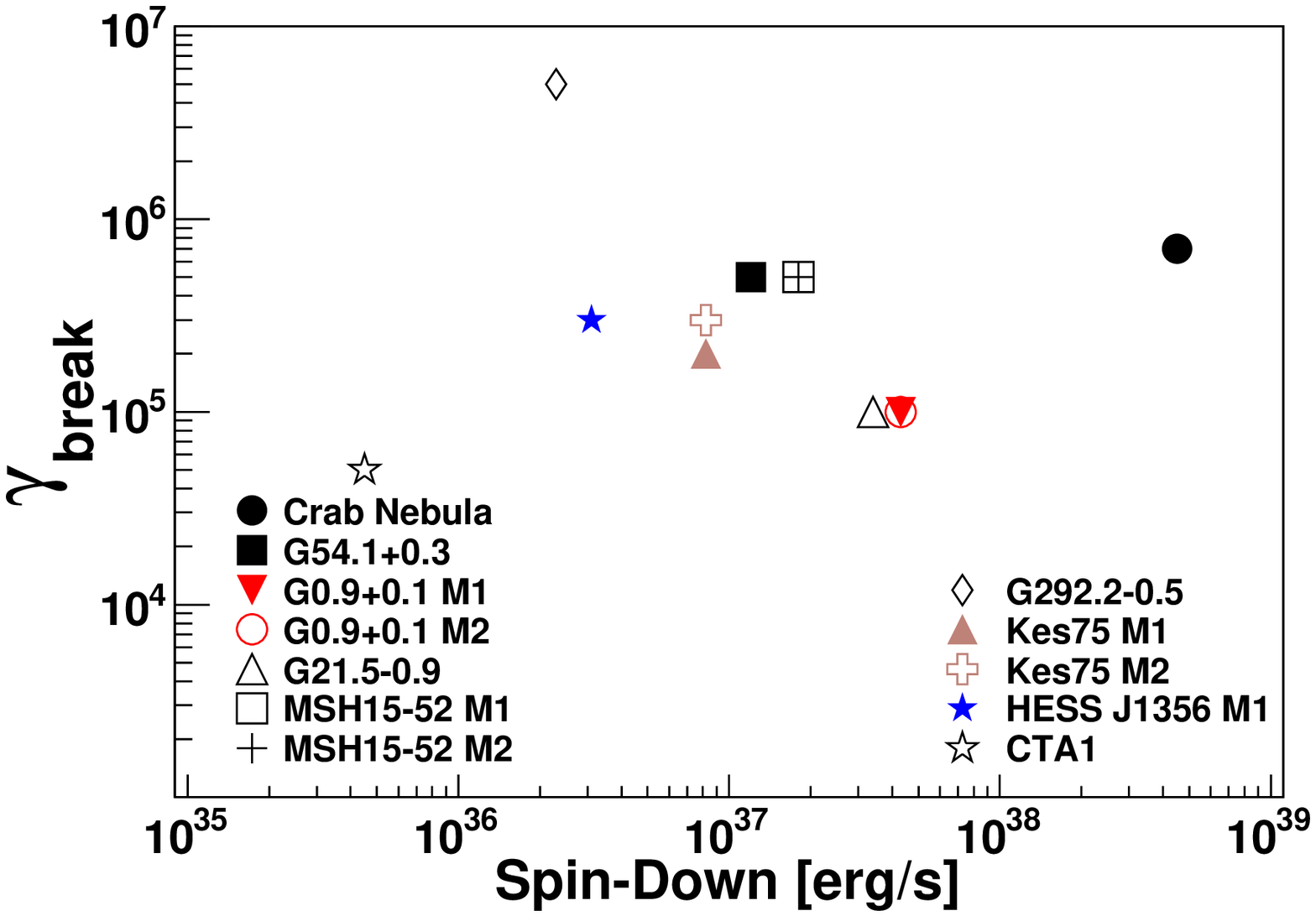}  \hspace{-0.3cm}
\includegraphics[width=46mm]{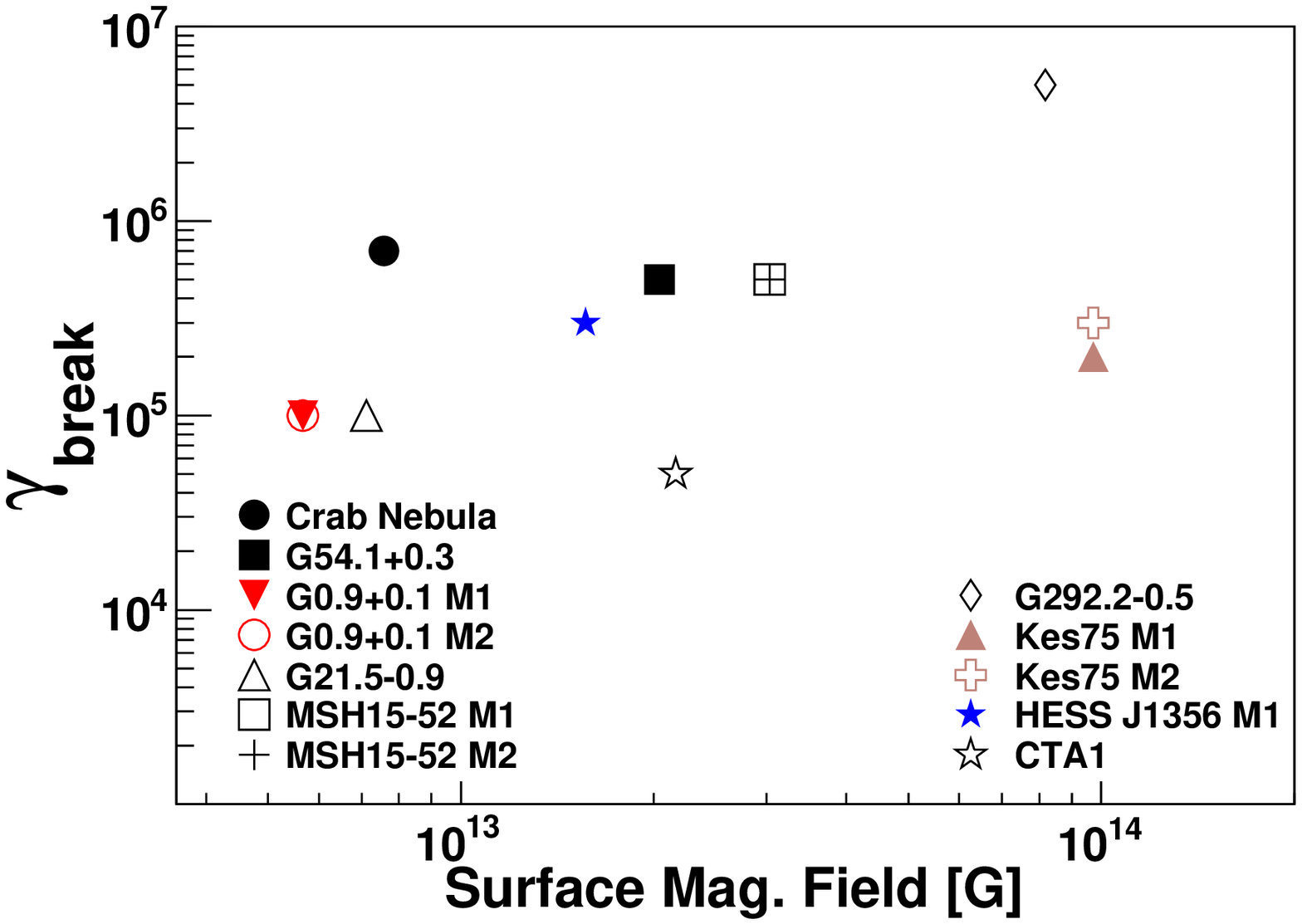}  \hspace{-0.3cm}
\includegraphics[width=46mm]{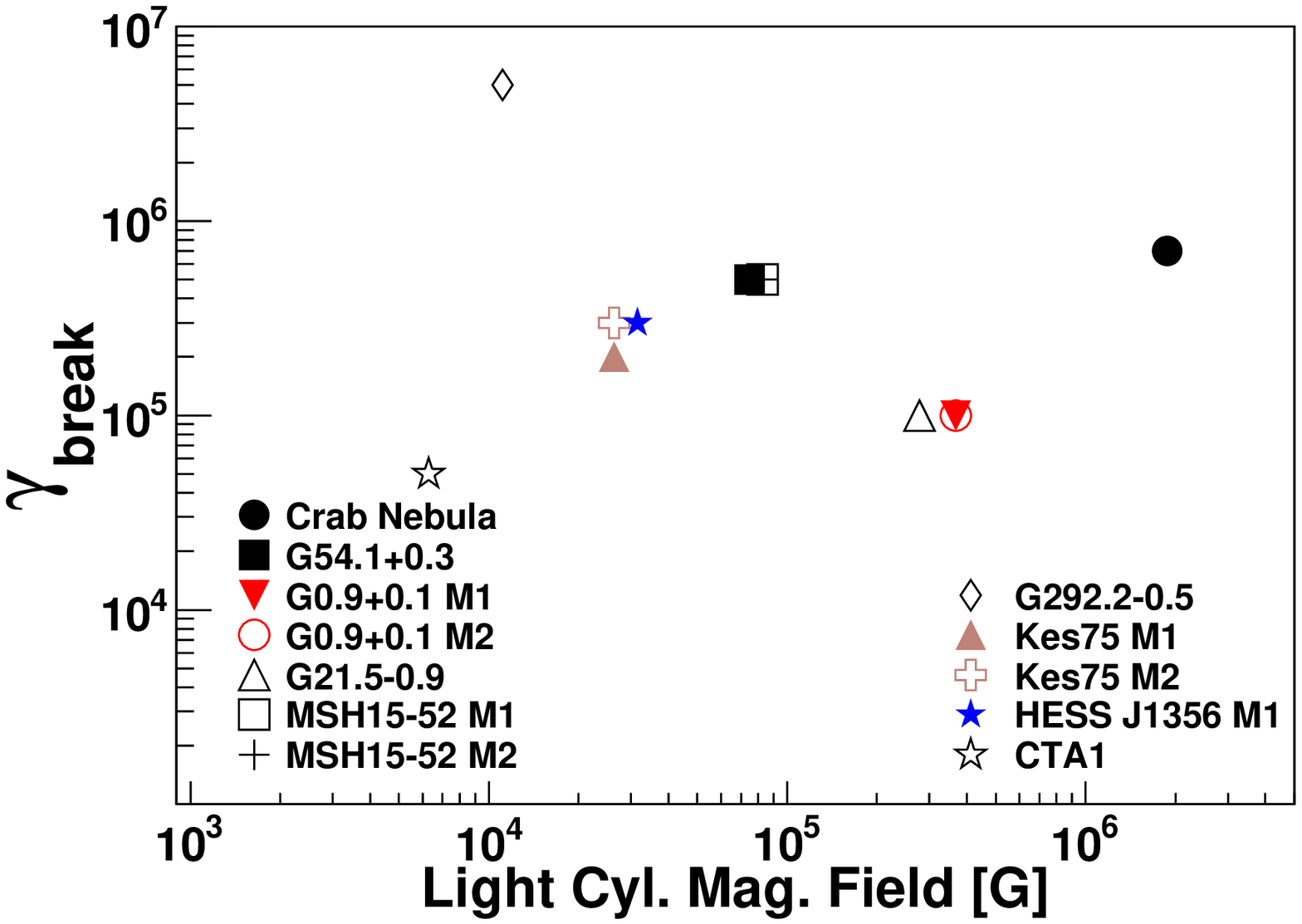}  \hspace{-0.3cm}
\includegraphics[width=46mm]{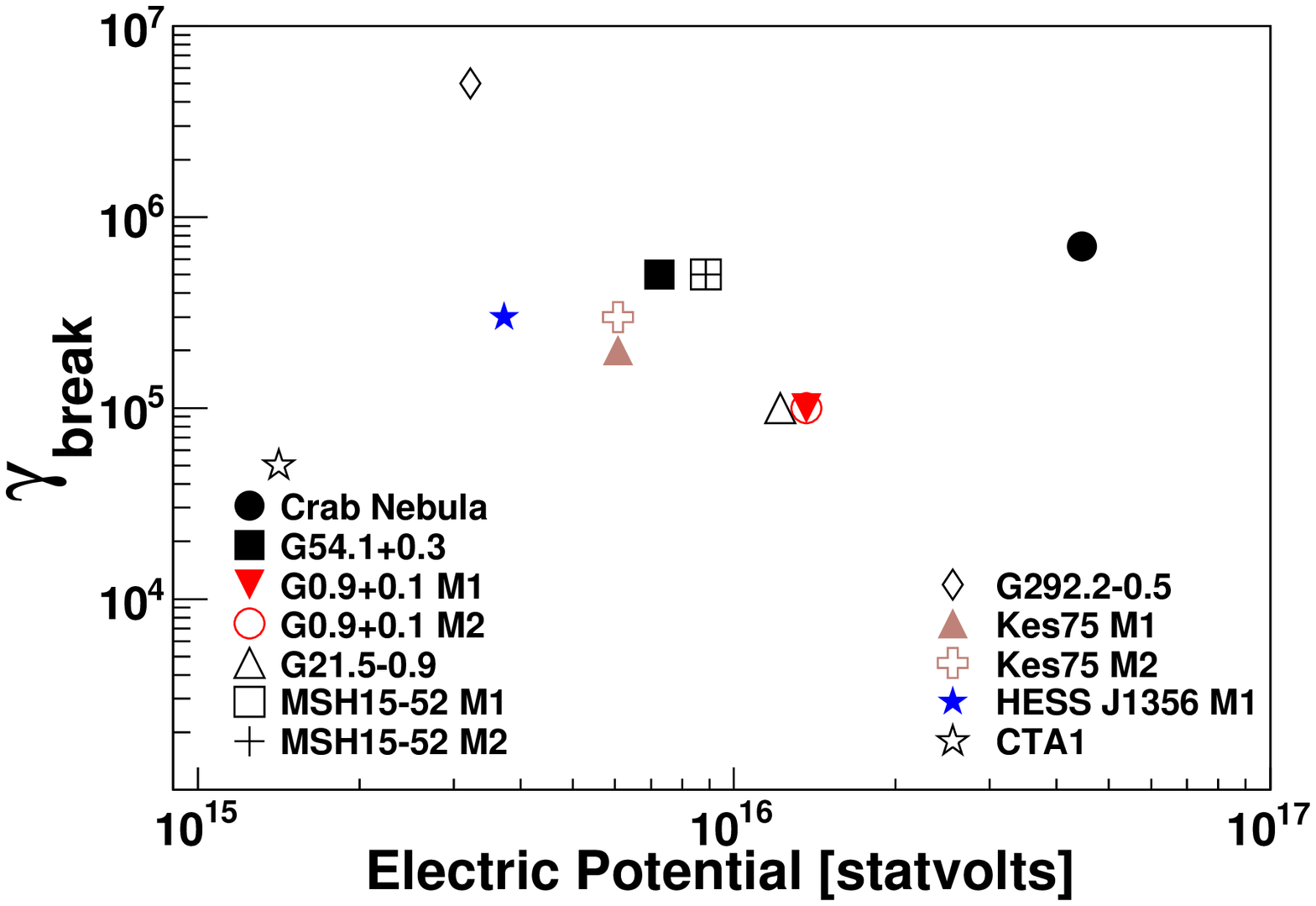} \\
\includegraphics[width=46mm]{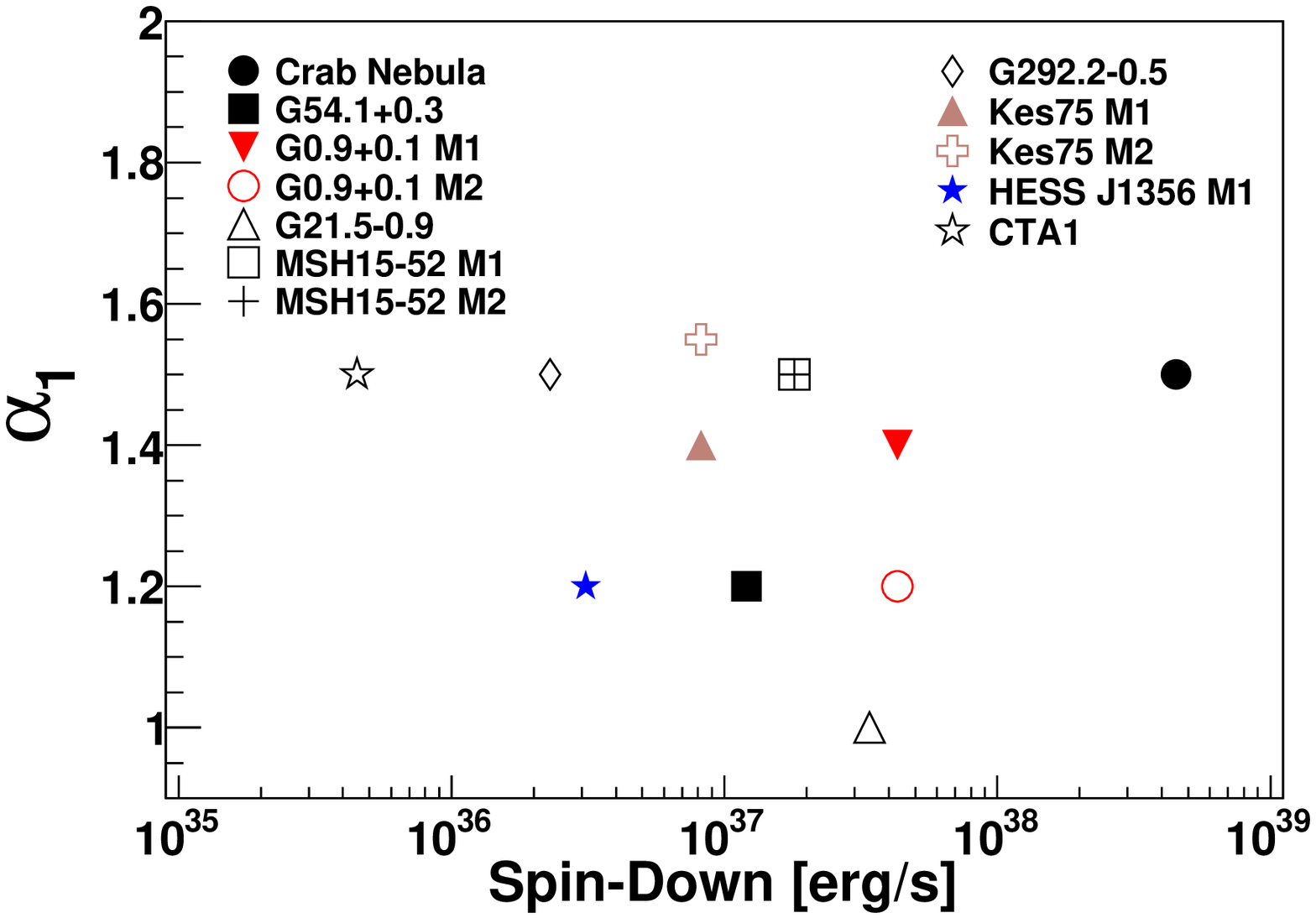} \hspace{-0.3cm}
\includegraphics[width=46mm]{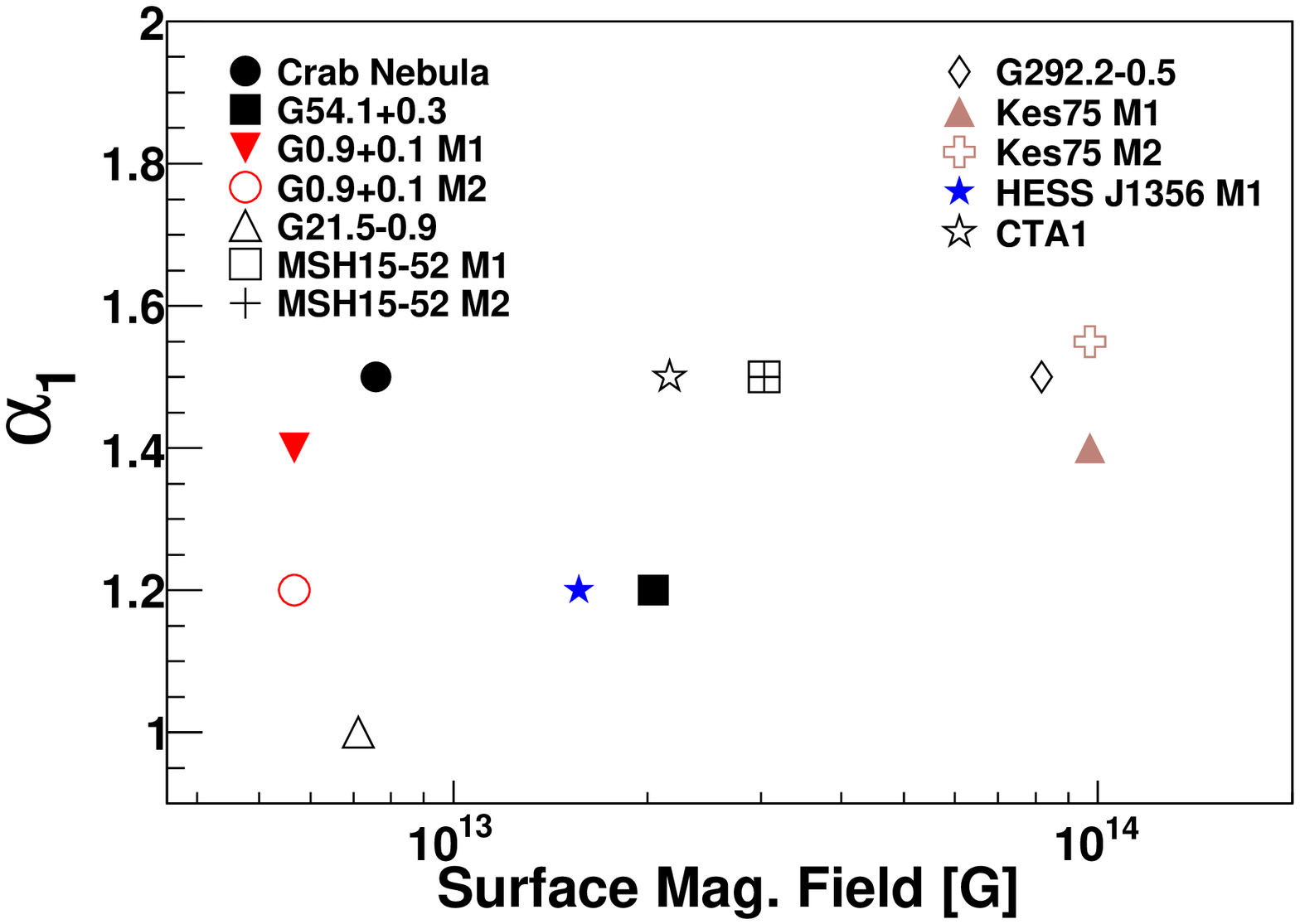} \hspace{-0.3cm}
\includegraphics[width=46mm]{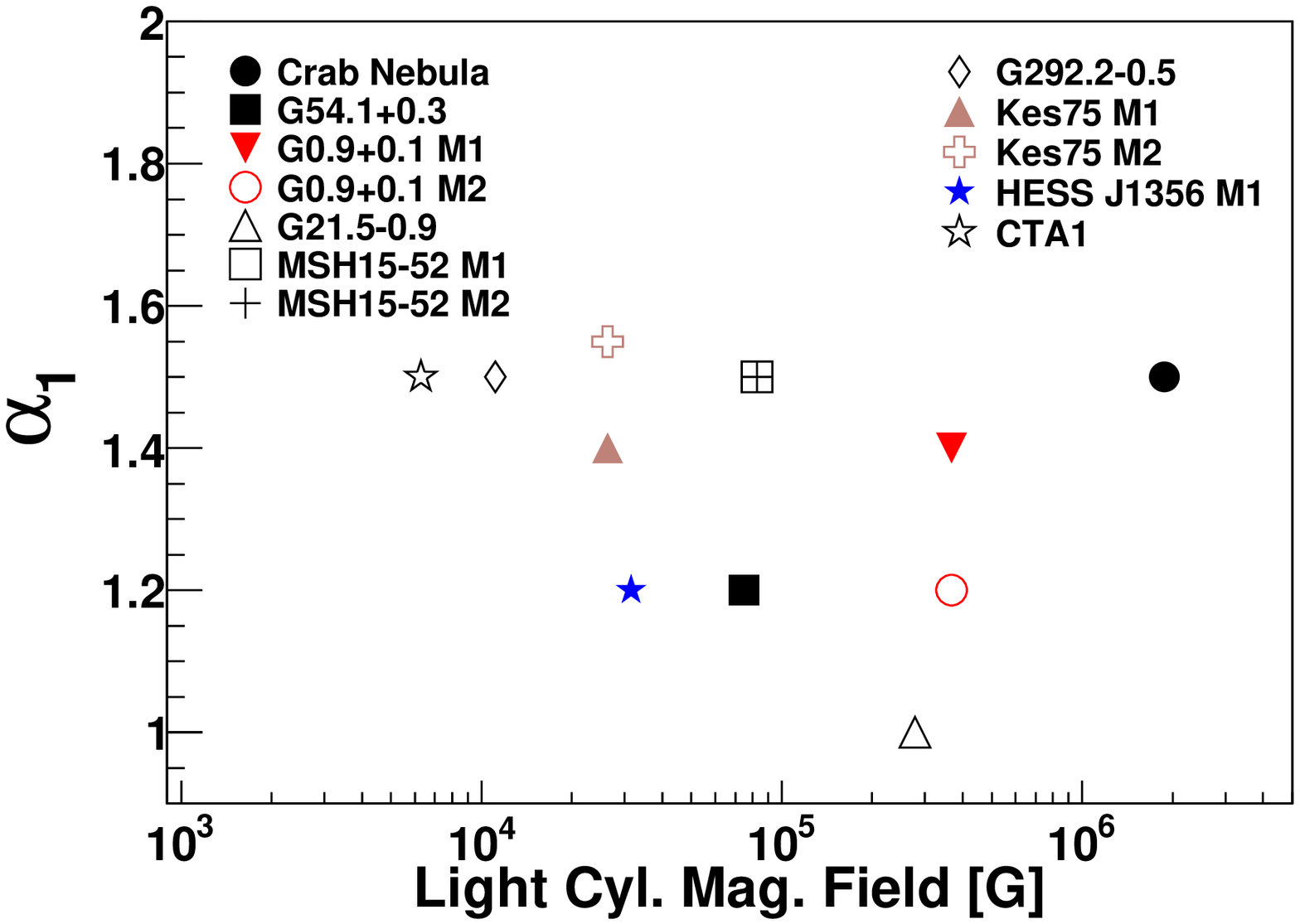} \hspace{-0.3cm}
\includegraphics[width=46mm]{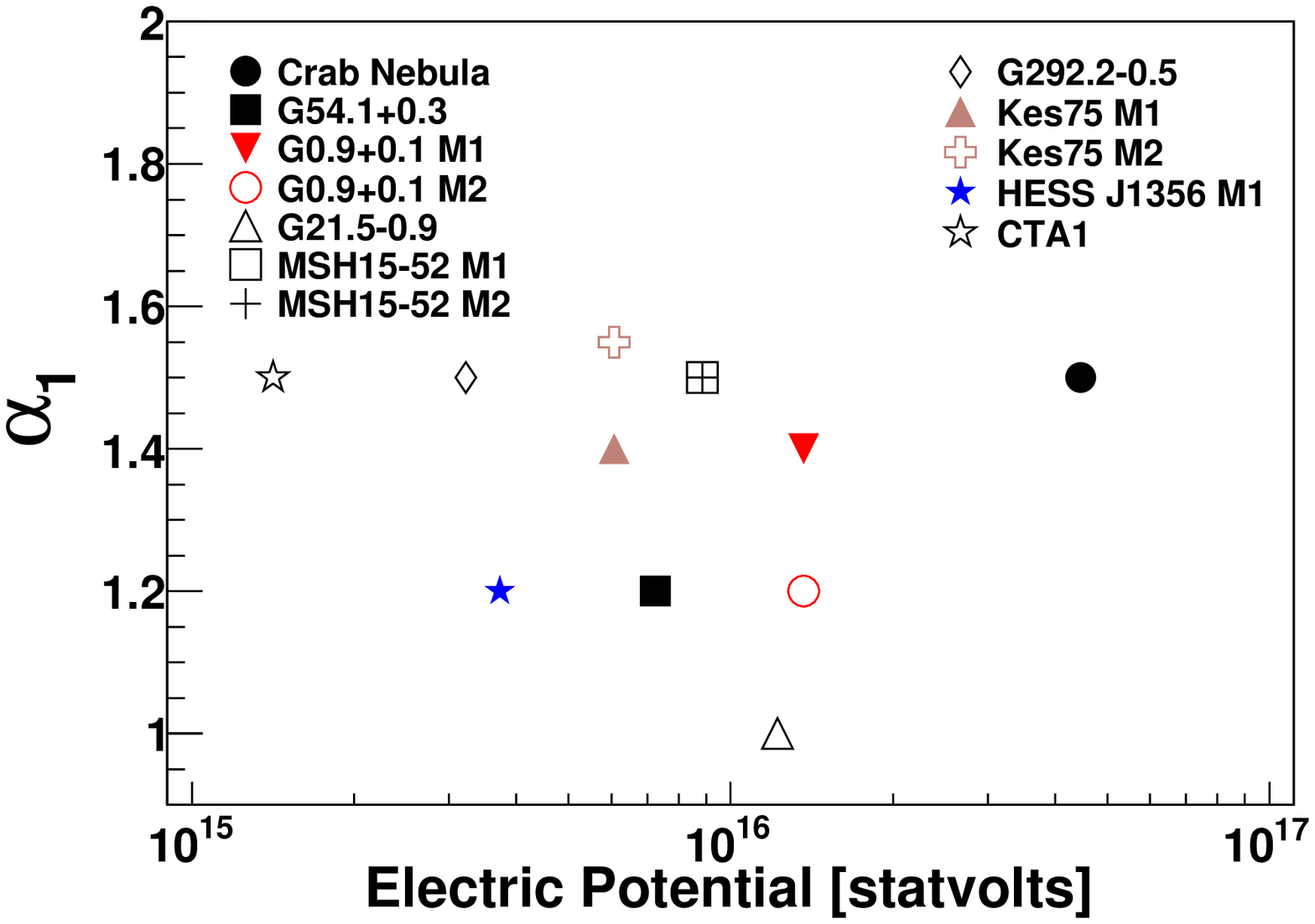} \\
\includegraphics[width=46mm]{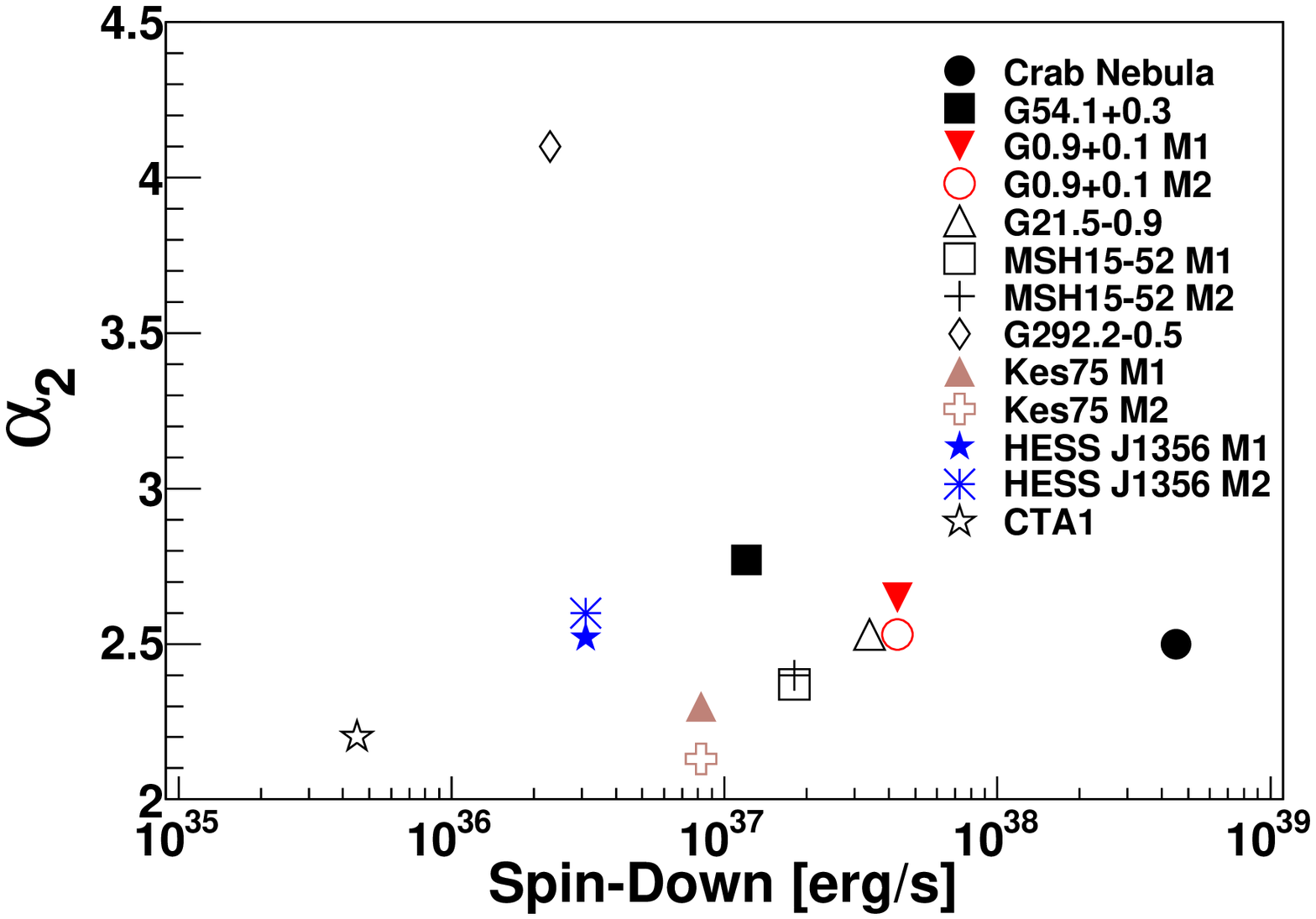} \hspace{-0.3cm}
\includegraphics[width=46mm]{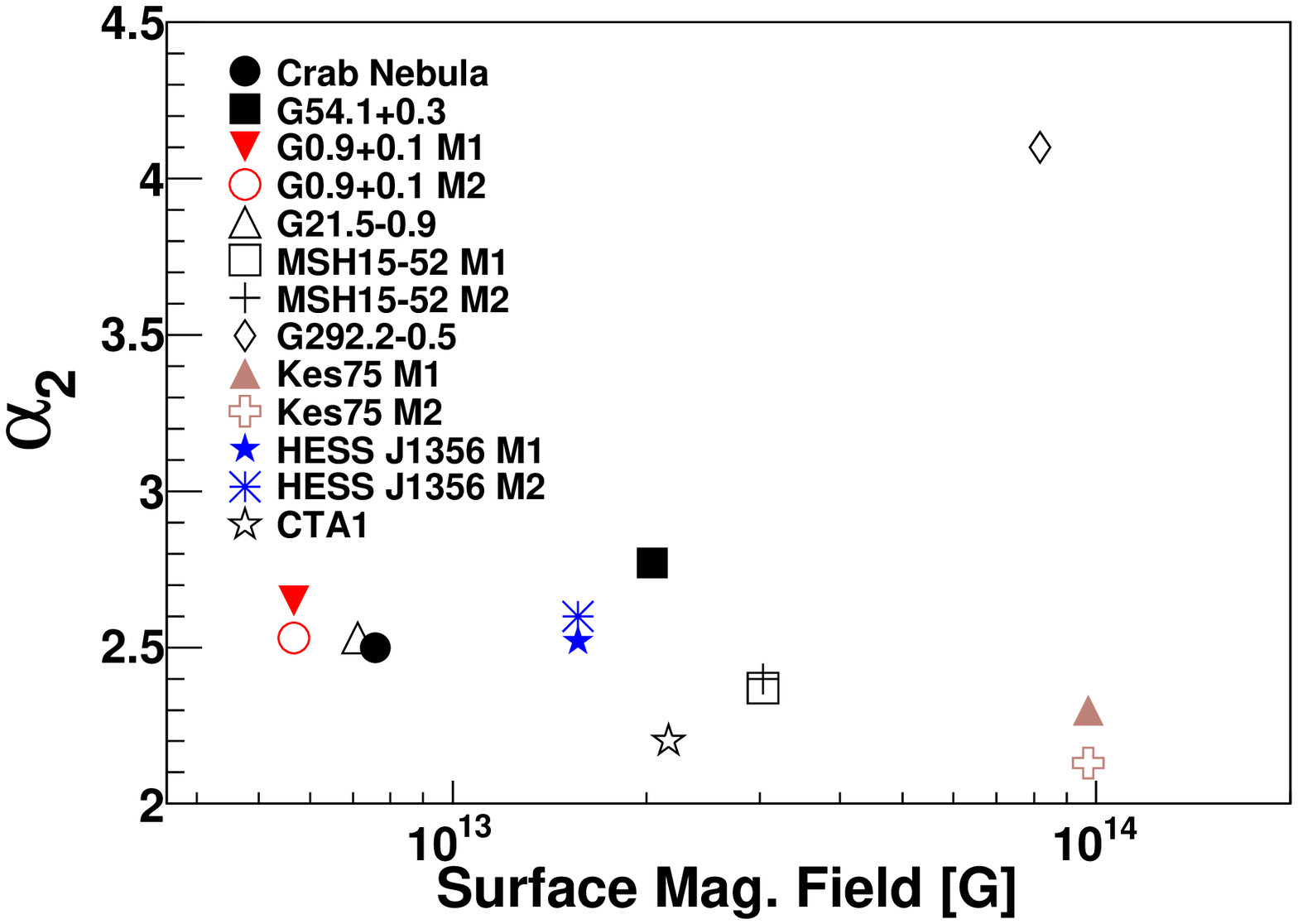} \hspace{-0.3cm}
\includegraphics[width=46mm]{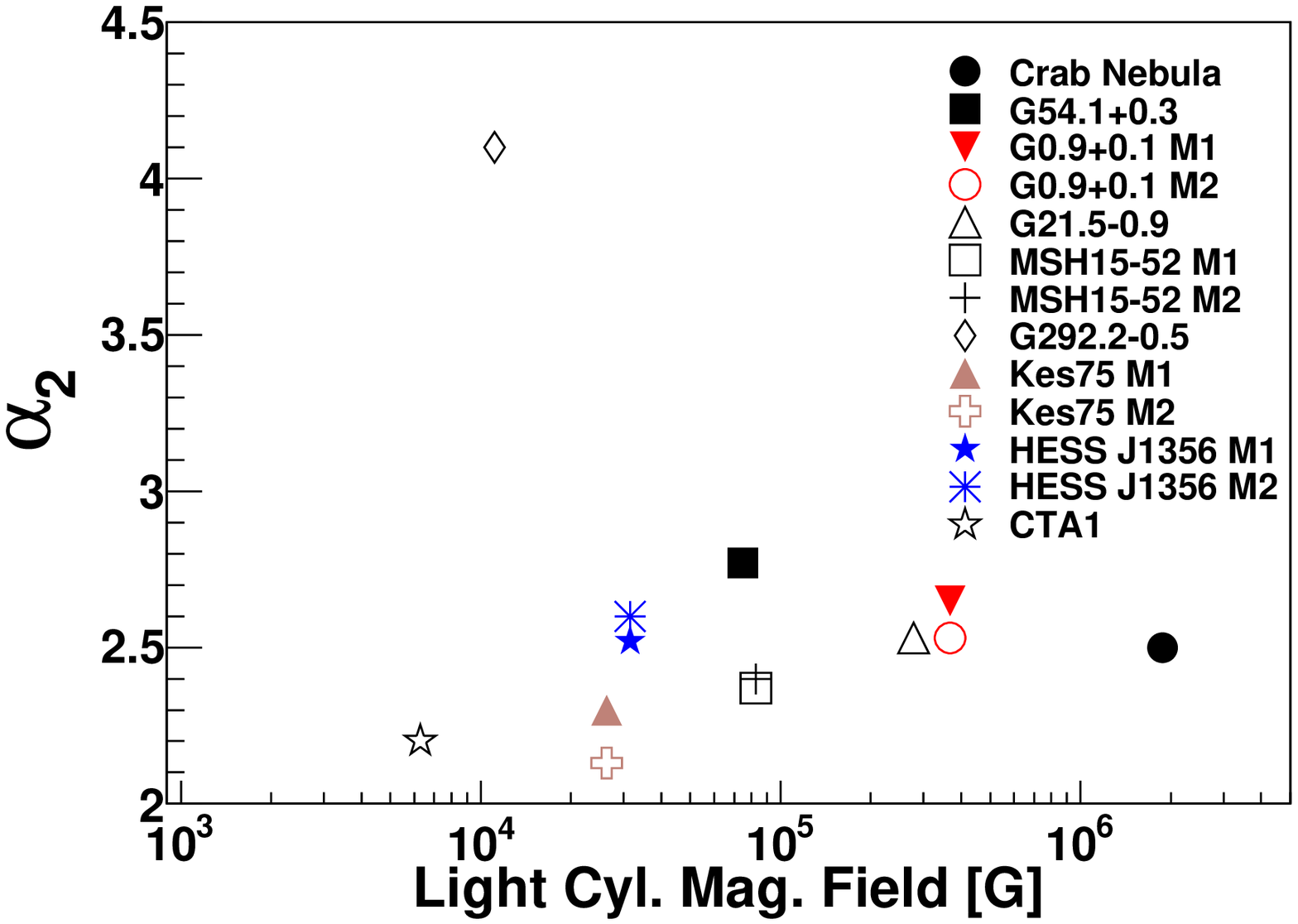} \hspace{-0.3cm}
\includegraphics[width=46mm]{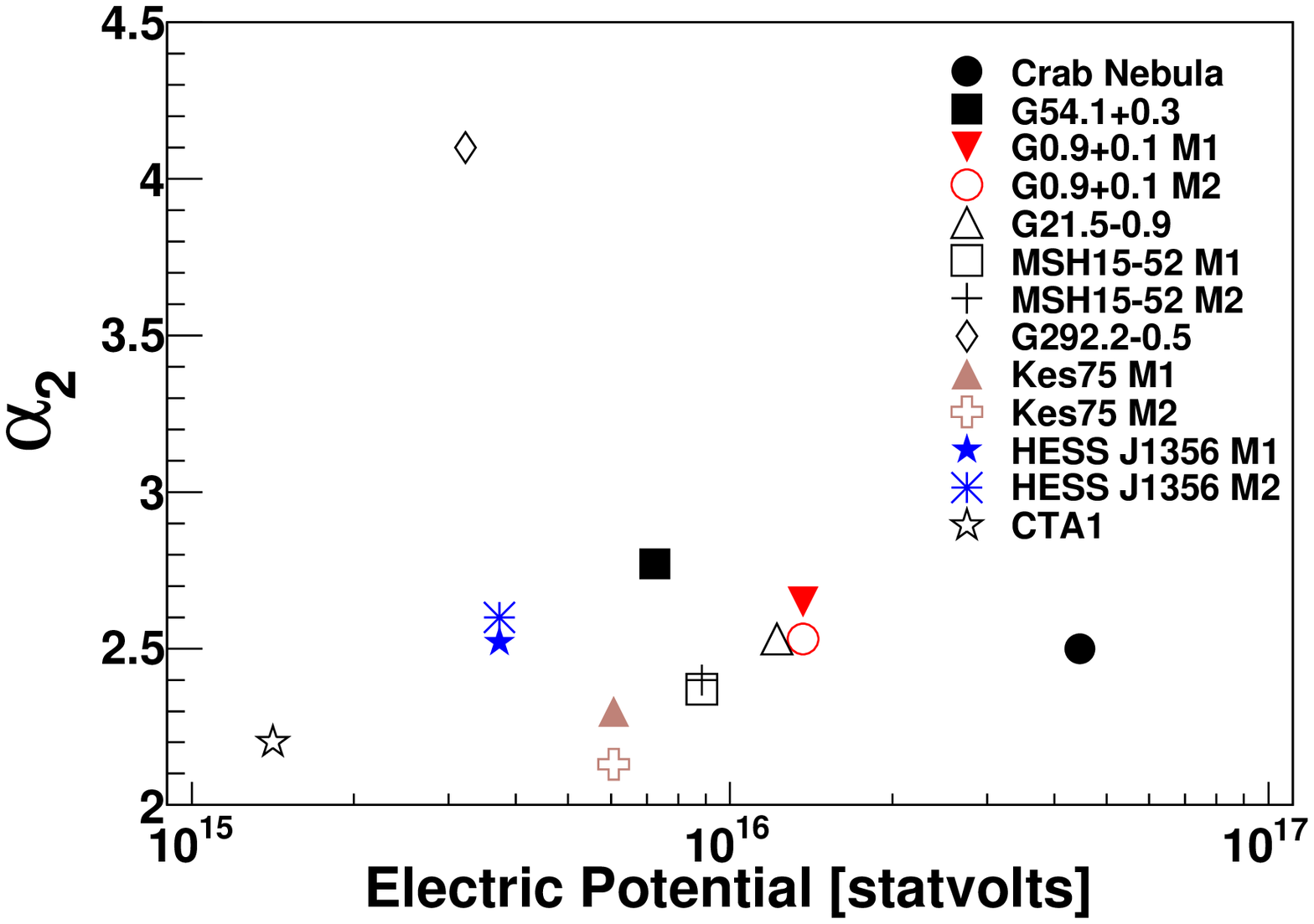} \\
\includegraphics[width=46mm]{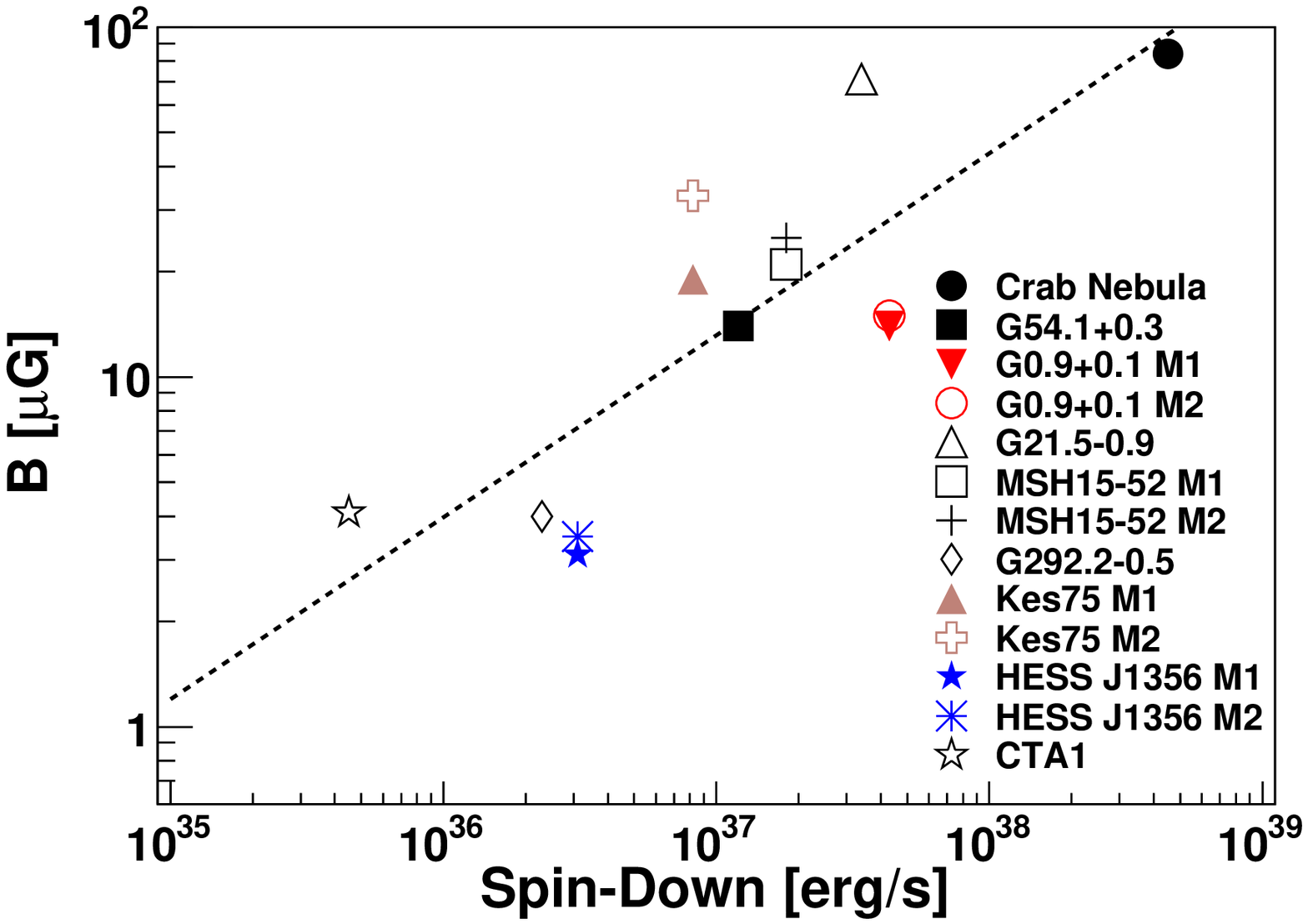} \hspace{-0.3cm}
\includegraphics[width=46mm]{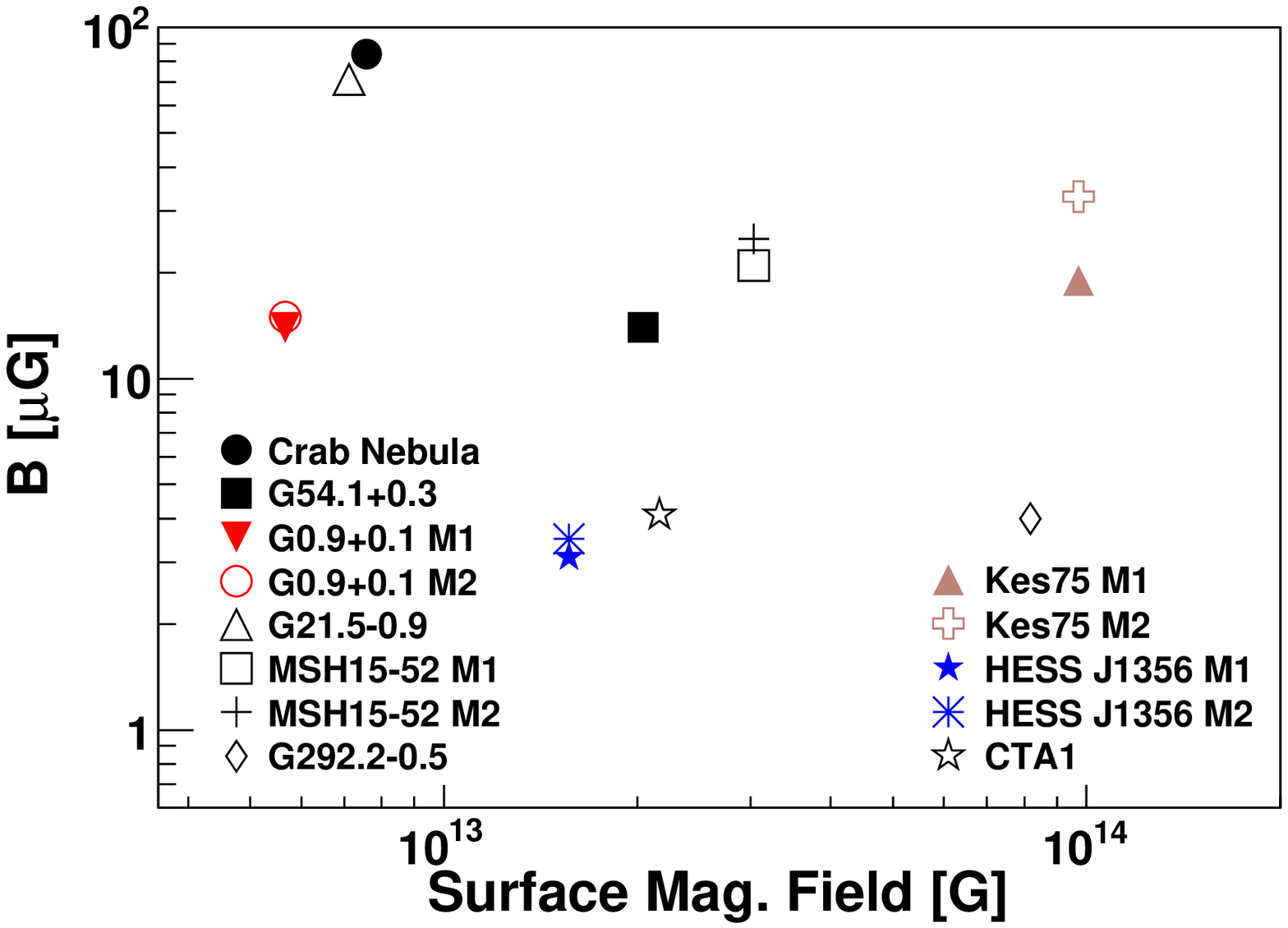} \hspace{-0.3cm}
\includegraphics[width=46mm]{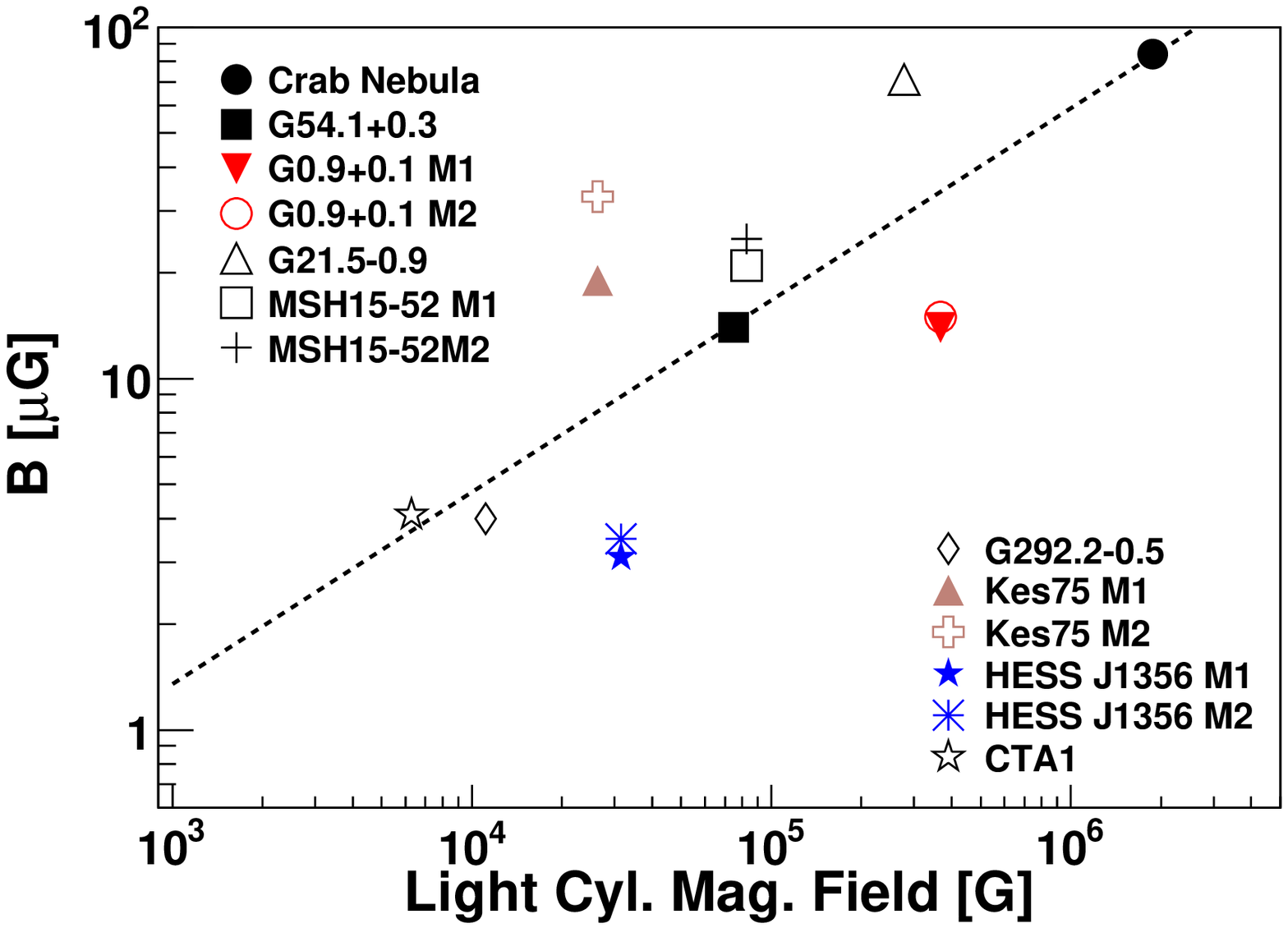} \hspace{-0.3cm}
\includegraphics[width=46mm]{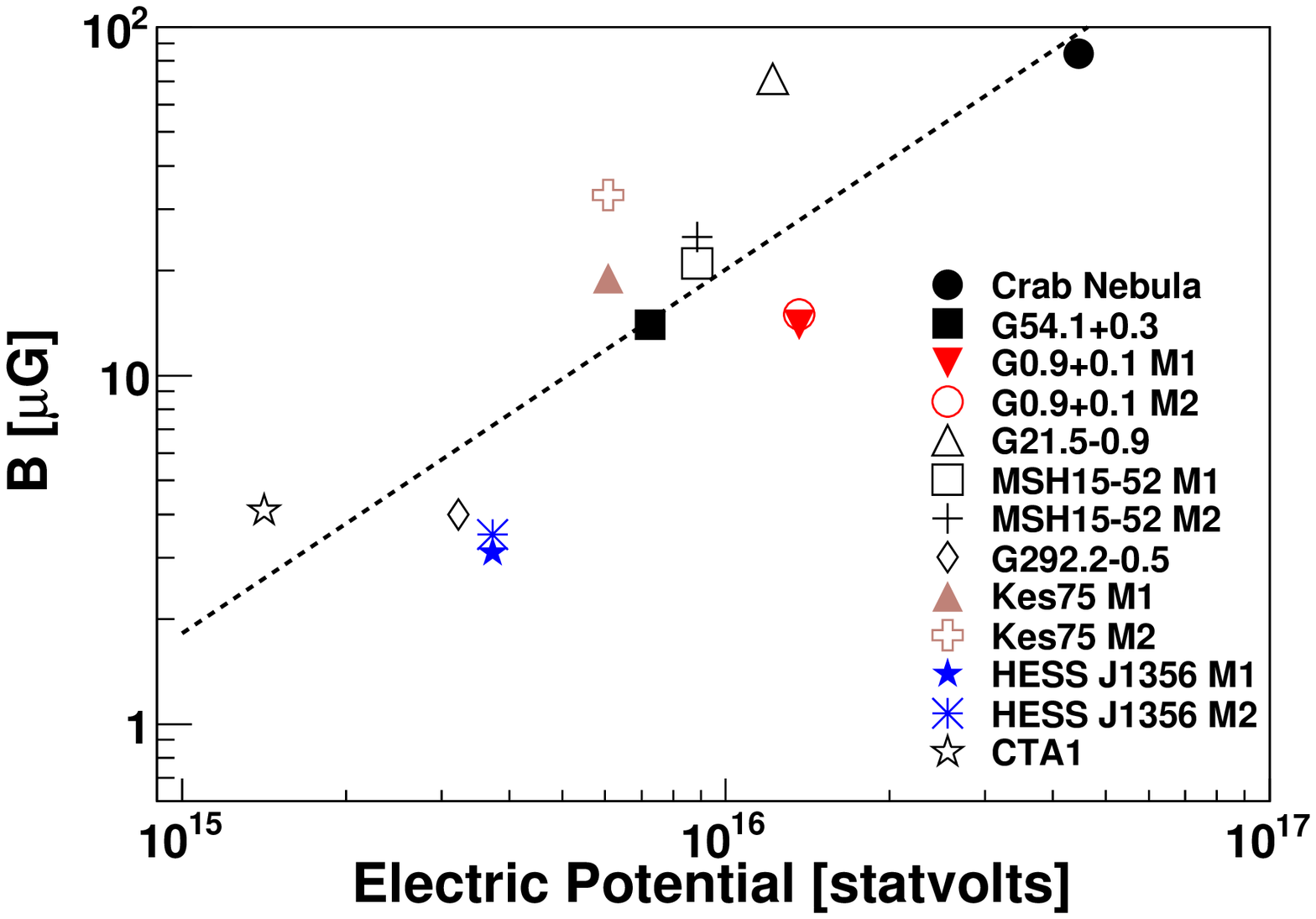} \\
\includegraphics[width=46mm]{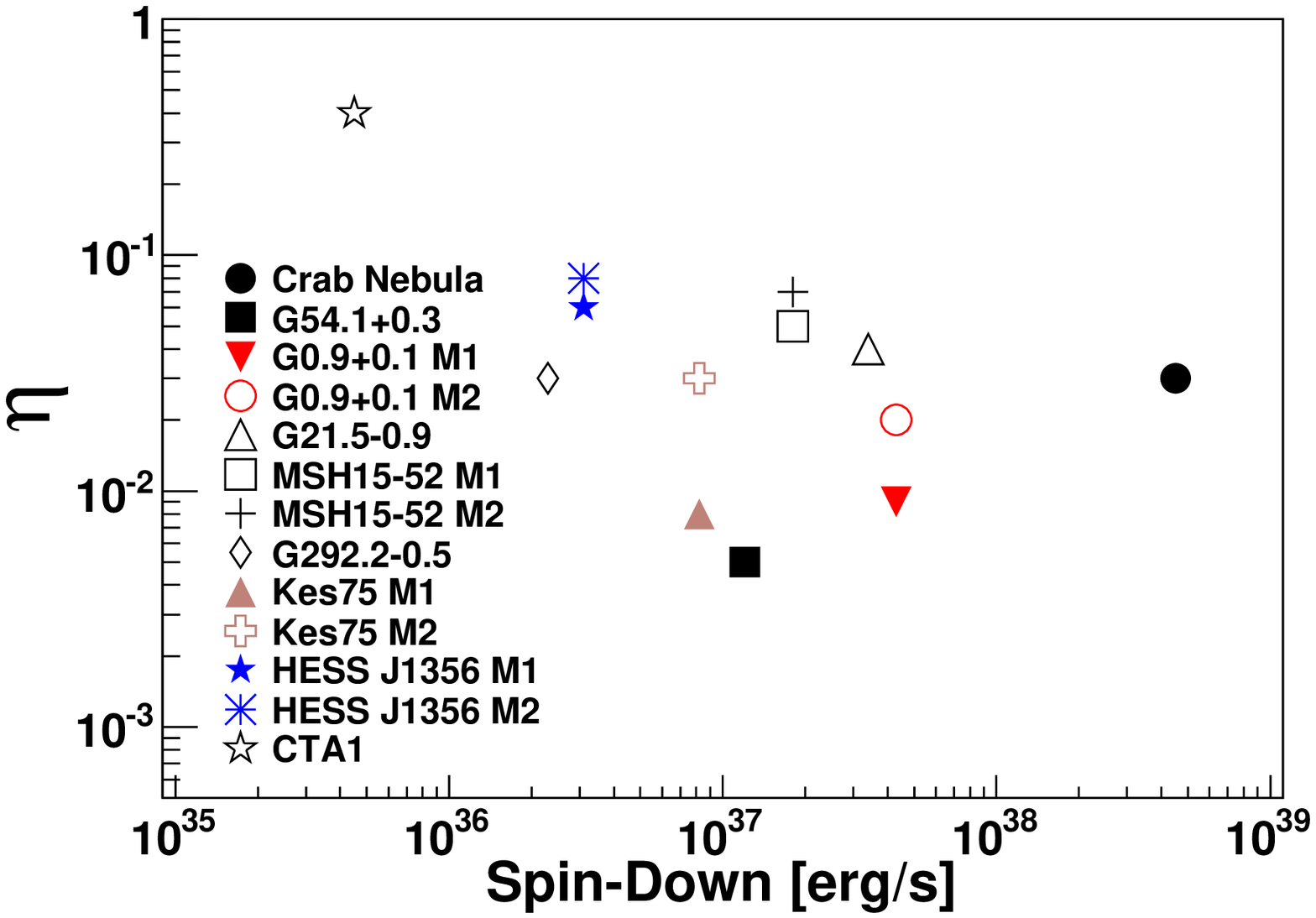} \hspace{-0.3cm}
\includegraphics[width=46mm]{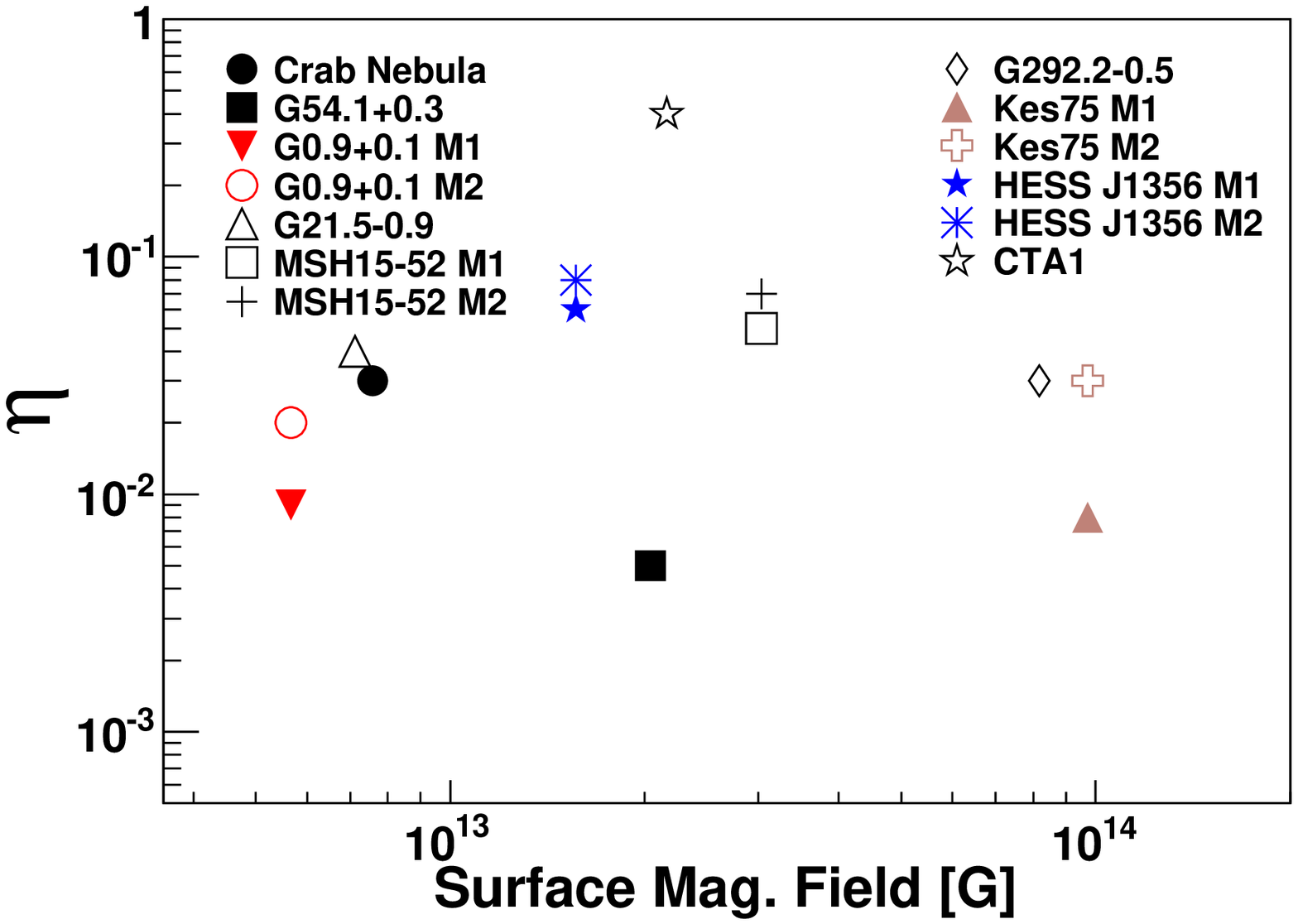} \hspace{-0.3cm}
\includegraphics[width=46mm]{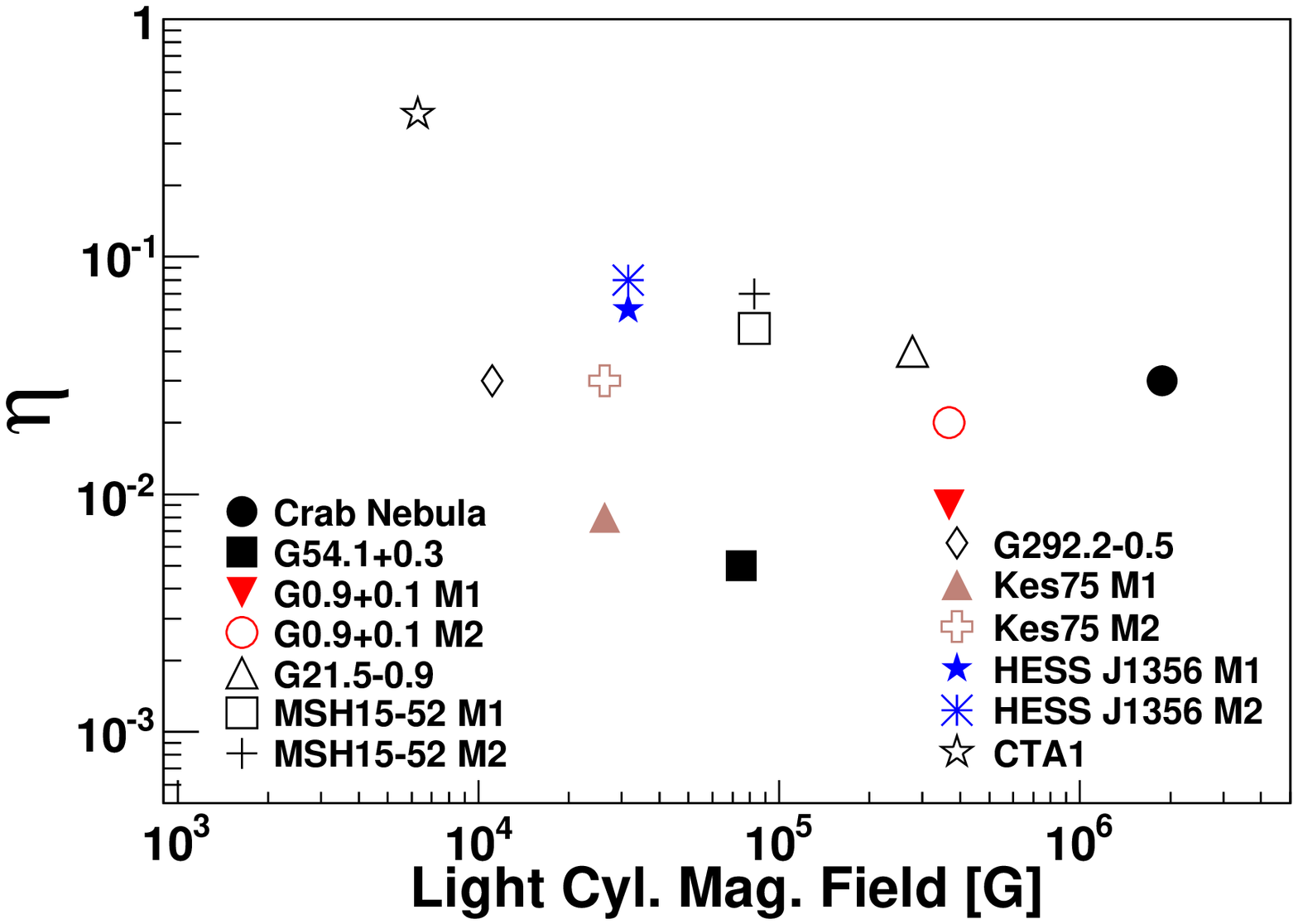} \hspace{-0.3cm}
\includegraphics[width=46mm]{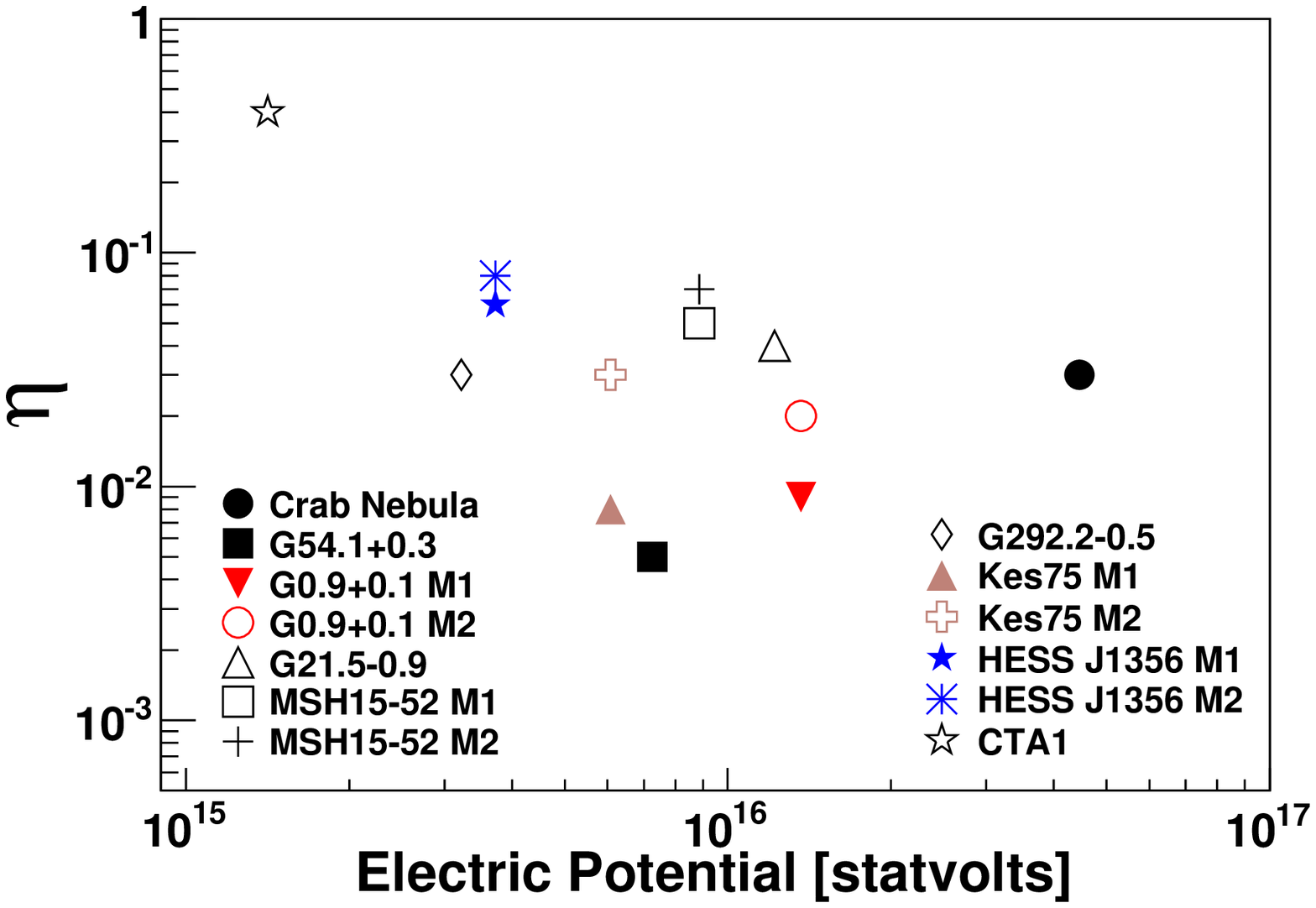} \\
\caption{
PWNe properties  {
{in the y-axis of all plots}} 
as a function of pulsar properties {
 {in the x-axis of all plots}}. The values of all magnitudes refer to the current time. 
 %so that each pulsar has a different age in this plot.
From left to right, we plot the obtained values of
$\gamma_{max}$, $\gamma_b$, $\alpha_1$, $\alpha_2$, $B(t_{age})$, and $\eta$, 
%and $\kappa$  
as a function of (from top to bottom) 
spin-down, surface magnetic field, light cylinder magnetic field, and potential.
}
\label{PWN-PSR}
\end{figure*}
%%%%%%%%%%%%%%%%%%%%%%%%%%%%%%%%%%%%%%%%%%%%

%gamma-max, alpha1,2, gamma_b
The first four
rows of Fig.  \ref{PWN-PSR}
plot the spectral parameters of the injected electrons 
%(all assumed to be time-independent in our model) 
as a function of
the pulsar properties.
We find no correlation of the slopes $\alpha_1$ and $\alpha_2$, or $\gamma_b$ with the pulsar properties. 
In the case of $\alpha_2$, this is true even disregarding the outlier, G292.2--0.5.

We do find a correlation of the 
the maximum Lorentz factor with the spin-down power (and thus the magnetic field at the light cylinder, and the pulsar electric potential).
The significance of the correlation surpasses 95\% CL. For the surface magnetic field, the significance we obtain is the level of 94\%, and this is why we do not quote this fit in Table \ref{fits} although
we show it in the corresponding plot for visual inspection. If this trend is considered, the $\gamma_{max}$ value is anti-correlated with the surface $B$ field of the pulsar. 
On the contrary, the larger is the spin-down power (or the
magnetic field at the light cylinder or the electric potential), the larger is the Lorentz factor of electrons  in the nebulae.
The maximum energy to which electrons are accelerated in the nebulae depends on the injected electrons at the bottom of the wind zone.  This correlation 
is to be expected via Eq. \ref{g1} and the fact that the dispersion that we find in the two other free parameters appearing in there, $\varepsilon$ and $\eta$, 
is relatively not large for most of the sample.

%{\red { 3000 years}}
Looked at the same age (at 3000 years) the $\gamma_{max}$ -- surface magnetic field anti-correlation is confirmed better than the 95\% level, whereas the results for the other parameters are very similar.

%magnetic properties
The magnetic field in the nebulae is also correlated with the pulsar properties. Also here, the larger the spin-down power (or the
magnetic field at the light cylinder or the electric potential) the larger the nebula magnetic field, but this too can be ascribed to 
the way we define the magnetic field in the model (see Section 2.2).
The magnetization, however, is a free parameter in the fit, and with the confidence cut imposed, we see no relation between $\eta$ and any of the pulsar characteristics. Take as an example the Crab nebula: it is the pulsar with the largest spin-down power and nebular $B$ (today magnitudes) but its magnetization is similar to that of the remaining PWNe. 

%{\red { 3000 years}}
Taking the PWNe at the same age of 3000 years, we find that the PWN magnetic field correlation with the spin-down  power (or the
magnetic field at the light cylinder or the electric potential) is lost. 
The nebular magnetic field and the spin-down power are both decreasing with the age of the system, thus looking for its relationship at the same age increases 
the dispersion.

%multiplicity
The multiplicity of the models studied is correlated (but only better than 94\% of CL) with the pulsar parameters, presenting positive correlations with
the spin-down power (or the
magnetic field at the light cylinder or the electric potential) and negative correlation with the surface magnetic field (albeit the scatter of the data points in this latter case seems to be worse). A caveat in this case is that the $\kappa$ parameter is already making use of the $P$ and $\dot P$ values
to normalize the injected electrons (see Eq. \ref{kappa2}), and in fact, because of its definition $Q$ itself is obviously correlated with 
the spin-down. %Fig. \ref{PWN-PSR} and 
%Table \ref{fits} shows these fits for completeness. 

\section{Concluding remarks}

The aim of this paper was to present numerical models of the TeV-detected, young PWNe along more than 20 decades
of frequencies; using a radiatively complete, time-dependent numerical approach. For the first time, we have a coverage of many such PWNe analyzed under the same framework, adopting similar assumptions, which allows for a more meaningful parameter comparison. Despite the caveats of the model used, for instance, the simplicity of having adopted a uniform magnetic field, a free expansion phase, and of disregarding morphological features,
we find that one-zone, leptonic-only generated radiation provides a reasonably good fit to the multifrequency data for 
PWNe detected at TeV. Here we summarize our findings.

\begin{itemize}

\item We favor a non-PWN origin for the radiation detected from HESS J1813--178. For the remaining 9 TeV sources studied,
we find a plausible PWN origin of the mutiwavelength emission. 

\item For all the TeV sources  plausibly related with a PWN, only the Crab nebula is SSC dominated. All the remaining PWNe except for HESS J1356--645 and CTA 1 are IC-FIR dominated. The dominance of the FIR contribution to IC is always significant.

\item The  FIR energy densities that we found is needed to fit the PWN high-energy emission are generally larger
than what is obtained from GALPROP (usually by up to a factor of a few).

\item The efficiencies of emission are $\sim 10^{-6 \div 7}$ in radio, $\sim 10^{-2 \div 3}$ in X-rays, 
and $\sim 10^{-3 \div 4}$ in gamma-rays, with only one outlier in the sample presenting very low X-ray fluxes (G292.2--0.5). 

\item The electron population can be described by a broken power law in all cases. The parameters of the injection cluster in relatively narrow ranges, especially, the break Lorenz factor, which is around $5 \times 10^5$. The high energy spectral slope is found to be in the range 2.2 -- 2.8 (except for the steeper case of G292.2--0.5, which also present a higher energy break). The low energy part is instead much harder, with the low energy index in the range 1.0 -- 1.6.

\item All PWNe have large multiplicities, in general in excess of $10^5$. The population of low-energy electrons is large by number,
and generate a low medium energy per particle in the spectrum in all cases.

\item All the nebulae except CTA~1 have low values of magnetization, of only a few percent. CTA~1 presents the largest magnetization of our sample, and reaches almost to equipartition. All the other PWNe are heavily particle dominated. This result is found to be stable against uncertainties.

\item We do not find significant correlations between the efficiencies of emission at different frequencies and the magnetization, implying that the specific
environment and the injection effects play a dominant role in determining, e.g.,  the gamma-ray luminosity. 

\item Comparing SEDs of the PWNe as observed today mixes pulsars of different spin-down power and age, and generates a variety of distributions. 
A normalized comparison of the SEDs (e.g., with the corresponding spin-down flux) at the same age significantly reduces the dispersion.

\item We do not find clear correlations between the pulsar's characteristic ages and the radio and X-ray luminosities. 
The gamma-ray luminosity seems to be anti-correlated with the characteristic age. On the other hand, we do find correlations of the radio and X-ray (and at a slightly lower confidence also 
gamma-ray luminosities) with the spin-down,  and an anti-correlation of the ratios of IC to synchrotron luminosities with the spin-down.

\item The injection parameters do not appear to be correlated with the pulsar properties, except for the maximum Lorentz factor and the magnetic field in the nebula 
which are correlated with the spin-down power 
(or the magnetic field at the light cylinder or the electric potential),
but these cases can be ascribed to 
the model properties. 

\item We do not find a significant correlation of any PWN parameter with the surface magnetic field of the pulsars.

\end{itemize}

%%%%%%%%%%%%%%%%%%%%%%%%%%%%%%%%%%%%%%%%%%%%
%%%%%%%%%%%%%%%%%%%%%%%%%%%%%%%%%%%%%%%%%%%%
\subsection*{Acknowledgments}
%%%%%%%%%%%%%%%%%%%%%%%%%%%%%%%%%%%%%%%%%%%%
%%%%%%%%%%%%%%%%%%%%%%%%%%%%%%%%%%%%%%%%%%%%

This work has been done in the framework of  the
grants AYA2012-39303,
SGR2009-811, and iLINK2011-0303. 
DFT was additionally supported by a Friedrich Wilhelm Bessel Award 
of the Alexander von Humboldt Foundation. ANC acknowledges CONICET, 
Consejo Nacional de Investigaciones Cient\'ificas y T\'ecnicas, for financial support (through PIP 305/2011).
ANC and DFT further acknowledge the grant PICT Ra\'ices 2012-0605.
We thank Sabrina Casanova, for
her help in extracting photon background values from GALPROP, Dieter Horns and Manuel Meyer
for discussions on Crab nebula modeling, and Daniele Vigan\'o  \& Nanda Rea for comments.

%%%%%%%%%%%%%%%%%%%%%%%%%%%%%%%%%%%%%%%%%%%%
%%%%%%%%%%%%%%%%%%%%%%%%%%%%%%%%%%%%%%%%%%%%
\section*{Appendix:
TeV detections and PWNe not included in the study}

\subsection*{HESS J1023--575}

HESS J1023--575 was discovered by H.E.S.S. \cite{rei2007}. Its
spectrum is fitted by a power law of the form $dN/dE=4.5 \times 10^{-12}(E/{\rm TeV} )^{-2.53}$ TeV$^{-1}$ cm$^{-2}$ s$^{-1}$, which
implies an integrated flux above 380 GeV of 1.3 $\times 10^{-11}$ cm$^{-2}$ s$^{-1}$. The closest central source
is PSR J1022--5746, but the association of these
two objects is uncertain due to the large distance between them, 0.28 degrees, assuming 8 kpc, and the proximity to Westerlund 2, which provides other candidates for the origin the radiation
\citep{abra2011}. As far as we are aware there is no synchrotron PWN detected for PSR J1022--5746, leaving any possible fit of the TeV emission quite unconstrained. 

\subsection*{HESS J1616--508}

HESS J1616--508 is one of the brightest sources in the HESS catalog \citep{aha2006}. It is located near RCW 103 
(SNR G332.4--0.4) and Kes 32 (G332.4+0.1) and has an extension of 16 arcmin. Its spectrum is fitted by a power law with an index of 
2.35$\pm$0.06 and its flux between 1 and 30 TeV is 2.1 $\times 10^{-11}$ erg cm$^{-2}$ s$^{-1}$. PSR J1617--5055 was discovered as a radio pulsar by Kaspi et al. (1998). This pulsar was also detected with INTEGRAL \citep{landi2007}, and it was argued that
PSR J1617--5055 was the power engine of HESS J1616--508 (e.g., Mattana et al. 2009).
However, there is still some controversy due to the lack of detection in other wavelengths and the position of the PSR in later observations with Chandra \citep{kar2008}. The latter authors discovered an X-ray PWN
surrounding PSR J1617-5055, with a total luminosity between 0.5 and 8 keV of 3.2 $\times 10^{33}$ erg s$^{-1}$ assuming a distance of
6.5 kpc. The X-ray efficiency is very low for a young PWN ($L_{PWN}/\dot{E} \sim 2 \times 10^{-4}d^2_{6.5 kpc}$) as is also for the ratio
between luminosities ($L_{PWN}/L_{PSR} \sim 0.18$). When compared with the TeV source, the size of the putative X-ray nebulae and the TeV emission has one of the largest mismatches. %since the TeV source is almost 0.2 degrees wide, and the PWN is of only 1 arc sec.
Due to the controversy in the connection with HESS J1616--508 and the lack of data in
the multiwavelength spectrum for the X-ray underluminous PWN, we do not include this source in our study.

\subsection*{HESS J1640-465}

HESS J1640$-$465 is one of the sources discovered by H.E.S.S. during its Galactic Plane survey (Aharonian et al. 2006).
The source is extended with a width of 2.7 $\pm$ 0.5 arcmin. Its spectrum is well fitted with a power law with an index of $\sim$ 2.4 and a total integral flux above 200 GeV of 2.2$\times10^{-11}$ erg cm$^{-2}$ s$^{-1}$. The source is partially coincident with the known radio  SNR G338.3$-$0.0 \citep{Whiteoak1996}. XMM-Newton observations \citep{Funk2007b} showed a hard-spectrum X-ray emitting object at the center of the HESS source, within the shell of the SNR, most likely a PWN associated with G338.3-0.0 and the counterpart of HESS J1640--465. Chandra observations \citep{Lemiere2009} constraint the distance and age of the system between 8 and 13 kpc  and 10 and 30 kyr, respectively.  For a distance of 10 kpc, the luminosity of the pulsar and PWN in the range 2-10 keV were estimated as $L_{PSR}\sim1.3\times10^{33}d_{10}^{2}$  erg s$^{-1}$ and $L_{PWN}\sim3.9\times 10^{33} d_{10}^{2}$ erg s$^{-1}$ ($d_{10}=d/$10 kpc), respectively. The region of HESS J1640-465 was also detected in Fermi data \citep{Slane2010}. No pulsations were found in the Chandra data of this system.  Multifrequency radio continuum observations toward SNR G338.3-0.0 were not able to detect pulsed emission up to a continuum flux density of 2.0 and 1.0 mJy at 610 and 1280 MHZ, respectively; no PWN was detected in the region of the X-ray PWN was detected \citep{Castelletti2011}. 
The lack of the observational data of the period and period derivative of the pulsar that could be associated with the PWN 
makes not possible to perform the fit in our model in the same setting as the others PWNe considered, and thus we do not consider this source in our analysis.

\subsection*{HESS J1834--087} 

The pulsar we quote being positionally correlated in Table 1 is a magnetar and unlikely related to the TeV emission unless having an unusually high spin-down power conversion into TeV photons, of the order of 10\% (orders of magnitude larger than typical values we found in Table \ref{models}). HESS J1834--087 is spatially coincident with the supernova remnant (SNR) G23.3--0.3 (W41) and was detected in the Galactic Plane survey (Aharonian et al. 2006). The MAGIC telescope also observed the source, confirming these results (Albert et al. 2006b). The TeV emission seems to have two components, a central source and an extended region surrounding it (see Mehault et al. 2011, Castro et al. 2013). The latter authors have also reported the GeV detection of this region, with a comparable intrinsic extension and a hard SED between 1 and 100 GeV, of 2.1$\pm$0.1, somewhat atypical
for a PWN spectrum, which smoothly join with the TeV detection. Only a single component is found at GeV energies; the compact TeV emission
is not separately seen by Fermi-LAT. The TeV emission region correlates with a local enhancement of molecular material of about 10$^5$ M$_\odot$ (see Albert et al. 2006, Tian et al. 2007), what makes possible that TeV emission is in fact hadronically produced in this cloud, similarly to the models explored in Gabici et al. (2009), or Torres et al. (2010). However,
details of the comparison between the CO  intensity tracing the mass and the TeV morphology are not perfectly matching. 
A new pulsar candidate has been identified by 
Misanovic et al. (2011), CXOU J183434.9--084443, but its $P$ and $\dot P$, if indeed a pulsar, are unknown. 
These uncertainties suggest that we could not consider this source on a par with the others in our sample.

\subsection*{HESS J1841--055}

This source is one of the largest and most complex detected by H.E.S.S., with an extension of approximately 1 degree \citep{aha2008}. It would appear that there are several emission peaks within the detection, and thus it is likely that HESS J1841--055 could have multiple origins. In particular, SNR Kes 73, the pulsar within Kes 73, 1E 1841--45, and also the High Mass X-Ray Binary AX 184100.4--0536 could all plausibly play in a role in partially generating the TeV emission (see e.g.,
Sguera et al. 2009). In addition, the pulsar we have proposed in Table 1 as a plausible connection to HESS J1841--055. PSR J1838-0537, 
was discovered by Fermi (Pletsch et al. 2012), and can also play a role in producing the TeV source, particularly when a PWN was detected in GeV gamma-rays (Acero et al. 2013). However, the plethora of possible origins of the TeV emission, the difficulty in separating the possible contributors if more than one, and the lack of multiwavelength detections of the PSR J1838-0537 nebula at lower frequencies preclude us to consider it further in our analysis.

\subsection*{Boomerang}

The Boomerang PWN (G106.6+2.9) is associated with the pulsar PSR J2229+6114. This pulsar is surrounded by an incomplete radio shell (Halpern et al. 2002) and 
it is unique due to its extremely flat spectrum in radio ($\alpha=0.0$). Its distance is not clear, and estimates range
from 3, e.g. see Pinneault et al. (2000) or Abdo et al. (2009), to only 0.8 kpc, see e.g., Kothes et al. (2006). The period of the central source is 51.6 ms and the
period derivative is 7.8 $\times 10^{-14}$ s s$^{-1}$ (Halpern et al. 2001). The inferred characteristic age is thus 10460 yr, 
and the spin-down
luminosity is 2.2 $\times 10^{37}$ erg s$^{-1}$.
%
%This PWN has been observed in radio (\cite{kothes2006}) by Effelsberg telescope. The shell was observed previously in radio by \cite{pin2000} and \cite{kothes2001}. The radio flux of the PWN is fitted by a broken power law with a lower index of -0.11 and an upper index of -0.59. The break is observed between 4 and 5 GHz. Regarding the structure of the nebula, it has an unusual half-shell structure and the peak of the emission is not coincident with the position of the pulsar, but displaced to the north (\cite{kothes2006}).
%
%PSR J2229+6114 has been detected also at X-rays (\cite{halp2001a}; \cite{halp2001b}) and $\gamma$-rays (\cite{halp2002}). \cite{halp2001a}and \cite{kar2008} give an X-ray luminosity between 0.5 and 8 keV for the PWN of 8.7 $\times 10^{32}$ erg s$^{-1}$ assuming a distance of 3 kpc. {\it FERMI} obtained an upper limit, but is was not detected (\cite{acker2011}) At very high energies, {\it MILAGRO} detected a possible $\gamma$-ray counterpart of the PWN (\cite{abdo2007}). {\it HESS} (\cite{abdo2009}) and {\it VERITAS} (\cite{acciari2009}) detected VHE extended emission from Boomerang PWN. The spectrum was fitted by a power law with index 2.29 and a flux of 1.15 $\times 10^{-13}$ cm$^{-2}$ s$^{-1}$ TeV$^{-1}$ at 3 TeV.
%
The PWN seems to have been displaced by the reverse shock of the SNR already.
%, which would explain the low radio emission. 
Kothes et al. (2001) observed that the forward shock of the SNR has
been expanding to the north-east where there is a dense HI medium. As a result of the interaction of the forward shock with the dense medium,
a strong reverse shock was created and crushed with the PWN. After the passage of the reverse shock, the pulsar created another PWN with less
luminosity than the first one, explaining the low radio flux of the nebula considering the spin-down power of the pulsar. The south-west area
is almost empty and the PWN is expanding freely. 
Kothes et al. (2006) have also studied the 
 the nature of the break in the spectrum at radio frequencies and inferred an age of 3900 yr since the crush with the reverse shock
and a magnetic field of 2.6 mG from the lifetime of the electrons. 
Due to the interaction with the reverse
shock, we do not consider this PWN in our analysis.

%%%%%%%%%%%%%%%%%%%%%%%%%%%%%%%%%%%%%%%%%%%%
%%%%%%%%%%%%%%%%%%%%%%%%%%%%%%%%%%%%%%%%%%%%

\begin{landscape}
 \begin{table}
 \label{datpsr}
\centering
\scriptsize
\caption{Pulsars in the ATNF 
catalog with known period $P$, and period derivative $\dot P$, and
less than 10000 years of characteristic age $(\tau)$. The first few columns are taken from ATNF data. The column 
``TeV Obs.?" answers whether the pulsar has been observed in TeV range, and, if so, by which telescope (noting H for H.E.S.S., M for MAGIC, and V for VERITAS).
The column ``TeV PWN?" indicates whether there has been a detection of a PWN or in general a TeV source spatially co-located with the pulsar. This information comes
from published literature. 
The last 
three columns represent, respectively, the age of Crab (assuming today's braking index) at which it would have the same
characteristic age than the corresponding pulsar ($T^{Crab}_{\tau}$), the Crab's spin-down power at that age ($ L_{sd}^{Crab}(T^{Crab}_{\tau}) $), and the spin-down power
of the pulsar in terms of percentage of $ L_{sd}^{Crab}(T^{Crab}_{\tau} )$, which we refer to as CFP (Crab fractional power).
Sources maked with $\dag$ are magnetars, which low rotational power is not expected to contribute significantly to the corresponding TeV sources (marked in red). Names of the TeV sources shown in blue are the ones studied in this work. }
\vspace{0.2cm}
\begin{tabular}{l r r r r c c c l l c c l}
\hline
Name & $P$ & $\dot P$ & $D$ & $\tau$ & $B_d$ & $L_{sd}$ & $L_{sd} / D^2 $ & TeV &  TeV                & $T^{Crab}_{\tau}$ &$ L_{sd}^{Crab}(T^{Crab}_{\tau}) $ & CFP\\
 J\ldots           & s     & s s$^{-1}$ & kpc & yrs & G & erg s$^{-1}$ & erg s$^{-1}$ kpc$^{-2}$ & Obs.?  & source & yrs                            &           erg s$^{-1}$      & \%   \\
\hline            
1808$-$2024  $\dag$  &     7.5559                  &     $5.49\times 10^{-10}$            &     13.0 &  218  &    2.0 $\times 10^{15}$ &  5.0 $\times 10^{34}$ &  3.0 $\times 10^{32}$ & H & {\red {J1809-194/G11.0+0.08}} & \ldots & \ldots  & \ldots   \\
1846$-$0258    &     0.3265     &     $7.10\times 10^{-12}$   &      5.8  & 728   &   4.9 $\times 10^{13}$ &  8.1 $\times 10^{36}$ & 2.4 $\times 10^{35}$ & H & {\blue {Kes 75}} & 238 & 1.6 $\times 10^{39}$ & 0.5\\
1907+0919 $\dag$   &     5.1983                &     $9.20\times 10^{-11}$               &      \ldots     &  895    &  7.0 $\times 10^{14}$   &  2.6 $\times 10^{34}$ & \ldots & H & {\red {J1908+063/G40.1-0.89} }& 459 & 1.0 $\times 10^{39}$ & 0.003  \\
1714$-$3810 $\dag$    &     3.8249                &     $5.88\times 10^{-11}$             &      \ldots     &  1030  & 4.8 $\times 10^{14}$ &  4.1 $\times 10^{34}$ &  \ldots & H & {\red {J1718--385/CTB37A}} & 638 & 7.2 $\times 10^{38}$  & 0.006\\
0534+2200   &     0.0334     &      $4.21\times 10^{-13}$         &   2.0  &  1258   &  3.8 $\times 10^{12}$ & 4.5 $\times 10^{38}$ & 1.2 $\times 10^{38}$ & HMV & {\blue {Crab nebula}} & 940 & 4.5 $\times 10^{38}$ & 100\\
1550$-$5418    &     2.0698              &     $2.32\times 10^{-11}$            &     9.7  & 1410   & 2.2 $\times 10^{14}$ & 1.0 $\times 10^{35}$ & 1.1 $\times 10^{33}$ & H & \ldots & 1141 & 3.5 $\times 10^{38}$  & 0.03\\
1513$-$5908    &     0.1512   &     $  1.53\times 10^{-12}$   &       4.4  &  1560  & 1.5 $\times 10^{13}$ & 1.7 $\times 10^{37}$ & 9.0 $\times 10^{35}$ & H & {\blue {J1514--281/MSH 15--52}} & 1340 & 2.8 $\times 10^{38}$ &6 \\
1119$-$6127    &     0.4079            &     $ 4.02\times 10^{-12}$       &      8.4  &  1610  & 4.1 $\times 10^{13}$  & 2.3 $\times 10^{36}$ &  3.3 $\times 10^{34}$ & H & {\blue {J1119-6127/G292.1--0.54}} & 1406 & 2.6 $\times 10^{38}$ & 0.9  \\
0540$-$6919    &     0.0504        &    $ 4.79\times 10^{-13}$      &      53.7 &  1670  & 5.0 $\times 10^{12}$ & 1.5 $\times 10^{38}$ & 5.1 $\times 10^{34}$ & H & \ldots & 1486 & 2.4 $\times 10^{38}$  & 63\\ 
0525$-$6607    &     8.0470                   &    $ 6.50\times 10^{-11}$              &   \ldots     & 1960  & 7.3 $\times 10^{14}$ & 4.9 $\times 10^{33}$ & \ldots & \ldots & \ldots & 1871 & 1.6 $\times 10^{38}$  & 0.003\\
1048$-$5937    &     6.4520               &   $  3.81\times 10^{-11}$         &     9.0 &  2680  & 5.0 $\times 10^{14}$ & 5.6 $\times 10^{33}$ & 6.9 $\times 10^{31}$ & H & \ldots & 2825 & 7.8 $\times 10^{37}$ & 0.007\\
1124$-$5916    &     0.1354         &  $   7.52\times 10^{-13}$    &     5.0  & 2850  & 1.0 $\times 10^{13}$ & 1.2 $\times 10^{37}$ & 4.8 $\times 10^{35}$ & H & \ldots & 3050 & 6.8 $\times 10^{37}$ &18 \\
1930+1852   &     0.1368         &  $   7.50\times 10^{-13}$       &     7.0  & 2890  & 1.0 $\times 10^{13}$ & 1.2 $\times 10^{37}$ & 2.4 $\times 10^{35}$ & V & {\blue {J1930+188/G54.1+0.3}} & 3103 & 6.6 $\times 10^{37}$ & 18 \\
1622$-$4950    &     4.3261                   &      $1.70\times 10^{-11}$              &     9.1 &  4030  & 2.7 $\times 10^{14}$ &  8.3 $\times 10^{33}$ & 9.9 $\times 10^{31}$ & H & \ldots & 4614 & 3.0 $\times 10^{37}$  & 0.03\\
1841$-$0456    &     11.7789             &   $ 4.47\times 10^{-11}$           &   9.6 &  4180  & 7.3 $\times 10^{14}$ & 1.1 $\times 10^{33}$ & 1.2 $\times 10^{31}$ & H & \ldots & 4813 & 2.8 $\times 10^{37}$ & 0.004\\
1023$-$5746    &     0.1115        &  $   3.84\times 10^{-13}$        &     {8.0}    &  4600  &  6.6 $\times 10^{12}$ &  1.1 $\times 10^{37}$ & 1.7 $\times 10^{35}$ &H & J1023+575 & 5370 & 2.2 $\times 10^{37}$   & 50\\
1833$-$1034    &     0.0618         &  $   2.02\times 10^{-13}$          &     4.10 &  4850  & 3.6 $\times 10^{12}$ &  3.4 $\times 10^{37}$ &  2.0 $\times 10^{36}$ &H & {\blue {J1833--105/G21.5--0.9}} & 5701 & 2.0 $\times 10^{37}$ & 170 \\
1838$-$0537    &     0.1457             &   $ 4.72\times 10^{-13}$           &    \ldots    &  4890  & 8.4 $\times 10^{12}$ &  6.0 $\times 10^{36}$ &\ldots &H & J1841--055 & 5754 & 1.9 $\times 10^{37}$ & 32\\
0537$-$6910    &     0.0161         &  $   5.18\times 10^{-14}$   &     53.7 & 4930  & 9.2 $\times 10^{11}$ &  4.9 $\times 10^{38}$ &  1.7 $\times 10^{35}$ &H & N157B (in the LMC) & 5807 & 1.9 $\times 10^{37}$ & 2579 \\
1834$-$0845 $\dag$    &     2.4823                &    $ 7.96\times 10^{-12}$          &    \ldots    &  4940  & 1.4 $\times 10^{14}$ & 2.1 $\times 10^{34}$ & \ldots &H & {\red {J1834--087/W41}} & 5820 & 1.9 $\times 10^{37}$ & 0.1\\
1747$-$2809    &     0.0521              &    $ 1.55\times 10^{-13}$      &     17.5   & 5310  & 2.9 $\times 10^{12}$ & 4.3 $\times 10^{37}$ & 1.4 $\times 10^{35}$ &H & {\blue {J1747--281/G0.9+0.1}} & 6311 & 1.6 $\times 10^{37}$ & 269 \\
0205+6449   &     0.0657     &    $ 1.94\times 10^{-13}$  &     3.2  & 5370  & 3.6 $\times 10^{12}$ & 2.7 $\times 10^{37}$ & 2.6 $\times 10^{36}$ &MV & \ldots & 6390 & 1.6 $\times 10^{37}$  & 169 \\
1813$-$1749    &     0.0446             &    $  1.26\times 10^{-13}$           &      \ldots   &   5600  & 2.4 $\times 10^{12}$ & 5.6 $\times 10^{37}$  & \ldots &H & {\blue {J1813--178/G12.8--0.02}} & 6695 & 1.4 $\times 10^{37}$ &400  \\
0100$-$7211    &     8.0203                 &     $1.88\times 10^{-11}$            &     62.4 & 6760  & 3.9 $\times 10^{14}$ & 1.4 $\times 10^{33}$ & 3.7 $\times 10^{29}$ & \ldots & \ldots & 8233 & 9.1 $\times 10^{36}$ & 0.02\\
1357$-$6429    &  0.1661         &     $3.60\times 10^{-13}$    &     4.1 &  7310  &  7.8 $\times 10^{12}$ &  3.1 $\times 10^{36}$ & 1.9 $\times 10^{35}$ &H & {\blue {J1356--645/G309.9--2.51}} & 8962 & 7.6 $\times 10^{36}$ &41 \\
1614$-$5048   &   0.2316            &   $4.94\times 10^{-13}$        &  7.2  & 7420 & 1.1 $\times 10^{13}$ & 1.6 $\times 10^{36}$ & 3.0 $\times 10^{34}$ &H &  \ldots & 9107 & 7.3 $\times 10^{36}$ & 22\\
1734$-$3333   &     1.1693             & $2.28\times 10^{-12 }    $     &     7.4  & 8130   & 5.2 $\times 10^{13}$ & 5.6 $\times 10^{34}$ & 1.0 $\times 10^{33}$ &H & \ldots & 10048 & 5.9 $\times 10^{36}$ & 0.9\\
1617$-$5055    &     0.0693              & $    1.35\times 10^{-13 }  $       &     6.4  &  8130 & 3.1 $\times 10^{12}$ &  1.6 $\times 10^{37}$ &  3.8 $\times 10^{35}$ &H & J1616-508 & 10048 & 5.9 $\times 10^{36}$ & 271 \\
2022+3842    &     0.0242         &$4.32\times 10^{-14 }$      &     10.0 & 8910 & 1.0 $\times 10^{12} $& 1.2 $\times 10^{38}$ & 1.2 $\times 10^{36}$ &\ldots & \ldots & 11082 & 4.8 $\times 10^{36}$ & 2500 \\
1708$-$4009 $\dag$    &     11.0013              &     $1.93\times 10^{-11}      $     &     3.8 &  9010 & 4.7 $\times 10^{14}$ &  5.7 $\times 10^{32}$ & 4.0 $\times 10^{31}$ &H &  {\red {J1708-443/G343.1--2.69}} & 11215 & 4.7 $\times 10^{36}$ & 0.01\\
1745$-$2900 $\dag$ & 3.76356 & $6.5 \times 10^{-12}$ & 8.0 & 9170 & 1.6 $\times 10^{14}$ &  $4.8 \times 10^{33}$ & 7.5 $\times 10^{31}$ & HM & {\red {(in the Galactic Center)}} & 11427 & 4.4 $\times 10^{36}$ & 0.99\\
\\

\hline
\end{tabular}
%\end{minipage}
\end{table}
\end{landscape}

\begin{table*}
\scriptsize
\centering
  \caption{Examples of radiative time-dependent models used to fit observations of young PWNe.}
  \begin{tabular}{  l l  ll ll l l ll l   }
  \hline
    
    & \rotatebox{90}{Tanaka  \& Takahara 2011}  & \rotatebox{90}{Zhang et al. 2008}  & \rotatebox{90}{Bucciantini et al. 2011}&  \rotatebox{90}{Fang \& Zhang 2010} & 
    \rotatebox{90}{Qiao, Zhang, \& Fang 2009} & \rotatebox{90}{Li, Chen, \& Zhang 2010}&  \rotatebox{90}{Venter \& de Jager 2007} & \rotatebox{90}{{ {This work}}}\\
     
    \hline
    
     Crab nebula & X  & X & X & --  & -- & -- & -- & X \\

      G54.1+0.3 & X & -- & --  & -- & -- &  X & -- & X\\
     G0.9+0.1 & X & -- & X & X & X & -- & X & X \\
     G21.5--0.9 & X & -- & -- & -- & -- & -- &-- & X \\
     MSH 15--52 & -- & X & X &   X & -- & -- &-- & X\\
     G292.2--0.5 & -- &-- &-- & -- & -- & -- &-- & X \\
     Kes 75 & X & X & X &-- & -- & -- &-- & X \\
   HESS J1356--645 & -- &-- &-- & -- & -- & -- &-- &X \\ 
   CTA~1 & -- &-- &-- & -- & -- & -- &-- & X \\
    HESS J1813--178    & -- &-- &-- & -- & -- & -- &-- &X \\ 
%      Boomerang & -- &-- &-- & -- & -- & -- &-- & X\\

 \hline
 \hline

\end{tabular}
\label{thmodels}
\end{table*}

%%%%%%%%%%%%%%%%%%%%%%%%%%%%%%%%%%%%%%%%%%%%
%\begin{landscape}
%\pagestyle{empty}
\begin{table*}[t!]
%\centering
\scriptsize
  \caption{Physical magnitudes. The dot symbols are used to represent the same value of the corresponding left column.
}

   % Moment of inertia (g cm$^{2}$), $I$          		     &   $10^{45}$    & $10^{45}$ \\
    % SN explosion energy (erg),  $E_{0}$                          & $10^{51}$            & $10^{51}$\\
    % T_CMB =2.73 K
    % w_CMB=0.25 eV / cm^3

  \begin{tabular}{  l l  ll ll l l ll l   }
  \hline
                               &{\bf \red {Crab nebula}} & {\bf \red { G54.1+0.3}}           &  {\bf \red { G0.9+0.1}} & \ldots  &  {\bf \red { G21.5$-$0.9}}  &  {\bf \red { MSH 15--52}} 
                                &  \ldots   \\
                               &                       &         & Model 1     & Model 2   &    &    Model 1  & Model 2 \\
 \hline
{\blue {Pulsar \& Ejecta}}  \\
 \hline
 
$P(t_{age})$ (ms) & 33.40 & 136  & 52.2 & \ldots & 61.86 & 150 &  \ldots \\

$\dot{P} (t_{age})$ (s s$^{-1}$)  	& 4.2$\times 10^{-13}$ &  7.5$\times 10^{-13}$ & 1.5$\times 10^{-13}$ & \ldots & 2.0$\times 10^{-13}$ & 1.5$\times 10^{-12}$  &   \ldots \\

$\tau_{c}$   (yr)  & 1296   &  2871 & 5305 &\ldots &  4860  & 1600  &  \ldots \\

$L(t_{age})$  (erg/s) 	& 4.5 $\times 10^{38}$  &  1.2$\times 10^{37}$  & 4.3$\times 10^{37}$ & \ldots & 3.4$\times 10^{37}$ &  1.8 $\times 10^{37}$  &    \ldots  \\

$n$ &  2.509 &   3  & 3 &\ldots & 3 & 2.839 &  \ldots\\

$t_{age}$  (yr) & 940  &  1700 & 2000 & 3000 & 870 &  1500 &  \ldots \\        

$D$  (kpc)  & 2.0   &  6  & 8.5 & 13   & 4.7 & 5.2  &    \ldots \\             

$\tau_{0}$ (yr) & 730  & 1171   & 3305 & 2305 & 3985 &   224    &  \ldots \\          
  % & From eq. 5 of \citet{Martin2012}\\

$L_{0}$   (erg/s)     	& 3.1 $\times 10^{39}$  & 7.2$\times 10^{37}$  & 1.1$\times 10^{38}$ & 2.3$\times 10^{38}$ &5.0$\times 10^{37}$ & 1.3$\times 10^{39}$  
&  \ldots \\ 
 %& From eq. 3 of \citet{Martin2012}\\

%$P_0$ (ms) &  &  \\  

$M_{ej}$ ($M_{\odot}$)                    	&  9.5  &    20 &  11 & 17 & 8  & 10  &  \ldots  \\

$R_{PWN}(t_{age})$ (pc)             & $2.1$      		&1.4      & 2.5 & 3.8  & 0.9 & 3  &  \ldots \\

\hline
 { \blue {Environment }}\\
\hline

$T_{FIR}$ (K)                                  	& 70  	 &       20       & 30 & \ldots & 35 & 20 &  20 \\

$w_{FIR}$ (eV/cm$^{3}$)       	&   0.5      &    2.0       & 2.5 & 3.8 & 1.4  & 5 &   4 \\ 

$T_{NIR}$ (K)     	&   5000     &     3000      & 3000 & \dots  & 3500 &  3000 &  400 \\

$w_{NIR}$   (eV/cm$^{3}$)	&  1.0       	 &       1.1     & 25 & \dots & 5.0 & 1.4 & 20 \\ 

$n_{H}$    	&   1.0         &    10  & 1.0 &\ldots & 0.1 &  0.4 &   \ldots \\

 \hline
 {\blue { Particles and field }} \\
 \hline

$\gamma_{max} (t_{age})$ & $7.9\times10^9$  & $7.5\times10^8$   & $1.3 \times10^9$ & $1.9 \times10^9$ & $2.4 \times10^9$ & $1.9 \times10^9$ 
&   $2.3 \times10^9$ 
&   \\

$\gamma_{b}$                      & $7\times 10^5$    & $5\times 10^5$    & $1.0 \times10^5$ & $0.5 \times10^5$ & $1.0 \times10^5$ &  $5.0 \times10^5$ 
&  \ldots \\

$\alpha_{1}$              & $1.5$      	 	& 1.20          & 1.4 & 1.2 & 1.0 & 1.5  &  \ldots  \\

$\alpha_{2}$              & $2.5$      		& 2.8          & 2.7 & 2.5 & 2.5 & 2.4 &   \ldots \\

$\epsilon$                 & $0.2$      		& 0.3         & 0.2 & \ldots & 0.2 & 0.2 &  \ldots \\

$B(t_{age})$    ($\mu$G)          & $84$       		& 14   & 14 & 15 & 71 & 21 & 25    \\

$\eta$                         & $0.03$    		& 0.005       &0.01 & 0.02 & 0.04 & 0.05 & 0.07  \\

\hline
\end{tabular}
\label{param}
\end{table*}

%\end{landscape}
%%%%%%%%%%%%%%%%%%%%%%%%%%%%%%%%%%%%%%%%%%%%

%\begin{landscape}
%\pagestyle{empty}
%

\begin{table*}[t!]
%\centering
\scriptsize
  Continued.

   % Moment of inertia (g cm$^{2}$), $I$          		     &   $10^{45}$    & $10^{45}$ \\
    % SN explosion energy (erg),  $E_{0}$                          & $10^{51}$            & $10^{51}$\\
    % T_CMB =2.73 K
    % w_CMB=0.25 eV / cm^3

  \begin{tabular}{  l l  lll ll l l ll l   }
  \hline
                                & {\bf \red {G292.2--0.5}}  & {\bf \red { Kes 75}}  & \ldots    &  {\bf \red {HESS J1356--645  }}  & \dots  & {\bf \red {CTA~1}} \\
                                &   & Model 1      & Model 2   &   Model 1 & Model 2  & \\
 \hline
{\blue { Pulsar \& Ejecta }}  \\
 \hline
 
$P(t_{age})$ (ms) & 408 &  324 & \ldots   & 166 & \ldots  & 316.86 & \\

$\dot{P} (t_{age})$ (s s$^{-1}$)  	&   4.0$\times 10^{-12}$ & 7.1$\times 10^{-12}$ & \ldots &  3.6$\times 10^{-13}$ & \ldots  & 3.6$\times 10^{-13}$ \\

$\tau_{c}$   (yr)  &   1610& 724 & \ldots  & 7300 & \ldots & 13900 &  \\

$L(t_{age})$  (erg/s) 	&  2.3 $\times 10^{36}$ & 8.2$\times 10^{36}$ & \ldots &  3.1$\times 10^{36}$ & \ldots    &  4.5$\times 10^{35}$ &  \\

$n$  & 1.72 &  2.16  & \ldots & 3 & 2 & 3 &  \\

$t_{age}$  (yr) & 4200 &  700 & 800 & 6000 & 8000 & 9000 & \\        

$D$  (kpc)     &  8.4 & 6  & 10.6  & 2.4 & \ldots & 1.4 &  \\             

$\tau_{0}$ (yr)   & 270 & 547   & 447 & 1311 & 6622 & 4901 &  \\          
  % & From eq. 5 of \citet{Martin2012}\\

$L_{0}$   (erg/s)   &  9.3 $\times 10^{40}$ & 7.7$\times 10^{37}$ & 1.3$\times 10^{38}$ &  9.6$\times 10^{37}$ & 3.3$\times 10^{37}$ & 3.6$\times 10^{36}$ &  \\ 
 %& From eq. 3 of \citet{Martin2012}\\

%$P_0$ (ms) &  &  \\  

$M_{ej}$ ($M_{\odot}$)                & 35   	  &   6 & 7.5 & 10 & 12 & 10 &  \\

$R_{PWN}(t_{age})$ (pc)            &13		& 0.9     & 1.0 &  9.5 & 9.9 & 8.0 &  \\

\hline
 {\blue { Environment}}\\
\hline

$T_{FIR}$ (K)                                 & 70 	  	 &       25      & \ldots  & 25 & \ldots & 70 & \\

$w_{FIR}$ (eV/cm$^{3}$)       & 3.8	      &    2.5       & 5.0  & 0.4 & \ldots & 0.1 & \\ 

$T_{NIR}$ (K)    &4000 	 &     5000    & \ldots   & 5000 & \ldots & 5000 &  \\

$w_{NIR}$   (eV/cm$^{3}$)	&1.4       	 &       1.4    & 1.4 & 0.5 & \ldots & 0.1 &   \\ 

$n_{H}$    	&0.02 &      1.0 & \ldots  & 1.0 & \ldots & 0.07 &  \\

 \hline
{\blue { Particles and field} }\\
 \hline

$\gamma_{max} (t_{age})$ & $8.0\times10^8$   & $5.2\times10^8$   & $4.9 \times10^8$  & $8.8\times10^8$ & $1.5\times 10^{9}$ & $8.6\times 10^{8}$ &  \\

$\gamma_{b}$          &                  $5.0\times 10^6$    & $2.0\times 10^5$    & $1.0 \times 10^5$ & $3.0 \times 10^5$ & -- & $0.8\times 10^{5}$ & \\

$\alpha_{1}$                   	& 1.5 	& 1.4          & 1.6  & 1.2 & -- & 1.5 & \\

$\alpha_{2}$                   &4.1  		& 2.3          & 2.1 & 2.52 & 2.6 & 2.2 &  \\

$\epsilon$                    &  0.3 		& 0.2          & 0.1  & 0.2 & 0.3 & 0.2 &  \\

$B(t_{age})$    ($\mu$G) & 4                		& 19    & 33  & 3.1 & 3.5 & 4.1 & \\

$\eta$                         &0.03   		& 0.008       &0.03  & 0.06 & 0.08 & 0.4 &  \\

\hline
\end{tabular}
\label{param2}
\end{table*}

%\end{landscape}
%%%%%%%%%%%%%%%%%%%%%%%%%%%%%%%%%%%%%%%%%%%%

%%%%%%%%%%%%%%%%%%%%%%%%%%%%%%%%%%%%%%%%%%%%

\begin{table*}[t!]
\scriptsize
%\centering
  \caption{Properties of the fitted models. For an explanation of all the columns, see the text.
  }
  \begin{tabular}{  l l  lc cc c c cc c   }
  \hline
    
  PWN  & 1$^{st}$ & 2$^{nd}$ & ratio & $L_r$ & $L_X$ & $L_\gamma$ & $f_r$ & $f_X$ & $f_\gamma$  \\
    & cont. & cont. & (1--10 TeV)  & (1.4 GHz) & (1--10 keV) & (1--10 TeV) & \\
      
    \hline
    
     Crab nebula       &  SSC   & IC-FIR & 1.3 & $1.3\times 10^{33}$  & $1.4\times 10^{37}$& $3.4\times 10^{34}$  & $2.8\times 10^{-6}$ & $3.2\times 10^{-2}$ & $7.5\times 10^{-5}$    \\
     
     G54.1+0.3         & IC-FIR & IC-CMB & 5.3 & $5.0\times 10^{30}$  & $3.0\times 10^{34}$& $6.4\times 10^{33}$ & $4.2\times 10^{-7}$ &  $2.5\times 10^{-3}$ &  $5.3\times 10^{-4}$  \\
     
     G0.9+0.1 (M1)& IC-FIR & IC-NIR & 4.1 & $5.0\times 10^{31}$  & $6.9\times 10^{34}$& $1.4\times 10^{34}$ & $1.2\times 10^{-6}$ & $ 1.6\times 10^{-3}$ & $3.2  \times 10^{-4}$  \\ 
     
     G0.9+0.1 (M2)& IC-FIR & IC-CMB & 6.6 & $1.2\times 10^{32}$  & $1.6\times 10^{35}$& $3.0\times 10^{34}$ & $2.9\times 10^{-6} $& $3.7 \times 10^{-3}$   & $7.1 \times 10^{-4}$ \\
     
     G21.5--0.9        & IC-FIR & IC-CMB & 3.6 & $5.1\times 10^{31}$  & $3.9\times 10^{35}$& $2.0\times 10^{33}$ & $1.5\times 10^{-6} $& 
     $1.2\times 10^{-2}$ & $5.8\times 10^{-5}$  \\
     
     MSH 15--52  (M1)      & IC-FIR & IC-CMB & 10.1 & $2.8\times 10^{31}$  & $3.9\times 10^{35}$& $5.0\times 10^{34}$& $1.5\times 10^{-6}$ & $2.2 \times 10^{-2}$ & $2.7\times 10^{-3}$    \\
     MSH 15--52    (M2)    & IC-FIR & IC-NIR & 1.3 & $3.4\times 10^{31}$  & $3.8\times 10^{35}$& $5.2\times 10^{34}$& $1.9\times 10^{-6}$ & $2.1 \times 10^{-2}$ & $2.9 \times 10^{-3}$    \\

     G292.2--0.5       & IC-FIR & IC-NIR & 31.1 & $1.1\times 10^{31}$ & $1.1\times 10^{32}$& $8.4\times 10^{33}$& $5.0\times 10^{-6}$ & $4.8\times 10^{-5}$ & $3.7 \times 10^{-3}$  \\
     
     Kes 75 (M1)  & IC-FIR & IC-CMB & 4.1 & $4.2\times 10^{30}$ & $1.3\times 10^{35}$& $7.4\times 10^{33}$& $5.1\times 10^{-7} $& 
     $1.5\times 10^{-2}$ & $9.0\times 10^{-4}$  \\
     
     Kes 75 (M2)  & IC-FIR & IC-CMB & 8.5 & $1.3\times 10^{31}$ & $3.7\times 10^{35}$& $1.5\times 10^{34}$& $1.5\times 10^{-6} $& $4.5 \times 10^{-2}$& $1.8\times 10^{-3}$ \\		
     
     HESS J1356--645 (M1)   & IC-CMB & IC-FIR & 1.3 & $1.6\times 10^{30}$ & $7.1\times 10^{33}$& $5.7\times 10^{33}$& $5.0\times 10^{-7} $& $2.3\times 10^{-3}$& $1.8\times 10^{-3}$   \\
     
     HESS J1356--645 (M2)   & IC-CMB & IC-FIR & 1.3 & $1.6\times 10^{30}$ & $6.0\times 10^{33}$& $4.0\times 10^{33}$& $5.2\times 10^{-7} $& $1.9\times 10^{-3}$& $1.3 \times 10^{-3}$  \\
     
     CTA~1             & IC-CMB & IC-FIR & 14.2 & $2.7\times 10^{29}$ & $4.1\times 10^{33}$& $8.6\times 10^{32}$& $6.1\times 10^{-7}$& 
     $9.1 \times 10^{-3}$& $1.9 \times 10^{-3}$ \\
     
%     HESS J1813--178   & IC-FIR & IC-CMB & 1.3 & $4.8\times 10^{29}$ & $8.8\times10^{34}$ & $2.2\times 10^{34}$& $8.6\times 10^{-9}$& $1.6\times 10^{-3}$ & $4.0  \times 10^{-4}$\\

 \hline
 \hline

\end{tabular}
\label{models}
\end{table*}

%%%%%%%%%%%%%%%%%%%%%%%%%%%%%%%%%%%%%%%%%%%%

\begin{table*}[t!]
\scriptsize
\centering
  \caption{Goldreich \& Julian estimation and multiplicity computed from our models (an upper limit). See the description in the text.
  }
  \begin{tabular}{  l l  lc cc c cc c   }
  \hline
    
 PWN  & $\dot N$ & $Q$ & $\kappa$ &$\gamma_w$\\
   & s$^{-1}$ & s$^{-1}$ &  \\      
    \hline
    
     Crab nebula       &  7.6$\times 10^{33}$      &    3.2$\times 10^{41}$ & $ 4.2\times 10^{7}  $ & $1.7 \times 10^3$  \\
     
     G54.1+0.3         & 1.2$\times 10^{33}$    &    7.4$\times 10^{38}$     & $ 6.2\times 10^{5}  $  & $2.0 \times 10^4$ \\
     
     G0.9+0.1 (M1)& 2.3$\times 10^{33}$     &    4.0$\times 10^{40}$ & $ 1.8\times 10^{7}  $ & $1.3 \times 10^3$ \\ 
     
     G0.9+0.1 (M2) & 2.3$\times 10^{33}$    &    1.3$\times 10^{40}$ & $ 5.6\times 10^{6}  $ & $4.0 \times 10^3$\\
     
     G21.5--0.9        & 2.1$\times 10^{33}$    &    1.7$\times 10^{39}$ & $ 8.0\times 10^{5}  $ & $2.4 \times 10^4$ \\
     
     MSH 15--52  (M1)      &   1.5$\times 10^{33}$    &    1.3$\times 10^{40}$ & $ 8.6\times 10^{6}  $ & $1.6 \times 10^3$\\
     
     MSH 15--52    (M2)    &   1.5$\times 10^{33}$    &    1.3$\times 10^{40}$ & $ 8.7\times 10^{6}  $ & $1.6 \times 10^3$\\
     
     G292.2--0.5       &  5.5$\times 10^{32}$    &    9.8$\times 10^{38}$ & $ 1.8\times 10^{6}  $ & $2.8 \times 10^3$\\
     
     Kes 75 (M1)  & 1.0$\times 10^{33}$    &    3.5$\times 10^{39}$ & $ 3.5\times 10^{6}  $ & $2.9 \times 10^3$\\
     
     Kes 75 (M2)  &1.0$\times 10^{33}$    &    1.4$\times 10^{40}$ & $ 1.4\times 10^{7}  $ & $7.2 \times 10^2$\\        
     
     HESS J1356--645 (M1)   & 6.4$\times 10^{32}$    &    2.2$\times 10^{38}$ & $ 3.4\times 10^{5}  $ & $1.6 \times 10^4$  \\
     
     HESS J1356--645 (M2)   & 6.4$\times 10^{32}$    &    1.3$\times 10^{37}$ & $ 2.1\times 10^{4}  $ & $2.7 \times 10^5$\\
     
     CTA~1             & 2.4$\times 10^{32}$    &    3.8$\times 10^{38}$    & $1.6 \times 10^6$ & $8.8 \times 10^2$\\
     
%     HESS J1813--178   & 2.7$\times 10^{33}$    &    1.4$\times 10^{41}$ & $2.3 \times 10^4$ & $1.1 \times 10^6$\\

 \hline
 \hline

\end{tabular}
\label{kappa}
\end{table*}

\begin{table*}[t!]
\scriptsize
\centering
  \caption{Comparison between modeled ($w$,$T$) and  GALPROP ($w^G$,$T^G$) energy densities and temperatures. When the parameters ($w$,$T$) in the model are the same as
  the extracted from GALPROP we quote \ldots 
  }
  \begin{tabular}{  l c  ccccccc   }
  \hline    
   PWN  & $w_{FIR}$ & $T_{FIR}$ & $w_{NIR}$ & $T_{NIR}$ & $w^G_{FIR}$ & $T^G_{FIR}$ & $w^G_{NIR}$ & $T^G_{NIR}$ \\
    & (eV cm$^{-3}$) & (K) & (eV cm$^{-3}$) & (K) & (eV cm$^{-3}$) & (K) & (eV cm$^{-3}$) & (K)  \\

    \hline
    
     Crab nebula & 0.5 & 70 & 1.0 & 5000 & 0.2 & 25 & 0.6 & 3500\\
     G54.1+0.3 & 2.0 & 20 & 1.1 & 3000 & 0.8 & 25 & 1.1 & 3000\\
     G0.9+0.1 (M1) & 2.5 & 30 & 25 & 3000 & 1.4 & 35 & 10.5 & 3500\\
     
     G0.9+0.1 (M2) & 3.8 & 30 & 25 & 3000 & 1.7 & 30 & 3.4 & 3200\\
     G21.5--0.9 & \ldots & \ldots & \ldots & \ldots & 1.4 & 35 & 5.0 & 3500\\
     MSH 15--52 (M1)  & 5 & 20 & 1.4 & 3000 & 1.2 & 30 & 2.2 & 3000\\
     MSH 15--52 (M2) Ê& 4 & 20 & 20 & 400 & 1.2 & 30 & 2.2 & 3000\\

     G292.2--0.5 & 3.8 & 70 & 1.4 & 4000 & 0.3 & 25 & 0.7 & 3300\\
     Kes 75 (M1) & 2.5 & 25 & 1.4 & 5000 & 1.5 & 30 & 4.4 & 3500\\
     Kes 75 (M2) & 5.0 & 25 & 1.4 & 5000 & 1.6 & 30 & 2.2 & 3000\\
   HESS J1356--645 (M1) & 0.4 & 25 & 0.5 & 5000 & 0.6 & 25 & 1.2 & 3100\\
   HESS J1356--645 (M2) & 0.4 & 25 & 0.5 & 5000 & 0.6 & 25 & 1.2 & 3100\\
      
      CTA~1 & 0.1 & 70 & 0.1 & 5000 & 0.3 & 25 & 0.6 & 3000\\
    
%    HESS J1813-178    & 0.4 & 70 & 0.8 & 5000 & 1.64 & 30 & 3.37 & 3000\\

 \hline
 \hline

\end{tabular}
\label{ISRF}
\end{table*}

\begin{table*}[t!]
\scriptsize
\centering
  \caption{Fits shown in the figures.    We use $y = p_1 x + p_0$, where variables can be in logarithmic scale, as shown in the corresponding figures.
  % $\log_{10} y = p1 \times \log_{10} x + p_0$. When magnitudes are plotted linearly, we use $y = p_1 x + p_0$. 
  Numbering of panels goes alphabetically, from left to right and top to bottom. Unless otherwise clarified 
  we used all PWNe for fitting (in cases where we have two models, we use Model 1). We show the Pearson's correlation coefficient $r$ and
  the non-directional significance implied  by it.}
  \begin{tabular}{lllllcccc}
  \hline
  $x$-Magnitude &  $y$-Magnitude & Fig. & $p_0$ & $p_1$ & Pearson's $r$ & $P$ \\ % & Comments \\
        \hline

 \hline
         
%%%%%%%%%%%%%%%%%%%%%%%%%%%%%%%%%%%%%%%%%
%Fig. 17 today, magnetization efficiencies
%%%%%%%%%%%%%%%%%%%%%%%%%%%%%%%%%%%%%%%%%

% eta & $f_{r}$             & Fig. \ref{Magnetization} - panel a & $   -5.85935\pm  0.398977 $ & $ 0.0636587 \pm 0.246237 $  & 0.10& 0.80 &\\
% eta & $f_{X}$             & Fig. \ref{Magnetization} - panel b & $  -2.03781   \pm 0.920621$ & $ 0.214868  \pm 0.56818 $  & 0.14 & 0.72 &\\
% eta & $f_{/gamma}$        & Fig. \ref{Magnetization} - panel c & $  -2.70356\pm 0.684616 $ & $ 0.312234  \pm  0.422525 $  & 0.27 & 0.48  &\\

%%%%%%%%%%%%%%%%%%%%%%%%%%%%%%%%%%%%%%%%%
%Fig. 17  magnetization efficiencies - 3000 years
%%%%%%%%%%%%%%%%%%%%%%%%%%%%%%%%%%%%%%%%%

%  eta & $f_{r}$             & Fig. \ref{Magnetization} - panel a & $ -5.45543 \pm 0.414994 $    & $  0.407204  \pm 0.256122 $  & 0.52  & 0.16 &\\
    $\eta$ & $f_{X}$             & Fig. \ref{Magnetization} - panel e & $   -0.75 \pm 0.89$ & $   1.35 \pm  0.55 $  & 0.68  & $4.3 \times 10^{-2}$&\\
%eta & $f_{/gamma}$        & Fig. \ref{Magnetization} - panel c & $ -3.61038\pm 0.549034  $  & $ -0.458742\pm 0.338848 $  & -0.46 & 0.2 &\\

\hline
%%%%%%%%%%%%%%%%%%%%%%%%%%%%%%%%%%%%%%%%%
%today PWN, Edot and tau
%%%%%%%%%%%%%%%%%%%%%%%%%%%%%%%%%%%%%%%%%

 Spin-down & $L_{r}$       & Fig. \ref{mattana} - panel a & $  -11.50\pm  5.67  $ & $  1.15  \pm  0.15 $ & 0.94 &  $1.4 \times 10^{-4}$ \\
 Spin-down & $L_{X}$       & Fig. \ref{mattana} - panel b & $ -17.67 \pm   12.78 $ & $ 1.41 \pm 0.34   $ & 0.84 & $4.5 \times 10^{-3}$ \\
 %Spin-down & $L_{\gamma}$  & Fig. \ref{mattana} - panel c & $   18.7069\pm 6.82285  $ & $ 0.409453  \pm 0.184041 $ & 0.64 & $6.2 \times 10^{-2}$\\
 
% $\tau$    & $L_{r}$       & Fig. \ref{mattana} - panel d & $  35.4196  \pm 2.93812  $ & $  -1.24178  \pm 0.840688$ & -0.49 & 0.18\\
% $\tau$    & $L_{X}$       & Fig. \ref{mattana} - panel e & $  38.5198 \pm 4.4101 $ & $  -1.10063  \pm   1.26187     $ & -0.31 & 0.41   \\
 $\tau$    & $L_{\gamma}$  & Fig. \ref{mattana} - panel f & $   36.95 \pm 1.31   $ & $  -0.88  \pm 0.38 $ & -0.67 & $4.8 \times 10^{-2}$\\

% Spin-down & $L_{X}/L_{r}$       & Fig. \ref{mattana} - panel g & $ -6.1765  \pm 16.652 $ & $  0.263515 \pm 0.449175  $ & 0.22   & 0.58 \\
 Spin-down & $L_{\gamma}/L_{r}$  & Fig. \ref{mattana} - panel h & $  30.20 \pm 7.69 $ & $ -0.74 \pm  0.21 $ & -0.80 &$9.6 \times 10^{-3}$\\
 Spin-down & $L_{\gamma}/L_{X}$  & Fig. \ref{mattana} - panel i & $   36.38  \pm 15.76   $ & $ -1.00 \pm 0.42 $ & -0.67 & $4.8 \times 10^{-2}$\\
 
  \hline

%%%%%%%%%%%%%%%%%%%%%%%%%%%%%%%%%%%%%%%%%
% PWN, Edot and tau; not in plots; correlations appear but with flat parameters and are not visual
%%%%%%%%%%%%%%%%%%%%%%%%%%%%%%%%%%%%%%%%%

% $\tau$    & $f_{r}$       & Fig. \ref{mattana} - panel j & $  -4.82  \pm 1.13 $  & $ -0.32 \pm 0.32  $ & $-0.75$ & $1.9  \times 10^{-2}$ &\\
% $\tau$    & $f_{X}$       & Fig. \ref{mattana} - panel k & $   -1.80 \pm 2.75 $  & $ -0.16 \pm 0.79  $ & $-0.73$ & $2.5  \times 10^{-2}$&\\
%$\tau$    & $f_{X}$       & Fig. \ref{mattana} - panel k & $   0.09  \pm 1.37 $  & $ -0.63 \pm 0.39  $ & $-0.77$ & 0.024309  &without  G292.2$-$0.5\\
%$\tau$    & $f_{\gamma}$  & Fig. \ref{mattana} - panel l & $  -3.44  \pm 2.08 $  & $ 0.07  \pm 0.60  $ & $-0.73$ & $2.5  \times 10^{-2}$&\\

%%%%%%%%%%%%%%%%%%%%%%%%%%%%%%%%%%%%%%%%%
%figure 21, PWN-PSR properties
%%%%%%%%%%%%%%%%%%%%%%%%%%%%%%%%%%%%%%%%%

\hline

Spin-down power            & $\gamma_{max}$ & Fig. \ref{PWN-PSR}  - panel a & $ -2.85 \pm 3.53 $ & $ 0.32 \pm 0.10 $ & 0.79 &  $1.0  \times 10^{-2}$ &\\
%Surf. Mag. Field     & $\gamma_{max}$ &  Fig. \ref{PWN-PSR} - panel b & $ 15.87  \pm 3.08 $ & $ -0.51\pm  0.23 $ &-0.64 &0.06 &\\
Mag. Field Light at LC & $\gamma_{max}$ &  Fig. \ref{PWN-PSR} - panel c & $  7.31 \pm 0.47 $ & $ 0.37 \pm 0.10 $ & 0.82 & $6.8  \times 10^{-3}$&\\
Electric Potential   & $\gamma_{max}$ &  Fig. \ref{PWN-PSR} - panel d & $ -1.16 \pm 3.03 $ & $ 0.65 \pm 0.19 $ & 0.78 & $1.3  \times 10^{-2}$ &\\

Spin-down power           & $B$ & Fig. \ref{PWN-PSR} - panel p & $  -18.13 \pm 4.28 $ & $ 0.52 \pm 0.11 $  & 0.86 & $2.9  \times 10^{-3}$ &\\
%Surf. Mag. Field     & Mag. Field  & Fig. \ref{PWN-PSR} - panel q & $  7.95   \pm 5.31 $ & $-0.51 \pm 0.40 $  & -0.44& 0.24   &\\
 Mag. Field Light at LC & $B$  & Fig. \ref{PWN-PSR} - panel r & $ -1.51  \pm 0.70 $ & $ 0.55 \pm 0.14 $  & 0.83 & $6.0  \times 10^{-3}$ &\\
 Electric Potential   & $B$  & Fig. \ref{PWN-PSR} - panel s & $ -15.40  \pm 3.68 $ & $ 1.04 \pm 0.23 $  & 0.86 &   $2.9  \times 10^{-3}$& \\

%spin-down            & $\eta$      & Fig. \ref{PWN-PSR} - panel t & $  -0.88  \pm  $ & $ -0.02  \pm  $  &  -0.25  & 0.51   & without CTA1 \\
%Surf. Mag. Field     & $\eta$      & Fig. \ref{PWN-PSR} - panel u & $  -0.25  \pm  $ & $ -0.11  \pm  $  &  -0.005 & 0.99   & without CTA1\\
%Mag. Field Light Cyl.& $\eta$      & Fig. \ref{PWN-PSR} - panel v & $  -1.71  \pm  $ & $  0.008 \pm  $  & -0.19   & 0.62   & without CTA1\\
%Electric Potential   & $\eta$      & Fig. \ref{PWN-PSR} - panel w & $  -1.02  \pm  $ & $ -0.041 \pm  $  & -0.25   & 0.51   & without CTA1\\

% Spin-down power            & $\kappa$ & Fig. \ref{PWN-PSR} - panel  x & $ -13.81 \pm 3.58 $ & $ 0.52 \pm 0.10 $  & 0.67  & $4.9 \times 10^{-2}$  &\\
% Surf. Mag Field      & $\kappa$ & Fig. \ref{PWN-PSR} - panel  y  & $ 14.65  \pm 4.34 $ & $-0.70 \pm 0.33 $  & -0.67 & $4.9 \times 10^{-2}$  &\\
% Mag. Field Light at LC & $\kappa$ & Fig. \ref{PWN-PSR} - panel z & $  2.54  \pm 0.49 $ & $ 0.57 \pm 0.10 $  & 0.73  &  $2.5  \times 10^{-2}$ &\\
% Electric Potential   & $\kappa$ & Fig. \ref{PWN-PSR} - panel  aa & $ -11.07 \pm 3.10 $ & $ 1.03 \pm 0.20 $  & 0.67  & $4.9 \times 10^{-2}$  &\\

\hline
\hline
\end{tabular}
\label{fits}
\end{table*}

\begin{table*}[t!]
\scriptsize
\centering
  \caption{Definitions used in search of correlations, as a function of $P$ and $\dot P$ and values for the pulsars associate with the PWNe
  considered in the study. Note that the dipolar field definition uses here an inclination angle $\alpha$ such that $\sin \alpha=1/2$ for all pulsars (i.e., there is a factor
  of 2 difference between the field here and that used in the ATNF catalog). }
  \begin{tabular}{  l l  ll ll l l ll l   }
  \hline
  PSR associated with &  Surface Magnetic field & Light Cylinder Radius & Magnetic field at Light Cylinder & Electric Potential   \\
        \hline
 & $B_s = \left( 3c^3 I P \dot P / (2  \pi^2 R^6  ) \right)^{1/2}$    & $R_{LC} = ( c  P ) / ( 2 \pi )$ & $B(LC)=B_s (R_s/R_{LC})^3$ & $\Delta V=2\pi^2 B_s R^3/ (c^2P^2)$ \\
 &  $=  6.4\times 10^{19} (P \dot P / {\rm s})^{1/2}$ & $= 4.77 \times 10^9 (P / {\rm s}) $ & $ = 5.9 \times 10^8 (P / {\rm s})^{-5/2} (\dot P/ {\rm s \, s^{-1}})^{1/2}$ & $=4.2 \times 10^{20} (\dot P P^{-3})^{1/2}$\\
 & G & cm & G & statvolts \\
     \hline

     Crab                & $7.58\times 10^{12} $& $1.59\times 10^8  $&$ 1.88\times 10^6 $&$ 4.46\times 10^{16}$ \\
     G54.1+0.3           & $2.04\times 10^{13} $& $6.49\times 10^8  $&$ 7.49\times 10^4 $&$ 7.25\times 10^{15}$\\
     G0.9+0.1 (M1/M2)  & $5.66\times 10^{12} $& $2.49\times 10^8  $&$ 3.67\times 10^5 $&$ 1.36\times 10^{16}$ \\
     G21.5--0.9          & $7.12\times 10^{12} $&$ 2.95\times 10^8  $&$ 2.77\times 10^5 $&$ 1.22\times 10^{16}$ \\
     MSH 15--52 (M1/M2)          & $3.04\times 10^{13} $&$ 7.16\times 10^8  $&$ 8.29\times 10^4 $&$ 8.85\times 10^{15}$\\
     G292.2--0.5         & $8.18\times 10^{13} $&$ 1.95\times 10^9  $&$ 1.11\times 10^4 $&$ 3.22\times 10^{15}$ \\
     Kes 75 (M1/M2)    & $9.71\times 10^{13} $&$ 1.55\times 10^9  $&$ 2.63\times 10^4 $&$ 6.07\times 10^{15}$\\
      HESS J1356--645    & $1.56\times 10^{13} $&$ 7.92\times 10^8  $&$ 3.15\times 10^4 $&$ 3.73\times 10^{15}$\\
      CTA 1              & $2.16\times 10^{13} $&$ 1.51\times 10^9  $&$ 6.27\times 10^3 $&$ 1.41\times 10^{15}$ \\
 %     HESS J1813--178    & $4.81\times 10^{12} $&$ 2.13\times 10^8  $&$ 4.97\times 10^5 $&$ 1.58\times 10^{16}$\\

 \hline
 \hline

\end{tabular}
\label{def2}
\end{table*}

\end{document}